\documentclass[aps,prb,showpacs,onecolumn,notitlepage,superscriptaddress]{revtex4-1}
\usepackage{amsmath}
\usepackage{amssymb,amscd,bm,dsfont,wasysym,mathrsfs,latexsym,psfrag,accents,mathtools}
\usepackage{graphicx}
\usepackage{multirow}
\usepackage{dcolumn}
\usepackage{float}
\usepackage{color}
\usepackage{xcolor}
\usepackage{comment}

\newcolumntype{L}[1]{>{\raggedright\arraybackslash}p{#1}}
\newcolumntype{C}[1]{>{\centering\arraybackslash}p{#1}}
\newcolumntype{R}[1]{>{\raggedleft\arraybackslash}p{#1}}

			\newcommand{\e}[1]{\begin{align}{#1}\end{align}}	
		
		\newcommand{\f}[2]{\frac{#1}{#2}}
		\newcommand{\tf}[2]{\tfrac{#1}{#2}}
		
		\newcommand{\p}[2]{\frac{\partial #1}{\partial #2}}

		\newcommand{\la}[1]{\label{#1}}

		\newcommand{\q}[1]{Eq.\ (\ref{#1})}
		\newcommand{\qq}[2]{Eqs.\ (\ref{#1})-(\ref{#2})}
		\newcommand{\s}[1]{Sec.\ \ref{#1}}
		\newcommand{\fig}[1]{Fig.\ \ref{#1}}		
		\newcommand{\app}[1]{App.\ \ref{#1}}				
		\newcommand{\tab}[1]{Tab.\ \ref{#1}}
		\newcommand{\ocite}[1]{Ref.\ \onlinecite{#1}}

		\newcommand{\sq}[1]{\left[\;#1\;\right]}

		\newcommand{\iwith}{\ins{with}}
				
		\newcommand{\iand}{\ins{and}}
		
		\newcommand{\ifor}{\ins{For}}

		\newcommand{\sgn}{\text{sgn}}

		\newcommand{\eq}{=&\;}
		\newcommand{\condeq}[1]{\as \substack{\sma{#1}\\=}\as}
		\newcommand{\condprop}[1]{\as \substack{\sma{#1}\\ \propto}\as}
		\newcommand{\limit}[1]{\substack{\text{lim}\\#1}\;}		
		\newcommand{\appr}{\approx &\;}
		\newcommand{\prop}{\propto&\;}
	
		\newcommand{\R}{\mathbb{R}}
		
		\newcommand{\C}{\mathbb{C}}
		\newcommand{\bbs}{\mathbb{S}}

	\newcommand{\eikr}{e^{i\bk \cdot \br}}
	\newcommand{\eikR}{e^{i\bk \cdot \bR}}
	\newcommand{\emikr}{e^{-i\bk \cdot \br}}

\newcommand{\curl}{\nabla \times}

\newcommand{\nabr}{\nabla_{\boldsymbol{r}}}
\newcommand{\nabk}{\nabla_{\boldsymbol{k}}}

\newcommand{\bohr}{\mu_{\sma{B}}}
\newcommand{\gfac}{g_{\sma{0}}}
\newcommand{\lmt}{l^{\mt}}
\newcommand{\lmf}{l^{\text{-}4}}
\newcommand{\lmo}{l^{\mo}}

\newcommand{\om}{orbital moment }
\newcommand{\omm}{orbital moment}
\newcommand{\zf}{Zilberman-Fischbeck }
\newcommand{\checkgp}{\check{g}^{\sma{\perp}}}
\newcommand{\bt}{$BT_{\sma{\perp}}$}

\newcommand{\var}{\varepsilon}

\newcommand{\enk}{\varepsilon_{n,\bk}}

\newcommand\as{\;\;\;\;}

\newcommand{\hp}{\hat{p}}
\newcommand{\hbp}{\hat{\bp}}
\newcommand{\hbr}{\hat{\br}}

\newcommand{\hH}{\hat{H}}

\newcommand{\hbPi}{\hat{\boldsymbol{\Pi}}}
\newcommand{\hPi}{\hat{{\Pi}}}

\newcommand{\ba}{\boldsymbol{a}}
\newcommand{\bb}{\boldsymbol{b}}
\newcommand{\bc}{\boldsymbol{c}}
\newcommand{\bd}{\boldsymbol{d}}

\newcommand{\bff}{\boldsymbol{f}}
\newcommand{\bg}{\boldsymbol{g}}

\newcommand{\bk}{\boldsymbol{k}}
\newcommand{\bkp}{\boldsymbol{k}^{\sma{\perp}}}

\newcommand{\bp}{\boldsymbol{p}}
\newcommand{\bq}{\boldsymbol{q}}
\newcommand{\br}{\boldsymbol{r}}

\newcommand{\bv}{\boldsymbol{v}}

\newcommand{\bA}{\boldsymbol{A}}
\newcommand{\bB}{\boldsymbol{B}}

\newcommand{\bG}{\boldsymbol{G}}
\newcommand{\bK}{\boldsymbol{K}}
\newcommand{\bM}{\boldsymbol{M}}
\newcommand{\bL}{\boldsymbol{L}}

\newcommand{\bQ}{\boldsymbol{Q}}
\newcommand{\bR}{\boldsymbol{R}}

\newcommand{\bze}{\boldsymbol{0}}

\newcommand{\bPi}{\boldsymbol{\Pi}}

\newcommand{\bsigma}{\boldsymbol{\sigma}}

\newcommand{\bdelta}{\boldsymbol{\delta}}

\newcommand{\btau}{\boldsymbol{\tau}}

\newcommand{\frake}{\mathfrak{e}}
\newcommand{\frako}{\mathfrak{o}}

\newcommand{\mx}{\mathfrak{X}}
\newcommand{\mxx}{\mathfrak{X}^x}
\newcommand{\mxy}{\mathfrak{X}^y}
\newcommand{\bmx}{\boldsymbol{\mathfrak{X}}}

\newcommand{\orb}{\boldsymbol{\mathfrak{A}}}

\newcommand{\tbPi}{\tilde{\bPi}}
\newcommand{\tPi}{\tilde{\Pi}}
\newcommand{\tbv}{\tilde{\bv}}
\newcommand{\tv}{\tilde{v}}
\newcommand{\tbmx}{\tilde{\bmx}}
\newcommand{\tmxx}{\tilde{\mx}^x}

\newcommand{\tmx}{\tilde{\mx}}
\newcommand{\tH}{\tilde{H}}
\newcommand{\tbsigma}{\tilde{\bsigma}}

\newcommand{\W}{{\cal W}}
\newcommand{\A}{{\cal A}}

\newcommand{\breveg}{\breve{g}}

\newcommand{\inv}{\mathfrak{i}}
\newcommand{\mir}{\mathfrak{r}}
\newcommand\glide{\mathfrak{g}}
\newcommand\rot{\mathfrak{c}}
\newcommand\scr{\mathfrak{s}}
\newcommand\tra{\mathfrak{t}}

\newcommand{\vectwo}[2]{\begin{pmatrix} {#1}\\{#2} \end{pmatrix}}
\newcommand{\matrixtwo}[4]{\begin{pmatrix} #1 & #2 \\ #3 & #4 \end{pmatrix}}
\newcommand{\diagmatrix}[2]{\begin{pmatrix} #1 & 0 \\ 0 & #2 \end{pmatrix}}

\newcommand{\sx}{\sigma_{\sma{1}}}
\newcommand{\sy}{\sigma_{\sma{2}}}
\newcommand{\sz}{\sigma_{\sma{3}}}

\newcommand{\tx}{\tau_{\sma{1}}}
\newcommand{\ty}{\tau_{\sma{2}}}
\newcommand{\tz}{\tau_{\sma{3}}}

\newcommand{\ins}[1]{\;\;\text{#1}\;\;}

\newcommand{\cala}{{\cal A}}

\newcommand{\calf}{{\cal F}}
\newcommand{\calg}{{\cal G}}
\newcommand{\calh}{{\cal H}}
\newcommand{\tcalh}{\tilde{\cal H}}

\newcommand{\call}{{\cal L}}

\newcommand{\caln}{{\cal N}}

\newcommand{\calr}{{\cal R}}
\newcommand{\cals}{{\cal S}}
\newcommand{\calt}{{\cal T}}

\newcommand{\calv}{{\cal V}}

\newcommand{\caly}{{\cal Y}}

\newcommand{\noi}[1]{\noindent (#1)}
\newcommand{\imp}{\;\;\Rightarrow\;\;}
\newcommand{\mo}{\text{-}1}
\newcommand{\mt}{\text{-}2}
\newcommand{\minus}{\text{-}}

\newcommand{\braket}[2]{\big\langle #1 \big| #2 \big\rangle}
\newcommand{\ketbra}[2]{\big|  #1  \big\rangle \big\langle #2 \big| }
\newcommand{\braopket}[3]{\big\langle #1 \big| #2 \big| #3 \big\rangle}
\newcommand{\bra}[1]{\big\langle#1\big|}
\newcommand{\ket}[1]{\big|#1\big\rangle}

\newcommand{\lin}{\notag \\}

\newcommand{\ab}{\alpha\beta}

\newcommand{\low}{L$\ddot{\text{o}}$wdin\;}

\newcommand{\bpm}{\begin{pmatrix}}
\newcommand{\epm}{\end{pmatrix}}

\newcommand{\bal}{\begin{align}}

\newcommand{\dg}[1]{#1^{\scriptstyle{\dagger}}}

\newcommand{\sma}[1]{\scriptscriptstyle{#1}}

\newcommand{\Z}{\mathbb{Z}}

\begin{document}

\title{Semiclassical theory of Landau levels and magnetic breakdown in topological metals}
 \author{A. Alexandradinata} \affiliation{Department of Physics, Yale University, New Haven, Connecticut 06520, USA}  
  \author{Leonid Glazman} \affiliation{Department of Physics, Yale University, New Haven, Connecticut 06520, USA}  
  

\begin{abstract}
The Bohr-Sommerfeld quantization rule lies at the heart of
the semiclassical theory of a Bloch electron in a magnetic field. This rule is predictive of Landau levels and de Haas-van Alphen oscillations for conventional metals, as well as for a host of topological metals which have emerged in the recent intercourse between band theory, crystalline symmetries and topology. The essential ingredients in any quantization rule are connection formulae that match the semiclassical (WKB) wavefunction across regions of strong quantum fluctuations. Here, we propose (a) a multi-component WKB wavefunction that describes transport within degenerate-band subspaces,  and (b) the requisite connection formulae for  saddlepoints and type-II Dirac points, where tunneling respectively occurs within the same band, and between distinct bands. (a-b) extend previous works by incorporating phase corrections that are subleading in powers of the field; these corrections include the geometric Berry phase, and account for the orbital magnetic moment and the Zeeman coupling. A comprehensive symmetry analysis is performed for such phase corrections occurring in closed orbits, which is applicable to solids in any (magnetic) space group.  We have further formulated a graph-theoretic description of semiclassical orbits. This allows us to systematize the construction of quantization rules for a large class of closed orbits (with or without tunneling), as well as to formulate the notion of a topological invariant in semiclassical magnetotransport -- as a quantity that is invariant under continuous deformations of the graph.  Landau levels in the presence of tunneling are generically quasirandom, i.e., disordered on the scale of nearest-neighbor level spacings but having longer-ranged correlations; we develop a perturbative theory to determine Landau levels in such quasirandom spectra. 
\end{abstract}
\date{\today}




\maketitle

{\tableofcontents \par}


\section{Introduction}

The Peierls-Onsager-Lifshitz semiclassical theory of Bloch electrons in a weak magnetic field is the bridge  that connects experimentally-accessible, field-induced oscillations to properties of a metal at zero field.\cite{peierls_substitution,onsager,lifshitz_kosevich,lifshitz_kosevich_jetp,luttinger_peierlssub} This theory underlies the phenomenological construction of the Fermi surface of normal metals\cite{shoenberg,ashcroft_mermin} and superconductors\cite{champel_mineev} -- from measuring the oscillatory period of the magnetization\cite{dHvA} or resistivity.\cite{SdH} These de Haas-van Alphen (dHvA) oscillations are generically disrupted by field-induced quantum tunneling  between semiclassical orbits. Such tunneling, known as magnetic breakdown, occurs wherever semiclassical orbits intersect at  saddlepoints (in the energy-momentum dispersion)\cite{azbel_quasiclassical} or at band-touching points.\cite{slutskin} The experimental discovery of breakdown in magnesium\cite{shoenberg} sparked an extension of the semiclassical theory to incorporate tunneling.\cite{cohen_falicov_breakdown,blount_effham,azbel_quasiclassical,pippard1,pippard2,chambers_breakdown,slutskin,kaganov_coherentmagneticbreakdown,proshin_breakdownwithspinflip} \\



The semiclassical theory has been further extended to incorporate two modern concepts: a wavepacket that orbits around the Fermi surface accumulates a geometric Berry phase,\cite{berry1984,mikitik_berryinmetal} as well as a second phase associated to the orbital magnetic moment of a wavepacket around its center of mass.\cite{chang_niu_hyperorbit} Both phases were first derived from the effective-Hamiltonian theory pioneered in the 1960's;\cite{kohn_effham,rothI,rothII,blount_effham,Wilkinson_semiclassicallimits} analogs of these phases appear ubiquitously in the asymptotic theory of coupled-wave equations,\cite{littlejohn_short,littlejohn_long,Emmrich_geometry,panati_spaceadiabatic} which apply in a much wider variety of physical contexts than the present study. Only more recently have the physical consequences of the geometric phase and the orbital magnetic moment been explored for solids\cite{rothII,Mikitik_quantizationrule,topofermiology}
 -- especially in the complementary theory of wavepackets.\cite{chang_niu_hyperorbit,sundaram1999,chang2008,culcer_multibandwavepacket}\\



Both of the above phases are  evaluated on  semiclassical orbits uniquely determined by Hamilton's equation. On the other hand, semiclassical orbits are no longer unique in the presence of breakdown. The challenge is to resolve this tension. Recently, we have synthesized the geometric phase, the orbital moment, and tunneling -- into a single, generalized Bohr-Sommerfeld quantization rule.\cite{AALG_breakdown} This rule is not only predictive of Landau levels and de Haas-van Alphen oscillations  for conventional metals, but it is also critically relevant to describe a host of topological metals which have emerged in the recent intercourse between band theory, crystalline symmetries and topology. Such topological metals have intrinsically unremovable geometric phase, owing to the presence of Dirac-Weyl points where conically-dispersing bands touch;\cite{Nielsen_ABJanomaly_Weyl,wan2010,burkov2011,Bradlyn_newfermions,Murakami2007B,zhijun_3DDirac,Young_diracsemimetal2D,tiantianAA_doubleweylphonons} their Fermi surfaces are twisted into unusual topologies\cite{Serbyn_LandauofTCI,obrien_breakdown,koshino_figureofeight} such that breakdown is also unavoidable. \\

The essential ingredients in any quantization rule are connection formulae that match the semiclassical (WKB) wavefunction across regions of strong quantum fluctuations. The main subject of this work is the derivation of these ingredients and the systematic construction of quantization rules for a large class of closed orbits -- with and without breakdown. Our results are summarized in the following section.

\section{Summary and organization of results}

The semiclassical theory is a method to approximate the wavefunctions and energy levels of a Bloch electron in a magnetic field. These approximations become increasingly accurate in the limit where a classical action function -- that characterizes the solid at zero field -- is much larger than  a parameter characteristic of the magnetic field. These quantities are simplest to exemplify for a semiclassical orbit on a Fermi surface having the topology of a sphere; an orbitting wavepacket evolves according to Hamilton's equation of motion:
\e{ \hbar \dot{\bk}\eq -\f{|e|}{c\hbar} \nabk \var \times \bB. \la{hamiltoneom}}
The semiclassical approximation is valid where the the area $|S|$ (in $\bk$-space) of this orbit is much larger than $1/l^2$, with
 the magnetic length  defined as
\e{ l = \sqrt{\f{\hbar c}{|e|B}}. \la{definemagneticlength}} 
To simplify notation, we henceforth adopt a coordinate system where the magnetic field $\bB=-B\vec{z}$, and the semiclassical orbit is therefore a band contour at fixed energy $E$ (and wavevector $k_z$, for 3D solids), equipped with an orientation from \q{hamiltoneom}.  The Bohr-Sommerfeld quantization rules of the semiclassical theory are derived in the effective-Hamiltonian formalism,\cite{peierls_substitution,luttinger_peierlssub,kohn_effham,wannier_fredkin,blount_effham,rothI,zak_dynamicsofelectrons,nenciu_review} which we briefly review  in \s{sec:revieweffham}. Certain notations used in this review, and throughout the text, are collected  in \s{sec:preliminaries} for easy reference.\\

We begin properly in \s{sec:simple} by deriving the quantization rules for closed orbits in the absence of breakdown; these rules are summarized in \qq{rule3a}{definenonabelianunitary}, and their consequences for Landau levels and de Haas-van Alphen oscillations are discussed in \s{sec:quantumoscillations}. These rules are equivalent to a continuity condition of the WKB solution to the above-reviewed effective Hamiltonian. The single-component WKB wavefunction\cite{zilberman_wkb,fischbeck_review} (as applied to a nondegenerate band) is reviewed in \s{sec:singlebandwkbwf}, and we generalize this to a multi-component wavefunction in \s{sec:multibandwkbwf} (as applied to bands of arbitrary degeneracy $D$). As this wavefunction is continued around a closed orbit, it accumulates a phase proportional to $1/B$ and  the oriented area of the orbit; the subleading-in-$B$ variation of the wavefunction is described by a $D \times D$ unitary propagator $\cala$ [cf.\ \q{definenonabelianunitary}], which is generated by a one-form that includes the Berry connection (non-abelian for $D>1$), the orbital magnetic moment, and the Zeeman coupling. Each eigen-phase (i.e., phase of each eigenvalue) of this propagator enters the quantization rule as an $O(1)$ phase correction. In addition, there is a subleading Maslov correction originating from turning points on the orbit where the WKB solution is invalid. A general method to determine Maslov corrections is described in \s{sec:turningpoint}, which is applicable to twisted Fermi surfaces whose orbits intersect at points. We emphasize that there are no further $O(1)$ corrections to the quantization rule.\\


\s{sec:symmetry} is an exposition of the effects of symmetry (in any space or magnetic space group) in the quantization condition. By a symmetry analysis of the propagator $\cala$ [cf.\ \q{definenonabelianunitary}], we ascertain how symmetry constrains the degeneracy and energetic offsets of the Landau levels, as well as phase offsets in the de Haas-van Alphen oscillations. In addition, we provide a general symmetry analysis of the orbital magentic moment and Zeeman coupling in \s{sec:symmetrysinglebandom}; this may be applied to $\bk$-resolved measurements of the orbital magnetic moment, e.g., through circular dichroism in photoemission.\cite{YaoWang_dichroism_valley}\\

In \s{sec:intraband}, we describe quantization rules which are applicable to orbits which intersect at saddlepoints in the energy-momenta dispersion. Saddlepoints are the nuclei of Lifshitz transitions in the Fermi-surface topology, as exemplified by the surface states of topological crystalline insulators.\cite{Serbyn_LandauofTCI,Hourglass} The vicinity of saddlepoints are regions of strong intraband tunneling where the WKB solutions lose their validity. WKB wavefunctions away from the saddlepoint are patched together by  a connection formula that we derive in \s{sec:connectionintraband}. The generalized quantization rule is equivalent to the continuity of patched-up WKB wavefunctions over the intersecting orbit; the general algorithm for constructing such rules is presented in \s{sec:quantizationintraband}. This algorithm is then applied to two case studies: a Weyl metal near a metal-insulator phase transition [cf.\ \s{sec:doublewell}], and the surface states in the SnTe-class of topological crystalline insulators\cite{Hsieh_SnTe} [cf.\ \s{sec:tci}]. \\

A qualitatively distinct type of breakdown occurs where orbits intersect at a touching point between two bands which are otherwise nondegenerate at generic wavevectors. In this work, we focus on touching points for which the nearby band dispersion is conical. From a general classification of Fermi surfaces near conical touching points,\cite{WTe2Weyl} we identify the orbit intersection as a type-II Dirac point (in short, a II-Dirac point),\cite{isobe_nagaosa_IIDirac,Bergholtz_typeII,LMAA} which might be viewed as an over-tilted version of the conventional, rotationally-symmetric Dirac point.\cite{Novoselov_graphene} Due to the discontinuity of the Bloch wavefunction across the II-Dirac point,\cite{Zak_diracpoint} the effective Hamiltonian that was reviewed in \s{sec:revieweffham} is not applicable. What we require is a different representation for the effective Hamiltonian, where the basis functions evolve smoothly across the II-Dirac point. Inspired by the basis functions proposed by Slutskin,\cite{slutskin} we formulate such an effective Hamiltonian in \s{sec:effhamgen}, as summarized in \qq{effhambanddeg}{effhambanddeglinearkx}. This effective Hamiltonian extends previous formulations\cite{slutskin,chambers_breakdown} by: (i) being applicable to any band-touching point (of any degeneracy and dispersion, e.g., Weyl\cite{Nielsen_ABJanomaly_Weyl,wan2010,burkov2011} and multi-Weyl\cite{chen_multiweyl} and spin-1 Weyl points,\cite{Bradlyn_newfermions} four-fold-degenerate Dirac points\cite{Murakami2007B,zhijun_3DDirac,Young_diracsemimetal2D} and charge-2 Dirac points\cite{tiantianAA_doubleweylphonons}), and (ii) by accounting for subleading-in-$B$ corrections, which includes the multi-band orbital magnetic moment at the band-touching point.\\

The solution of the above effective Hamiltonian -- particularized to a II-Dirac point -- affords us a connection formula presented in \s{sec:connectioninter}. This rule is a crucial ingredient in quantization rules for orbits that intersect at a II-Dirac point. We demonstrate how to construct such rules for orbits surrounding an isolated, over-tilted Weyl point in \s{sec:casestudysingleIIDirac}. The Landau-level spectrum in the presence of interband breakdown (and also intraband breakdown in low-symmetry metals) is generically quasirandom, i.e., disordered on the scale of nearest-neighbor level spacings but having longer-ranged correlations. A perturbative theory to determine Landau levels in quasirandom spectra is presented in  \s{sec:perturbation} and applied to our case study.\\

Throughout the text, we will employ a graph-theoretic description of orbits that is summarized in \s{sec:graph} for easy reference. Such a description is not only useful in systematizing the construction of quantization rules (with or without breakdown), it allows us to define an equivalence class of Fermi surfaces -- through the homotopy equivalence of their corresponding graphs. This allows us to precisely define a topological invariant in semiclassical magnetotransport: as a quantity that is invariant under continuous deformations of the Hamiltonian that preserves the homotopy class of the graph. The generalization to symmetry-protected topological invariants is simply described in \s{sec:graph}. Examples of such topological invariants have been presented in our previous companion works: \ocite{topofermiology} and \ocite{AALG_breakdown}.

\section{Preliminaries}\la{sec:preliminaries}

We review the exact Hamiltonian of a Bloch electron, with and without a magnetic field, to establish notation that would be used throughout this paper.

\subsection{Bloch Hamiltonian in the crystal momentum representation}

In materials with light elements and consequently weak spin-orbit coupling, we would apply the field-free Schrodinger Hamiltonian, defined as
\e{\hat{H}_0 =\f{ \bp^2}{2m}+V(\br); \la{fieldfreeschrodinger}}
otherwise, we apply the Pauli Hamiltonian:
\e{\hat{H}_0 \eq  \f{1}{2m}\left( \bp + \f{\hbar}{4mc^2}\bsigma \times \nabla V \right)^2+V(\br), \la{fieldfreepauli}}
which is accurate to order $E/mc^2$. We use the same symbol $\hat{H}_0$ for both Hamiltonians, and unless otherwise stated in the context, we assume that expressions with $\hat{H}_0$ apply to both types of Hamiltonians. Each eigenstate of $\hat{H}_0$ may be expressed as a Bloch function
\e{\psi_{n\bk} =\eikr u_{n\bk},}
where $u_{n\bk} = u_{n\bk}(\br)$ in the Schrodinger case, and $=u_{n\bk}(\br,s)$ with additional spin index $s$ in the Pauli case. 
In both cases, $u_{n\bk}$ is periodic with respect to Bravais-lattice translations $\br \rightarrow \br +\bR$, and shall henceforth be referred to as cell-periodic functions. It is convenient to define the cell-periodic position coordinate $\btau$ with the equivalence $\btau \sim \btau+\bR$, as well as the variable $\alpha$, which is a flexible shorthand for $\btau$ in the Schrodinger case, and for $(\btau,s)$ in the Pauli case. It is well known that cell-periodic functions form an orthonormal set which is complete with respect to the space of $\alpha$:
\e{ &\sum_{\alpha} u^*_{m\bk}(\alpha)u_{n\bk}(\alpha) = \delta_{mn}, \iand \lin
  &\sum_{m} u_{m\bk}(\alpha)u^*_{m\bk}(\beta)=\delta_{\ab}. \la{cellperiodicbasis}}
Here and henceforth, we employ the Dirac notation: $\braket{u}{v} = \sum_{\alpha}u^*(\alpha)v(\alpha),$ where
$\sum_{\alpha}$ should be interpreted as an integration of $\btau$ over the unit cell (normalized multiplicatively by the volume of the Brillouin torus), and possibly also a sum over the spin indices. Analogously, $\delta_{\ab}$ denotes the Dirac delta function $\delta(\btau-\btau')$, possibly multiplied with a Kronecker delta function in spin space.  When there is no topological obstruction to constructing Wannier functions ($W_n$), we will find it useful to expand the cell-periodic function in terms of Wannier functions as
\e{ u_{n\bk}(\btau,s) = \f1{\sqrt{N}}\sum_{\bR}e^{-i\bk \cdot (\btau-\bR)}W_n(\btau-\bR,s).\la{expandcellperiodicfunc0}}
The Hamiltonian acts on cell-periodic functions as
\e{ \hat{H}_0(\bk) = e^{-i\bk \cdot \hbr}\hat{H}_0e^{i\bk \cdot \hbr};}
we will refer to $\hat{H}_0$ as the Hamiltonian, and $\hat{H}_0(\bk)$ as the Bloch Hamiltonian. The velocity operator is defined by
\e{ \hbPi = -\f{i}{\hbar}[\hbr,\hat{H}_0]= \nabla_{\bp}\hH_0,\la{definevelocity0}}
and it acts on cell-periodic functions as
\bal
\hbPi(\bk) :\eq\emikr\,\hbPi\,\eikr= \hbPi+\f{\hbar \bk}{m}= \nabla_{\bp}\hH_0(\bk)\lin
\eq  \nabk \hat{H}_0(\bk)  =\begin{cases} \f{\hp +\hbar\bk}{m}, \\
                                                                          \f{\hp +\hbar \bk}{m} -\f{\bohr}{2e mc}\bsigma \times \nabla V,\end{cases}
																																					\la{definevelocity}
\end{align}
with the Bohr magneton $\bohr ={|e|\hbar}/{2mc}$. The Bloch Hamiltonian may always be expanded around a chosen wavevector $\bk_0$ as
\e{ \hH_0(\bk) = \hH_0(\bk_0)+ \hbar(\bk-\bk_0)\cdot \hbPi(\bk_0)+ \f{\hbar^2(\bk-\bk_0)^2}{2m}. \la{expandBlochham}}

Any operator which acts on functions of $\br$ (and possibly also on spin index $s$) are denoted with a hat, as exemplified in \q{fieldfreeschrodinger}-\ref{definevelocity}; the same operator in the basis of cell-periodic functions is a matrix denoted by the same symbol with a tilde. Unless specified otherwise, we will usually employ a basis of cell-periodic functions which correspond to energy bands, i.e., for which the Hamiltonian matrix is diagonal 
\e{ \tH_0(\bk)_{mn} = \braopket{u_{m\bk}}{\hH_0(\bk)}{u_{n\bk}}=\var_{n\bk}\delta_{mn}. \la{definehamiltonianmatrix}}
Another example is the velocity matrix
\e{\tbPi (\bk)_{mn} = \braopket{u_{m\bk}}{\hbPi(\bk)}{u_{n\bk}}, \la{definevelocitymatrix}}
which may be identified, in the basis of energy eigenstates, as
\e{ \hbar \tbPi(\bk)_{mn} = \nabk \var_n \delta_{mn} + i\tbmx(\bk)_{mn}(\var_{m\bk}-\var_{n\bk}). \la{idenvel}}
Here, we have introduced
\e{\tbmx(\bk)_{mn}=i\braket{u_{m\bk}}{\nabk u_{n\bk}}, \la{defineberryconnection}}
which occurs as part of the matrix elements of the position operator in the crystal-momentum representation.\cite{Blount} It is also useful to define the diagonal component of the velocity matrix as
\e{ \tbv(\bk)_{mn} = \tf1{\hbar}\nabk \var_n \delta_{mn}, \la{definediagonalvelocitymatrix}}
as well as the spin-half matrix
\e{\f{\hbar}{2}\tbsigma(\bk)_{mn} = \f{\hbar}{2} \braopket{u_{m\bk}}{\hat{\bsigma}}{u_{n\bk}}. \la{definespinhalf}}
While these matrices are formally infinite-dimensional, we are often interested in the physics of a finite number of (possibly degenerate) bands, projected by
\e{P(\bk) = \sum_{n=1}^D \ketbra{u_{n\bk}}{u_{n\bk}}, \la{defineprojP}}
with $D$ the dimension of  said subspace at each wavevector. Bands not in $P$ are henceforth labelled with an extra bar: $\bar{m},\bar{n}$, and their corresponding projection is
\e{Q(\bk) = \sum_{\bar{n}} \ketbra{u_{{\bar{n}}\bk}}{u_{{\bar{n}}\bk}}=I-P(\bk).\la{defineprojQ}}
Let us then define the restriction of any matrix in the infinite basis to the subspace projected by $P$ as 
\e{ P:\;\{\tbPi, \tbmx, \tbv, \tbsigma\} \rightarrow \{\bPi, \bmx, \bv, \bsigma\}. \la{restriction}}
These finite-dimensional matrices are distinguished notationally by having no accents. A case in point is $\bmx$, which is the Berry connection\cite{berry1984} for solids.\cite{zak1989} This connection manifests whenover one differentiates operators represented in the $D$-dimensional cell-periodic basis: for any $\hat{O}(\bk)$,
\e{ \nabk& O(\bk)_{mn}= \nabk \braopket{u_{m\bk}}{\hat{O}(\bk)}{u_{n\bk}} \lin
\eq \braopket{u_{m\bk}}{\nabk\hat{O}}{u_{n\bk}} + i[\bmx(\bk),O(\bk)]_{mn}. \la{idenderivativek}}

\subsection{Gauge transformations in band theory}

We would often deal with $U(D)$ basis transformations in the cell-periodic functions in $P$:
\e{\ket{u_{n\bk}}\rightarrow \sum_{m=1}^D \ket{u_{m\bk}}V_{mn}(\bk), \as V^{\mo}=\dg{V}.\la{basistransf}}
We will refer to this as a gauge transformation within $P$. With respect to this transformation, certain objects are invariant (such as the projection $P$ itself); other objects transform covariantly, i.e., they change only in being conjugated by the unitary $V$ (e.g., the just-defined spin matrix  $\bsigma \rightarrow V^{-1}\bsigma V$); other objects have a more complicated tranformation rule, e.g., the non-abelian Berry connection transforms as\cite{wilczek1984} 
\e{\bmx \rightarrow V^{-1}\bmx V+iV^{-1}\nabk V. \la{gaugetransform}}

\subsection{Review of symmetry in Bloch Hamiltonians}\la{sec:symmetryinBloch}

Let $g$ denote a symmetry in the (magnetic) space group ($G$) of a solid; we use $\hat{g}$ to denote its representation in real space tensored with spin space. Its action on the position operator can always be decomposed as a point-group operation (an operation that preserves at least one point) and a translation:\cite{Lax}
\e{ \hat{g}^{\mo}\hat{r}_i\hat{g} = \check{g}_{ij}\hat{r}_j+\delta_i, \as \check{g}^{\mo}=\check{g}^t \in\R, \la{gactsonposition}}
which we shorten notationally as $\hat{g}^{-1}\hbr\hat{g} = \check{g}\hbr+\bdelta.$ Here, we have introduced a real, orthogonal matrix $\check{g}$ that represents the point-group component of $g$ acts in real space. For all symmetry elements in symmorphic space groups, a spatial origin may be chosen such that $\bdelta$ is a Bravais-lattice vector.\cite{Lax} To describe nonsymmorphic operations such as screw rotations and glide reflections, we allow $\bdelta$ to be a rational fraction  of a Bravais-lattice vector.\\

In addition to $g$ that transforms space, we also consider $g$ that reverses time. The time reversal operation $g=T$ acts trivially on space ($\check{g}=I, \bdelta=0$), and is represented by $\hat{T}=U_TK$, with $U_T$ a unitary transformation and $K$ the complex conjugation operation; $\hat{T}^2=(-1)^{\sma{F}}$, where $F=0$ for integer-spin representations ($U_T=I$), and $F=1$ for half-integer-spin representations ($U_T=-i\sigma_y$ in spinor space). It is useful to introduce a $\Z_2$ index that distinguishes between transformations which are purely spatial (and therefore have a unitary representation $\hat{g}$), and transformations which involve a time reversal, possibly composed with a spatial operation ($\hat{g}$ here is antiunitary):  
\bal
&g: \vectwo{\hbr}{t} \rightarrow \matrixtwo{\check{g}}{0}{0}{ (-1)^{s(g)}}\vectwo{\hbr}{t}+\vectwo{\bdelta}{0}; \lin
 &s(g)=\begin{cases}0, \as \hat{g} \; \ins{unitary,} \\ 1, \as \hat{g} \; \ins{antiunitary.}\end{cases} \la{definesg}
\end{align}
As a useful example, we apply \q{definesg} and  \q{gactsonposition} to derive
\e{ \hat{g}e^{i\bk\cdot \hbr}\hat{g}^{-1}=e^{i(-1)^{\sma{s(g)}}[\check{g}\bk]\cdot( \hbr-\bdelta)},\la{gactsoneikr}}
which implies that a Bloch function at wavevector $\bk$, when operated upon by $g$, transforms in the representation 
\e{ g \circ \bk:=(-1)^{\sma{s(g)}}\check{g}\bk. \la{definegcirck}}
 If $g$ is a symmetry of the Hamiltonian ($[\hat{g},\hH_0]=0$), then 
\e{ \hat{g}(\bk)&\hH_0(\bk)\hat{g}^{-1}(\bk) =\hH_0\big( g\circ \bk\big), \iwith\lin
 &\hat{g}(\bk):=e^{-i(g\circ\bk)\cdot \bdelta}\hat{g}. \la{symmetryofH0}}
This implies that if $\ket{u_{m\bk}}$ is an eigenstate of $\hH_0(\bk)$ with eigenvalue $\var_{m\bk}$, then ${\hat{g}(\bk)}\ket{u_{m\bk}}K^{s(g)}$ belongs to the eigenspace of $\hH_0( \;g\circ\bk\;)$ with the same energy $\var_{m\bk}$; this is expressed as 
\e{ {\hat{g}}(\bk)\ket{u_{m\bk}}K^{s(g)} \eq \ket{u_{n,g\circ \bk}}\breve{g}(\bk)_{nm},\la{gactsonu}}
where $\breve{g}$, a unitary matrix that is block-diagonal with respect to the energy eigenspaces, expresses the ambiguity in our choice of basis vectors within each energy eigenspace. To clarify a possible source of confusion, $\braopket{\alpha}{\hat{g}(\bk)}{u_{m\bk}}K^{s(g)}$ is just a complex number -- where $s(g)=1$, there are two $K$ operators in this expression: one explicit, and the other implicit in $\hat{g}$.\\

We refer to $\breveg(\bk)$ colloquially as the `sewing matrix', owing to its function in `sewing' together the cell-periodic functions by symmetry. Sewing matrices are the basic objects that encode symmetry constraints in the crystal-momentum representation, and they will play a prominent role in constraining the effective Hamiltonian. These matrices may be understood from a group-cohomological perspective;\cite{shiozaki_review} the winding number of the sewing matrix over the Brillouin torus also plays a role in the topological classification of band insulators.\cite{hughes_inversionsymmetricTI,Chen_bulktopologicalinvariants} \\

For our purpose of determining the symmetry constraints on the effective Hamiltonian, we will need to review a few properties of sewing matrices. Depending on the presence of spin-$SU(2)$ symmetry, $\{\breveg(\bk)\}$ forms either an integer- or half-integer-spin representation of the space group.\cite{Cohomological,shiozaki_review} A simple example might convey this point: let $g$ be a glide operation ($\glide_{x,\vec{y}/2}$) that is composed of a reflection, that inverts $x\rightarrow -x$, and a translation by half a Bravais-lattice vector in $\vec{y}$ (denoted $\tra_{\vec{y}/2}$). In the space group, the multiplication rule for this element is: $\glide_{x,\vec{y}/2}^2 = \frake\,\tra_{\vec{y}}$, with $\frake$ a $2\pi$ rotation and $\tra_{\vec{y}}$ a full lattice translation; this is represented as\footnote{See related footnote just after \ocite{weinbergbook1}.}
\e{ &[\breve{\glide}_{x,\vec{y}/2}(-k_x,k_y,k_z)\breve{\glide}_{x,\vec{y}/2}(\bk)]_{mn}\lin
 \eq \sum_l \braopket{u_{m\bk}}{e^{-ik_y/2}\hat{\glide}_{x,\vec{y}/2}}{u_{l,(-k_x,k_y,k_z)}} \lin 
&\times\;\braopket{u_{l,(-k_x,k_y,k_z)}}{e^{-ik_y/2}\hat{\glide}_{x,\vec{y}/2}}{u_{n\bk}} \lin
\eq  e^{-ik_y} \braopket{u_{m\bk}}{\hat{\glide}_{x,\vec{y}/2}^2}{u_{n\bk}} = e^{-ik_y}(-1)^{\sma{F}}\delta_{mn}. \la{glidesquare}  }
In the last equality, we employed that a rotation by $2\pi$ produces a representation-dependent, $\pm 1$ factor, and also that $\tra_{\bR}$ has a trivial action on cell-periodic functions. \\

More generally, for any nontrivial $g$ which is not purely a translation, we may assign to $g$ an order $N(g)$, which is the smallest integer in $\{2,3,4,6\}$  such that
\e{ g^{N(g)}=\frake^{a(g)} \tra_{\bR(g)}, \as a(g) \in \{0,1\},\la{defineorderg}}
with  $\tra_{\bR}$ a translation by  a Bravais-lattice vector $\bR$ (possibly the zero vector) which depends on $g$. We have introduced a $\Z_2$ index $a(g)$ that equals $0$ (resp.\ $1$) if $g^N$ is proportional to an odd (resp.\ even) multiple of a $2\pi$ rotation $(a=1)$. In the case of $g=\glide_{x,\vec{y}/2}^2$ in \q{glidesquare}, $N=2,a=1$ and $\bR=\vec{y}$; other representative examples are summarized in \tab{tab:exampleproj}. If $g$ reverses time, its order must be even:
\e{ s(g)=1 \imp N(g)\in 2\Z. \la{simpliesN}}
This follows because  $g^N$ by assumption does not invert time [cf.\ \q{defineorderg}], and on the other hand it is the composition of $T^N$ with a spatial transformation. \\

For any $\bk_1$, we define
\e{ &g\text{-orbit of}\;\bk_1:= \{\bk_i\}_{i=1}^N, \iwith \lin
&\bk_{i+1}:=g^i\circ \bk_1:=\bk_{i+N+1},\la{definegorbit}}
which is not to be confused with Hamilton's semiclassical orbit; we are guaranteed that $\bk_{i}=\bk_{i+N}$ owing to \q{defineorderg}. \q{defineorderg} is represented with the sewing matrices as
\e{ \breve{g}_{i}:=\breve{g}(\bk_i), \as \breve{g}_{\sma{N}}K^s\, \ldots \;\breve{g}_{\sma{2}}K^s\; \breve{g}_{\sma{1}}K^s =(-1)^{Fa}e^{-i\bk \cdot \bR}.\la{orderNrepsewingmatrix}}
When this equation is particularized to $g$ which is unitarily represented, and to $\bk = g\circ \bk$, we obtain
\e{   \breve{g}\big(\bk)^N =(-1)^{Fa}e^{-i\bk \cdot \bR}.\la{orderNginv}}
The $N$ possible eigenvalues of $\breve{g}$ (at $g$-invariant wavevectors), corresponding to the $N$ roots of $e^{i\pi Fa-i\bk\cdot \bR}$, label the different representations of $g$. More examples of sewing matrices are provided in the second column of Tab. \ref{tab:magnetization}.\\

\begin{table}[ht]
	
\centering
		
\begin{tabular} {|c|c|c|c|c|c|} \hline
			
$g$ &$N$&  $a$ & $\bR$ &$m$& $p$ \\  \hline \hline 
		  
$\inv$ & $2$& $0$ &  $\bze$& $1$ & $1$        \\ \hline			 
$T$ & $2$& $1$ & $\bze$& $1$&  $1$        \\ \hline			 
$\rot_{nz}$ & $n$& $1$ & $\bze$&$1$&  $1$        \\ \hline			 
$T\rot_{3z}$ & $6$& $1$ & $\bze$&$1$&  $5$        \\ \hline			 
$T\rot_{4z}$ & $4$& $1$ & $\bze$&$1$&  $3$        \\ \hline			 
$T\rot_{6z}$ & $6$& $1$ & $\bze$&$2$&  $4$        \\ \hline			 
$\glide_{x,\vec{y}/2}$ & $2$& $1$ & $\vec{y}$&$1$&  $1$        \\ \hline			 
\end{tabular}
		
\caption{ Examples of symmetries $g$ of order $N$. $a$ and $\bR$ are defined through \q{defineorderg}. $m$ and $p$ are quantities that are introduced later in \s{sec:classIIA}: $m$ is the number of cycles in the $g$-orbit, and $p \sim p+N$ labels inequivalent extensions (by quasimomentum loop translations) of the point group generated by $g$. We have chosen the convention that $\frako$ is clockwise-oriented, and that $\rot_{nz}$ induces an anticlockwise rotation in  $\bk$ space; if both $\frako$ and $\rot_{nz}$ are anticlockwise-oriented, then the above values of $p$ should be inverted in sign.  
	\label{tab:exampleproj}}
\end{table}

Finally, we consider how the sewing matrix transforms under basis transformations of the form in \q{basistransf}.
From \q{gactsonu}, we derive 
\e{ \breve{g}(\bk) \rightarrow   \dg{V}\big(g\circ \bk\big)\breve{g}(\bk)K^{s(g)}V(\bk)K^{s(g)}. \la{generalbasistransformsew}}
For $g$-invariant wavevectors (defined through $\bk = g\circ \bk$ modulo a reciprocal vector), \q{generalbasistransformsew} particularizes to
\begin{align}
\breveg \rightarrow \begin{cases} \dg{V}\breveg V, &\text{for unitary} \;g \\
                                      \dg{V}\breveg V^*, &\text{for anti-unitary} \;g.\end{cases}\la{basistransformsew}
																			\end{align}
This distinction between unitary and anti-unitary symmetries becomes relevant when we consider the symmetry constraints of the \om in \s{sec:symmetrysinglebandom}.\\

For future reference, we employ the following notation for symmetry operations: $T$ denotes time reversal, $\tra_{\vec{z}/2}$ a real-space translation by half a Bravais-lattice vector parallel to $\vec{z}$; $T\tra_{\vec{z}/2}$ is the composition of $T$ and $\tra_{\vec{z}/2}$. $\inv$ denotes spatial inversion. $\mir_{\alpha}$ ($\glide_{\alpha, \vec{\beta}/2}$) is normal (glide) reflection that inverts the spatial coordinate $\alpha$; the glide operation includes an additional translation by $\vec{\beta}/2$, which is half a Bravais-lattice vector in the $\beta$ direction. $\rot_{nz}$ is an $n$-fold rotation about $\vec{z}$ ($n \in \{2,3,4,5\})$, $\scr_{nz,m}$ is a screw rotation that satisfies ${\scr_{nz,m}}^n=\tra_{m\bG}$ with $\bG$ the smallest reciprocal vector parallel to $\vec{z}$ and $m\in \{0,1,\ldots,n-1\}$. To describe half-integer-spin representations, we will employ the double-group formalism that identifies a $2\pi$ rotation with a group element ($\frake$) that is distinct from and squares to the identity.

\subsection{Hamiltonian of a Bloch electron in a magnetic field}

We study a Bloch electron immersed in a spatially homogeneous magnetic field $\bB$, with corresponding  vector potential
\e{\bB = \curl \bA, \as \ba=\f{|e|}{c}\bA,}
The field-on Schrodinger Hamiltonian is defined as
\e{ &\hH = \f{ (\bp+\ba(\br))^2}{2m}+V(\br);\la{definefieldonhamiltonian}}
this is distinguished notationally from the zero-field Schrodinger Hamiltonian ($\hat{H}_0$) by having no subscript. Analogously,
the field-on Pauli Hamiltonian is
\e{\hH \eq  \f{1}{2m}\left( \bp +\ba+ \f{\hbar}{4mc^2}\bsigma \times \nabla V \right)^2\lin
&\;+V(\br)+\f{\gfac}{2}\bohr \bsigma \cdot \bB, \la{definefieldonpaulihamiltonian}}
with the free-electron g-factor $\gfac \approx 2$. The semiclassical equation of motion for a Bloch electron in a magnetic field is
\e{ \hbar \dot{\bk}^{\sma{\perp}}\bigg|_{\bk}\eq   l^{-2} \epsilon^{\ab}\vec{\alpha}v^{\beta}(\bk), \as \alpha, \beta \in \{x,y\},\la{hamilton2}}
which particularizes \q{hamiltoneom} to the case $\bB =-B\vec{z}$. We refer to $\dot{\bk}^{\sma{\perp}}|_{\bk}$ as the \emph{orbit velocity} at $\bk$, which is distinguished from the field-independent band velocity $\bv(\bk)$.

\subsection{Field-free Bloch Hamiltonian in the Luttinger-Kohn representation}\la{sec:reviewluttkohn}

This subsection reviews a set of basis functions which are more convenient to employ near a conical band touching -- this would be useful when we derive the effective Hamiltonian near a band degeneracy in \s{sec:effhamgen}, and derive the Bohr-Sommerfeld quantization conditions in the presence of interband breakdown in \s{sec:interband}.\\

The Bloch functions are not an ideal basis for application near conical band touchings, owing to their discontinuity with respect to $\bk$ at the touching point,\cite{Zak_diracpoint} which we set by convention to $\bze$. Here, it is convenient to employ  the Luttinger-Kohn functions $\{{u}_{nk_x0}(\br)\eikr\}$, which are known to form a complete and orthonormal set of basis functions,\cite{Luttinger_Kohn_function} and are analytically better-behaved at $\bk=\bze$.  To clarify the terminology we employ, Luttinger and Kohn (LK) considered in similar spirit the functions $\{u_{n\bze}\eikr\}$;\cite{Luttinger_Kohn_function} we take the liberty of referring to $\{\tilde{u}_{nk_x0}\eikr\}$ as LK functions -- the proof of completeness and orthonormality for $\{\tilde{u}_{nk_x0}\eikr\}$ is nearly identical to that presented in Ref.\ \onlinecite{Luttinger_Kohn_function}.\\

The Bloch Hamiltonian in the LK representation has the form (with $\hbar{=}1$) 
\e{&\braopket{{u}_{m,k_x,0}}{\hat{H}_0(\bk)}{{u}_{n,k_x,0}} \lin
\eq \tilde{H}(k_x,0)_{mn}  +k_y{\tilde{\Pi}}^y_{mn}(k_x,0)+\f{k_y^2}{2m}\delta_{mn},}
where we have applied the expansion \q{expandBlochham} around the $k_y=0$ line, which intersects the band touching point at $\bk=\bze$.
It is convenient to choose ${u}_{nk_x0}$ to be eigenfunctions of the Bloch Hamiltonian $\hat{H}_0(k_x,0)$ (i.e., such that $\tilde{H}(k_x,0)$ is diagonal); this choice for ${u}_{nk_x0}$ will be emphasized notationally by adding a tilde: ${u}_{nk_x0}\rightarrow \tilde{u}_{nk_x0}$.\\

Let us restrict $\{\tilde{u}_{nk_x0}\}_{n=1}^D$ to the D-dimensional subspace projected by $P$, and further assume that this subspace of bands is D-fold degenerate at $\bk=\bze$; we further set the origin of energy such that $H_0(\bze)=0$. Applying the identity of \q{idenderivativek} to $\nabk^xH_0\big|_{\bze}$, the Hamiltonian to linear order in $k_i$ simplifies to
\e{\braopket{\tilde{u}_{m,k_x,0}}{\hat{H}_0(\bk)}{\tilde{u}_{n,k_x,0}} \eq k_x\Pi_{mn}^x(\bze) +k_y{\Pi}^y_{mn}(\bze), \la{LKhamlinear}}
with $\Pi^x$ and $\Pi^y$ D-by-D diagonal matrices.\\

It would be useful to transform between the crystal-momentum and Luttinger-Kohn representations by the overlap matrix $\tilde{S}$ defined as
\e{ u_{n\bk} \eq \sum_{l=1}^{\infty}\tilde{u}_{l,k_x,0} \tilde{S}_{ln}(k_x,0,\bk) \lin
\eq \sum_{l=1}^D\tilde{u}_{l,k_x,0} {S}_{ln}(k_x,0,\bk)+O(k_y/G_y), \lin
&\iwith \tilde{S}(k_x,0,\bk)_{mn}=\braket{\tilde{u}_{mk_x0}}{u_{n\bk}},}
and $G_y$ a reciprocal period. The projection of $\tilde{S}$ into the $D$-dimensional subspace is approximated, to an accuracy of $O(k_y/G_y)$, by a unitary matrix $S$ defined by the eigenvalue equation:   
\e{ \sum_{l=1}^D\left[k_x\Pi^x(\bze)  +k_y{\Pi}^y(\bze)\right]_{ml} S_{ln} = \enk S_{mn}, \la{truncatedLKeigenvalueeq}}
where $u_{n\bk}$ diagonalizes $H_0(\bk)$ with energy eigenvalue $\enk$.

\subsection{Graph-theoretic description of orbits}\la{sec:graph}

With the eventual goal of formulating quantization conditions (with and without breakdown) and their topological invariants, we will find it useful to formulate a graph-theoretic description of orbits. This section is written for easy reference of graph-theoretic terminology that we will eventually employ, and the reader may skip this on a first reading, and refer back to it when necessary. \\

\begin{figure}[ht]
\centering
\includegraphics[width=8.4 cm]{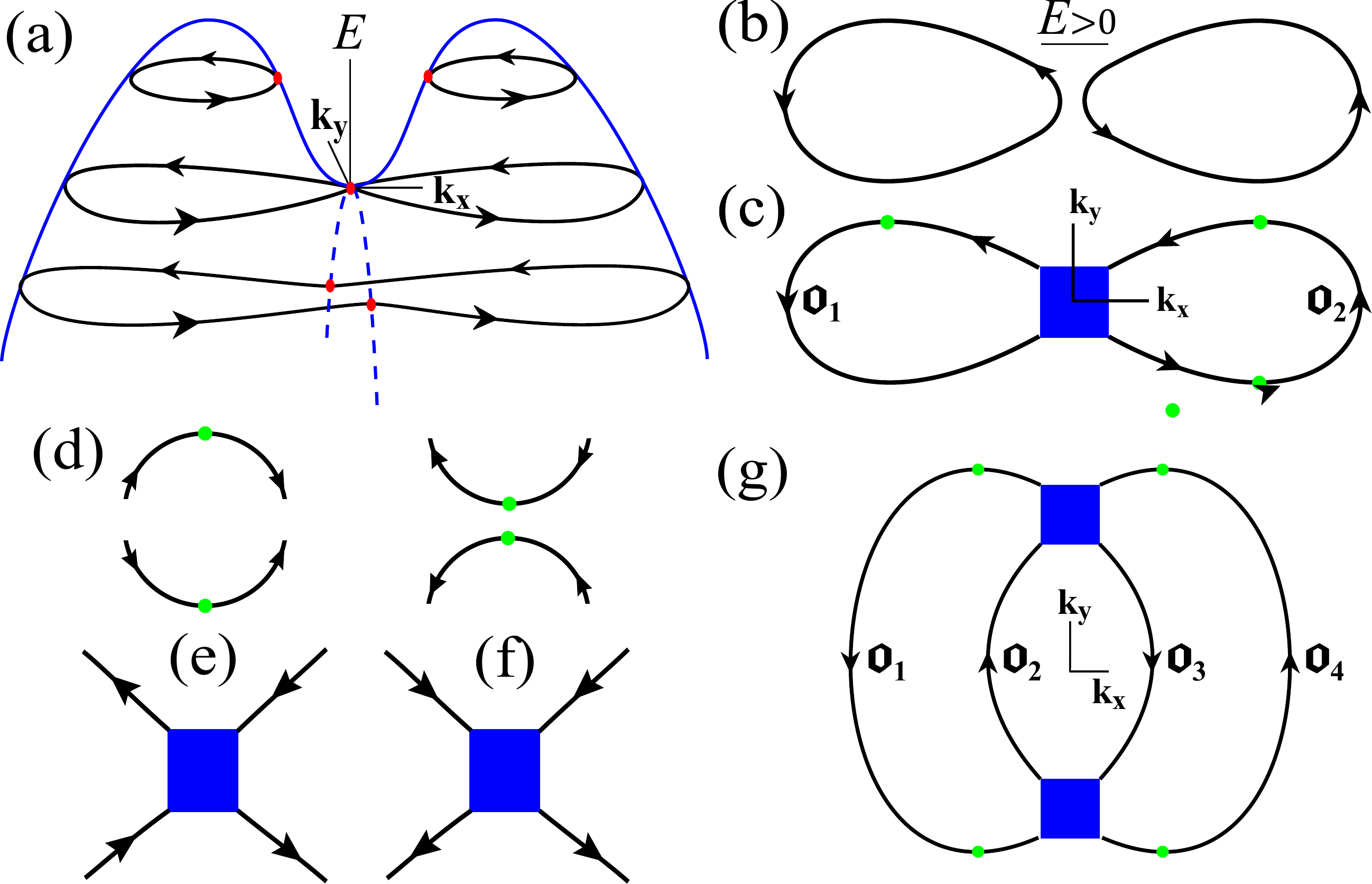}
\caption{(a) An example of a zero-field band dispersion with a saddlepoint. (b) A constant-energy band contour of (a). (c) The graph interpretation of (b); we refer to this as the `double-well' graph. (d) Turning points are degree-two vertices. (e) Intraband-breakdown vertex. (f) Interband-breakdown vertex. (g) The `butterfly' graph.}\label{fig:graphs}
\end{figure}

Any zero-field bandstructure [as exemplified in \fig{fig:graphs}(a)], when considered at fixed energy (and fixed $k_z$ for 3D solids) [see \fig{fig:graphs}(b)], may be represented as a directed graph [\fig{fig:graphs}(c)]. A directed graph is composed of directed edges and vertices. A directed edge is a continuous line with an orientation -- its beginning and end points are referred to as vertices. In our context, a directed edge is a section [black, arrowed trajectory in \fig{fig:graphs}(c)] of a constant-energy band contour, and the vertices are either turning points [green dots], where the $y$-component of the wavepacket velocity vanishes,  or breakdown regions [blue squares] where quantum tunneling between orbits is signficant. \\

Each vertex is associated with a degree, which is the number of edges connected to the vertex: a turning vertex has degree two, and a breakdown vertex has degree four. Note that a breakdown region typically has dimension of order $1/l$, but its assumed smallness compared to the size of a typical orbit justifies our use of the term `breakdown vertex'. From the orientation of the edges connected to the vertex, we might describe a turning vertex as one-in-one-out, and the breakdown vertex as two-in-two-out. It is useful to assign an orientation to each turning vertex, which might be clockwise or anticlockwise [\fig{fig:graphs}(d)] -- this determines the phase ($-i$ and $+i$ respectively) acquired by a wavepacket as it turns, as we elaborate in \s{sec:turningpoint}. It is also useful to assign an orientation to distinguish two classes of breakdown vertices: in the case of the intraband-breakdown vertex [\fig{fig:graphs}(e)], where tunneling occurs between orbits in the same band, the two incoming edges are parallel and lie on the same diagonal; for the interband-breakdown vertex [\fig{fig:graphs}(f)], where tunneling occurs between different bands,  the incoming edges lie on distinct diagonals. Intraband and interband breakdown is described respectively in \s{sec:intraband} and \s{sec:interband}. For exemplification, \fig{fig:graphs}(c) illustrates a `double-well' graph composed of six edges, four turning vertices, and a single intraband-breakdown vertex. \fig{fig:graphs}(g) illustrates a `butterfly' graph composed of eight edges, four turning vertices, and two intraband-breakdown vertices. The quantization conditions for these two graphs will be studied in \s{sec:doublewell} and \ref{sec:tci} respectively. \\

Two further comments regard the type of directed graphs that are relevant to the Bloch electron in a magnetic field. Firstly, we are generally interested in directed multigraphs, which means that we allow for two vertices to be connected by more than one edge, e.g., the two breakdown vertices in the butterfly graph are connected by two edges lying in the middle of the graph.  We also insist that our graphs are two-toroidal, by which we mean it may be drawn/embedded on a two-torus (here, the Brillouin torus) such that no edges cross. \\


The notion of connected components is intimately related to quantum tunneling. A connected component is maximal connected subgraph -- each vertex and edge belongs to exactly one connected component, and any two vertices in a connected component can be linked by a path. The appropriate description of orbits which are not linked by tunneling is a disconnected graph with multiple connected components; when the minimal separation in $\bk$-space between two neighboring components is of order $1/l$, it becomes appropriate to connect the two components by an intraband- or interband-breakdown vertex.  In the presence of tunneling, we may define a broken orbit in the following way: it is an oriented subgraph, composed only of directed edges and turning vertices, that forms a continuous path beginning and ending at a breakdown vertex. The beginning and ending vertex may be identical (as for the double well, which is composed of two broken orbits linked by a single breakdown vertex), or distinct (as for the butterfly graph, which is composed of four broken orbits linked by two vertices). The two-in-two-out rule for each breakdown vertex implies there are always two broken orbits which shoot out from the vertex, and another two broken orbits which terminate at the same vertex. The orientation of a broken orbit is determined from Hamilton's equation, and it generally comprises $N_s$ number of edges  and $(N_s-1)$ number of turning vertices, with $N_s \geq 1$; if an edge $\nu$ [resp.\ turning vertex $p$] belongs to a broken orbit $\frako_i$, we denote this by $\nu \in \frako_i$ [resp.\ $p \in \frako_i$].\\

For our  formulation of a topological invariant in the quantization condition, it is useful to formulate a class of homotopically-equivalent graphs. Two equivalent graphs may be continuously deformed into each other, given three rules for what is meant by `continuous': (i) one can neither break apart a connected component, nor merge two connected components into one, (ii) a breakdown vertex is movable in $\bk$-space, but unremovable from the graph, and (iii) the total number of turning vertices is not invariant, as explained in \s{sec:turningpoint}, however the net circulation of all turning vertices on a connected path is invariant.        \\

\noindent \emph{Definition} A topological invariant in magnetic transport is a quantity that is invariant under continuous deformations of the zero-field Hamiltonian that preserve the homotopy class of the graph.  A symmetry-protected topological invariant in magnetic transport is a quantity that is invariant under continuous deformations of the zero-field Hamiltonian that: (i) preserve the homotopy class of the graph, as well as (ii) respects the symmetry of the zero-field Hamiltonian.

\section{Review of effective Hamiltonian in the absence of interband breakdown } \la{sec:revieweffham}

We are interested in semiclassical approximations to the exact Hamiltonians of a Bloch electron in a magnetic field, as shown in \q{definefieldonhamiltonian} and \q{definefieldonpaulihamiltonian}; such approximations will be referred to as effective Hamiltonians. In this section, we particularize to cases where interband breakdown is negligible; the effective Hamiltonian that is valid near an interband degeneracy takes a different form that is described in \s{sec:interband}. \\

In the presence of a field along $\vec{z}$, $\bk^{\sma{\perp}}=(k_x,k_y)$ is no longer a conserved quantity for the Bloch electron. In the lowest-order approximation, the effective Hamiltonian is obtained by the Peierls substitution:\cite{peierls_substitution}  
\e{ H_0(\bk) \as \longleftrightarrow \as H_0(\bK),}
which describes the unique, Weyl correspondence between a function of commuting variables $(k_x,k_y)$, and a function of noncommuting variables:
\e{ \bK  =\bk+\ba(i\nabk).\la{defineK}}
We refer to $\bK$ as the kinetic quasimomentum operators, and their noncommutivity is manifest in 
\e{ \bK \times \bK = -i\f{e}{\hbar c}\bB; \as e<0.}
Generally, a one-to-one correspondence exists between a classical `symbol' ($A(\bk)$) and an operator ($A(\bK)$), if $A(\bk)$ is a Fourier-invertible function of commuting variables, with the Fourier transform $\check{A}(\br)$:
\e{A(\bk) = \int d\br e^{i\bk \cdot \br} \check{A}(\br). \la{fourierinversion}}
Many of the functions we deal with, including the matrix $H_0(\bk)$, are periodic in reciprocal-lattice translations: $\bk \rightarrow \bk+\bG$, in which case \q{fourierinversion} particularizes to a Fourier-series expansion. The operator to which the symbol corresponds is
\e{A(\bK):=[A(\bk)]:=  \int d\br e^{i\bK \cdot \br} \check{A}(\br). \la{replace}}
To make this definition rigorous, one assumes certain regularity conditions on $A$, and checks that the integral converges in some suitable sense.\cite{Tao} The lowest-order effective Hamiltonian $H_0(\bK)$ was first derived in a tight-binding approximation;\cite{luttinger_peierlssub} its form may be argued from general principles of electromagnetic gauge invariance.\cite{lifshitz_pitaevskii_statphys2} However, we cannot appeal to gauge invariance to predict the form of higher-order corrections, which may be organized in an asymptotic\cite{Zak_exactsymmetry,fischbeck_review} expansion:
\e{\calh(\bK) = H_0(\bK)+H_1(\bK)+H_2(\bK)+\ldots, \la{effhamasymptoticexpansion}} 
where each term in the expansion corresponds to the symbol $H_j(\bk)=O(l^{-2j})$. By $O(l^{-2j})$, we mean that $H_j$ is of the order $(a/l)^{2j}$, where $a$ is a typical lattice period. $H_j$ is obtained systematically\cite{rothI} by expanding an eigenstate of $\hat{H}$, defined by
\e{ (\hH -E)\Psi_E =0,}
in a complete\cite{MISRA_completenessofroth,nenciu_review} basis of field-modified Bloch functions
\e{ \Psi_E(\br) = \sum_{n\bk}g_{n\bk E}\phi_{n\bk}(\br), \la{modifiedblochfunctions}}
such that, in an asymptotic sense,
\e{  \sum_n(\calh(\bK)_{mn}-E\delta_{mn}) g_{n\bk E}=0. \la{eigenvalueeq}}
Note that by $\sum_{\bk}$ we really mean a continuous integral over the Brillouin torus. There are several different proposals for the best basis functions to formulate an effective Hamiltonian,\cite{kohn_effham,wannier_fredkin,rothI,blount_effham,nenciu_review}
but all these proposals agree\cite{ZAK_questionmagneticenergybands} to lowest order in $\lmt$:\cite{harper_magneticgaugetransformations,luttinger_peierlssub,zilberman_wkb} 
\e{ \phi_{n\bk}(\br) = \eikr u_{n,\bk+\ba(\br)}(\br).\la{zerothrothfunction}}
This form of $\phi$ manifests the semiclassical intuition that for a slowly-varying vector potential, the ordinary Bloch function is modified locally in space, but only through the wavevector dependence of the cell-periodic component: $u_{n\bk}\rightarrow u_{n,\bk+\ba},$ which is no longer periodic in Bravais-lattice translations; we provide a further argument that motivates the form of $\phi$ in App.\ \ref{app:fieldmodBlochfunc}.
The Fourier transform of $\phi$ is a real-space function that may be obtained from applying the magnetic translation\cite{Brown_magnetictranslation} to a  Wannier function;  in this real-space basis, the effective-Hamiltonian equation is a finite-difference equation for a wavefunction defined on a  lattice,\cite{chambers_breakdown} as famously exemplified by the Harper equation.\cite{hofstadter_butterfly}  \\

The value of \q{eigenvalueeq} is that, in many cases of interest, the matrix elements between  a single band and its complement (i.e. all other bands) have been removed perturbatively in the parameter $\lmt$; this decoupling of bands is asymptotic and fails if inter-band gaps become too small.\cite{fischbeck_review} Assuming otherwise, we may truncate $\sum_n$  and solve for $(\calh(\bK)_{nn}-E) g_{n\bk E}=0$; in this sense we say $\calh(\bK)_{nn}$ is a one-band effective Hamiltonian. The Weyl correspondence thus provides the link between the magnetic problem [given by $\calh(\bK)_{nn}$] and band properties at zero field: $[H_0]_{nn}(\bk)$ describes the dispersion of a single band, which is renormalized\cite{Wilkinson_semiclassicallimits,wilkinson_exactrenormalisation,wilkinson_generalizedwannierfunction} by the higher-order $\{H_j(\bk)\}$ as we eliminate degrees of freedom in the other bands. In other applications of \q{eigenvalueeq}, we may utilize a multi-band effective Hamiltonian to describe a degenerate band subspace. \\


The effective-Hamiltonian theory has been 
rigorously justified for an energetically-isolated nondegenerate band;\cite{nenciu_review} the justification for a finite family of (possibly magnetic) bands was achieved only recently.\cite{freund_peierlsmagnetic} While these impressive works go a long way in solving `one of the few unsolved problems of one particle quantum mechanics',\cite{Schellnhuber} they  rely on the assumption of a strictly isolated band (or family of bands), i.e., that there exists a direct energy gap above and below the band(s) in question. The complicated nature of bands in naturally-occuring crystals often implies indirect gaps are as common as direct gaps, except in the extreme tight-binding limit; this problem is especially severe in highly-symmetrical crystals with many band touchings, of both the immovable\cite{Bradlyn_newfermions} and movable kinds.\cite{connectivityMichelZak} We further highlight a class of metallic systems where the nonexistence of a gap is guaranteed from topological principles -- the surface states of certain topological insulators robustly interpolate between conduction and valence bands;\cite{kane2005A,Hsieh_SnTe,Hourglass} this phenomenon of spectral flow is familiar from the integer quantum Hall effect.\\ 

From physical grounds, one may expect that the effective Hamiltonian is valid for bands which are not energetically isolated over the Brillouin torus. That is, the existence of well-defined semiclassical orbits (at some energy $E$) at least validates $\calh(\bK)$ in a local neighborhood of the orbit and at that energy $E$. Let us exemplify our perspective for graphene in a magnetic field -- we would apply a single-band $\calh$ at wavevectors sufficiently far (on the scale of $l$) from the Dirac point, even though it is impossible to symmetrically separate graphene's two bands (not counting spin). This impossibility is enforced by symmetry, i.e., the $p_z$ bands of graphene form  an elementary band representation\cite{Zak_bandrepresentations,Evarestov_bandrepresentations,Bacry_bandrepresentations,TQC} with two branches. Generally, the existence of semiclassical orbits leads to discrete magnetic energy levels (henceforth referred to as Landau levels) which may be determined by Bohr-Sommerfeld quantization rules;\cite{onsager,lifshitz_kosevich,lifshitz_kosevich_jetp,zilberman_wkb,Wilkinson_semiclassicallimits,fischbeck_review} this method has been verified numerically for simple models,\cite{Butler_modelmagneticband} and is predictive\cite{topofermiology} of de Haas-van Alphen oscillations in metals.\cite{dHvA,SdH} We further substantiate our perspective  for the single-band effective Hamiltonian in \s{sec:singlebandeffham}, and for the multi-band case in \s{sec:multibandeffham}.

\subsection{Single-band effective Hamiltonian}\la{sec:singlebandeffham}

 The applicability of the single-band $\calh_{nn}(\bK)$   generically depends on the wavevector and energy in question, and is contingent on the cell-periodic functions (in this one-band subspace) being smooth enough.  By smooth enough, we mean that $\partial_ku=O(a)$ with $a$ a typical lattice period; there may be isolated regions in $\bk$-space where such smoothness cannot hold, e.g., where two bands touch at a conical degeneracy (a Dirac point), the Berry connection (with $\nabk$ in the azimuthal direction) diverges.\cite{Zak_diracpoint} Even so, we may apply $\calh_{nn}(\bK)$ at wavevectors sufficiently far (on the scale of $\lmo$) from the conical degeneracy.  For notational convenience, we henceforth drop the subscript $\calh(\bK)_{nn}\rightarrow \calh(\bK)$. These are common scenarios in which we might consider a single-band effective Hamiltonian:\\

\noi{i} Low-symmetry wavevectors where all bands are  nondegenerate, e.g., a generic wavevector for a spin-orbit-coupled system without spacetime-inversion symmetry. By spacetime inversion, we mean a simultaneous inversion of both space and time (denoted $T\inv$ in later sections), which is known to result in spin-degenerate bands. \\

\noi{ii} Spin-degenerate bands in the absence of spin-orbit and Zeeman couplings. Since $S_z$ (spin component in the $\vec{z}$ direction) is conserved, electron dynamics in a field is effectively constrained within a single band. \\

Explicit expressions for the single-band effective Hamiltonian have been derived up to $H_2=O(l^{-4})$;\cite{rothI} in this work, we derive the quantization conditions for the truncated $H_0+H_1$, where 
\e{H_0(\bk)= \var_{n\bk}, \la{Peierlsonsageroneband}}
is the energy-momentum dispersion of a band (labelled by an integer $n$), and $H_1$ may be split into gauge-dependent ($H_1^B$) and -independent ($H_1^R$, $H_1^Z$) terms as\cite{rothI,Panati_effectivedynamics}
\e{ H_1(\bk) \eq H_1^B(\bk)+H_1^R(\bk)+H_1^Z(\bk), \la{H1}\\ 
H_1^B(\bk) \eq    \lmt \epsilon^{\ab} \mx^{\beta} v_n^{\alpha}, \la{H1berry}\\
H_1^R(\bk) \eq   \tf1{2l^2} \epsilon^{\ab} \sum_{m}{\tmx}_{nm}^{\beta}(\tPi^{\alpha}-\tv^{\alpha})_{mn}, \la{H1orbmag}\\
H_1^Z(\bk) \eq  - \f{g_0 \hbar^2}{4ml^2} \sigma^z.  \la{H1zeeman}
 }
Here, band indices $m$ and $n$ are not summed over unless explicitly stated, $\epsilon^{\ab}$ is the Levi-Cevita tensor with $\epsilon^{xy}=1=-\epsilon^{yx}$, $\bv_n := \nabk \var_n$,  $\tbPi$, $\tbmx$, $\tbv$, $\bv$, $\sigma^z$ and $\bmx$ are $\bk$-dependent matrices defined in \q{definevelocitymatrix}, \q{defineberryconnection}, \q{definediagonalvelocitymatrix} and \q{restriction}, respectively; in particular, $\bmx=\tbmx_{nn}$ in this context. By gauge dependency, we refer to a phase ambiguity in the cell-periodic functions of band $n$ [cf. \q{basistransf} with $D=1$], and hence also of the field-modified Bloch functions which form our basis; cf.\ \q{modifiedblochfunctions}]; this results in $H_1^B$ being not uniquely defined:
\e{ &\ket{u_{n\bk}}\rightarrow \ket{u_{n\bk}}e^{i\phi_n(\bk)}\lin
 &\imp H_1^B \rightarrow H_1^B+\lmt \epsilon^{\ab} \partial_{\alpha}\phi_n v_n^{\beta}. \la{h1bnotunique}}
We will shortly demonstrate that the quantization condition is nevertheless gauge-invariant. If a symmetry (e.g., $T\inv$) constrains the Berry curvature to vanish, a basis may be found such that the Berry connection $\bmx_{nn}(\bk)$ (hence also $H_1^B$) vanishes at any $\bk$; this basis may be continuously defined over the Brillouin torus unless there is a topological obstruction, which may originate from a Dirac point in 2D,  or a line node in 3D.\cite{mikitik_berryinmetal}\\

On the other hand, $H_1^R$ may be expressed in a manner that manifests its gauge invariance:
\e{ \sum_m\tmx^{\alpha}_{nm}(\tPi^{\beta}-\tv^{\beta})_{mn}\eq i\braopket{u_{n\bk}}{(\partial_{\alpha}Q)\hPi^{\beta}}{u_{n\bk}}, \la{expressH1Osingleband}}
where $Q$ is the gauge-invariant, cell-periodic projections defined in \q{defineprojQ}, for $D=1$. As we will demonstrate in  \s{sec:wkbwavefunction}, the WKB wavefunction of $\calh=H_0+H_1$ includes multiplicatively a geometric Berry phase factor\cite{berry1984} that originates from the gauge-dependent $H_1^B$ (henceforth called the Berry term), and a non-geometric phase factor that originates from the gauge-independent $H_1^R$. We interpret $H_1^R$ as a coupling ($-\bM_n\cdot \bB$) of the field to the band orbital moment, defined for band $n$ as
\e{ M(\bk)_n^{\alpha} \eq  -\f{|e|}{2\hbar c}\epsilon^{\ab \gamma}\sum_{\bar{m}} \tmx_{n{\bar{m}}}^{\beta}(\tPi^{\gamma}-\tv^{\gamma})_{{\bar{m}}n} \lin
\eq  i\f{|e|}{2\hbar c}\epsilon^{\ab \gamma}\sum_{\bar{m}}\f{\tPi^{\beta}_{n{\bar{m}}}\tPi^{\gamma}_{{\bar{m}}n}}{(\var_n-\var_{\bar{m}})} \lin
\eq -i\f{|e|}{2\hbar c}\epsilon^{\ab \gamma} \braopket{\partial_{\beta}u_n}{\hH_0(\bk)-\var_{n\bk}}{\partial_{\gamma}u_n}.
\la{orbitalmagnetization29}}
Here, we have used $\bar{m}$ to label bands which are orthogonally complement to band $n$ [cf.\ \q{defineprojQ}].  These equivalent expressions for the \omm in \qq{expressH1Osingleband}{orbitalmagnetization29} are derived in \app{sec:equivalentexpressionsorbitalmag}. \\

An expression identical to \q{orbitalmagnetization29} appears in the correction to the energy of a wavepacket in a Bloch band.\cite{chang_niu_hyperorbit,sundaram1999,chang2008} We, however, disagree with a claim in Ref.\ \onlinecite{chang_niu_hyperorbit} that the orbital moment is absent in non-magnetic Bloch bands (i.e., eigenstates of $\hH_0$ without spontaneous time-reversal-symmetry breaking); we substantiate this point by a comprehensive symmetry analysis of $H_1$ in \s{sec:symmetry}. The derivation of the semiclassical equations of motion, as corrected by $H_1$, was accomplished in Ref.\ \onlinecite{Panati_effectivedynamics}. $H^R_1$ is sometimes referred to as the Rammal-Wilkinson term, and has been alternatively derived from a purely algebraic approach,\cite{rammal_algebraicapp} as well as in a semiclassical treatment of the Harper-Hofstadter model.\cite{Wilkinson_semiclassicallimits,Gat_semiclassicalHofstadter} Finally, we remark that terms analogous to $H_1^R$ and $H_1^B$ appear ubiquitously in the asymptotic theory of coupled-wave equations (i.e., multi-component WKB theory)\cite{littlejohn_short,littlejohn_long,Emmrich_geometry}  as well as in space-adiabatic perturbation theory,\cite{panati_spaceadiabatic} which apply in a much wider variety of physical contexts than the present study.

\subsection{Multi-band effective Hamiltonian}\la{sec:multibandeffham}

Consider a multi-band effective Hamiltonian that describes a $D$-fold degenerate band subspace projected by $P$ [cf. \q{defineprojP}]. A common example of $D=2$ arises in spin-orbit-coupled  solids with spacetime-inversion symmetry --  bands are spin-degenerate at generic wavevectors, and dynamics in a magnetic field is described by a two-band effective Hamiltonian  [$\calh(\bK)$]. $\calh(\bK)$ loses its applicability near (on the scale of $1/l$) four-fold-degenerate band touchings which might occur in various contexts, e.g.: (i) a 3D Dirac point, which is the critical point of a topological phase transition between trivial and topological insulators,\cite{Murakami2007B} or (ii) a symmetry-protected degeneracy that can be found in nonsymmorphic space groups.\cite{Young_diracsemimetal2D,Bradlyn_newfermions}  \\

For any $D$, the multi-band generalization of \q{H1} is\cite{rothI}
\e{H_1 \eq  {l^{-2}\epsilon_{\ab}}\left[ \f1{2}\{\tPi^{\alpha}-\tv^{\alpha},{\tmx^{\beta}}\} + \mx^{\beta}v^{\alpha}\right] -\f{\gfac \hbar^2}{4ml^2} \sigma^z \lin
\eq H_1^R+H_1^B+H_1^Z. \la{H1multiband2}}
where $\{a,b\}=ab+ba$, and we consider only matrix elements of $H_1$ within the $P$ subspace.  $H_1^B\propto \mx^{\alpha}v^{\beta}$ is just the product of two $D\times D$ matrices; in contrast, since $\bv$ is the diagonal component of $\bPi$ [cf.\ \ref{definevelocitymatrix}], the first term in \q{H1multiband2} involves \emph{only} matrix summations between $P$ and $Q$ subspaces:
\e{ [(\tPi^{\beta}-\tv^{\beta}){\tmx^{\alpha}}]_{mn} =\sum_{\bar{l}}[\tPi^{\beta}-\tv^{\beta}]_{m\bar{l}}\tmx^{\alpha}_{\bar{l}n}.}
While $H_1^Z$ has the advantage of looking more symmetric with respect to $\alpha$ and $\beta$, the following alternative expressions reveal a closer resemblance to the one-band $H_1^Z$ in \q{H1}: 
\e{\f{\epsilon_{\ab}}{2}\{\tPi^{\beta}-\tv^{\beta},{\tmx^{\alpha}}\}\eq  \epsilon_{\ab}\tmx^{\alpha}(\tPi^{\beta}-\tv^{\beta})\lin
\eq  -\epsilon_{\ab}(\tPi^{\alpha}-\tv^{\alpha})\tmx^{\beta}.\la{degorbitalmagmatrixelement}}	
The multi-band \omm, defined by $H_1^R=-\bM \cdot \bB$, therefore has a very similar form to \q{orbitalmagnetization29}:
\e{ M(\bk)_{mn}^{\alpha} \eq  -\f{|e|}{2\hbar c}\epsilon^{\ab \gamma}\sum_{\bar{l}} \tmx_{m\bar{l}}^{\beta}(\tPi^{\gamma}-\tv^{\gamma})_{\bar{l}n} \lin
\eq  i\f{|e|}{2\hbar c}\epsilon^{\ab \gamma}\sum_{\bar{l}}\f{\tPi^{\beta}_{m\bar{l}}\tPi^{\gamma}_{\bar{l}n}}{(\var_m-\var_{\bar{l}})}.\la{definemultibandom}}
We stress that the multi-band orbital moment nontrivially affects the Landau levels, which motivates a comprehensive symmetry analysis of $H_1^R$ in \s{sec:symmetrysinglebandom}. Through $H_1^B$, the energy levels are also sensitive to the non-abelian gauge structure in the subspace $P$, as we will demonstrate in the next section [\s{sec:simple}].



\section{Quantization conditions for closed orbits without breakdown}\la{sec:simple}

As motivated in the last paragraphs of \s{sec:revieweffham}, we are interested in determining Landau levels from Bohr-Sommerfeld quantization rules. In this section we derive the rules for \emph{closed} orbits, by which we mean orbits that do not extend beyond one unit cell in $\bk$-space. These clearly do not exhaust all possible orbits,\cite{kaganov_coherentmagneticbreakdown} but they are sufficient to exemplify the results of this work; we will briefly remark on generalizations beyond closed orbits in \s{sec:singlebandquant}. We further particularize to isolated orbits whose closest distance to any other orbit (if they exist) is much greater than $1/l$, with $l$ the magnetic length. If this condition is violated, tunneling between orbits must be accounted for; generalized quantization conditions that incorporate tunneling are presented in \s{sec:intraband} and \s{sec:interband}. \\     

Let us first summarize our results, which we will derive in the subsequent subsections. For a closed  orbit ($\frako$) corresponding to a nondegenerate band (labelled $n$) of the Pauli Hamiltonian [\q{fieldfreepauli}], the Bohr-Sommerfeld quantization rule is
\e{ & l^2 S[\frako]+\phi_M+ \oint_{{\frako}}({\bmx}+\orb)\cdot d\bk \lin 
&+Z\oint_\frako \sigma^z \f{dk}{v^{\sma{\perp}}}\bigg|_{E=E_j,k_z} 
=  2\pi j +O(\lmt), \as j\in \Z.\la{rule3a}}
The left-hand side of \q{rule3a} comprises five terms which we define in their order of appearance:\\

\noi{i} The first term is a dynamical phase that is proportional to the $\bkp$-space area $S$ bounded by ${\frako}_j$, with  $S$ being positive (resp.\ negative) for a clockwise-oriented (resp.\ anticlockwise) orbit. \\

\noi{ii} The second term is a Maslov phase,\cite{keller1958} e.g., $\phi_M=\pi$ for orbits which are deformable to a circle, and equals $0$ for figure-of-eight orbits, as elaborated in \s{sec:turningpoint}. To leading order in the field, $l^2S+\phi_M=2\pi(j+1/2)$ is a  well-known result by Onsager and Lifshitz.\cite{onsager,lifshitz_kosevich,lifshitz_kosevich_jetp} The Landau-level degeneracy ($\caln$) may be obtained from the following semiclassical phase-space argument:\cite{pathria} $\caln$ equals the phase-space density of states [$(2\pi)^{-2}$ for two spatial dimensions, which we assume in this paragraph for simplicity], multiplied by the phase-space volume ($\delta \calv$) in between two constant-energy hypersurfaces; these hypersurfaces corresponding to nearest-neighbor Landau levels indexed by adjacent integers in the quantization rule, hence 
\e{ \caln=\f{\delta \calv}{(2\pi)^2}=\f{\bar{A} \delta S}{2\pi^2}=\f{\bar{A}}{2\pi l^2},}
with $\bar{A}$ the real-space area of the 2D solid. This degeneracy simply reflects that a single-particle state undergoing localized cyclotron motion occupies an average area of $2\pi l^2$ in the semiclassical limit.  \\

\noi{iii} Beyond Onsager-Lifshitz, $\oint \bmx$  is the single-band Berry phase acquired over a single cyclotron period [$\bmx=\tbmx_{nn}$ as defined in \q{defineberryconnection}]. We might utilize Stoke's theorem to combine terms (i) and (iii) as 
\e{ &l^2S[\frako]+ \oint_{\sma{\frako}}{\bmx}\cdot d\bk = l^2\tilde{S}, \lin
& \tilde{S}:=\int |d^2\bk| \big(1 - \lmt \calf^z(\bk) \big),\la{phasespace}}
 with the Berry curvature defined by $\calf^z=\epsilon_{\ab}\nabk^{\alpha}  \mx^{\beta}$. One may therefore interpret $(2\pi)^{-2}(1 - \lmt \calf^z(\bk))$ as the Berry-corrected, phase-space density of states for a 2D solid immersed in a spatially-homogeneous field, i.e., a single-particle state occupies a volume in phase space that is modified by the coupling of the magnetic field to the Berry curvature. This correction to the phase-space density of states may alternatively be derived from a different route: through the semiclassical equations of motion,\cite{dixiao_berryphasecorrection,Duval_comment,duval_berryphasecorrection} which ultimately also derives from the effective-Hamiltonian formalism.\cite{zak_dynamicsofelectrons,lifshitz_pitaevskii_statphys2}\\

\noi{iv} The fourth term is the line integral of a one-form that encodes the orbital magnetic moment: 
\e{ \orb \cdot d\bk :=  \f{[\tmxx(\tPi^y-\tv^y)]_{nn}}{2{v}^y_n} dk_x+ (x \leftrightarrow y), \la{defineorbitaloneform2}}
where $\tmxx(\tPi^y-\tv^y)$ is formally the product of two infinite-dimensional matrices expressed in \q{expressH1Osingleband}. $\oint \orb \cdot d\bk$ is not a geometric phase because it depends on the rate at which the orbit is traversed [recall that the orbit velocity is related to the band velocity through \q{hamilton2}]. Analogous expressions of $\oint \orb\cdot d\bk$ have been called, in various contexts, the `no-name' phase,\cite{yabana_coupledchannel,littlejohn_short,littlejohn_long}, and sometimes the Ramal-Wilkinson phase. However, we will refer to it as the Roth phase to honor its first discoverer\cite{rothI,rothII} in the context of Bloch electrons in a magnetic field.\\

\noi{v} Finally, $Z\oint \sigma^z/v^{\sma{\perp}}dk$ is the Zeeman energy of the nondegenerate band  integrated over the orbit; the $\bk$-dependence of the Zeeman energy originates from spin-orbit coupling. Note $Z:=g_0\hbar/4m$, $dk=|d\bk|$, $v^{\sma{\perp}}:=(v_x^2+v_y^2)^{\sma{1/2}}$, $\sigma^z(\bk):=\sigma_{nn}^z(\bk)=\braopket{u_{n\bk}}{\hat{\sigma}^z}{u_{n\bk}} \in \R$. \\

\noindent All of (i-v) may be evaluated knowing the band structure at zero field; (i) and (ii-v)  depend continuously on the energy of the orbit $E$; in three spatial dimensions, they depend additionally on the wavevector ($k_z$) parallel to the field. \q{rule3a} leads to discrete, macroscopically-degenerate Landau levels labelled as $E_j$; more details about the spectrum, as well as consequences in dHvA oscillations, are described in \s{sec:quantumoscillations}. \\

 In the absence of spin-orbit coupling, a spin-degenerate band results in spin-split Landau levels obtained from the two quantization conditions:
\e{  &l^2 S[\frako]+\phi_M+ \oint_{{\frako}}({\bmx}+\orb)\cdot d\bk \lin
&\pm\pi \f{g_0}{2}\f{m_c[\frako]}{m}\bigg|_{E=E_{\pm,j},k_z} = 2\pi j +O(\lmt), \as j\in \Z,\la{singlequantrulewithzeeman}} 
 where $m_c{:=}(\hbar^2/2\pi)\partial S/\partial E$ is the cyclotron mass for the orbit $\frako$. Despite the notational similarity  of \q{rule3a} and \q{singlequantrulewithzeeman}, we remind the reader that the velocity matrix $\tbPi$ is defined differently when there is no spin-orbit coupling [cf. \q{definevelocity}]. In spite of the spin degeneracy of the bands at zero field, we analyze this case under the heading of `single-band' [e.g., in \s{sec:singlebandwkbwf}
] because the field-on Hamiltonian may be block-diagonalized with respect to the spin quantum number $S_z=\pm \hbar/2$; all `single-band' statements are then understood to apply to either of $S_z=\pm \hbar/2$, and all symmetries that we consider preserve $S_z$.\\


 \q{rule3a} and \q{singlequantrulewithzeeman} may be derived from the condition of continuity of the WKB wavefunction around the closed orbit.\cite{chambers_breakdown,fischbeck_review} This wavefunction is derived in \s{sec:wkbwavefunction}, where we also demonstrate that the Berry and Roth phases are respectively generated from $H_1^B$ and $H^R_1$. The additional phase of $\pi$ on the right-hand-side of \q{rule3a} [and also of \q{singlequantrulewithzeeman}] is a Maslov correction that we derive in \q{sec:turningpoint}; here, we argue that previous derivations\cite{chambers_breakdown,fischbeck_review} of the Maslov correction introduces an uncertainty of $O(l^{\sma{-2/3}})$, which we reduce to $O(\lmt)$ in an improved derivation. We combine these results in \s{sec:singlebandquant} to finally derive \q{rule3a}, and further discuss experimental signatures in quantum oscillations in \s{sec:quantumoscillations}.\\

For a closed orbit ($\frako$) corresponding to a $D$-fold degenerate band subspace, the quantization condition is
\e{ & l^2 S(E_{a,j},k_z)+ \phi_M+ \lambda_a(E_{a,j},k_z)\lin
\eq 2\pi j +O(l^{\sma{-2/3}}), \as j\in \Z, \;a \in \Z_D,\la{rule3b}}    
where $\{e^{i\lambda_a}\}_{a=1}^D$ is the spectrum of the unitary propagator
\e{\A[\frako] = \overline{\exp}\left[i\oint_{\sma{\frako}} \left\{(\orb+\bmx) \cdot d\bk +  Z(\sigma^z/v^{\sma{\perp}})\,dk \right\} \right]. \la{definenonabelianunitary}}
Here, $\overline{\exp}$ denotes a path-ordered exponential, and we employ the same symbol $\bmx$ for both the abelian [as in \q{rule3a}] and non-abelian [as in \q{rule3b}] Berry connection. The non-abelian generalization of the abelian Roth one-form [in \q{defineorbitaloneform2}] is 
\e{ (\orb \cdot d\bk)_{mn} =  \f{[\tmxx(\tPi^y-\tv^y)]_{mn}}{2{v}^y} dk_x+(x\leftrightarrow y),  \la{definerothoneform}}
with $m,n=1,2,\ldots, D$. Due to the assumed degeneracy within $P$, the band velocity $\bv_1=\ldots=\bv_D:=\bv$. \q{rule3b} leads to $D$ sets of Landau levels (labelled by the $a$ subscript on $\{E_{a,j}\}$). Landau levels within each set are locally periodic, i.e., the difference between two adjacent Landau levels ($E_{a,j+1}-E_{a,j}$) is approximately $2\pi/l^2(\partial S/\partial E)$ evaluated at $E_{a,j}$, as elaborated in \s{sec:quantumoscillations}. \\


The quantizations rule in \q{rule3b}-(\ref{definenonabelianunitary}) may be compared with previous works. For $T\inv$-symmetric, spin-orbit-coupled systems ($D=2$), two-band  quantization conditions have been derived\cite{Mikitik_quantizationrule} with an `equation-of-motion' method,\cite{rothII} which leads to formulating $\{e^{i\lambda_a}\}$ as eigenvalues of a complex Ricatti equation.\cite{Mikitik_quantizationrule} Their method presupposes a special basis for the Bloch functions (i.e., a special gauge)  in which the matrix exponent in \q{definenonabelianunitary} is traceless, as elaborated in \s{sec:AIIwithinversion}; note that $\{e^{i\lambda_a}\}$ are gauge-independent, so in principle their and our methods should converge to the same quantization rule for this symmetry class. In other formulations of the quantization rule for spin-degenerate bands, the Berry phase and/or orbital moment have either been neglected explicitly,\cite{proshin_breakdownwithspinflip} or derived in a form that is difficult for comparison.\cite{falkovskii,Gorbovitskii} On the other hand, \q{rule3b}-(\ref{definenonabelianunitary}) represents the quantization condition in its most general form, which would apply to any symmetry class, and to bands of any energy degeneracy ($D$). Since no special gauge was assumed in our expressions, they are useful for numerical computations where gauge fixing is often troublesome. One further contribution we make is a comprehensive, group-theoretic analysis of the propagator in \q{definenonabelianunitary}, which determines in complete generality the symmetry constraints on the Landau levels [see \s{sec:symmetryunitarygenbyH1}]. In a complementary perspective, multi-band wavepacket theory has been derived in \ocite{culcer_multibandwavepacket}, and reviewed in \ocite{chang2008} with notation that is closer to ours. They derived an equation of motion for a multi-component wavepacket that is also sensitive to the non-abelian gauge structure; however, their dynamical equations are nontrivially coupled, and it is unclear to us if a non-abelian quantization rule can be derived in their approach.\\



Our results may plausibly be applied to charge-neutral, cold-atomic systems (e.g., optical lattices of bosonic cold atoms, degenerate Fermi gases) described by the field-on Schrodinger Hamiltonian [\q{definefieldonhamiltonian}] with an artificially-induced gauge field. It is possible to mimic a magnetic field by: (i) rotation of a Bose-Einstein condensate,\cite{Schweikhard_rotateBECLandaulevel} (ii) coupling neutral fermionic atoms to slow light,\cite{Juzeliunas_neutralfermigasLandaulevels,JuzeliunasII} and (iii) by laser-assisted tunneling in an optical lattice.\cite{Jaksch_magneticfieldopticallattice} In the limit of weak interactions, this leads to the quantization of energy levels which are analogous to Landau levels.\cite{jacob_landaulevelscold}\\

We remark on one caveat to the above discussion: energy quantization of closed orbits is never strictly correct in a solid. While the magnetic field tends to quantize electronic motion and form discrete levels, the crystalline potential tends to form bands. From the perspective of semiclassical orbits in $\bk$-space, there generally exists a nonzero tunneling probability between closed orbits in distinct Brillouin zones. This leads to broadening of the Landau levels that cannot be accounted for with the above quantization rules. While this broadening is usually exponentially small in the field,\cite{blount_effham,kohnII} it cannot be neglected in narrow energy ranges where the separation of orbits is of order $1/l$; this frequently occurs at saddlepoints at the Brillouin-zone edge.\cite{azbel_energyspectrum,hsu_falicov}



\subsection{WKB wavefunction of effective Hamiltonians}\la{sec:wkbwavefunction}

\subsubsection{Single-band WKB wavefunction}\la{sec:singlebandwkbwf}

We look for an eigenfunction of the single-band $\calh(\bK)=H_0+H_1$ with the WKB ansatz
\e{ g_{\bk}=e^{-i\psi_{\bk}} \iwith \psi=\psi_{-1}+\psi_0+\psi_1 +\ldots, \la{assumefwkb}}
where the subscript denotes the order in the WKB parameter $\lmt$. Any function that is asymptotically expandable as \q{assumefwkb} will be called a WKB function. In the classically-allowed regions, $\psi_{-1} \in \R$ is the integral of a classical action, while higher-order $\psi_j \in \C$. In the Landau gauge $\bA=(By,0,0)$, the kinetic quasimomentum operators are
\e{ K_x=k_x+ il^{-2}\partial_{y},\as K_y=k_y,\la{defineKyrepresentation}}
and we would look for wavefunctions over the circle parametrized by $k_y$, with  $k_x$ a good quantum number. We shall refer to this as the wavefunction in the $(K_x,k_y)$ representation. In this representation, $\calh$ may be solved as
\e{ &(H_0(\bK)+H_1(\bK)-E)g^{\nu}_{\bk E}=O(\lmf), \la{eigenproblemsingleband}\\
&g^{\nu}_{\bk E}= \f1{\sqrt{|v^x_{\nu}|}}e^{ik_xk_yl^2}e^{-il^2\int \big(k_x^{\nu}-H^{\nu}_1(v^x_{\nu})^{\mo}\big)dk_y}. \la{zilbermanfischbeck}} 
Here, all quantities carrying a $\nu$ superscript or subscript depend on $k_y$ and $E$; they also depend, in three spatial dimensions, on the wavevector $k_z$, but we shall henceforth omit this notationally. As a case in point, $k_x^{\nu}(k_y,E)$  should be  distinguished from the continuous parameter $k_x$. $k_x^{\nu}$ describes an oriented \emph{edge} (labelled by ${\nu}$) of the zero-field band contour ($\frako$) at fixed energy $E$, with the orientation prescribed by Hamilton's equation; each $k_x^{\nu}$ corresponds to a single-valued solution of 
\e{ H_0\big( k_x^{\nu}(k_y,E),k_y \big)=E. \la{zerothorderkxnu}}
The constant-energy contour of a single band may be divided into multiple edges, e.g., a closed contour has at least two edges. A more elaborate, graph-theoretic description of edges is provided in \s{sec:graph}. For $s_{\nu}$ corresponding to a band index $n$, we further define 
\e{& v^x_{\nu}(k_y,E):= v_n^x(k_x^{\nu}(k_y,E),k_y)\lin
 &\ins{and} H^{\nu}_1(k_y,E):=H_1(k_x^{\nu}(k_y,E),k_y),\la{definenuquantities}}  
as the band velocity and the first-order Hamiltonian [cf.\ \q{H1}] evaluated on the edge $\nu$. The single-band WKB wavefunction of $H_0$ was first derived by Zilberman;\cite{zilberman_wkb} Fischbeck later derived the corrections due to  $H_1$ and $H_2$,\cite{fischbeck_review} of which we have shown only the first-order correction in \q{zilbermanfischbeck}. 
We will therefore refer to \q{zilbermanfischbeck} as the Zilberman-Fischbeck function; the same expression without the $H_1$ correction will be referred to as the Zilberman function. With sufficient hindsight, we may now identify the Roth, Berry and Zeeman phases as being generated, respectively, by $H^R_1$, and $H^B_1$ and $H^Z_1$:
\e{l^2\int H_1 \f{dk_y}{v^x} \eq l^2\int (H^R_1+H^B_1+H^Z_1)\f{dk_y}{v^x} \lin
\eq \int (\orb+\bmx)\cdot d\bk + \sigma^z \f{Z\,dk}{v^{\sma{\perp}}}; \la{rbzphase}}
this expression is understood to be evaluated on a certain edge. To derive the last equality, we combine the definitions in \q{H1} with the identity $0=v^x \,dk_x+v^y\,dk_y$ (which is valid on a constant-energy contour); Hamilton's equation in \q{hamilton2} is also useful in identifying $-dk_y/v^x = dk/v^{\sma{\perp}}$.  \\

Let us derive the single-band, Zilberman-Fischbeck wavefunction. This serves a pedagogical purpose, but also warms us up for the slightly more complicated derivation of the multi-band WKB wavefunction in \s{sec:multibandwkbwf}, which in its most general form has not been seen. \\


\noindent \emph{Proof of \q{zilbermanfischbeck}:} Applying the identity \q{letitactonidentity} [derived in \app{app:moyalidentities}] to \q{eigenproblemsingleband}, with the WKB ansatz $g=e^{-i\psi}$, we derive
\e{  E \eq H_0 + H_1+ l^{\minus 2}\left( \f{i}{2}\p{^2H_0}{k_x\partial k_y}+\psi_0'(k_y) \partial_xH_0\right) \lin
&+\; \f{i}{2}l^{\minus 4}\psi_{-1}''\partial^2_xH_0+O(l^{{-}4}), \la{Eequals}
}
where $H_j$ and its derivatives are evaluated at $(k_x+\psi_{-1}'/l^2,k_y)$, and $\psi_j'$ is the first derivative of $\psi_j$ with respect to $k_y$. A solution exists if $\psi_{-1}$ can be found that satisfies the zeroth-order relation:
\e{ E \eq H_0(k_x+\psi_{-1}'/l^2,k_y). \la{zerothordercondition}}
 For the purpose of deriving the quantization conditions, we will only need the WKB wavefunctions in the classically-allowed regions, where $\psi_{-1}' \in \R$. Generally, there might be multiple single-valued and real solutions, which we label with $\nu$ as $k_x+{\psi^{\nu}_{-1}}'/l^2:=k_x^{\nu}(k_y,E)$ [compare \q{zerothordercondition} with \q{zerothorderkxnu}]. This implies
\e{ \psi^{\nu}_{-1} = -l^2k_xk_y + l^2\int {k}^{\nu}_x(k_y,E)dk_y}
up to an irrelevant integration constant. Collecting the first-order terms in \q{Eequals}, and substituting the just-obtained expression for ${\psi^{\nu}_{-1}}''$,
\e{ 0\eq l^2H_1+\f{i}{2}\left(\p{^2H_0}{{k}_x\partial k_y} +\p{{k}^{\nu}_x}{k_y}\p{^2H_0}{{k}_x^2}  \right)\lin
&+{\psi^{\nu}_0}'(k_y) \p{H_0}{{k}_x} \bigg|_{\bk\rightarrow ({k}^{\nu}_x,k_y)}. \la{settozero}}
Let us separate $\psi_0=\psi_{0R}+i\psi_{0I}$ into real and imaginary parts. Setting the imaginary component of \q{settozero} to zero:
\e{ 0=\f{1}{2}  \p{{v}^x_{\nu}}{k_y}+{\psi^{\nu}_{0I}}' {v}^x_{\nu} \imp e^{\psi^{\nu}_{0I}} \propto \f1{\sqrt{|{v}^x_{\nu}|}}, \la{setimagtozero}}
with ${v}^x_{\nu}$ defined in \q{definenuquantities}. Setting the real component of \q{settozero} to zero:
\e{ &l^2H_1 +{\psi^{\nu}_{0R}}' \p{H_0}{{k}_x} \bigg|_{\bk\rightarrow ({k}^{\nu}_x,k_y)}=0 \lin
 &\imp \psi^{\nu}_{0R}=-l^2\int H^{\nu}_1(k_y,E)({v}^{\nu}_x)^{-1} dk_y, \la{psi0R}}
with $H^{\nu}_1$ defined in \q{definenuquantities}.


\subsubsection{Multi-band WKB wavefunction}\la{sec:multibandwkbwf}

Let us define the multi-band WKB wavefunction $\bff$ as the eigenfunction of
\e{   (H_0(\bK)+H_1(\bK)-E)\bff^{\nu}_{\bk E} =O(l^{-4}); \la{definemultibandwkbwf}}
matrix summation is implicit in this expression, and $\bff$ is a vector-valued function with as many components as the number ($D$) of bands  in the degenerate subspace $P$. We would like to demonstrate that
\e{ \bff^{\nu}_{\bk E}= \cala^{\nu}_{\bk E}\bff^{0\nu}_{\bk E}  \la{definegvector} }
with $\bff^0$ the product of an as-yet-undetermined, $k_y$-independent vector  $\boldsymbol{c}$ with  the Zilberman function:
\e{ f_a^{0\nu} =c_a^{\nu}\f1{\sqrt{|v^x_{\nu}|}}e^{ik_xk_yl^2}e^{-il^2\int k_x^{\nu}dk_y}; \as c_a^{\nu} \in \C,\la{defineg0nu}}
with $a=1,\ldots,D$, and $\cala$ a unitary propagator defined as the path-ordered exponential  
\e{\cala_{\bk E}^{\nu} =  \overline{\exp}\left[il^2\int H^{\nu}_1 (v^x_{\nu})^{-1} dk_y\right]. \la{definepropagator}}
Despite being a simple extension of the single-band wavefunction, we have not seen a multi-band ansatz for the Roth effective Hamiltonian in the literature. \\



\noindent \emph{Proof:} The assumed band degeneracy within $P$ implies $[H_0(\bk)]_{mn} =\delta_{mn}[H_0(\bk)]$, and therefore 
\e{(H_0(\bK)-E)\bff^{0\nu}=O(l^{-4}), \la{zerothbg0nu}}
 with $\bff^{0\nu}$ defined in \q{defineg0nu}, as a special case of \q{zilbermanfischbeck} with $H_1=0$. We propose the ansatz $\bg=\cala \bff^0$ with  
 $\A_{\bk}$ a $D\times D$ matrix that is differentiable with respect to $k_y$. Each matrix element $\A_{ab} \in \C$ is assumed to be of order one. The following identity is useful: 
\e{  &H_0(\bK) \A_{ab} = \sum_{\bR}\check{H}_0(\bR) e^{i\bk\cdot \bR-iR_xR_yl^2}\lin
&\times\bigg\{\A_{ab} -l^{-2}R_x\partial_y\A_{ab}+O(l^{-4})\bigg\}e^{-l^{-2}R_x\partial_y}, \la{toeavl2}}
which is derivable from \q{Hactsonorderone}.  Letting \q{toeavl2}, a matrix operator, act on the vector $\bff^{0\nu}$, we obtain
\e{  &\sum_{c}H_0(\bK) \A_{bc}f_c^{0\nu} = \sum_c \A_{bc} H_0(\bK)f^{0\nu}_c \lin
&+il^{-2}\sum_c\partial_y\A_{ac}v^x_{\nu}f^{0\nu}_c+O(l^{-4}),\la{toeavl3}}
with help from \q{uluright}. Here, we have introduced the band velocity $v^x_{\nu}$ on the edge labelled ${\nu}$. By similar manipulations with the $H_1$-term, we derive
\e{ \sum_{bc} H_1(\bK)_{ab} \A_{bc} f^{0\nu}_c 
\eq  \sum_{bc} ({H}^{\nu}_1)_{ab}\A_{bc} \bff^{0\nu}_c+O(l^{-4}). \la{ah1a}}
 Inserting \qq{toeavl3}{ah1a} into \q{definemultibandwkbwf}, the zeroth-order terms cancel owing to \q{zerothbg0nu}; after factoring out a common multiplicative factor (the Zilberman function), what remains is 
\e{  \sum_bil^{-2}\partial_y\A_{ab}v^x_{\nu} c^{\nu}_b+ \sum_{bd}({H}^{\nu}_1)_{ab}\A_{bd}c^{\nu}_d=O(l^{-4}).\notag}
We would like this equation to be true for arbitary $\bc^{\nu}$, hence we are led to a simplified differential equation
\e{&\partial_y\A  = il^2(v^x_{\nu})^{-1}{H}^{\nu}_1\A,}
which is solved by \q{definepropagator}. 




\subsection{Maslov correction from turning points}\la{sec:turningpoint}

\begin{figure}[ht]
\centering
\includegraphics[width=8 cm]{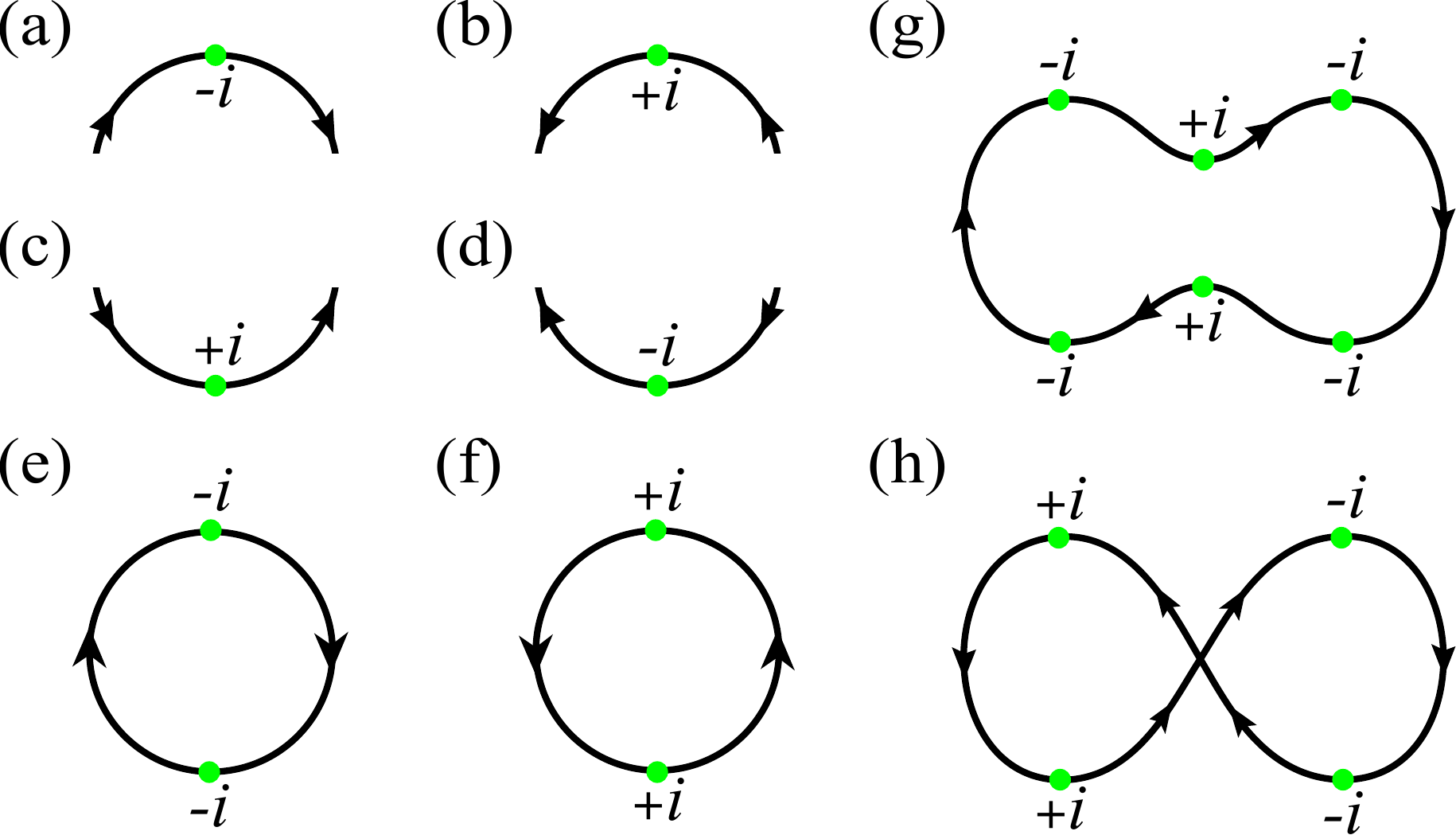}
\caption{Illustration of various turning points. Each turning point may be assigned a sense of circulation, which will be indicated by the sign of $\pm i$ next to each green dot. Each turning point may be divided into four classes illustrated by (a-d); these classes are distinguished by: (i) the sign in $k_y \sim \pm k_x^2$ [i.e., whether the band contour forms an upright ($\smile$) or inverted ($\frown$) parabola], as well as (ii) an orientation determined by the direction of a semiclassical wavepacket. If the band dispersion is expanded around each of the turning point as $\var_{\bk}=u_yk_y+k_x^2/2m_x$, then in (a) $u_y,m_x>0$, (b) $u_y,m_x<0$, (c) $u_y>,m_x<0$, and in (d) $u_y<0,m_x>0$. The correspondence, between the sign of $\pm i$ and the four classes (a-d) of turning points, is derived in \app{app:reviewturningpoint}.  (e-g) illustrate simple, closed orbits which are deformable to a circle, and (h) a nonsimple, closed orbit in the shape of a figure-of-eight.  \label{fig:turning_point}}
\end{figure}

For any closed orbit, as exemplified in \fig{fig:turning_point}(e-h), there are at least two turning points for which, in their vicinity, the Zilberman-Fischbeck (ZF) wavefunction loses its validity due to strong quantum fluctuations. In the graph-theoretic language introduced in \s{sec:graph}, a turning point is a  \emph{vertex} which is connected to two edges; alternatively stated, the beginning and end points of edges are vertices, and a turning point exemplifies a degree-two vertex. \\

It is well-known from the theory of caustics\cite{keller1958} that in passing around a turning point the WKB wavefunction \emph{effectively} picks up a phase $\phi_r$. For us, $\phi_r$ describes the phase difference between incoming and outgoing single-band ZF wavefunctions, which are valid sufficiently far from the turning point;\cite{berry_mount_review} `incoming' and `outgoing' are interpretive characterizations of different edges of the ZF wavefunction -- we may uniquely assign an orientation to each edge from Hamilton's equation[cf.\ \q{hamilton2}].  In analogy with a 1D Schrodinger particle reflecting off a wall, we might interpret the semiclassical wavepacket for a Bloch electron as being reflected in the coordinate $k_y$ -- we therefore refer to $\phi_r$ as a reflection phase.\\

For a turning point in the orbit of a single band, we determine that $\phi_r=\pm \pi/2 +O(\lmt)$, where the sign of $\pi/2$  is determined by the  sense of circulation when passing the turning point: plus for anticlockwise, and minus for clockwise. We should clarify that this orientation  is assigned locally to each turning point, and in a manner independent of the shape and orientation of the rest of the orbit. We may imagine minimally extending the parabolic contour at each turning point into a circle [e.g., $\curvearrowleft \rightarrow \circlearrowleft$, $\curvearrowright \rightarrow \circlearrowright$]; we then assign the orientation by interpreting the circle as a clock face.  \\

$\phi_r$ is derived by a divide-and-conquer approach -- we approximately describe the turning region with an effective Hamiltonian that is linear in one momentum component and quadratic in the other. What distinguishes our approach from previous works\cite{zilberman_wkb,fischbeck_review,chambers_breakdown} -- the effective Hamiltonian we adopt to describe the turning point is not just the Peierls-Onsager Hamiltonian, but includes the first-order correction by $H_1$. For simplicity, we consider a hard-wall boundary condition at the turning point, i.e., we ignoring tunneling between closed orbits. We then match the asymptotic wavefunction of this small-momentum, effective Hamiltonian with the incoming and reflected WKB functions; the proof is completed in \app{app:proofmaslov}.   \\

The multi-band analog of the  calculation in  \app{app:proofmaslov} is more involved. One may, however, avoid this calculation if one is willing to accept an uncertainty of $O(l^{\sma{-2/3}})$ in the quantization condition; this viewpoint seems to be implicitly adopted in past works,\cite{zilberman_wkb,fischbeck_review,chambers_breakdown,falkovskii,Gorbovitskii} though no attempt was made to quantify this uncertainty. To clarify, by exploiting the smallness of the turning regions relative to the rest of the semiclassical orbit, we might neglect the effect of $H_1$ in the turning regions, but account for it everywhere else on the orbit. In practice, this just means applying the zeroth-order Maslov correction ($\pi$ for a closed orbit) to a quantization condition which already includes first-order corrections through the propagator of \q{definepropagator}. Let us estimate the uncertainty in this approach. (i) If our asymptotic expansion of the quantization conditions is at all valid, we expect $H_j$ to make an $O(l^{2-2j})$ contribution to the quantization condition; in particular, $H_1$ makes an $O(1)$ contribution. (ii) The length of the orbit lying within the turning region is of $O(l^{\sma{-2/3}})$, as shown in \app{app:proofmaslov}; generically, the length of the semiclassical orbit is of the order of the reciprocal period. The ratio of the turning length to that of the entire orbit is then of $O(l^{\sma{-2/3}})$. Combining (i) and (ii), we expect that neglecting $H_1$ in the turning regions introduces an uncertainty of $O(l^{\sma{-2/3}})$. For the multi-band case, we therefore argue that the Maslov correction for a closed orbit is $\pi+O(l^{\sma{-2/3}})$, for each of the $D$ sets of sub-Landau levels [indexed by $a$ in \q{rule3b}].

\subsection{Quantization conditions} 

\subsubsection{Single-band quantization condition for closed orbits}\la{sec:singlebandquant}

Let us illustrate how to formulate the continuity condition  for the circular closed orbit, which is composed of two edges (labelled by $\nu=\pm$) which touch at two turning points, as illustrated for two orientations in \fig{fig:turning_point}(e-f). We shall focus on the orbit that circulates as $\circlearrowright$ in \fig{fig:turning_point}(e), which is expected of an electron pocket at the Fermi level. The quantization condition for an energy eigenstate at energy $E$ and wavevector $(k_x,k_z)$ is the continuity (with respect to $k_y$) of the wavefunction in the $(K_x,k_y)$-representation.\\


The continuity condition on $f$ may be formulated in a manner that emphasizes a semiclassical motion along the orbit; this motion is described by the time evolution of certain scalar amplitudes that we define for each edge:
\e{ a_{\nu,E}(t_{\nu}):=e^{-il^2\int \big(k_x^{\nu}-H^{\nu}_1(v^x_{\nu})^{\mo}\big)(dk_y/dt_{\nu}) dt_{\nu}}\bigg|_E. \la{definescalarampsection2}}
This amplitude is simply the phase component of the Zilberman-Fischbeck wavefunction $g^{\nu}_{\bk E}$[cf. \q{zilbermanfischbeck}], except without $e^{ik_xk_yl^2}$, which is trivially continuous over any closed orbit. The triviality of $e^{ik_xk_yl^2}$ would not be true of open orbits, as we will substantiate in \s{sec:beyondsimpleorbits}. 
We have parametrized each amplitude in \q{definescalarampsection} by a time-like variable $t_{\nu}\in [0,1]$, which increases along the orbit in a direction consistent with Hamilton's equation. The end points [$t_{\nu}=0$ and $1$] correspond to distinct turning points that bound the edge $\nu$. We may loosely interpret  $\bk(t_{\nu})$  as the wavevector of the `moving wavepacket' at time $t_{\nu}$; we caution the reader that there are no turning points in conventional wavepacket theory,\cite{ashcroft_mermin} but the language of a `moving wavepacket' offers a convenient and visually appealing metaphor for the continuity and patching of wavefunctions in WKB theory.\\

 With this caveat in mind, we would interpret  
\e{ e^{i\theta_{\nu}(E,l^2)}\eq \f{a_{\nu,E}(1)}{a_{\nu,E}(0)} \lin
:\eq e^{-il^2\int_0^1 \big(k_x^{\nu}-H^{\nu}_1(v^x_{\nu})^{\mo}\big)(dk_y/dt_{\nu}) dt_{\nu}}\bigg|_E \la{ratioofamplitude2} }
as the semiclassical phase acquired by a wavepacket as it traverses the edge $\nu$. As the wavepacket approaches a turning point along the edge $\nu'$, it is reflected onto a distinct edge $\nu$ and picks up an additional phase of $\pm \pi/2$; the sign depends on the sense of circulation of the turning point, as we have illustrated in \fig{fig:turning_point}(a-d). We implement this reflection phase in the boundary condition  
\e{ a_{\nu}(0)= e^{i\phi_r^{\nu'}} a_{\nu'}(1), \as e^{i\phi_r^{\nu'}}=\pm i. \la{bcturning}}
For any closed orbit, the set of equations in \q{ratioofamplitude2} and \q{bcturning} (for all $\nu,\nu'$), may be combined into a single equation that is parametrized by $E$ (and $k_z$ in 3D solids). For our case study of the simplest closed orbit $\circlearrowright$, this single equation may be expressed in a manner that emphasizes its motional interpretation:
\e{    1= (-i)e^{i\theta_-}(-i)e^{i\theta_+}\bigg|_{E,k_z,\lmt}. \la{singleeq}}
Reading from right to left, a wavepacket that begins at $t_+=0$ first accumulates the semiclassical phase $e^{i\theta_+}$ along edge $+$, is then reflected unto edge $-$ and accumulates $e^{i\theta_-}$ in travelling along this edge; a second reflection closes the loop, and the quantization rule states that the net phase (acquired by the wavepacket around the loop) is an integer multiple of $2\pi$. \q{singleeq} may equivalently be expressed as in \q{rule3a}, or as
\e{-1\eq \text{exp}\bigg[ il^2S +i \oint_{\frako} (\orb+\bmx)\cdot d\bk \lin
&+i Z\oint_{\frako} (\sigma^z/v^{\sma{\perp}}) dk\bigg]_{E_j,k_z}, \la{phasecontinuity}}
 with $S$ defined as the oriented area of the orbit:
\e{S[\frako]:= -\int^1_{0}k_x^{\nu}(dk_y/dt_{\nu}) dt_{\nu}.} 
The other terms in the exponent are, collectively, the Roth-Berry-Zeeman phase originating from  the $H_1$ term in $e^{i\theta_{\nu}}$ [cf.\ \q{rbzphase}].  The gauge ambiguity in the definition of $H^B_1$[recall \q{h1bnotunique}] is reflected in the above equations by the ambiguity in the Berry connection $\bmx$. However, it is known\cite{berry1984} that the exponentiated loop integral of $i\bmx$, as appears in  \q{phasecontinuity}, is gauge-invariant.\\

Being independent of $k_x$, \q{phasecontinuity} defines a set of discrete energy levels $\{E_j\}$ which are each macroscopically degenerate -- we refer to them as Landau levels.  The Zeeman term in \q{phasecontinuity} may be further simplified if spin-orbit coupling is absent -- we might then replace $\sigma^z \rightarrow \pm 1$ and \q{phasecontinuity} reduces to \q{singlequantrulewithzeeman} with the identification $Z\int dk/v^{\sma{\perp}}=\pi (g_0/2)(m_c/m)$, with $m_c$ the cyclotron mass.\\

The extension of \q{singleeq} to the most general  closed orbit, composed of $N_s \in 2\Z$ edges and an equal number of turning points, is 
\e{ 1= \prod_{\nu=1}^{N_s} e^{i\phi_{r}^{\nu}} e^{i\theta_{\nu}}\bigg|_{E,k_z,\lmt},\la{generalclosedorbit}}
where we identify the Maslov phase
\e{\phi_M:=\prod_{\nu=1}^{N_s}\phi_{r}^{\nu},} 
as the net reflection phase of all turning points. To derive the same quantization condition from more conventional means (i.e., continuity and patching of wavefunctions), we refer the interested reader to \app{app:justifyrules}. \q{generalclosedorbit} applies to any simple closed orbit, which we define as orbits that are deformable to a circle, e.g., \fig{fig:turning_point}(g). \q{generalclosedorbit} also applies to nonsimple closed orbits which are homotopically \emph{in}equivalent to a circle, a case in point being the figure-of-eight illustrated in \fig{fig:turning_point}(h). A figure-of-eight pinches together an electron-like pocket with a hole-like pocket, and has recently been studied in the context of over-tilted Weyl/Dirac fermions.\cite{obrien_breakdown,koshino_figureofeight} The four turning points in a figure-of-eight have cancelling circulations [as indicated by the sign of $\pm i$ in \fig{fig:turning_point}(h)], and therefore there is no Maslov correction, contrary to a claim in \ocite{obrien_breakdown}; we shall extend our analysis of the over-tilted Weyl/Dirac fermion  to include interband breakdown in \s{sec:interband}.\\

The net reflection phase ($\prod_{\nu=1}^{N_s} e^{i\phi_{r}^{\nu}}$) of a closed orbit is invariant under continuous deformations of the orbit trajectory -- we will argue for this by locally deforming the band contour near a turning point, while maintaining a closed orbit. If we invert the parabolic contour associated to a single turning point, we  necessarily introduce a Mexican-hat wiggle with two additional turning points, e.g., \fig{fig:turning_point}(g) is a deformed version of \fig{fig:turning_point}(e). Since the net circulation [a notion we  make precise in \s{sec:turningpoint}] of the final three points is always equal to that of the original point, there is no net change in the reflection phase. Having argued that the combined reflection phase is topologically invariant for a closed orbit, we may therefore evaluate this quantity for the simplest, homotopically-equivalent representative. For the simple closed orbits, this is a circle [\fig{fig:turning_point}(e-f)], which has a reflection phase of $\pi$ -- this accounts for the $\pi$ Maslov correction to the quantization condition. One implication of this argument: while \q{phasecontinuity} has been derived for a circular orbit, its final expression is generally valid for any simple closed orbit.

\subsubsection{Multi-band quantization condition for closed orbits}\la{sec:multibandquant}

In the multi-band case with $P(\bk)$ having rank $D$, we may analogously define a vector-valued amplitude for each edge (labelled by $\nu$) as
\e{ \ba_{\nu E}(t_{\nu}):=e^{-il^2\int^{t_{\nu}}_0 k_x^{\nu}(d{k}_y/dt'_{\nu}) dt'_{\nu}} \;\cala_{\bk(t_{\nu}) E}^{\nu}\; \ba_{\nu E}(0) \bigg|_E. \la{definevecampmultiband}}
As defined in \q{definepropagator}, $\A$ is a $D\times D$ unitary matrix acting on an as-yet-unspecified, constant vector $\ba(0)$. $t_{\nu}\in [0,1]$ and $\bk(t_{\nu})$ have the same meaning as for the single-band case, as described below \q{definescalarampsection2}.  
We implement the boundary condition
\e{\ba_{\nu E}(0)=e^{i\phi_r^{\nu'}}\ba_{\nu'E}(1), \as e^{i\phi_r^{\nu'}}=\pm i, \la{bcvector}}
for every two edges $(\nu$ and $\nu'$) that touch at a turning point. \\

For a closed orbit ($\frako$) comprising of $N_s$ edges (and an equal number of turning points), \q{definevecampmultiband} and (\ref{bcvector}) may be combined into a system of linear equations with $D$ variables. The quantization condition is then equivalent to solving this system of equations; a solution exists upon satisfaction of the following determinantal equation:
\e{     \det \left[ \bigg(\prod_{\nu=1}^{N_s}e^{i\phi_r^{\nu}}\bigg)\;e^{il^2S}\;\cala[\frako] -I\right]\bigg|_{E=E_{a,j},k_z}=0,\la{multibandquantizationcondition}}
with $a \in \Z_D, \; j \in \Z$. $\cala[\frako]$ is the propagator of \q{definenonabelianunitary} defined over the full orbit. Its solution corresponds to $D$ sets of equidistant, macroscopically-degenerate Landau levels. For a simple closed orbit, $\prod_{\nu=1}^{N_s}e^{i\phi_r^{\nu}}=-1$, and we are led directly to the multi-band quantization conditions in \q{rule3b}-(\ref{definenonabelianunitary}). In \s{sec:gaugecovariancepropag}, we show that \q{multibandquantizationcondition} is invariant under the $U(D)$ gauge transformations [cf.\ \q{basistransf}]; the transformation of the propagator under symmetry is further investigated in \s{sec:symmetrycovariancepropag}, where we prove certain symmetry constraints for the Landau levels. \\

As an example, consider a spin system where $T\inv$ symmetry imposes a two-fold degeneracy ($D=2$) in the zero-field band dispersion. In the presence of a field, the same symmetry imposes $\lambda_1=-\lambda_2$ mod $2\pi$, as elaborated in \s{sec:symmetrycovariancepropag}. If spin-orbit coupling is negligible, $|\lambda_1-\lambda_2|$ just equals  the free-electron Zeeman splitting [$\pi (g_0/2)(m_c/m)$ from \q{singlequantrulewithzeeman}].

\subsubsection{Beyond closed orbits} \la{sec:beyondsimpleorbits}

We briefly comment on the quantization conditions for open orbits with negligible breakdown. One example would be a noncontractible orbit which extends across the Brillouin torus in single direction -- in the extended-zone scheme, these orbits traverse across different Brillouin zones. Since the phase factor $e^{ik_xk_yl^2}$ is not single-valued when $\bk$ is advanced by a reciprocal vector,  this phase cannot be neglected when one imposes continuity on the wavefunction in the $(K_x,k_y)$-representation. When this phase is accounted for, it introduces a $k_x$-dependence to the quantization condition, and consequently a loss of the (exponentially-accurate) macroscopic degeneracy that characterizes closed orbits. We refer the reader to \ocite{kaganov_coherentmagneticbreakdown} for a more extensive discussion of open orbits.

\subsection{Landau levels and de Haas-van Alphen oscillations for closed orbits} \la{sec:quantumoscillations}

\subsubsection{Single-band case}\la{sec:singlebandoscillations}

When the single-band quantization condition [\q{rule3a}]  is viewed at a fixed field (and a fixed wavevector $k_z$ for a 3D solid), the energy difference between  adjacent Landau levels is locally periodic as  
\e{ E_{j+1}-E_j =  \f{2\pi }{l^2\partial S/\partial E}\bigg|_{E=E_j,k_z} +O(l^{-4}). \la{orderedspectrum}}
This follows from the assumption that the area of the orbit ($S$), as well as the Roth-Berry-Zeeman (RBZ) phase, collectively defined as
\e{\lambda(E,k_z):=\oint_{\frako} (\orb+\bmx)\cdot d\bk+Z\oint_{\frako}(\sigma^z/v^{\sma{\perp}})dk,}
are smooth functions of energy on the scale of $E_{j+1}-E_j=O(\lmt)$. Equivalently stated, we assume $\partial S/\partial E$ and $\partial \lambda/\partial E$ are $O(l^0)$ quantities.  The quantity $(\hbar^2/2\pi)\partial S/\partial E$ that determines Landau-level differences has been referred to as the cyclotron mass; it coincides, in the free-electron limit ($V{=}0$), with the free-electron mass. Supposing $E_n^0$ are zeroth-order solutions of the quantization condition, the $H_1$-correction to $E_n^0$ is 
\e{\delta E_n = -l^{-2}\f{\lambda}{\partial S/\partial E}\bigg|_{E_n^0} +O(l^{-4}).\la{H1correctenergy}}

If we view \q{rule3a} at fixed energy (e.g., the Fermi energy $E_F$) and variable field, then the quantization condition is satisfied for a discrete set of fields (indexed by integer $j$), with corresponding magnetic lengths satisfying
 \e{ &l_{j+1}^2-l_j^2 = \f{2\pi}{S(E_F,k_z)}+O(\lmt);\lin
 &l_j^2 = \frac{2\pi j -\phi_M -\lambda}{S}\bigg|_{E=E_F,k_z}+O(\lmt).\la{simpleorbitoscillation}}
The first equation forms the basis of quantum oscillatory phenomena of the de Haas-van Alphen (dHvA) type -- they reflect how quasi-periodic Landau levels successively become equal to the Fermi energy as the reciprocal magnetic field is changed. Such $l_j^2$ is henceforth referred to as a dHvA level, and the set of all dHvA levels is referred to as the dHvA spectrum.  The period in $l_j^2$ is not affected by the RBZ phase ($\lambda$), in accordance with the conventional theory of metals.\cite{shoenberg} On the other hand, one may look to the phase offset   of the dHvA oscillation to extract $\lambda$. In 3D metals, the curvature of the Fermi surface results in an additional Lifshitz-Kosevich correction\cite{lifshitz_kosevich,lifshitz_kosevich_jetp} to this phase offset, which in sum equals   
\e{\gamma:= l_j^2S +\phi_{LK} \;\text{mod}\; 2\pi \equiv -\lambda-\phi_M+\phi_{LK}.\la{defgammaint}}
with $\phi_{LK}=\pm \pi/4$ depending on whether the orbit is maximal or minimal. Restated from the perspective of measurement, $\gamma$ is the phase offset of the oscillations of the magnetization of a solid.\cite{champel_mineev} In the experimental literature,  $\gamma$ is often  viewed graphically as the intercept of an extrapolated line connecting the discrete values of $l^2$ where the magnetization is peaked; we shall therefore refer to the quantity defined in \q{defgammaint} as the $\gamma$-intercept.\\

A comprehensive symmetry analysis of the RBZ phase is performed in \s{sec:symmetryunitarygenbyH1}; here we illustrate two highlights: \\

\noi{i} Orbits which are mapped to themselves, up to a reversal in orientation, are said to be self-constrained. Graphene provides a paradigmatic example, for which an orbit encircling the Dirac point is invariant under $T\rot_{2z}$ symmetry -- this leads to the vanishing of the Roth moment at each wavevector, as well as the quantization of the Berry phase to $\pi$, which cancels the Maslov correction in the $\gamma$-intercept. \\

\noi{ii} Just as relevant are orbits which are mapped to distinct orbits by a symmetry -- two related orbits are said to be mutually constrained. In a toy model of spinless graphene, two orbits which encircle different valley centers are mutually constrained by $T$ symmetry; since each orbit does not encircle a $T$-invariant wavevector, it is not self-constrained by $T$ symmetry. If the spatial symmetry ($\rot_{2z}$) is further broken (plausibly by epitaxial growth on certain substrates\cite{Zhou_gapingraphene_SiC,hBN_gap}), each valley-centered orbit develops an orbital moment, i.e., when integrated over the orbit, this moment results in a nontrivial Roth phase. Owing to $T$ symmetry, the sum of Roth-Berry phases in two mutually-constrained orbits cancel modulo $2\pi$, but individually each phase should be measurable from the $\gamma$-intercept of its corresponding orbit. Alternatively stated, two distinct but mutually-constrained harmonics should appear in the magnetization oscillations. To make this toy model of graphene more realistic, one must incorporate the Zeeman effect, as described in \ocite{topofermiology}.\\ 

The equations in this section directly apply to spin-orbit-coupled systems with nondegenerate bands; for Zeeman-coupled systems with negligible spin-orbit coupling, the above equations would apply to either of the two spin species, with $\sigma_z$ replaced by $\pm 1$; for  intrinsically spinless systems described by a Schrodinger Hamiltonian, the above equations apply without the Zeeman term (i.e., set $Z=0$).



\subsubsection{Multi-band case} \la{sec:multibandquantization}

Considering \q{rule3b} at fixed field and $k_z$, we obtain, for each eigenvalue $e^{i\lambda_a}$ of the $D\times D$ propagator [\q{definenonabelianunitary}], a set of discrete energy levels $\{E_{a,j}\}_{j\in \Z}$. For fixed $a$, the energy difference between adjacent Landau levels,  is to leading order,
\e{ E_{a,j+1}-E_{a,j} \approx  \f{2\pi }{l^2\partial S/\partial E}\bigg|_{E=E_{a,j},k_z} .}
We have assumed here that $\partial S/\partial E$ and $\partial \lambda_a/\partial E$ are both $O(l^0)$. When \q{rule3b} is viewed at fixed energy ($E_F$) and varying field, the quantization condition is satisfied for a discrete set of fields corresponding to
\e{ &l_{a,j}^2 \approx \frac{2\pi j-\phi_M -\lambda_a}{S}\bigg|_{E=E_F,k_z}; \lin
& l_{a,j+1}^2-l_{a,j}^2 \approx \f{2\pi}{S(E_F,k_z)}.\la{laj}}
All equations in this section are accurate to $O(l^{-8/3})$, due to the $O(l^{-2/3})$ uncertainty in the multi-band Maslov correction [derived in \s{sec:turningpoint}]. The dHvA spectrum therefore divides into $D$ sets of levels indexed by $a=1,\ldots,D$; each set corresponds to a harmonic in the magnetization oscillations, with corresponding intercept $\gamma_a:=l_{a,j}^2S +\phi_{LK}$ mod $2\pi$. \\

Particularizing to spin-orbit-coupled bands with $T\inv$ symmetry ($D=2$), \q{laj} implies the existence of two harmonics in the magnetic oscillations. Absent any other symmetries, these harmonics are generally distinct, with $\lambda_1=-\lambda_2$ mod $2\pi$, as proven in the paragraph surrounding \q{Tinvconstraint} in \s{sec:symmetryunitarygenbyH1}. A more comprehensive symmetry analysis of $\lambda_a$ is performed in the next section [\s{sec:symmetry}].

\section{Symmetry in the first-order effective Hamiltonian theory}\la{sec:symmetry}

This section describes the effects of symmetry in the first-order effective theory; our analysis covers all possible symmetries that occur in crystals, i.e., in any space group or magnetic space group.  We first identify in \s{sec:symmetryoforbits} the symmetries that are relevant to semiclassical orbits;   we then describe how symmetry constrains the orbital magnetic moment  and the Zeeman coupling [\s{sec:symmetrysinglebandom}], the first-order effective Hamiltonian ($H_1$) [\s{sec:symmetryH1}], and the propagator that is generated by $H_1$ [\s{sec:symmetryunitarygenbyH1}] over an orbit. 
The eigen-phases of this propagator enters the quantization conditions, from which one may determine the symmetry constraints on the Landau levels and dHvA oscillations. Our symmetry analysis is simplified by the classification of closed orbits into ten (and only ten) symmetry classes. These ten symmetry classes were first introduced and exemplifed in \ocite{topofermiology}; in \s{sec:symmetryunitarygenbyH1}, we provide a more detailed derivation which focuses on the possible types of  symmetry representations.  
In addition, \s{sec:symmetrysinglebandom} may be used to analyze $\bk$-resolved measurements of the orbital magnetic moment, e.g., through circular dichroism in photoemission.\cite{YaoWang_dichroism_valley}


\subsection{Symmetries of semiclassical orbits} \la{sec:symmetryoforbits}

We encourage the reader to scan through \s{sec:symmetryinBloch}, where we reviewed how symmetry constrains Bloch functions at zero field. We assume the reminder is familiar with certain notations for symmetry transformations that was introduced therein. As a reminder, $g$ denotes a symmetry in the (magnetic) space group ($G$) of a solid, and its representations in various contexts [cf.\ \s{sec:symmetryinBloch}] are denoted by $\hat{g},\check{g},\breve{g}$.\\

We would like to particularize to symmetries which are relevant to Bloch electrons in a field. Assuming that the field is oriented along $\vec{z}$, all semiclassical orbits are contained in quasimomentum planes orthogonal to $\vec{z}$, and we are interested in symmetries which relate one such orbit to another (or possibly an orbit to itself, up to a reversal in orientation). For 3D solids, $g$'s action in $\bk$-space may be block-diagonalized as
\e{g:& \bk \rightarrow\; g\circ \bk := (-1)^{s(g)}\check{g} \bk\lin
\eq (-1)^{s(g)}\bigg(\check{g}^{\sma{\perp}}\bk^{\sma{\perp}},(-1)^{t(g)} k_z\bigg),\as t(g) \in \{0,1\}, \la{kactionblockdiag}}
where $t(g)=0$ (resp. $1$) for symmetries whose point-group operation preserves (resp. inverts) the coordinate parallel to the field. We distinguish between $\bk$ which parametrizes the 3D Brillouin torus, and $\bkp=(k_x,k_y)$ which parametrizes a two-torus (\bt) perpendicular to the field. We shall sometimes refer to \bt as a plane; symmetry operations that act only in \bt are described as planar.\\

In  \q{kactionblockdiag}, we have also introduced   $\check{g}^{\sma{\perp}}$ as a real, orthogonal, two-by-two matrix; it represents the point-group operation that is restricted to \bt. The determinant of $\check{g}^{\sma{\perp}}$ defines a $\Z_2$ variable $u$ as 
\e{(-1)^{u(g)} :\eq \det[\check{g}^{\sma{\perp}}],\as u(g) \in \{0,1\},\lin
\det[\check{g}] \eq (-1)^{t(g)+u(g)}.\la{defineug}}
Let us demonstrate that $u(g)=0$ (resp. $1$) if $g$ preserves (resp. inverts) the orientation of the semiclassical orbit; to clarify, the orientation of an orbit is its sense of circulation, whether clockwise or anticlockwise, that is determined from Hamilton's equation [cf. \q{hamiltoneom}].  \\

\noindent The symmetry constraint on the band velocities at $\bk$ and $g\circ \bk$ [recall \q{kactionblockdiag}]
\e{\bv(\bk)\eq (-)^{s(g)} [\check{g}^T\bv(g\circ \bk)],\as \check{g}^T_{\ab} = \check{g}_{\beta \alpha}, \la{constraintbandvelocity}}
implies, through \q{hamilton2}, an analogous relation between the orbit velocities [defined in \q{hamilton2}]:
\e{\det[\check{g}^{\sma{\perp}}]\dot{\bk}^{\sma{\perp}}\bigg|_{g\circ \bk} =  (-)^{s(g)}\left[\check{g}^{\sma{\perp}} \dot{\bk}^{\sma{\perp}} \bigg|_{\bk}\right]. \la{relatetwotangentvectors}}
To interpret this equation, consider the map $\caly_g:\R^2 \rightarrow \R^2$ between two planar wavevectors related by symmetry $g$: 
\e{\caly_g(\bk^{\sma{\perp}}) =(-1)^s\check{g}^{\sma{\perp}}\bk^{\sma{\perp}}={(g\circ \bk)}^{\sma{\perp}}.} 
\q{relatetwotangentvectors} states that $\caly_g(\dot{\bk}^{\sma{\perp}}|_{\bk})$ is equal in magnitude to the orbit velocity at $g\circ \bk$, with a minus-sign difference iff $\det[\check{g}^{\sma{\perp}}]=(-1)^{u(g)}=-1.\blacksquare$\\

All symmetries, whose point-group operation block-diagonalizes as in \q{kactionblockdiag}, would henceforth be referred to as \emph{symmetries of the orbit configuration}. In deriving how these symmetries constrain the effective Hamiltonian, the following decomposition, valid for any symmetry of the orbit configuration, would be useful: 
\e{ g =T^{s(g)}\,\tra_{\bdelta}\,\mir_z^{t(g)}\,\mir_x^{u(g)}\,\rot_{n(g),z}^{v(g)}\,\mathfrak{e}^{w(g)}, \la{symmetrydecomposition}}
where $\tra_{\bdelta}, T, \mir,\rot,\frake$ are symmetry operations defined in \s{sec:symmetryoforbits};  $s,t,u \in \{0,1\}$ have been previously defined[\q{definesg}, (\ref{kactionblockdiag})-(\ref{defineug})], $n\in \{2,3,4,6\}$ labels the possible discrete rotations, and we introduce here $w\in \{0,1\}$ and $v \in \{0,1,\ldots,n-1\}$. \q{symmetrydecomposition} is really valid for a double group; for ordinary groups the same decomposition holds without the factor of $\frake^w$. \\

\noindent \emph{Proof of decomposition \q{symmetrydecomposition}:} \\

\noindent If $g$ inverts time, we decompose $g=Tg'$ such that $g'$ is a purely spatial operation; otherwise, $g=g'$. Our shorthand for this is $g=T^sg'$. We further decompose $g'$ into translational and point-preserving spatial transformations: $g'=\tra_{\bdelta}g''$. Applying \q{kactionblockdiag}, we find that $g''$  decomposes as $g''=\mir_z^{t}{g}^{\sma{\perp}},$ such that the $g^{\sma{\perp}}$ acts trivially on the coordinate orthogonal to the plane, i.e.,
${g}^{\sma{\perp}}: \bk \rightarrow (\check{g}^{\sma{\perp}}\bk^{\sma{\perp}},k_z).$ \\

To complete the proof, we would need to show that any planar, spatial transformation may be expressed as $g^{\sma{\perp}} =\mir_x^{u}\rot_{nz}^m \frake^w$. Any point group is built up of discrete rotations ($\rot$) and reflections ($\mir$), which  are the fundamental covering operations;\cite{tinkhambook} 2D point groups are built up from planar rotations ($\rot_{nz}$) and reflection-invariant lines contained in the plane (in short: planar reflections); the latter are exemplified by $\mir_x$ and $\mir_y$. It is useful to distinguish between planar-proper (det $\check{g}^{\sma{\perp}}=+1$) and planar-improper (det $\check{g}^{\sma{\perp}}=-1$) transformations; all planar reflections (resp.\ rotations) are planar-improper (resp.\ proper). Some properties of successive transformations will be needed:\cite{tinkhambook} (i) the product of two planar rotations is another planar rotation; (ii) the product of two planar reflections is a planar rotation, and (iii) the product of a planar rotation with a planar reflection is another planar reflection. It follows from (i-iii) that any planar-proper transformation is proportional to $\rot_{nz}^m$ for some integers $m$ and $n\in \{2,3,4,6\}$; the proportionality factor must act trivially in space, so it may be the identity operation or a $2\pi$-rotation. Therefore, if $g^{\sma{\perp}}$ is planar-proper, it is expressible as  $\rot_{nz}^m \frake^w$. Otherwise if $g^{\sma{\perp}}$ is planar-improper, it can always be expressed as the product of (a) an arbitrarily chosen reflection (e.g., $\mir_x$), with (b) a proper transformation ($\rot_{nz}^m \frake^w$) that depends on our choice in (a). In summary, we may say $g^{\sma{\perp}} =\mir_x^{u}\rot_{nz}^m \frake^w$ with $(-1)^{u} = \det[\check{g}^{\sma{\perp}}]$ [cf.\ \q{defineug}]. This completes the proof.$\blacksquare$\\

Combining \q{kactionblockdiag} with \q{symmetrydecomposition}, $g$ maps $\bkp \in$ \bt to
\e{ g\circ \bkp =(-1)^s\check{g}^{\sma{\perp}}\bkp, \iwith \check{g}^{\sma{\perp}}= \check{\mir}^u_x\, \check{\rot}_{nz}^v. \la{definecheckgperp}}
We say that $\bkp$ is $g$-invariant if $g\circ \bkp=\bkp$ up to a planar reciprocal vector. If $g$ acts as a planar reflection, then its order must be even:
\e{ u(g)=1 \imp N(g) \in 2\Z. \la{uimpliesN}}
 This follows because any odd power of a planar reflection is still a planar reflection, while by assumption $u(g^N)=0$ [cf.\ \q{defineorderg}].\\

It will be useful to classify symmetries according to the topology of the $g$-invariant points. Type-I symmetries are defined to leave every, generic $\bkp$ invariant, hence
\e{ (-1)^s\checkgp=I_2 \imp \det \checkgp = 1 \imp u(g)=0, \la{ugequal0}}
with $I_2$ the two-by-two identity matrix. We may further distinguish type-I symmetries by $s$: either $u=s=v=0$, or $u=0, s=1$ and $\check{\rot}_{nz}^v=\check{\rot}_{2z}$. \\

Type-II symmetries are symmetries for which the generic $\bkp$ is not invariant;  $g$-invariant $\bkp$ are isolated points if $u=0$; otherwise ($u=1$), they form  isolated lines. To prove the last claim, if type-II $g$ is planar-proper ($u=0$), then $g$ maps $\bkp$ to $\check{\rot}_{2z}^s\check{\rot}_{nz}^v\bkp$, where we have identified  $(-1)^s$ with $\check{\rot}_{2z}^s$. Being the product of two planar rotations, $\check{\rot}_{2z}^s\check{\rot}_{nz}^v$  must be again a planar rotation, whatever the values of $s,n,v$. Moreover, $\check{\rot}_{2z}^s\check{\rot}_{nz}^v$ cannot be the trivial rotation (identity transformation), due to our assumption that it is type-II. The only rotationally-invariant $\bkp$ are isolated points. Now if type-II $g$ is planar-improper, $g$ acts on $\bkp$ as the product of a planar reflection ($\check{\mir}_x$) and a planar rotation  ($\check{\rot}_{2z}^s\check{\rot}_{nz}^v$); any such product is a planar reflection, possibly with the reflection-invariant line rotated. \\

Our classification of type-I and -II symmetries is extended to a classification of symmetric orbits in \s{sec:symmetryunitarygenbyH1}.



\subsection{Symmetry constraints of the orbital magnetic moment and the Zeeman coupling} \la{sec:symmetrysinglebandom}

We would like to identify the symmetry classes which allow for a $\bk$-dependent orbital moment/Zeeman coupling. If the  average of the orbital moment/Zeeman coupling over an orbit is nonzero, then the Landau levels are nontrivially affected, as explained in \s{sec:symmetrycovariancepropag}. Our symmetry analysis is further motivated  by recent experiments that are able to probe the \om (at each wavevector) through circular dichroism in momentum-resolved photoemission.\cite{YaoWang_dichroism_valley} We would like to determine if dichroism is allowed by symmetry, and where dichroism may be found in the Brillouin torus.\\

Consider the \om for a subspace of bands projected by $P(\bk)$ in \q{defineprojP}, where, again, $D$ the dimension of  the subspace at each wavevector. For $D=1$, we refer to the single-band $\bM_n$ defined in \q{orbitalmagnetization29} for band $n$, but henceforth we would drop the $n$ subscript; for $D>1$, we refer to the multi-band $\bM$ defined in \q{definemultibandom}, which is a $D\times D$ Hermitian matrix. To derive the symmetry constraints on the orbital moment, it is convenient to begin with its expression with the velocity matrix $\bPi$ [second line of \q{definemultibandom}], since the current operator $\hbPi=-i[\hbr,\hH_0]/\hbar$ transforms simply under a symmetry ($g$). From the action of $g$ on the position operator [cf. \q{gactsonposition}], and the symmetry contraint on the cell-periodic functions [cf. \q{gactsonu}], we obtain 
\e{ \Pi^{\alpha}_{mn}\bigg|_{g\circ\bk}  \eq (-1)^{s(g)}\check{g}_{\ab}K^{s(g)}[\breve{g}^* \,\Pi^{\beta}\, \breve{g}^{T}]_{mn}K^{s(g)}\bigg|_{\bk}. \la{gactsonvelocitymatrixfinal}}
Inserting this into \q{definemultibandom} and after a little gymnastics [detailed in \app{app:symmetryom}], 
\e{ \bM\bigg|_{g\circ\bk} = (-1)^{s(g)}\det[\check{g}] \,\breve{g}\, K^{s(g)} \, (\check{g}\bM)\, K^{s(g)} \,\breve{g}^{-1}\bigg|_{\bk}. \la{gactsonMmultiband}}
While this expression is valid for any number of bands, it simplifies in the single-band case owing to: (i) $\bM$, being a Hermitian one-by-one matrix, is a real number, and (ii) $\breveg$, being a unitary one-by-one matrix, is a commuting phase factor that cancels with its Hermitian adjoint $\breveg^{\mo}$. Therefore, the single-band \om satisfies
\e{ \bM \bigg|_{g\circ\bk} = (-1)^{s(g)}\det[\check{g}][\check{g}\bM]\bigg|_{\bk}. \la{gactsonM}}
To interpret this equation, recall that each $g$ corresponds to a certain action in spacetime, which may be decomposed as a point-preserving transformation and a translation [cf. \q{definesg}]; $(-1)^{s(g)}\det[\check{g}]$ is the determinant of the matrix corresponding to the point-preserving transformation. Therefore, $\bM$ transforms like the spatial components of a (3+1)-dimensional pseudovector, in addition to the transformation of its argument $\bk$. \q{gactsonM} is the full generalization (to any symmetry) of well-known constraints on the single-band moment with $T$ and/or $\inv$ symmetry.\cite{sundaram1999,rothII,lifshitz_pitaevskii_statphys2} We exemplify \q{gactsonMmultiband} and (\ref{gactsonM}) with some naturally-occuring, but certainly not exhaustive, symmetries in Tab.\ \ref{tab:magnetization}; there, we employ certain notation for symmetries that have been introduced in \s{sec:symmetryinBloch}.   \\


\begin{table}[H]
	
\centering
		
\begin{tabular} {|c|c|c|c|c|} \hline
			
$g$ &Space-group rule& Single-band $\bM$ & Multi-band $\bM$ & Two-band constraints ($F{=}1$)  \\  \hline \hline 
		  
$T$ & $\breveg\big|_{-\bk}\breveg^*\big|_{\bk}=(\mo)^{\sma{F}},$& $\bM\big|_{-\bk}=-\bM\big|_{\bk}$& $\bM\big|_{-\bk}=-\breve{g}\bM^*\breve{g}^{\mo}\big|_{\bk}$ & For $\bk=-\bk,$  \\ 
&$\breveg\big|^T_{\bk=-\bk}=(\mo)^{\sma{F}}\breveg\big|_{\bk=-\bk}$&&&$\breveg=-i\sy,\;\text{Tr}[\bM]=0.$\\ \hline			 
			 
$T\inv$& $\breveg\big|_{\bk}\breveg^*\big|_{\bk}=(\mo)^{\sma{F}},$ & $\bM\big|_{\bk}=0$&  $\bM\big|_{\bk}=-\breve{g}\,\bM^*\,\breve{g}^{\mo}\big|_{\bk}$  & For all $\bk$,\\
& $\breveg\big|^T_{\bk}=(\mo)^{\sma{F}}\breveg\big|_{\bk}$ &&&$\breveg  =-i\sy,\;\text{Tr}[\bM]=0.$\\ \hline

$T\rot_{2z}$& $\breveg\big|_{-\check{\rot}_{2z}\bk}\breveg^*\big|_{\bk}=I,$   & $\bM\big|_{-\check{\rot}_{2z} \bk} =  -\check{\rot}_{2z}\bM\big|_{\bk},$ & $\bM\big|_{-\check{\rot}_{2z} \bk} =  -\breve{g}\,[\check{\rot}_{2z}\bM^*]\,\breve{g}^{\mo}\big|_{\bk}$ & For $k_z=0$ or $\pi,$ \\ 
&$\breveg\big|^T_{\bk=-\check{\rot}_{2z}\bk}=\breveg\big|_{\bk=-\check{\rot}_{2z}\bk}$ & $M^z\big|_{\bk=-\check{\rot}_{2z} \bk} =0$  & &$\breveg=I,\; M^z \propto \sy.$\\ \hline

$T\tra_{\vec{z}/2}$ & $\breveg\big|_{-\bk}\breveg^*\big|_{\bk}=(\mo)^{\sma{F}}e^{\minus ik_z},$& $\bM\big|_{-\bk}=-\bM\big|_{\bk}$& $\bM\big|_{-\bk}=-\breve{g}\bM^*\breve{g}^{\mo}\big|_{\bk}$ & (i) For $\{k_z=0,\bk^{\sma{\perp}}=-\bk^{\sma{\perp}}\},$ \\
&$\breveg\big|^T_{\bk=-\bk}=(\mo)^{\sma{F}}e^{\minus ik_z}\breveg\big|_{\bk=-\bk}$&&& $\breveg=-i\sy,\;\text{Tr}[\bM]=0.$ \\   
&&&&(ii) For $\{k_z=\pi,\bk^{\sma{\perp}}=-\bk^{\sma{\perp}}\},$ \\
&&&&$\breveg=I,\;\bM \propto \sy.$\\ \hline	
			 
$\inv$ & $\breveg\big|_{-\bk}\breveg\big|_{\bk}=I$& $\bM\big|_{-\bk}=\bM\big|_{\bk}$ &  $\bM\big|_{-\bk}=\breveg\bM\breveg^{\mo}\big|_{\bk}$, & $\text{Tr}[\bM]\big|_{-\bk}=\text{Tr}[\bM]\big|_{\bk}$\\ 

&&& $[\breveg,\bM]\big|_{\bk=-\bk}=0$ & \\ \hline

$\mir_x$ &$\breveg\big|_{\check{\mir}_x\bk}\breveg\big|_{\bk}=(\mo)^{\sma{F}}$& $\bM\big|_{\check{\mir}_x\bk} =-[\check{\mir}_x\bM]\big|_{\bk},$ &  $\bM\big|_{\check{\mir}_x\bk} =-\breveg[\check{\mir}_x\bM]\breveg^{\mo}\big|_{\bk}$, &  $\text{Tr}[\bM]\big|_{\check{\mir}_x\bk} =-\check{\mir}_x\text{Tr}[\bM]\big|_{\bk}$\\ 

&& $M^{\alpha}\big|_{\check{\mir}_x\bk=\bk} =0, \; \alpha {\in} \{y,z\}$& $[\breveg,M^x]\big|_{\check{\mir}_x\bk=\bk}=0$ &\\ \hline

$\glide_{x,\vec{y}/2}$ & $\breveg\big|_{\check{\mir}_x\bk}\breveg\big|_{\bk}=(\mo)^{\sma{F}}e^{\minus ik_y}$ & $\bM\big|_{\check{\mir}_x\bk} =-[\check{\mir}_x\bM]\big|_{\bk},$ &  $\bM\big|_{\check{\mir}_x\bk} =-\breveg[\check{\mir}_x\bM]\breveg^{\mo}\big|_{\bk}$, & $\text{Tr}[\bM]\big|_{\check{\mir}_x\bk} =-\check{\mir}_x\text{Tr}[\bM]\big|_{\bk}$\\ 

&& $M^{\alpha}\big|_{\check{\mir}_x\bk=\bk} =0, \; \alpha {\in} \{y,z\}$&$[\breveg,M^x]\big|_{\check{\mir}_x\bk=\bk}=0$ &\\ \hline

$\rot_{nz}$ & $ \breveg\big|_{\bk^{\sma{(n\mo)}}} \ldots \breveg\big|_{\bk^{\sma{(1)}}}\breveg\big|_{\bk}=(\mo)^{\sma{F}}$ & $\bM\big|_{\check{\rot}_{nz}\bk} =[\check{\rot}_{nz}\bM]\big|_{\bk}$  &   $\bM\big|_{\check{\rot}_{nz}\bk} =\breveg[\check{\rot}_{nz}\bM]\breveg^{\mo}\big|_{\bk}$, & $\text{Tr}[\bM]\big|_{\check{\rot}_{nz}\bk} =\check{\rot}_{nz}\text{Tr}[\bM]\big|_{\bk}$\\ 

&&&$[\breveg,M^z]\big|_{\check{\rot}_{nz}\bk=\bk}=0$ &\\ \hline

$\scr_{nz,m}$ &$ \breveg\big|_{\bk^{\sma{(n\mo)}}} \ldots \breveg\big|_{\bk^{\sma{(1)}}}\breveg\big|_{\bk}=(\mo)^{\sma{F}}e^{\minus imk_z}$& $\bM\big|_{\check{\rot}_{nz}\bk} =[\check{\rot}_{nz}\bM]\big|_{\bk}$  &   $\bM\big|_{\check{\rot}_{nz}\bk} =\breveg[\check{\rot}_{nz}\bM]\breveg^{\mo}\big|_{\bk}$, & $\text{Tr}[\bM]\big|_{\check{\rot}_{nz}\bk} =\check{\rot}_{nz}\text{Tr}[\bM]\big|_{\bk}$\\ 

&&&$[\breveg,M^z]\big|_{\check{\rot}_{nz}\bk=\bk}=0$ &\\ \hline

\end{tabular}
		
\caption{The first column lists some commonly-found symmetries in crystals, and we have employed the notation for symmetries that was introduced in \s{sec:symmetryinBloch}; their corresponding sewing matrices form a representation of the space group, as described in the second column. The third and fourth columns describe general constraints on the \om for the single- and multi-band cases. In the last column, we describe the constraints which are specific to two-fold spin-degenerate bands in a spin-orbit-coupled system, for which $\breveg$ transforms in a half-integer-spin representation of the symmetry $g$; no restrictions of the symmetry representations have been made in the other columns; note that the multi-band constraints in the fourth column also apply to spin-degenerate bands. A relation with the qualifier $|_{\bk}$ applies to all wavevectors; $|_{\bk=-\check{\rot}_{2z} \bk}$ applies to any wavevector in the planes defined by $k_z=0$ and $\pi$; finally, $|_{\bk=-\bk}$ applies to wavevectors which are invariant under inversion, modulo reciprocal-lattice translations. If the last two rows, we employ the notation $\bk^{\sma{(i)}}:=\check{\rot}^{i}_{nz}\bk$. In the last column, certain canonical choices for the sewing matrices are displayed, i.e., a basis may always be found where $\breveg$ assumes the displayed forms, assuming that $\breveg$ transforms in the half-integer-spin representation; by $M^z \propto \sy$, we mean that it is proportional to the Pauli matrix $\sy$ with a real, $\bk$-dependent proportionality constant. As explained in the main text, the form of a symmetry constraint on $M^z$ applies also to the Roth and Zeeman Hamiltonians ($H_1^R$ and $H_1^Z$); in the latter case, we should fix $F=1$ for half-integer-spin representations. 
	\label{tab:magnetization}}
\end{table}


The single-band \omm, in a certain direction $\alpha$, vanishes at a specific $\bk$, if there exists a symmetry that inverts $M^{\alpha}\rightarrow -M^{\alpha}$, and simultaneously maps said $\bk$ to itself. This vanishing may occur at  isolated points, e.g., $\bM=0$ at inversion-invariant wavevectors (where $\bk=-\bk$) in systems with only time-reversal symmetry.  In $T\inv$-symmetric systems, $\bM=0$ in the entire torus.  For further exemplification, $M^z=0$ for high-symmetry planes in systems with $T\rot_{2z}$ symmetry, and assuming no other symmetry we should not expect $M^x$ or $M^y$ to likewise vanish. For these examples we list the symmetry constraints on the single-band moment in the third column.  \\



In the multi-band case, symmetry cannot enforce that $M^{\alpha}$ vanishes as a matrix, but the analogous constraint is that its trace vanishes, as shown in the second-to-last column.  Let us particularize the following discussion to spin-orbit-coupled systems, where bands are two-fold spin-degenerate, and transform in a half-integer-spin representation ($F=1$). Any of $T$, $T\inv$ or $T\rot_{2z}$ symmetries constrains the trace of $M^{\alpha}$ to vanish for at least one $\alpha$, i.e., $M^{\alpha}$ depends on  (at most) three real parameters. The traceless condition for $T\inv$ symmetry has previously been observed by \ocite{cohen_blount_gfactorBismuth}. For $T\rot_{2z}$, the constraint on $M^z$ is comparatively stronger, leading to $M^z$ only depending on one real parameter. This distinction originates from $(T\rot_{2z})^2$ being a $4\pi$ rotation, and $(T\inv)^2=T^2$ being a $2\pi$ rotation, as we proceed to explain.\\ 

\noi{i} \emph{Antiunitary representations that square to minus one.} Since $(T\inv)^2=-I$ for a half-integer-spin representation, the corresponding sewing matrices satisfy $\breveg(\bk)K\breveg(\bk)K=\breveg(\bk)\breveg^*(\bk)=-I$ which, in combination with the unitary of $\breveg$, implies that  $\breveg$ is skew-symmetric. From physical grounds, we expect that $T\inv$ inverts the spin and should map any state to an orthogonal state. Mathematically, we understand this from the impossibility of finding a basis where the sewing matrix is diagonal, i.e., the effect of a basis transformation is to conjugate the sewing matrix by a complex, orthogonal matrix [cf.\ \q{basistransformsew}], but no skew-symmetric matrix can be diagonalized by an orthogonal transformation. In the case of two-fold degenerate bands, we employ that any two-by-two, unitary, skew-symmetric matrix is proportional to the Pauli matrix $\sy$; the proportionality factor is an irrelevant phase factor that depends on the basis choice [cf. \q{basistransf}]. The constraint on the \omm: $\bM=-\sy \bM^* \sy$ [from the fourth column] then implies $\bM$ is traceless over the entire torus. A similar story unfolds for $T$ symmetry at the inversion-invariant wavevectors.\\

\noi{ii} \emph{Antiunitary representations that square to one.} $(T\rot_{2z})^2=I$ implies that the corresponding $\breveg$ (in high-symmetry planes) is symmetric, and may be diagonalized by an orthogonal transformation. By phase redefinitions of the cell-periodic functions, it is always possible to find a basis where $\breveg=I$ (the two-by-two identity matrix); in this basis, $M^z=-M^{z*}$ [from the fourth column] implies that $M^z$ is proportional to the Pauli matrix $\sy$ with a real proportionality constant. One may verify that in whatever basis is chosen, $M^z$ only depends only on one real parameter.\\

For any antiunitary representation that squares to a phase factor, the associativity of symmetry representations guarantees that this phase factor is either one [henceforth called type (ii)] or minus one [type (i)]. (Indeed, if $\breveg^2=e^{i\phi}, \breveg^3=e^{i\phi}\breveg = \breveg e^{i\phi} \imp e^{i\phi} \in \R$.) In symmorphic space groups, all order-two symmetries that invert time may be classified, by their corresponding sewing matrices, into types (i) and (ii).  This statement must be refined in nonsymmorphic, magnetic space groups where the multiplication rules for sewing matrices are wavevector-dependent. A case in point is a spin-orbit-coupled system with $T\tra_{\vec{z}/2}$ symmetry, which arises in layered, antiferromagnetic compounds where the layers are stacked in the ${z}$ direction; the ferromagnetic alignment in each layer alternates between every adjacent layer (separated by half a lattice vector: $\vec{z}/2$). Since $(T\tra_{\vec{z}/2})^2$ is the composition of a $2\pi$ rotation and a full lattice translation, the sewing matrix satisfies $\breveg\big|_{-\bk}\breveg^*\big|_{\bk}=-e^{\minus ik_z}$, which acts like a symmetry of type (i) where $k_z=0$, and of type (ii) where $k_z=\pi$; this leads to wavevector-dependent constraints on the moment, as detailed in the last column.\\

The last class of symmetries are completely spatial and unitarily represented. Under basis transformations of the cell-periodic functions, the sewing matrices at high-symmetry points or lines may always be diagonalized by a unitary transformation [cf.\ \q{basistransformsew}]; this is superfluously true for the single-band sewing matrix. The possible eigenvalues are discrete in phase, and they are determined by space-group rules in the second column. Alternatively stated, if a completely-spatial $g$ belongs to the group of the wavevector $\bk$,\cite{tinkhambook} the eigenvalues of the sewing matrix label the representations of the bands at $\bk$. Depending on the symmetry, different components of the multi-band \om may be simultaneously diagonalized at the high-symmetry wavevectors: $\bM$ for spatial inversion $\inv$, $M^x$ for reflection $\mir_x$ and glide $\glide_{x,\vec{y}/2}$, $M^z$ for rotations $\rot_{nz}$ and screws $\scr_{nz,m}$.\\

Let us add one final remark regarding the utility of Tab.\ \ref{tab:magnetization}. Since the Roth Hamiltonian is defined by $H_1^R:=-M^z B^z$, constraints that act on $M^z$ apply directly to $H_1^R$. A case in point: since $T$ symmetry imposes $M^z(-\bk)=-M^z(\bk)$ in the single-band case [second column], it follows immediately that $H_1^R(-\bk)=-H_1^R(\bk)$. For half-integer-spin representations ($F=1$), the spin matrix $\sigma^z$ transforms in the same way as $M^z$, as derived in \app{app:symmetryH1}. Consequently, symmetry constraints on the Zeeman coupling $H_1^Z \propto \sigma^zB^z$ may also be deduced from the table, if particularized to $F=1$.

\subsection{Symmetry of the first-order effective Hamiltonian} \la{sec:symmetryH1}

As derived in \app{app:symmetryH1}, the first-order effective Hamiltonian transforms under a symmetry ($g$) of the orbit configuration as 
\e{ &H_1\bigg|_{g\circ\bk} = (-1)^{s(g)+u(g)}\breve{g}K^{s(g)}H_1K^{s(g)}\breve{g}^{\mo}\lin
&+i(-1)^{u(g)}\lmt\epsilon_{\ab}\,\breve{g} \nabk^{\beta}\breve{g}^{\mo}v^{\alpha}\bigg|_{\bk}+\lmt \epsilon_{\ab}\delta^{\beta}v^{\alpha}\bigg|_{g\circ\bk}. \la{H1transformssymmetry}}
This expression may be applied to (a) spin-orbit-coupled systems, (b) spinful systems with negligible spin-orbit coupling, and (c) plausibly to charge-neutral systems with effective magnetic fields. However, the meaning of $H_1$ is slightly different in each context:\\

\noi{a} In spin-orbit-coupled systems, the Bloch functions form a half-integer-spin representations of the (magnetic) space group, and $H_1=H_1^R+H_1^B+H_1^Z$ is contributed by Roth, Berry and Zeeman [cf.\ \q{H1}]. In systems with spacetime-inversion $(T\inv)$ symmetry, bands are spin-degenerate and $H_1$ is a matrix with minimal dimension of two.  \\

\noi{b} $H_0$ is spin-$SU(2)$-symmetric in spinful systems with negligible spin-orbit coupling. It is therefore possible to work in a single-band basis that diagonalizes the Zeeman term, i.e., if the field points along $\vec{z}$, we work in the eigenbasis of $\hat{\sigma}^z$ where basis vectors are distinguished by the spin eigenvalue $s\in \pm 1$. In this basis, $H_1(\bk)$ is a diagonal two-by-two matrix: 
\e{H_1(\bk)\eq\diagmatrix{H_1^{+}}{H_1^-},\lin
 H_1^{\pm}(\bk)\eq H_1^R+H_1^B \mp \f{g_0\hbar^2}{4ml^2}.\la{diagH1}}
The symmetry analysis, when restricted to the $s=+1$ eigenspace (we could also have picked the $-1$ subspace, it matters not), is considerably simplified. The spin-restricted set of Bloch functions $\psi_{\bk}(\br,s=+)$ form an integer-spin representation of the (magnetic) space group; the scalar Hamiltonians $H_1^R$ and $H_1^B$ are defined, just as in \qq{H1berry}{H1orbmag}, but with respect to Bloch functions in one spin eigenspace. Since we are ignoring the full spinor structure of the Bloch functions $\psi_{\bk}(\br,s\in \pm )$, we might colloquially refer to $\psi_{\bk}(\br,+)$ as `single-spin' Bloch functions, and $H_1^R+H_1^B$ as the `spin-independent' first-order-effective Hamiltonian. A symmetry operation in the (magnetic) space group that preserves the eigenvalue of $\hat{\sigma}^z$ is described as a single-spin symmetry. For example, while  time reversal flips spin and is represented by $\hat{T}=-i\hat{\sigma}_y K$ satisfying $\hat{T}^2=-I$, we may define an single-spin time-reversal operator $T'$ that preserves $S_z$ by composing $T$ with a $\pi$-spin rotation about the $\vec{y}$ axis: $\hat{T}'=K$ squares to identity. While $T$ constrains the full spin-dependent effective Hamiltonian [$H_1$ in \q{diagH1}], it is $T'$ that constrains the spin-independent Hamiltonian $H_1^R+H_1^B$. Rather than carry around two symbols ($T'$ and $T$) for time reversal, it is simpler to talk about a single time reversal ($T$) which is represented on integer or half-integer spins, which we distinguish by $F\in \{0,1\}$: $\hat{T}^2=(-1)^{\sma{F}}$.\\  

\noi{c} Charge-neutral, cold-atomic systems are characterizable by the Berry phase and the Roth orbital moment: $H_1=H_1^R+H_1^B$; we shall leave out of this discussion the Zeeman effect.   Bloch functions of bosonic atoms (in optical lattices) form an integer-spin representations of the (magnetic) space group.\\


In all cases (a-c), the form of \q{H1transformssymmetry} may be motivated from the following two arguments: \\

\noi{i} A field-free Bloch Hamiltonian having the same symmetry ($g$) transforms in nearly the same way to the \emph{first} term on the right-hand-side of \q{H1transformssymmetry} [recall \q{symmetryofH0}]; the only difference is an additional factor of  $(-1)^{s+u}$ (which may be trivial) in \q{H1transformssymmetry}. To understand this phase factor, let us consider an alternative definition of a  symmetry of the orbit configuration ($g$) which is consistent with the original definition in \s{sec:symmetryoforbits}: it is an element of the space group (or magnetic space group) which induces a coordinate transformation where the magnetic field $\bB \rightarrow  (-1)^{s+u}\bB$. In fact, the field-on Hamiltonian [\q{definefieldonhamiltonian}-(\ref{definefieldonpaulihamiltonian})] is invariant under $g$, if $g$ acts not only on the electronic degrees of freedom in the solid through \q{definesg}, but also on the magnetic field.\cite{sakurai_modern} Since \q{H1transformssymmetry} describes a symmetry relation between electronic wavefunctions of the solid at a fixed field, we expect a compensating factor of $(-1)^{s+u}$.\\

\noi{ii} The second term originates from the transformation of the Berry term, which we recall from \q{H1multiband2} as being proportional to $\epsilon_{\ab}\mx^{\beta}v^{\alpha}$. Applying \q{gactsonu} to the definition of the non-abelian Berry connection in \q{defineberryconnection}, 
\e{ \mx^{\alpha}\bigg|_{g\circ\bk} 
\eq  \check{g}_{\ab}\, \bigg(  \breve{g}K^{s(g)}\mx^{\beta}K^{s(g)}\breve{g}^{\mo}\lin
 &+ i(-1)^{s(g)}\,\breve{g} \nabk^{\beta}\breve{g}^{\mo}\bigg)\bigg|_{\bk} +\delta^{\alpha}. \la{symmetryconstraintberryconn}}
The derivation of the above equation is aided by two identities [\q{gactsonu3} and (\ref{ginverseactsonu2})] proven in \app{app:symmetryinbloch}.  The first (resp. second) term in \q{symmetryconstraintberryconn} contributes to the first (resp. second) term in \q{H1transformssymmetry}. \\



 Supposing the second and third terms in \q{H1transformssymmetry} were absent, we say that $H_1$ transforms covariantly under $g$. Given that the sewing matrix for a symmetry $g$ depends on the basis chosen for the cell-periodic functions [cf.\ \qq{generalbasistransformsew}{basistransformsew}], one may ask if a basis exists where $H_1$ transforms covariantly under $g$ for all $\bk$ in the Brillouin torus. If the answer is yes, such a basis may be exploited to derive symmetry constraints on the Landau levels with relative ease. In short, the answer is no for a large class of band subspaces; we devote \app{app:no} to a self-contained elaboration of `no', which originates from an obstruction in topologically-nontrivial band subspaces. Some well-known obstructions forbid the construction of exponentially-localized Wannier functions\cite{TKNN} (i.e., global sections of the vector bundle), or symmetry-invariant Wannier functions.\cite{alexey2011} In \app{app:no}, we would describe a novel type of obstruction -- to symmetry covariance of $H_1$. The reader who is more interested in quantization conditions may transit immediately to the next section [\s{sec:symmetryunitarygenbyH1}].

\subsection{Symmetry of the first-order effective propagator} \la{sec:symmetryunitarygenbyH1}

In the last section we dealt primarily with the symmetry and gauge transformations of the first-order effective Hamiltonian $H_1$, and argued that $H_1$ generically does not transform non-covariantly. A related observation is that the eigenspectrum of $H_1$ has no gauge-invariant meaning. This reflects how the effective-Hamiltonian description of a Bloch electron in a magnetic field is fundamentally a nondynamical gauge theory; in gauge theories, a known source of gauge-invariant observables comes from the spectrum of Wilson-loop operators.\cite{wilczek1984} In our context, we identify the analogous operator as 
the propagator $\cala$, which we defined in \q{definenonabelianunitary} as the unitary generated by $H_1$ over the cyclotron period.\\

We would show that, unlike $H_1$, $\cala$ behaves nicely under gauge and symmetry transformations. Precisely, we  would show that $\cala$ transforms covariantly under the $U(D)$ gauge transformation of the type \q{basistransf} [in \s{sec:gaugecovariancepropag}], and covariantly under symmetry transformations of the type \q{H1transformssymmetry} [in \s{sec:symmetrycovariancepropag}]. One motivation for investigating these transformation behavior is that $\cala$ encodes the subleading corrections to  the Bohr-Sommerfeld quantization conditions, as derived in \q{sec:singlebandquant}. As we will show, the gauge covariance of $\cala$ implies the gauge invariance of the Landau levels determined from the quantization conditions; the symmetry covariance of $\cala$ implies certain symmetry constraints for the Landau levels that we will prove below.\\

We will use the same symbols ($\cala$ and $H_1$) in a variety of contexts which are not necessarily mutually exclusive:  (i) nondegenerate subspaces ($D=1$), in which case $\cala$ is a unimodular phase factor, (ii) degenerate subspaces ($D>1$), in which case $\cala$ is a $D\times D$ unitary, (iii) spin-orbit-coupled systems, (iv) spinful systems with negligible spin-orbit coupling, and (v) charge-neutral particles coupled to effective magnetic fields. We remind the reader that $H_1$ has a slightly different meaning in each of (iii-v), as detailed below \q{H1transformssymmetry}. Unless $D$ or the symmetry representation is explicitly specified in an equation, the reader may safely assume that the equation applies to all of (i-v). \\

We highlight one potentially confusing case where it is useful to have two related notions of $\cala$: this is the case of spin-degenerate ($D=2$) bands in solids with negligible spin-orbit coupling. We define the spin-dependent $\cala_{F=1}$ as the unitary generated by the spin-dependent $H_1$ [the two-by-two matrix in \q{diagH1}], and the spin-independent $\cala_{F=0}$ as the unitary generated by the spin-independent, scalar $H_1^R+H_1^B$ [defined in \q{diagH1}]. Both notions are related as
\e{ &\cala_{\sma{F=1}} = \diagmatrix{\cala_{\sma{F=0}}e^{i\pi  \f{g_0}{2}\f{m_c[\frako]}{m}}}{\cala_{\sma{F=0}}e^{{-}i\pi  \f{g_0}{2}\f{m_c[\frako]}{m}}};\lin
 &\det \cala_{F=1}= \cala_{F=0}^2, \la{spindependentH1}}   
where the only spin-dependent component of $\cala_{F=1}$ originates from the Zeeman coupling [$g_0, m_c,m$ are defined in \q{singlequantrulewithzeeman}]. The right equation in \q{spindependentH1}  expresses how the determinant of $\cala_{F=1}$ is fully determined by the spin-independent $H_1^R+H_1^B$, and is not affected by the Zeeman splitting. Supposing $g$ is a symmetry of the orbit configuration, it is represented differently when it acts on $\cala_{F=1}$ vs $\cala_{F=0}$, e.g., time-reversal is represented as $\hat{T}^2=(-1)^F$, as explained below \q{diagH1}.

\subsubsection{Gauge-covariance of the first-order propagator} \la{sec:gaugecovariancepropag}

One motivation to prove that the first-order propagator transforms covariantly: it follows that the spectrum obtained from the multi-band quantization condition in \q{multibandquantizationcondition} is gauge invariant; we remind the reader that the gauge invariance of the single-band quantization condition has been proven with less effort in \s{sec:singlebandquant}. \\

To substantiate the multi-band claim, we remind the reader that a matrix transforms covariantly if it is conjugated by the unitary $V$ which reshuffles the bands within $P$ [cf.\ \q{basistransf}], as exemplified by the Roth and spin matrices [defined in \q{definerothoneform}, \q{definespinhalf} and \q{restriction}]
\e{ \orb \rightarrow V^{-1}\orb V, \as \bsigma \rightarrow V^{-1}\bsigma V.\la{rothspintransform}}
In the single-band case, $V$ is a commuting phase factor that cancels with $V^{-1}$, hence all covariant objects are also invariant. 
In contrast, the Berry connection transforms non-covariantly, as shown in \q{gaugetransform}. Nevertheless, we would show that the propagator around a loop transforms as  
\e{\cala[\frako] \rightarrow V(\bk(0))^{\mo}\cala[\frako]V(\bk(0)), \la{covarianceprop}}
 with $\bk(0)$ the base point of the loop $\frako$. It would follow that the quantization condition in \q{multibandquantizationcondition} is gauge-invariant:
\e{ &0= \det\left[ e^{il^2S}\cala[\frako]+I \right] \lin
&\rightarrow\; \det\left[ e^{il^2S}V(\bk(0))^{\mo}\cala[\frako]V(\bk(0))+I \right]\lin
\eq \det\left[V(\bk(0))^{\mo}\right]\det\left[ e^{il^2S}\cala[\frako]+I \right]\det\left[V(\bk(0))^{\mo}\right].\notag}
To prove \q{covarianceprop}, it is convenient to consider the propagator 
\e{ &\A[\bk \leftarrow \bk-d\bk] \lin
\appr \exp\left[i (\orb+\bmx) \cdot  d\bk +i\sigma^z(Z/v^{\sma{\perp}})dk\right] +O(dk^2)}
over an infinitesimal path along the orbit, ending at $\bk(t)$ and beginning at $\bk(t-\delta t)=\bk(t)- d\bk(t)$; in short, we call such objects infinitesimal propagators. Applying \q{rothspintransform} and (\ref{gaugetransform}), the infinitesimal propagator transforms as 
\e{&\A [\bk \leftarrow \bk-d\bk] \lin
\rightarrow&\; e^{i V^{-1}(\orb+\bmx)V \cdot d\bk +iV^{-1}\sigma^z(Z/v^{\sma{\perp}})Vdk - V^{-1}\nabk V \cdot d\bk}\lin
\eq V^{-1}(\bk)e^{i(\orb+\bmx)\cdot d\bk +i\sigma^z(Z/v^{\sma{\perp}})dk}V(\bk) e^{-V^{-1}\nabk V \cdot d\bk}\lin
\eq  V(\bk)^{-1}\A[\bk \leftarrow \bk-d\bk]  V(\bk-d\bk), \la{infAtransforms}}
to linear order in $dk$. Consider a path-ordered multiplication of these infinitesimal propagators around a closed orbit beginning and ending at $\bk(0)$; every $V$ matrix that is not evaluated at $\bk(0)$ is multiplied with its inverse. What remains of this path-ordered product, after taking the limit  $\delta t\rightarrow 0$, is the right-hand-side of \q{covarianceprop}.

\subsubsection{Symmetry-covariance of the first-order propagator}\la{sec:symmetrycovariancepropag}

For a system having a symmetry ($g$) of the orbit configuration, we consider the infinitesimal propapagator centered at wavevector $g\circ\bk$ on an orbit, which is related through \q{H1transformssymmetry} to the infinitesimal propagator centered at  $\bk$:
\e{ e^{-iH_1\delta t/\hbar}\bigg|_{g\circ\bk} = \bigg(\breve{g} K^{s} e^{-i(-1)^{u}H_1\delta t/\hbar} K^{s} \breve{g}^{-1}\bigg)\bigg(e^{(-1)^{u}\lmt\epsilon_{\ab}\,\breve{g} \nabk^{\beta}\breve{g}^{\mo}v^{\alpha}\delta t/\hbar}\bigg)\bigg|_{\bk}e^{-i\lmt \epsilon_{\ab}\delta^{\beta}v^{\alpha} \delta t/\hbar}\bigg|_{g\circ\bk} +O(\delta t^2). \la{U1transformssymmetry}}
Hamilton's equation of motion [\q{hamiltoneom} particularized to $\bB=-B\vec{z}$] informs us that $\delta t/\hbar l^2=-\delta k_y(\bk)/v^x(\bk)=\delta k_x(\bk)/v^y(\bk)$ [let us also define $\delta \bk^{\sma{\perp}}(\bk)=(\delta k_x,\delta k_y)$], hence the above equation simplifies to
\e{ e^{-iH_1\delta t/\hbar}\bigg|_{g\circ\bk} \eq \bigg(\breve{g} K^{s} e^{-i(-1)^{u}H_1\delta t/\hbar} K^{s} \breve{g}^{-1}\bigg)\bigg(e^{-(-1)^{u}\breve{g} \nabk \breve{g}^{\mo} \cdot \delta \bk^{\sma{\perp}}}\bigg)\bigg|_{\bk}e^{i\bdelta \cdot \delta \bkp}\bigg|_{g\circ\bk}+O(\delta t^2)\lin
\eq  \bigg(\breve{g} K^{s} e^{-i(-1)^{u}H_1\delta t/\hbar} K^{s}\bigg)\bigg|_{\bk} \breve{g}^{-1}\bigg|_{\bk-(-1)^{u}\delta \bkp(\bk)}e^{i\bdelta \cdot \delta \bkp}\bigg|_{g\circ\bk} +O(\delta t^2).\la{U1transformssymmetry2}}
If $u(g)=1$, the infinitesimal-time propagator centered at $\bk$ [the right-hand-side of the above equation] is reversed in orientation with respect to the semiclassical orbit, and vice versa. \q{U1transformssymmetry2} forms the basis to derive the symmetry constraints for any configuration of orbits. \\

Let us translate the symmetry constraint on infinitesimal propagators into a constraint for finite-time propagators, which are analogous to Wilson lines. Suppose $\bk_i$ and $\bk_f$ are boundary points of a curved line segment ($\cals$) contained within an orbit; $\cals$ is equipped with an orientation such that $\bk_i$ [resp.\ $\bk_f$] is the initial point [resp.\ final point], which we will denote as $\cals: \bk_f \leftarrow \bk_i$. The propagator over $\cals$ is defined by
\e{ \cala[\cals:\bk_f\leftarrow \bk_i]:= \overline{\exp}\bigg\{-i\int_{t(\bk_i)}^{t(\bk_f)}H_1\big(\bk(t)\big) \f{dt}{\hbar}\bigg\}, \la{defwilsonline}}
with $\overline{\exp}$ a path-ordered exponential, and $\bk(t)$ and its inverse $t(\bk)$ determined by Hamilton's equation. We may define $-\cals$ as the same line segment as $\cals$ but with the opposite orientation; the corresponding propagator satisfies
\e{ \cala[-\cals:\bk_i \leftarrow \bk_f]= \cala[\cals:\bk_f\leftarrow \bk_i]^{-1}.}
Let us  define the symmetry-mapped segment ($g\circ  \cals$) as being bounded by initial point $g\circ  \bk_i$ and final point $g \circ  \bk_f$ [with $g\circ  \bk :=(-1)^s\check{g}\bk$]. The two corresponding segment propagators are related as:
\e{  \cala[g\circ \cals: g\circ  \bk_f \leftarrow g\circ  \bk_i]\eq\breve{g}(\bk_f)\,K^s\;\cala[\cals:\bk_f\leftarrow \bk_i]\;K^s\,\breve{g}^{-1}(\bk_i)\,e^{i\bdelta \cdot \int_{g\circ \cals}d\bkp}, \la{wilsonlineconstraint}}
which may be derived by a path-ordered multiplication of the infinitesimal propagators in \q{U1transformssymmetry2}, and taking the limit $\delta t\rightarrow 0$; in this process, every sewing matrix  (originating from the right-hand-side of \q{U1transformssymmetry2}) is multiplied with its inverse, except for the sewing matrices at the boundary points.\\

 Let us generalize \q{wilsonlineconstraint} to a relation between propagators over closed orbits. If $\frako$ is a closed orbit with base point $\bk_1$, then
\e{ \frako: \bk_1 \leftarrow \bk_1, \as  \cala[g\circ \frako]=\breve{g}(\bk_1)\,K^s\;\cala[\frako]\;K^s\,\breve{g}^{-1}(\bk_1). \la{wilsonloopconstraint}}
Note that $\int_{g\circ \frako}d\bkp=\bze$ for a closed orbit, hence the $\bdelta$-dependent phase factor on the right-hand-side of \q{wilsonlineconstraint} is trivial. When particularized to the case that $g\circ \frako$ and $\frako$ are identical orbits, up to a reversal in orientation that depends on $u(g)$: 
\e{g\circ \frako=(-1)^u\frako,  \as  \cala[\frako]^{(\mo)^{u}}=\breve{g}(\bk_1)\,K^s\;\cala[\frako]\;K^s\,\breve{g}^{-1}(\bk_1), \la{wilsonloopconstraintsameorbit}}
where $\cala^{\sma{(\mo)}^{\sma{u}}}$ equals $\cala$ [resp.\ $\cala^{\sma{\mo}}$] if $u=0$ [resp.\ $u=1$]. A topologically distinct possibility is that $g\circ \frako$ and $\frako$ are disconnected orbits. Let us define $\frako_1$ and $\frako_2$ as two disconnected orbits, whose orientations are determined by Hamilton's equation; $\frako_1$ is equipped with a base point $\bk_1$, and $\frako_2$ with a base point $\bk_2:=g\circ \bk_1$. Then a simple generalization of \q{wilsonloopconstraintsameorbit} provides us with
\e{&g\circ \frako_1=(-1)^u\frako_2,  \as \cala[\frako_2]^{(\mo)^{u}}=\breve{g}(\bk_1)\,K^s\;\cala[\frako_1]\;K^s\,\breve{g}^{-1}(\bk_1). \la{wilsonloopconstraintdifforbit}}

\subsubsection{Ten classes of closed, elementary orbits}\la{sec:tenfoldway}

In a Brillouin two-torus (\bt), any closed orbit configuration  possessing a symmetry $g$ may be divided into a set of elementary orbits ($\{E_i\}$). An elementary  orbit  $E_i$ is defined to be the smallest possible closed orbit configuration that is closed under $g$, i.e., it cannot further be divided into smaller configurations which are closed under $g$. To clarify two distinct notions, `closed orbits' do not wrap around \bt; if $E_i$ is `closed under $g$', we mean that for every $\bkp \in E_i$, $g \circ \bkp \in E_i$ as well. We remind the reader that $g$ maps $\bkp$ to $g \circ \bkp:=(-1)^{s(g)}\check{g}^{\sma{\perp}}\bkp$, with $\check{g}^{\sma{\perp}}$ the point-group component of $g$, as restricted to $BT_{\sma{\perp}}$.  Generally, $E_i$ is composed of one or more closed orbits.\\

   If there are multiple symmetries ($g,g',\ldots$) in the group of the orbit configuration, the same orbit configuration may be divided into two (or more) distinct sets of elementary orbits ($\{E_i\}$ and $\{E'_i\}$), which are closed under $g$ and $g'$ respectively; each elementary orbit is therefore defined by its closure under a \emph{single} symmetry, and we emphasize this by the paired notation $(g,E_i)$. The motivation for this $g$-centric organization is that distinct symmetries impose distinct constraints on the propagators, which we classify into ten (and only ten) classes.  In other words, any pair $(g,E_i)$ falls into one of ten classes, and the propagator(s) over closed orbits $(\in E_i)$ satisfy one of ten classes of contraints, which we summarize in \tab{tab:tenfold} below. These ten classes were first introduced in \ocite{topofermiology} and exemplified by many existing materials; here we present a more mathematically-oriented discussion that emphasizes the group-theoretic aspects of the ten classes. In addition, (i) a more thorough discussion is also provided for $g'$-symmetric orbit configurations that lie within mirror/glide-invariant planes, where $g'$ is an additional symmetry distinct from said mirror/glide. (ii) We also demonstrate how, with additional input about the symmetry representations of Bloch functions on the orbit, one may derive additional constraints on the spectrum of $\cala$ that go beyond \tab{tab:tenfold}. \\

\begin{table}[H]
	
\centering
		
\begin{tabular} {|r|c|c|l|l|l|l|} \cline{2-7}
			
\multicolumn{1}{c}{} &\multicolumn{1}{|c}{$u(g)$}&  \multicolumn{1}{|c}{$s(g)$} &  \multicolumn{1}{|c}{Constraint on $\cala$} & \multicolumn{1}{|c}{Spectrum of $\cala$} & \multicolumn{1}{|c}{Representation of $g$} & \multicolumn{1}{|c|}{Ex. of $g$} \\  \hline \hline 
		  
(I)$\as\as \forall \; \bkp,$ & $0$& $0$ &  $\cala=\breve{g}\cala\breve{g}^{\mo}$   & $\sigma(\cala){=}\sigma_{\sma{+}}{\sqcup}\,\sigma_{\sma{-}}$ &  $\breve{g}^{\sma{2}}{=}e^{\sma{i\pi Fa \minus i\bk_1 \cdot \bR}}$ & $\mir_z,\glide_{z,\vec{x}/2}$     \\ \cline{2-7}			 

$\bkp{=}g{\sma{\circ}}\bkp$ & $0$& $1$ &  $\cala=\breve{g}\cala^*\breve{g}^{\mo}$   & $\sigma(\cala)=\sigma(\cala)^*$ & $(\breve{g}K)^{\sma{2}}{=}e^{\sma{i\pi Fa \minus i\bk_1 \cdot \bR}}$  & $T\inv,T\rot_{2z}$     \\ \hline

\multicolumn{1}{|l|}{(II-A)} & $0$& $0$ &  $\cala=\tilde{g}\cala\tilde{g}^{\mo}$    & \multicolumn{1}{c|}{$-$}  &     $\tilde{g}^{\sma{N}}{=}\cala^{p}e^{\sma{i\pi Fa}}$    & $\inv,\rot_{nz}$     \\ \cline{2-7}			 
			 
$\bkp \in \frako,$  & $0$& $1$ &  $\cala=\tilde{g}\cala^*\tilde{g}^{\mo}$   &  $\sigma(\cala)=\sigma(\cala)^*$ &    $(\tilde{g}K)^{\sma{N}}{=}\cala^{p}e^{\sma{i\pi Fa}}$  & $T,T\rot_{6z}$     \\ 			 \cline{2-7}

$|\frako| = |g{\sma{\circ}}\frako|$ & $1$& $0$ &  $\cala =\breve{g}\cala^{\mo}\breve{g}^{\mo}$   & $\sigma(\cala)=\sigma(\cala)^*$ &  $\breve{g}^{\sma{N}}{=}e^{\sma{i\pi Fa \minus i\bk_1 \cdot \bR}}$  & $\mir_x,\mir_y$     \\ \cline{2-7}			 

 & $1$& $1$ &  $\cala =\breve{g}\cala^{t}\breve{g}^{\mo}$   & \multicolumn{1}{c|}{$-$}  &  $(\breve{g}K)^{\sma{N}}{=}e^{\sma{i\pi Fa \minus i\bk_1 \cdot \bR}}$   &  $T\mir_x,T\mir_y$     \\ \hline

\multicolumn{1}{|l|}{(II-B)} & $0$& $0$ &  $\cala_2=\breve{g}_{\sma{1}}\cala_1\breve{g}_{\sma{1}}^{\mo}$   & $\sigma(\cala_2)=\sigma(\cala_1)$ &  $\breve{g}_{\sma{N}}\ldots \breve{g}_{\sma{1}}{=}e^{\sma{i\pi Fa \minus i\bk_1 \cdot \bR}}$   & $\rot_{nz}$     \\ \cline{2-7}			 
			 
$\bkp \in \frako,$ & $0$& $1$ &  $\cala_2=\breve{g}_{\sma{1}}\cala_1^*\breve{g}_{\sma{1}}^{\mo}$   & $\sigma(\cala_2)=\sigma(\cala_1)^*$ &  $\breve{g}_{\sma{N}}K\ldots \breve{g}_{\sma{1}}K{=}e^{\sma{i\pi Fa \minus i\bk_1 \cdot \bR}}$  & $T$     \\ \cline{2-7}			 

$|\frako| \neq  |g{\sma{\circ}}\frako|$ & $1$& $0$ & $\cala_2=\breve{g}_{\sma{1}}\cala^{\mo}_1\breve{g}_{\sma{1}}^{\mo}$    & $\sigma(\cala_2)=\sigma(\cala_1)^*$ &     $\breve{g}_{\sma{N}}\ldots \breve{g}_{\sma{1}}{=}e^{\sma{i\pi Fa \minus i\bk_1 \cdot \bR}}$  &  $\mir_x,\mir_y$     \\ \cline{2-7}			 

										& $1$& $1$ &  $\cala_2=\breve{g}_{\sma{1}}\cala^t_1\breve{g}_{\sma{1}}^{\mo}$   & $\sigma(\cala_2)=\sigma(\cala_1)$ &   $\breve{g}_{\sma{N}}K\ldots \breve{g}_{\sma{1}}K{=}e^{\sma{i\pi Fa \minus i\bk_1 \cdot \bR}}$  &    $T\mir_x,T\mir_y$     \\ \hline

\end{tabular}
		
\caption{The first column distinguishes distinguishes between three topologically distinct mappings of $g:\bk \rightarrow g\circ \bk$, as summarized in \qq{caseequal}{casenotequal}. The second and third columns subdivide the three mapping classes according to two $\Z_2$ indices defined in \q{definesg} and (\ref{defineug}); this gives ten classes in total. Fourth column describes the constraints on the propagator: for class I and II-A (top six rows), $\cala$ is the propagator for a single, elementary orbit $\frako$; for class II-B (last four rows), $\{\cala_j\}_{j=1}^2$ is shorthand for $\cala[\frako_j]$, with $\frako_{j+1}=g\circ \frako_j$ being symmetry-related, closed orbits. $\breve{g}$, $\{\breve{g}_i\}_{i=1}^N$ and $\tilde{g}$ are representations of the point group generated by $g$ [cf.\ \qq{definegorbit}{orderNginv}], as summarized in the sixth column.  The fifth column describes the constraint imposed by $g$ on the spectrum of $\cala$, which we denote by $\sigma(\cala)$. We indicate the lack of a symmetry constraint with a $-$. $\sigma(\cala)=\sigma(\cala)^*$ means the spectrum is invariant under complex conjugation; it follows immediately that det$\,\cala=\pm 1$. In some cases, the sign of this determinant is fully determined by specifying the band degeneracy ($D$) and the symmetry representation of the Bloch functions (whether integer- or half-integer-spin).  Class-I symmetries have order two, and $\sigma(\cala)=\sqcup_{i\in \pm}\sigma_{i}$ indicates that $\cala$ is block-diagonal with respect to the two representations of the order-two symmetry. The last column lists some representative examples of the ten symmetry classes; the symbolic notation of various symmetries have been summarized in \s{sec:symmetryinBloch}.
	\label{tab:tenfold}}
\end{table}

The ten classes are partially distinguished by two $\Z_2$ indices which we have previously defined: $u(g)$ and $s(g)$. To remind the reader, $s(g)=1$ if $g$ contains a time reversal, and $0$ otherwise [cf.\ \q{definesg}]; $u(g)=0$ if the determinant of $\check{g}^{\sma{\perp}}$ [the point-group component of $g$, restricted to the $xy$-plane] equals $1$, and $u(g)=1$ if det$\,\check{g}^{\sma{\perp}}=-1$ [cf.\ \q{defineug}]; as explained in \s{sec:symmetryoforbits}, $u(g)=0$ (resp. $1$) if $g$ preserves (resp. inverts) the orientation of the semiclassical orbit. \\

We shall subdivide the ten classes according to three topologically distinct mappings of $g: \bkp \rightarrow g\circ \bkp$ as
\bal
 \ins{(I)} &\forall \;\bkp, \as \bkp =g \circ \bkp, \la{caseequal} \\
    \ins{(II)} &\text{Generically,} \as \bkp \neq g \circ \bkp; \as \bkp \in \frako, \as \begin{cases} \text{(II-A)} \as |\frako| = |g\circ \frako|;  \\  \text{(II-B)} \as |\frako| \neq |g\circ \frako|.\end{cases}  \la{casenotequal} 
\end{align}
In mappings of class I, all wavevectors in \bt are individually invariant under the symmetry, which implies that $u(g)=0$, as proven in \q{ugequal0}. There are therefore two classes of class-I elementary orbits   which we distinguish by $s(g)\in \Z_2$.  For mappings of class II, generic wavevectors are not invariant under $g$, but there exist closed submanifolds (isolated points/lines) of \bt which are invariant; we have shown in \s{sec:symmetryoforbits} that points occur iff $u=0$, and lines iff $u=1$. Suppose $\bkp$ is a point in a closed orbit $\frako \in E_i$; since $E_i$ is closed under $g$, $g\circ \bkp \in g\circ \frako \in E_i$. We further distinguish between mappings where $g\circ \frako$ is identical to $\frako$ up to orientation [class II-A], or they are disconnected orbits [class II-B]. We employ the notation that $\frako$ and $-\frako$ have opposing orientations, and $|\frako|$ as having no orientation. The defining characteristics of II-A and II-B may then be  expressed as in \q{casenotequal}. In class I and II-A, $E_i$ is composed of a single orbit $\frako$, and we may say that $\frako$ is self-constrained by $g$; in II-B, $E_i$ is composed of at least two closed orbits, which we say are mutually constrained by $g$. For class-II mappings, there are no constraints on $s$ or $u$ [as there was for class I in \q{ugequal0}], hence there are four classes for each of II-A and II-B. This gives ten classes of elementary orbits in total, whose defining characteristics are summarized in the first three columns of \tab{tab:tenfold}; representative examples of each class are given in the last column.\\

The rest of the table summarizes how the space-group symmetry $g$ constrains the propagators $\cala$; the operators (denoted by $\breve{g}$ in eight rows, and by $\tilde{g}$ in two) that constrain $\cala$ form a representation of the space-group symmetry $g$, as shown in column six.  $\breve{g}$  form either a linear or projective\cite{weinbergbook1} representation of the point group ($P_g$) generated by $g$,\footnote{The last four rows do not describe a standard representation of a point group, owing to the nontrivial action of $g$ in $\bk$-space. To map the equation to a more standard representation, we may collect  $\{\breve{g}_i\}_{i=1}^N$ as blocks in a single permutation matrix representing an $N$-cycle. This larger matrix indeed represents $g$ in $P_g$.  With slight abuse of language, we will anyway refer to $\breve{g}_i$ as a representation of $P_g$.} while $\tilde{g}$ forms necessarily a projective representation. To clarify this comment, $P_g$ is isomorphic to $\Z_N$ if $g$ has order $N$; generally, $P_g$ is a subgroup of the full point group of the space group. It is well-known that symmorphic (resp.\ nonsymmorphic) space groups are split (resp.\ unsplit) extensions of point groups by discrete spatial translations.\cite{cohomologyhiller} Unsplit extensions may contain nonsymmorphic elements of order $N$ -- the corresponding multiplication rule is represented by $\breve{g}^N \propto e^{-i\bk_1\cdot \bR}$. Double space groups are known to correspond to a further extension by a $2\pi$ spin rotation; the multiplication rule for an order-$N$ symmetry is represented by $\breve{g}^N \propto e^{iF\pi a}$. These two observations explain the form of the multiplication rules in all ten rows except for the third and fourth, where respectively $\tilde{g}^N$ and $(\tilde{g}K)^N$ are proportional to $\cala^p$, with $p \neq 0$ (mod $N$) and  depending on $(g,E_i)$. These two rules represent an unsplit extension of the point group by quasimomentum translations around the orbit $E_i$, which in the present context is a single loop; these translations are represented by the propagator $\cala$, which generates a normal subgroup (isomorphic to $\Z$) of the extension. Extensions by quasimomentum loop translations are one key result of this section, and occur for all self-constrained orbits having no $g$-invariant points -- this sharply delineates class-II-A orbits with $u=0$ from the remaining eight classes, which are all linearly represented with respect to $\cala$. To recapitulate, $\breve{g}$ (or $\tilde{g}$), $e^{i\bk_1\cdot \bR}$, $e^{i\pi}$ and $\cala$ generate a group; (i) the multiplication rules of this group [columns four and six], when combined with (ii) the spectral constraint on $\cala$ [column five], uniquely distinguishes each of the ten classes. In other words, given (i-ii), one may uniquely determine the corresponding mapping type (I,II-A,II-B), $u$ and $s$. 
We derive the table and discuss its implications in the following subsections, which are divided according to class-I mappings [\s{sec:classI}], class-II-A [\s{sec:classIIA}] and class-II-B [\s{sec:classIIB}]. For some (and only some) classes, the above spectral constraints are further strengthened when given additional data about the band degeneracy $D$ and the spin representation (whether integer or half-integer).\\

One last remark regards the application of \tab{tab:tenfold} beyond the semiclassical theory of magnetotransport. All constraints in \s{sec:tenfoldway}-\ref{app:suppargument} which are  tabulated or expressed in labelled equations remain valid if we substitute $\cala \rightarrow \W$, with $\W$ the purely-geometric component of $\cala$. That is, if we set the Roth and Zeeman Hamiltonians to zero, $\cala$ reduces to $\W$ -- a path-ordered exponential of the  Berry connection,\cite{berry1984} which is non-abelian for $D>1$. Though generically the spectra of $\cala$ and $\W$ are distinct, they satisfy the same type of constraints, e.g., if $\sigma(\cala)=\sigma(\cala)^*$ from \tab{tab:tenfold}, so would $\sigma(\W)=\sigma(\W)^*$.  $\W$ is the matrix representation of holonomy around the orbit $\frako$,\cite{barrysimon_holonomy} and has been called the Wilson loop of the Berry gauge field.\cite{wilczek1984}  The commonality between  $\W$ and $\cala$ originates from their identical transformation behavior under symmetry [cf.\ \qq{wilsonlineconstraint}{wilsonloopconstraintdifforbit}]. The Wilson loop is a basic geometric characterization of bands that is intimately related to the topology of wavefunctions over the Brillouin torus.\cite{barrysimon_holonomy} 

 

\subsubsection{Class-I elementary orbits}\la{sec:classI}

Let $g$ be a symmetry such that every wavevector ($\bkp$) in a Brillouin two-torus \bt is $g$-invariant. Common examples include $g=T\inv, T\rot_{2z},$ and $\mir_z$; for 3D $T\inv$-symmetric solids, any two-torus embedded in the 3D Brillouin zone is invariant under $T\inv$, while for 3D  solids  with either $T\rot_{2z}$ or $\mir_z$ symmetry, we would particularize to the high-symmetry planes ($k_z=0$ and $\pi$).\\

Since every $\bkp \in$ \bt is $g$-invariant, if $\bkp \in \frako$ (a single closed orbit), then $\frako=g\circ\frako$, which further implies $\frako$ is itself an elementary orbit [of class-I, as classified in \qq{caseequal}{casenotequal}]. If $g$ is the only symmetry of $\frako$, there is no contraint on the shape of $\frako$.  We have also proven that $u(g)=0$ in \q{ugequal0}, i.e., that class-I symmetries are orientation preserving. We further subdivide class-I orbits according to whether $g$ includes a time reversal or not [$s(g)=1$ or $0$ respectively]; $s$ distinguishes between two classes of constraints on the propagator $\cala[\frako]$ over the oriented $\frako$. In contexts where we are discussing a single orbit, we employ  $\cala$ as a shorthand for $\cala[\frako]$.\\

\begin{center}
\underline{Class-I elementary orbits with $s(g)=0$}
\end{center}

This occurs when $g$ is purely a spatial transformation; we ignore $g$ that is purely a spatial translation, because they trivially constrain the propagator. To leave every wavevector in \bt invariant, \bt must be a mirror ($g=\mir_z$) or a glide (e.g., $g=\glide_{z,\vec{x}/2}$) plane. In either case, $g$ is an order-two spatial symmetry [the order of a symmetry is defined in \qq{defineorderg}{orderNginv}], which implies that $g$ has two distinct representations. It is useful to block-diagonalize the Hilbert space ($\call$) with respect to the two representations of $g$; we shall denote this decomposition as $\call=\call_+\oplus \call_-$. The corresponding block-diagonalization of $\cala$ is denoted as $\cala=\cala_+\oplus \cala_-$ in the first row of \tab{tab:tenfold}. \\

Suppose there exists a distinct symmetry $g'$ in the group of the orbit configuration, whose operation preserves the decomposition $\call_+\oplus \call_-$. That is, if a Bloch function $\psi \in \call_{+}$, then the symmetry-mapped Bloch function $g'\circ \psi$ belongs also in $\call_+$. To analyze how $g'$ further constrains $\cala_{\pm}$, we would divide the orbit configuration into elementary orbits $\{(g',E_i')\}$; each of $\{(g',E_i')\}$ falls into one of the remaining nine classes.  We may then apply any of the results in the bottom nine rows of \tab{tab:tenfold}, with the understanding that $\cala$ (as denoted in the table) is the propagator restricted to $\call_{\pm}$. \\

Let us particularize to $g'$ that permutes the two representations of $g$, i.e., $g' \circ \call_{\pm}=\call_{\mp}$. Then if $\cala_+$ is the propagator for a closed orbit $\frako$, it is symmetry-related to $\cala_-$ which is the propagator for $g'\circ \frako$; in general $\frako \neq g'\circ \frako$. $\cala_+[\frako]$ and $\cala_-[g'\circ \frako]$ are mutually constrained in four possible ways, depending on the $\Z_2$ indices $u(g')$ and $s(g')$ which characterize $g'$ (not $g$); these constraints are summarized in \tab{tab:fourfold} below, which applies regardless of whether $\frako=g'\circ \frako$ or not. \tab{tab:fourfold} summarizes one new result of this work.  \\

\begin{table}[ht]
	
\centering
		
\begin{tabular} {|c|c|c|l|l|l|} \cline{2-6}
			
\multicolumn{1}{c}{} &\multicolumn{1}{|c}{$u'$}&  \multicolumn{1}{|c}{$s'$} &  \multicolumn{1}{|c}{Constraint on $\cala_{\pm}$} & \multicolumn{1}{|c}{Spectrum of $\cala_{\pm}$} & \multicolumn{1}{|c|}{$g'$} \\  \hline \hline

 & $0$& $0$ &  $\cala_+=\breve{g}'\cala_-({\breve{g}'})^{\mo}$   & $\sigma(\cala_+)=\sigma(\cala_-)$ & $\scr_{2z,\vec{z}/2}$     \\ \cline{2-6}			 
			 
$g' {\sma{\circ}} \call_{\pm}$ & $0$& $1$ &  $\cala_+=\breve{g}'\cala_-^*({\breve{g}'})^{\mo}$   & $\sigma(\cala_+)=\sigma(\cala_-)^*$ & $T,T\inv$     \\ \cline{2-6}			 

${=}\call_{\mp}$ & $1$& $0$ & $\cala_+=\breve{g}'\cala^{-1}_-({\breve{g}'})^{\mo}$    & $\sigma(\cala_+)=\sigma(\cala_-)^*$ & $\mir_x,\mir_y$     \\ \cline{2-6}			 

										& $1$& $1$ &  $\cala_+=\breve{g}'\cala^t_-({\breve{g}'})^{\mo}$   & $\sigma(\cala_+)=\sigma(\cala_-)$ & $T\glide_{x,\vec{z}/2}$     \\ \hline

\end{tabular}
		
\caption{ Table of constraints for solids with: (i) a class-I, unitarily-represented, order-two symmetry $g$, and (ii) an additional symmetry $g'$ that permutes the two representations of $g$. 
 The second and third columns classifies the constraints according to two $\Z_2$ indices [defined in \q{definesg} and (\ref{defineug})] that characterize $g'$ (not $g$). Fourth column describes the  constraints on propagators $\cala_{\pm}$ which are defined with respect to states in $\call_{\pm}$.  In the entire table, $\cala_+$ is short-hand for $\cala_+[\frako]$, and $\cala_-$ for $\cala_-[g\circ\frako]$. The sixth column lists some representative examples of $g'$, for the specific case of a half-integer-spin representation of $g=\mir_z$. For the nonsymmorphic examples of $g'$ [$\scr_{2z,\vec{z}/2}$ and $T\glide_{x,\vec{z}/2}$], $g'$ permutes the half-integer-spin representation of $\mir_z$ in the $k_z=\pi$ plane; for the remaining two symmorphic examples, this permutation occurs in both $k_z=0$ and $\pi$ planes. 
	\label{tab:fourfold}}
\end{table}

The four classes of $(g',E'_i)$ in \tab{tab:fourfold} are essentially identical to the four classes of class-II-B elementary orbits [bottom four rows of \tab{tab:tenfold}], if one relabels $\cala_{1,2} \leftrightarrow \cala_{\pm}$. The basic commonality is the existence of two distinct but symmetry-related vector bundles, each of which is defined over a 1D base space (embedded in \bt). In the case of $(g',E'_i)$, the two vector bundles are distinct because the fibres transform in different representations of the order-two symmetry $g$; in the case of class-II-B elementary orbits, the two vector bundles are distinct because their base spaces ($\frako$ and $g\circ \frako$) are distinct. Given this broader perspective, the derivation of the four classes of constraints listed in \tab{tab:fourfold} are essentially identical to those for class II-B, which may be found in \s{sec:classIIB} below.\\

\begin{center}
\underline{Class-I elementary orbits with $s(g)=1$}
\end{center}

If $g$ includes a time reversal, as exemplified by $g=T\inv$ and $T\rot_{2z}$, we  apply \q{wilsonloopconstraintsameorbit} to derive 
\bal
 \cala = \breve{g}\cala^* \breve{g}^{\mo} \imp \sigma(\cala)=\sigma(\cala)^* \imp \text{det}[\cala]= \pm 1 \la{Tinvconstraint}
\end{align}
The middle line states that the spectrum of $\cala$ is invariant under complex conjugation. That det$[\cala]=-1$ might seem surprising for a contractible orbit, especially when one recalls that the $U(1)$ Berry curvature $\calf^z(\bk)=\epsilon_{\ab}\nabk^{\alpha}$Tr$[\mx^{\beta}(\bk)]$ vanishes almost everywhere -- in the torus for the $T\inv$-symmetric case, and in the high-symmetry planes for the $T\rot_{2z}$-symmetric case. For $D=1$, the resolution is that the orbit must enclose a singularity in the curvature: the orbit is linked with an odd number of line nodes in the $T\inv$-symmetric case (a known result by Mikitik\cite{mikitik_berryinmetal}), and encircles an odd number of Dirac point in the $T\rot_{2z}$-symmetric case; the latter is exemplified by graphene, as we have substantiated in \s{sec:singlebandoscillations}. To complete the argument that det$[\cala]=-1$ in these cases, the conical dispersion around a Dirac point/line node guarantees that the velocity ($\nabk\var$) is finite at the singular point, hence the non-geometric one forms (Roth and Zeeman) negligibly contribute to $\cala$ in the limit where the area of the loop (that encircles the singular point) vanishes.\\

For spin-orbit-coupled solids with bands which are spin-degenerate ($D=2$) owing to $T\inv$ symmetry, we may rule out det$[\cala]=-1$ because all time-reversal-symmetric orbits can be continuously contracted to a point; the argument for this is presented in \s{sec:classIIATinv}. The implications of this determinantal constraint for the quantization conditions and magnetic oscillations have been discussed, around \q{rule3b} and (\ref{laj}) respectively. We remark that det$[\cala]=+1$ may be alternatively derived if $H_1$ is traceless, as we have discussed in \s{sec:AIIwithinversion}.

\subsubsection{Class-II-A elementary orbits}\la{sec:classIIA}

A class-II-A elementary orbit is a single closed orbit (denoted $\frako$), which is closed under $g$ (i.e., $g\circ \frako =\frako$). Just as for class-I orbits, we define $\cala[\frako]$ as the propagator over the oriented orbit $\frako$. At times  we may suppress the argument of $\cala$ notationally; in these cases $\cala$ should be understood as $\cala[\frako]$.

\begin{center}
\underline{Class-II-A elementary orbits with $u(g)=1$}
\end{center}

If $u(g)=1$, we have shown in \s{sec:symmetryoforbits} that $g$ acts on $\bkp$ as a planar reflection, and therefore $g$-invariant $\bkp$ form isolated lines. Since $\frako$ is closed as an orbit, it must intersect a $g$-invariant line at minimally two points. For simple, closed orbits (which are equivalent to circle), there are only two intersections, which we denote by  $\bk_a$ and $\bk_b$. There might be more intersections for nonsimple closed orbits (e.g., a figure-of-eight),  but we  shall identify the two intersection points that are furthest apart (on the $g$-invariant line) as $\bk_a$ and $\bk_b$. It is analytically convenient in derivations to let the base point of $\cala$ lie on one of these invariant wavevectors (say, $\bk_a$); we remark that the spectrum of $\cala$ is independent of the position of the base point.\cite{AA2014} Particularizing \q{wilsonloopconstraintsameorbit} to the present context, 
\e{ \cala = \breve{g} K^s\,\cala^{-1}\,K^s\, \breve{g}^{\mo} \imp \sigma(\cala)= K^{1+s}\sigma(\cala)K^{1+s},}
with the sewing matrix $\breve{g}$ evaluated at $\bk_a$. To clarify the above notation, $K^{\sma{1+s}}\sigma(\cala)K^{\sma{1+s}}=\sigma(\cala)^*$ iff $1+s$ is odd, and therefore $\sigma(\cala)$ is not constrained if $s(g)=0$. \\

To obtain another useful constraint, we might split the propagator into the product $\cala=\cala(\bk_b)\cala(\bk_a)$, where $\cala(\bk_a)$ propagates through half the orbit beginning from $\bk_a$ and ending at $\bk_b$, and $\cala(\bk_b)$ completes the orbit. The constraint between $\cala(\bk_a)$ and $\cala(\bk_b)$ in \q{wilsonlineconstraint} implies that
\e{&\cala = \cala(\bk_b)\cala(\bk_a) \lin
\eq e^{i\bdelta\cdot (g\circ \bk_a-g\circ\bk_b)}\breve{g}(\bk_a)\,K^s\,\cala(\bk_a)^{\mo}\,K^s\,\breve{g}^{\mo}(\bk_b)\,\cala(\bk_a).\la{halftohalf}}
This is an additional constraint that has not been included in \tab{tab:tenfold}. The spectra of unitaries with such a constraint have been studied by one of us in \ocite{AA2014,berryphaseTCI,JHAA}; a common theme in these works is that, for certain symmetries $\{g\}$, the spectrum of $\cala$ (or a subset thereof) may be robustly fixed to special values; the existence of such robust eigenvalues depends on the symmetry representations at the $g$-invariant wavevectors.\\

 To provide a simple illustration, we consider a simple closed orbit that is invariant under either the mirror symmetry $g=\mir_x$ ($u=1,s=0,\bdelta=\bze$). Since $\mir_x$ is an order-two symmetry, it has two distinct types of representations which we shall refer to as even and odd. For a nondegenerate band ($D=1$), \q{halftohalf} simplifies to $\cala= \breve{\mir}_x(\bk_a)\breve{\mir}_x^{\mo}(\bk_b)$, which equals $+1$ if the  representations at $\bk_a$ and $\bk_b$ are identical, and $-1$ if the two representations are distinct. $\cala=+1$ is exemplified by a band that is nondegenerate at all $\bkp$ bounded by $\frako$ -- due to continuity of the mirror representation along the $g$-invariant line, the representations at $\bk_a$ and $\bk_b$ must be identical. We may derive $\cala=+1$ from an alternative argument: the nondegeneracy at all $\bkp$ implies that $\frako$ is continuously contractible to a point. $\cala=-1$ occurs iff there is an odd number of band touchings along the segment of the mirror line contained within $\frako$ -- at each band touching (a Dirac point), the mirror representation flips discontinuously, and an odd number of flips implies that the representations at $\bk_a$ and $\bk_b$ are distinct. This is exemplified by the surface state of the SnTe-class\cite{Hsieh_SnTe} of topological crystalline insulators. Dirac cones protected by glide or screw symmetry are also characterized  by $\cala=-1$.\cite{LMAA}

\begin{center}
\underline{Class-II-A elementary orbits with $u(g)=0$}
\end{center}

If $u(g)=0$, we have shown in \s{sec:symmetryoforbits} that $g$ acts on generic $\bkp$ as a discrete rotation, while $g$-invariant (non-generic) $\bkp$ are isolated points. Given that $g\circ \frako=\frako$, and that $\frako$ is closed as an orbit, $\frako$ must  encircle a $g$-invariant point; however, $\frako$ itself contains no $g$-invariant points. In other words, $g$ maps every wavevector on $\frako$ to a distinct wavevector on the same orbit.   A commonly encountered example is   $g=T$ or $\inv$, which maps $\bk_1 \rightarrow -\bk_1$; for orbits that encircle an inversion-invariant point, $\{\bk_1,-\bk_1\}$ are distinct points lying on the same orbit.\\

Before stating the main result of this section, it would be useful to review and expand on the definition of order-$N$ symmetries $(g)$ and their corresponding $g$-orbits. For any $g$ which is not purely a translation, we may assign to $g$ an order $N(g) \in \{2,3,4,6\}$, a $\Z_2$ index $a(g)$, and a Bravais-lattice vector $\bR(g)$, such that \q{defineorderg} is satisfied. A case in point is $g=\inv$, where $\inv^2=I$ implies $N=2, a=0, \bR=\bze$, while $\rot_{nz}^n=\frake$ implies $N=n, a=1, \bR=\bze$. Further examples are provided in \tab{tab:exampleproj}. Let $\bk_1$ by an arbitrarily chosen base point in $\frako$, and define the $g$-orbit of $\bk_1$ as in \q{definegorbit}; in particular, the $g$-orbit of any  $\bk \in \frako$ also lies within $\frako$. For $g=\rot_{nz}$, there are $N$ distinct points in the $g$-orbit, which is a single cycle of length $N$. More generally, the $g$-orbit may contain $m(g)$   cycle(s) of length $L(g)=N/m \in \mathbb{N}$; $L$ is the smallest integer such that $g^L\circ \bkp=\bkp$ for all $\bkp$; $u=1{\imp}L\in 2\Z$ owing to $u(g^L)=0$. $m$ is a positive natural number that divides $N$, but is not equal to $N$;  the latter inequality follows from the assumption that $g$ is class-II ($m=N$ would imply that generic wavevectors are invariant under $g$). For example $g=T\rot_{6z}$ has order $N=6$, and its $g$-orbit is composed of $m=2$ cycles of length $L=3$; further examples are provided in \tab{tab:exampleproj}.  It will be useful to define $\breve{g}_i$ as the sewing matrix that relates the Bloch functions at $\bk_i$ to those at $\bk_{i+1}$: in more detail, $\breve{g}_i:=\breve{g}_{i+L}:= \breve{g}(\bk_i)$, as defined in \q{gactsonu}. It follows from \q{gactsonu} and (\ref{defineorderg}) that  the sewing matrices  form a representation of the space group, as shown in \q{orderNrepsewingmatrix}. \\

The main result of this section is that for every class-II-A symmetry $(g$) with $u(g)=0$, there exists an equivalence class of operators $[\tilde{g}K^s]$ that constrains the propagator as 
 \e{ 0 \eq \big[\,\tilde{g}K^s,\cala\,\big]. \la{definegtildegeneraller}}
$\tilde{g}$ is a unitary defined with the equivalence:
\e{ \tilde{g}^{\mo}=\dg{\tilde{g}}, \as \tilde{g}K^s \sim \cala \tilde{g}K^s. \la{equivgtilde}}
The motivation for this equivalence: if $\tilde{g}K^s$ were to be found that commutes with $\cala$, it follows trivially that $\cala \,\tilde{g}K^s$ would also commute with $\cala$. $\tilde{g}$ and $\cala$ are mutually constrained as
    \e{\big(\tilde{g}K^s\big)^N\eq  \cala^{p}\;(-1)^{Fa}; \as p(g) \sim p+N, \la{groupextensiongeneraller}}
where $s,a,p,N$ and $\bR$ are $g$-dependent. \q{definegtildegeneraller} and \q{groupextensiongeneraller} may be viewed as multiplication rules in a group generated by $\cala$, $\tilde{g}K^s$ and $(-1)^F$.\\

		
		Observe that the group relation for $\tilde{g}K^s$ in \q{groupextensiongeneraller} differs from the point-group relation for $g$ only by a multiplicative factor of $\cala^p$; we say that \q{groupextensiongeneraller} represents an extension of the point group by the loop propagators $\cala$. The exponent $p(g)$ is an integer defined with an equivalence $p \sim p+N$, which  reflects  $\tilde{g}K^s \sim \cala \tilde{g}K^s$; the values of $p$ for our list of representative symmetries are provided in \tab{tab:exampleproj}. Moreover, we prove in \s{app:suppargument} that 
\e{ [p(g)] = [\nu\,m]   \in \{[1],[2],\ldots,[N-1]\}, \as \nu(g)\in \{1,-1\} \la{relateptom}}		
where $m$ (as defined above) is the number of cycles in the $g$-orbit, and $\nu(g)= -1$ (resp.\ $+1$) if the $g$-orbit has the same (resp.\ opposite) orientation as $\cala$. Recall that $m$ is a positive natural number that divides $N$ but is less then $N$, and therefore $p$ is not $\sim 0$. This implies that \q{groupextensiongeneraller} represents an unsplit extension of the point group (generated by $g$) by  the group of loop translations (generated by $\cala$ and isomorphic to $\Z$). Equivalently stated, $\tilde{g}K^s \sim  \cala \tilde{g}K^s$  form an intrinsically projective representation\cite{weinbergbook1} of a point group; inequivalent projective representations are classified by the second group cohomology,\cite{cohomologysharifi} as we further  develop in  \s{app:suppargument}. In addition to this general group-theoretic discussion, we provide a more detailed case study of the order-two symmetries $T$ and $\inv$ in \s{sec:classIIATinv}.\\

The following constraint on the spectrum and determinant of $\cala$ follows directly from \q{definegtildegeneraller}:
\e{ \sigma(\cala)= K^s\sigma(\cala)K^s \substack{\imp} \as \text{if}\;s=1, \;\det \cala=\pm 1. \la{determinantvague}}
While the determinantal constraint (for $s=1$) is a general result that applies independent of the band degeneracy and the symmetry representation, we may further restrict the determinant once these additional data are specified; we shall exemplify this claim with $g=T$. For bands which are nondegenerate along $\frako$ ($D=1$), the determinantal constraint is merely a reality constraint on $\cala$, a unimodular phase factor. The sign of $\cala \in \R$ is determined by the symmetry representation as 
\e{ g=T,\as D=1,\as \cala=(-1)^{\sma{F}}.\la{detD1}}
$F=0$ corresponds to integer-spin representations, which include  single-spin bands in solids without spin-orbit coupling, and also charge-neutral bosonic systems. In the former case, $D=1$ should be interpreted as the energy degeneracy of bands restricted to one spin subspace, and the reality constraint applies to the spin-independent propagator $\cala_{F=0}$ defined in \q{spindependentH1}.\\

 Next, let us  consider spin-degenerate bands ($D=2$) which transform in a half-integer-spin representation ($F=1$) of time reversal ($T$). They may arise in (a) spin-orbit-coupled solids with $\inv$ symmetry (in addition to the assumed $T$ symmetry), and (b) solids with negligible spin-orbit coupling. In these two cases, the constraint \q{determinantvague} particularizes to:  
\e{ \text{For spin-degenerate bands,}\; D=2, F=1, \;\text{det} \,\cala=1. \la{determinantalspindeg}}
The proof for case (b) follows: we have already shown in \q{spindependentH1} how det$\, \cala$ is independent of the Zeeman effect, because the coupling to spin up exactly cancels the coupling to spin down. Consequently,  det$\, \cala$ is completely determined by the Roth-Berry phase, which characterizes the zero-field Hamiltonian $H_0$. Due to the spin-SU(2) symmetry of $H_0$, det$\, \cala$ equals the square of the Roth-Berry phase factor of the scalar (i.e., non-spinor, spinless) wavefunction [cf.\ \q{spindependentH1}]. To complete the proof, we utilize our general result in \q{determinantvague}, which applies in particular to spinless, nondegenerate $(D=1)$ bands: the Roth-Berry phase factor  is restricted to $\pm 1$, owing to time-reversal symmetry.\\ 

The proof of \q{determinantalspindeg} for case (a), as well as that of \q{detD1}, is more involved and will be deferred to \s{sec:classIIATinv}.






\subsubsection{Class-II-B elementary orbits}\la{sec:classIIB}

 Let $\cala_1:=\cala[\frako_1]$ and $\cala_2:=\cala[\frako_2]$ denote the propagators for two disconnected closed orbits related by $g\circ \frako_1=(-1)^{u}\frako_2$; the orientations of both $\frako_i$ are determined by Hamilton's equation. We denote the spectrum of $\cala_i$ by $\sigma(\cala_i) = \{$exp$\,{i\lambda_a^{\sma{(i)}}}\}_{\sma{a=1}}^{\sma{D}}$, where $D$ is the band degeneracy (and may equal $1$); $\{\lambda^{\sma{(1)}}_{\sma{a}}\}_{\sma{a=1}}^{\sma{D}}$ and $\{\lambda_{\sma{a}}^{\sma{(2)}}\}_{\sma{a=1}}^{\sma{D}}$ enter two independent quantization conditions having the same form as in \q{rule3b}. It follows from \q{wilsonloopconstraintdifforbit} that 
\e{ \cala_1 =\breve{g}K^{s}\cala_2^{(\mo)^{\sma{u}}}K^{s}\breve{g}^{\mo} \imp \sigma(\cala_1) =K^{s+u}\sigma(\cala_2)K^{s+u}.}

 For illustration, consider two disconnected  orbits related by time-reversal symmetry, but  neither orbit encircles a $T$-invariant point. We particularize to a spinless solid whose bands are nondegenerate ($D=1$) along both of $\frako_i$. The above equations then simplify to the mutual constraint $\cala_1=\cala_2^*$ or equivalently $\lambda_1=-\lambda_2$ mod $2\pi$. Since $\frako_i$ is not individually invariant under $T$, the average of the orbital moment over each orbit is generically nonzero -- this leads to a nonzero Roth contribution to each of $\lambda_i$.  The assumed absence of any stabilizing symmetry of $\frako_i$ implies that the Berry-phase contribution is not fixed to any special value. To recapitulate, there exists no constraints on individual values of $\lambda_i$; they satisfy only a mutual constraint. There are then two ladders of  sub-Landau levels corresponding to two uncoupled orbit. In energetic units (locally defined) where the separation between adjacent levels (within one sub-Landau ladder) is one, the offset between the two ladders is $2\lambda_1/2\pi$ mod $1$ [cf.\ \q{H1correctenergy}].  This splitting should be observable as two mutually-constrained harmonics in the dHvA oscillations, as exemplified by a toy model of distorted, spinless graphene in \s{sec:singlebandoscillations}.


\subsubsection{Class-II-A orbits with time-reversal or spatial-inversion symmetry}\la{sec:classIIATinv}

We provide a case study of class-II-A orbits with $T$  ($u=0,s=1$) or $\inv$ ($u=0,s=0$) symmetry. We may study each symmetry independently, without assuming that the solid simultaneously has both symmetries. Both $T$ and $\inv$ are order-two symmetries ($N=2$), and their corresponding $g$-orbits consists of a single cycle ($m=1$).\\

First, we will provide an elementary derivation of \qq{definegtildegeneraller}{relateptom} particularized to $N=2,m=1,[p]=[1]$. In the following proof, equations with the symbol $g$ applies to both $g=T$ and $g=\inv$; they are distinguished by $s(T)=1$ and $s(\inv)=0$. It is convenient to decompose the propagator as $\cala=\cala(-\bk_1)\cala(\bk_1)$, where $\cala(\bk_1)$ propagates through half the loop beginning from $\bk_1$ and ending at $-\bk_1$, and $\cala(-\bk_1)$ completes the loop. \q{wilsonlineconstraint} constrains the half-propagators as
\e{\cala(\pm \bk_1) \eq \breve{g}(\pm  \bk_1)K^s\cala(\mp \bk_1)K^s\breve{g}^{\mo}(\mp\bk_1), \la{constrainthalfprop}
}
where $\breve{g}(\bk)K^s$ forms a representation of the space group:
\e{ \ifor g\eq T,\as s(g)=1, \as \breve{g}(-\bk)\breve{g}(\bk)^*=(-1)^{\sma{F}};\lin
\ifor g\eq \inv,\;\as\,s(g)=0, \as \breve{g}(-\bk)\breve{g}(\bk)=I.\la{spacegroupinvtrs}}
The above equations are the particularization of \q{orderNrepsewingmatrix} for $N=2$, they respectively represent $T^2=\frake$ and $\inv^2=I$; $F$ distinguishes between integer-spin ($F=0$) and half-integer-spin representations ($F=1$). Owing to \q{constrainthalfprop}, the full propagator is constrained as
\e{  &\cala= \cala(-\bk_1)\cala(\bk_1)\lin
\eq     \breve{g}(-\bk_1)K^s\cala(\bk_1)\cala(-\bk_1)K^s\breve{g}^{\mo}(-\bk_1) \lin
          \eq \breve{g}(-\bk_1)K^s\cala^{\mo}(-\bk_1)\cala(-\bk_1)\cala(\bk_1)\cala(-\bk_1)K^s\breve{g}^{\mo}(-\bk_1) \lin
					\eq \tilde{g}K^s \cala K^s \tilde{g}^{\mo}. \la{constraintprojective}}
We have introduced the unitary matrix $\tilde{g}$:  					
\e{\tilde{g}:\eq\breve{g}(-\bk_1)K^s\cala^{\mo}(-\bk_1)K^s; \as \tilde{g}^{\mo}=\dg{\tilde{g}}, \la{definegtilde}}
which satisfies 
\bal
(\tilde{g}K^s)^2\eq\cala^{\mo}\,\breve{g}(-\bk_1)K^s\breve{g}(\bk_1)K^s \lin
\eq \begin{cases} (-1)^F\cala^{\mo}, &g=T;\\ \cala^{\mo}, &g=\inv. \as\as \blacksquare  \end{cases} \la{extension}
\end{align}

For $g=\inv$, \q{constraintprojective} implies that $\tilde{g}$ and $\cala$ are simultaneously diagonalizable, while  \q{extension} implies their eigenvalues are mutually correlated. A similar story occurs for $g=\rot_{nz}$: an operator $\tilde{\rot}_{nz}$ can be found that commutes with $\cala$ and satisfies the extended group relation $\tilde{\rot}_{nz}^n= \cala\,\frake.$ The mutual constraints between $\cala$ and $\tilde{g}$ do not constrain the spectrum of $\cala$ for a \emph{single} orbit; however, they may result in robust crossings in the spectra of a continuous family of rotationally-invariant orbits, which cover a 2D Fermi surface embedded in a 3D Brillouin torus. Incidentally, such crossings are already known to exist in the spectra  of a continuous family of  Wilson loops ($\W$) that cover a Fermi surface;  as mentioned earlier, $\cala$ and $\W$ are similarly constrained, i.e., the above equations are valid with $\cala$ replaced by $\W$. The presence of an odd number of crossings diagnoses the presence of a 3D Dirac point (protected by rotational symmetry\cite{zhijun_3DDirac}) enclosed by the Fermi surface.\cite{Gangli_3DDiracmetal,z2pack} \\

Let us particularize to $g=T$, for which \q{constraintprojective} implies  det$\,\cala= \pm 1$. As noted in \qq{detD1}{determinantalspindeg} of \s{sec:classIIA}, the determinant is completely determined by the following additional data: band degeneracy ($D$) at generic $\bkp$, the symmetry representation (whether integer- or half-integer-spin, as specified by $F$). In the subsequent subsections, we derive \q{detD1}, as well as \q{determinantalspindeg} for solids with both $T$ and $\inv$ symmetries.\\

\begin{center}
\underline{$D=1,F=0$}
\end{center}

By assumption, the $\bk$-space loop $\frako_0$ encloses a time-reveral-invariant point, which we denote by $\check{\bk}$. We first offer a simplified argument for $\cala=+1$ given two assumptions, which we will subsequently relax: our first assumption is that (a) the group of $\check{\bk}$ [denoted $G(\check{\bk})$] is only generated by $T$, hence  all irreducible representations (irreps, in short) are one-dimensional. It follows that the group of a generic wavevector enclosed by $\frako$ is trivial. We may therefore conclude that the minimal, symmetry-enforced degeneracy at any wavevector within $\frako$ is unity. Our second assumption is that, (b) at any $\bk$ within $\frako$, there are no accidental degeneracies between two one-dimensional irreps; we use `accidental' to generally describe degeneracies that are not enforced by symmetry, but require some fine-tuning of the Hamiltonian parameters. (a-b) imply that the band degeneracy is constant for all $\bk$ within $\frako_0$, and consequently there exists a family of time-reversal-symmetric loops ($\frako_s$, parametrized by $s\in [0,1]$) that interpolates between $\frako_0$ and a zero-area loop $\frako_1$ which encircles $\check{\bk}$; these loops are just the constant-energy contours of the assumed-nondegenerate band dispersion. Correspondingly, there exists also a family of time-reversal-symmetric propagators $\cala[\frako_s]$ which continuously interpolates between $\cala[\frako_0]$ to $\cala[\frako_1]$; in short, we say that $\cala$ is contractible $T$-symmetrically. Since $T$ is preserved throughout the interpolation, the sign of $\cala[\frako_s]$ is independent of $s$, from which follows that $\cala[\frako_0]=\cala[\frako_1]$. To complete the argument, we would demonstrate that $\cala[\frako_1]=+1$. Since the loop is of zero area and encloses no singularity in the Berry curvature, the Berry phase contribution to $\cala[\frako_1]$ vanishes. Such an argument cannot be applied to the non-geometric contributions (orbital moment and Zeeman coupling), owing to their inverse proportionality to the band velocity  -- which vanishes at the $T$-invariant point. Instead, by utilizing that time reversal inverts the angular momentum of states at $\pm \bkp$, we derive that the orbit-average of the non-geometric one-forms vanish. This completes the demonstration.\\ 

This result persists were we to relax our assumptions (a-b), as we proceed to explain.  Let us consider the case where the band, which is presumed to be nondegenerate along $\frako_0$, is continuously connected to a band touching point at $\check{\bk}$ enclosed by $\frako_0$. This touching point may be of three types: (i) it may correspond to a higher-dimensional irrep of $G(\check{\bk})$, that includes one or more point-group symmetries. (ii) The degeneracy might be an accidental touching between two one-dimensional irreps of $T$ symmetry. It is also possible that (iii) the touching is an accidental degeneracy between multiple irreps, one or more of which has dimension greater than one due to a point-group symmetry. Due to the presence of this band touching at $\check{\bk}$, we might question the $T$-symmetric contractibility of $\frako_0$. However, the reality constraint in \q{detD1} relies only on $T$ symmetry, hence any $T$-symmetric perturbation to $H_0$ cannot change the sign of $\cala$.
We may choose our $T$-symmetric perturbation to remove any accidental or point-group-symmetry-enforced degeneracy at $\check{\bk}$; in the latter case we would choose a perturbation that lowers the symmetry of $\check{\bk}$. Analogously, we may also remove any degeneracy at generic wavevectors within $\frako_0$. To clarify our argument, our perturbation to $H_0$ may be arbitrarily small in magnitude, and the energetic splitting of the degeneracy also arbitrarily small -- but strictly nonzero. We might define $\frako_0'$ as the band contour of the perturbed $H_0$, which lies  at the same energy as $\frako_0$; $\frako_0'\rightarrow \frako_0$ as the strength of the perturbation vanishes. The smallness of the perturbation guarantees that the topology of $\frako_0$ does not change discontinuously, i.e., in the sense of a Lifshitz transition; the reality condition guarantees that under such continuous deformations, $\cala[\frako_0]=\cala[\frako'_0]$. In this manner, we are once again able to construct the continuous, $T$-symmetric interpolation from $\cala[\frako_0] \rightarrow \cala[\frako_0'] \rightarrow +1$; in the second $\rightarrow$, the family of $T$-symmetric loops $\{\frako_s'|s\in [0,1]\}$ are just the constant-energy band contours of the perturbed $H_0$, with $\frako_1'$ the zero-area loop enclosing $\check{\bk}$.\\ 


\begin{center}
\underline{$D=1,F=1$}
\end{center}

Let us consider \q{detD1} for half-integer-spin representations ($F=1$); we restrict ourselves to solids without spatial inversion ($\inv$) symmetry -- only then are bands nondegenerate at generic wavevectors. Since $\frako_0$ encloses a Kramers-degenerate wavevector ($\check{\bk}$), $\cala[\frako_0]$ is not contractible  $T$-symmetrically. The linearly-dispersing band touching at $\check{\bk}$ contributes a Berry phase of $\pi$;\cite{berry1984} if linearly-dispersing touchings occur elsewhere within $\frako_0$, they come always in time-reversed pairs, hence the net Berry phase for all touchings remains $\pi$. Furthermore, the Roth and Zeeman phases individually vanish, since $T$ symmetry imposes $H_1^R(\bk)=-H_1^R(-\bk)$ and $H_1^Z(\bk)=-H_1^Z(-\bk)$ [cf.\ second column of Tab.\ \ref{tab:magnetization}]. We therefore conclude that $\cala=-I$.  Our assumption of a two-fold, Kramers degeneracy at $\check{\bk}$ may be challenged:  the degeneracy may be further enhanced by point-group symmetry\cite{Bradlyn_newfermions} and/or by fine-tuning of parameters in $H_0$. However, by $T$-symmetric perturbations which preserve the sign of $\cala$, we may always remove all point-group symmetries and accidental touchings, and recover the minimal scenario of a single Dirac touching at $\check{\bk}$.     \\

\begin{center}
\underline{$D=2,F=1$}
\end{center}

We may argue for this stronger constraint in case (ii-c-i) in two different ways. The first is based on the observation that the only nontrivial symmetry of a generic wavevector is the combined space-time inversion $T\inv$; its half-integer-spin irrep is two-dimensional. The group of $\check{\bk}$ (an inversion-invariant wavevector) is generated by $T$ and $\inv$ individually -- this group has only two inequivalent half-integer-spin irreps (corresponding to even and odd parities under $\inv$) which are both two-dimensional. Consequently, bands are two-fold-degenerate at every $\bk$ lying in $\frako_0$, absent accidental touchings and any other point-group symmetry (beyond $\inv$) that may enhance the two-fold degeneracy. As we have argued analogously above, these absences may be guaranteed by $T$- and $\inv$-symmetric perturbations that preserve the sign of det$[\cala]$. The constancy of band degeneracy at all $\bk$ within $\frako_0$ implies that $\cala[\frako_0]$ is $T$-symmetrically contractible to the two-by-two identity matrix, and therefore $\text{det}[\cala]= +1$. \\

In alternative argument, we may exploit the existence of  a continuous $T$-symmetric interpolation of the spin-degenerate subspace to a limit with vanishing spin-orbit coupling [case (ii-c-ii)]; det$[\cala]$, being fixed to either of $\pm 1$, is invariant throughout this interpolation. Since we have independently proven det$[\cala]=+1$ in case (ii-c-ii), we obtain a consistent result for (ii-c-i).\\

This unit determinant also applies to loops $\frako''$ that neither wrap around the Brillouin torus, nor enclose an inversion-invariant point ($\check{\bk}$). Absent other superfluous point-group symmetries, the group of any wavevector in $\frako''$ is generated by $T\inv$ and has a single inequivalent half-integer-spin irrep, which is two-dimensional -- we may then apply the perturb-then-contract argument to obtain the desired result.

\subsubsection{Group-theoretic analyis of class-II-A orbits with $u(g)=0$}  \la{app:suppargument}

One goal of this section is to derive \qq{definegtildegeneraller}{relateptom}. Before this, we shall elaborate on their implications on the group-theoretic structure of class-II-A orbits $(u=0)$. We have claimed that $\tilde{g}K^s \sim \cala\,\tilde{g}K^s$ reflects an intrinsic ambiguity in how we represent symmetries of the propagator $\cala$. The reader may be familiar with an analogous $U(1)$-phase ambiguity in the representation of symmetries of quantum Hamiltonians,\cite{weinbergbook1} which motivates the extension of groups by $U(1)$ phase factors. In different contexts,  these groups are known as ray or double groups, and have applications in magnetic translations\cite{Brown_magnetictranslation} and in describing half-integer-spin systems.\cite{weinbergbook1} Analogously, $\cala^p$ in \q{groupextensiongeneraller} originates from an  extension of the point group by quasimomentum loop translations. \\


To elaborate on this extension, let us define $G_{\frako}$ as the subgroup of the (magnetic) space group ($G$) that stabilizes a contractible  orbit $\frako$. $G_{\frako}$ is itself a (magnetic) space group, and its quotient with respect to its translational subgroup is a point group defined as $P_{\frako}$. Let $\cala \in G_{\sma{\cala}}:=\{\cala^z|z\in \Z\}$ represent a single translation around $\frako$. The action of $P_{\frako}$ on $G_{\sma{\cala}}$ is defined through
\e{ g\in P_{\frako},\as \tilde{g}K^s\,\cala\, K^s\tilde{g}^{-1}= \cala^{(-1)^{u}}, \la{Gorbacts}}
where $g$ is a representative element in $P_{\frako}$.
 $\tilde{g}K^s$ is defined to be an operator that maps the propagator to itself, up to a reversal in orientation that is determined by $u(g)$. If $\tilde{g}K^s$ is found that satisfies \q{Gorbacts}, it follows trivially that $\cala \tilde{g}K^s$ only satisfies \q{Gorbacts}. Therefore, $\tilde{g}K^s$ is only defined up to an equivalence $\tilde{g}K^s \sim \cala \tilde{g}K^s$, and we say that the equivalence class $[\tilde{g}K^s]$ forms a  (possibly projective) representation of $g\in P_{\frako}$. Alternatively stated, $\tilde{g}K^s$ and $\cala$ are elements of a group which is an extension of $P_{\frako}$ by $G_{\sma{\cala}}$; the  possible extensions are  classified by the second group cohomology:\cite{cohomologyhiller,cohomologysharifi} $H^2(G_{\frako},G_{\sma{\cala}})$.  Extensions of the point group by \emph{non-contractible} translations in $\bk$-space were first studied by one of us in \ocite{Cohomological}, to classify the symmetries of non-contractible Wilson loops that wrap around the Brillouin torus. The present program further demonstrates that extensions by \emph{contractible} $\bk$-space translations are needed in the group-theoretic description of closed orbits.\\

\noindent \emph{Proof of \qq{definegtildegeneraller}{relateptom}:}\\

Let us define $S_i \equiv S_{i+L}$ as the minimal-length, oriented line segment (contained within $\frako$) that begins at $\bk_i$ and ends at $\bk_{i+1}$; recall that $\bk_i$ are points on the $g$-orbit of $\bk_1$, and $\bk_{i+1}=g\circ \bk_i$, as defined in \q{orderNrepsewingmatrix}. For order-two symmetries such as $T$ or $\inv$, there are two equal-length segments connecting $\bk_1$ and $\bk_2:=-\bk_1$; in this case, either choice of segment is valid, and will not affect $[p(g)]$ in \q{relateptom}. We further define $\cala_i\equiv \cala_{i+L}$ as the propagator along $S_i$; in more detail, $\cala_i:=\cala[S_i:\bk_{i+1}\leftarrow \bk_i]$ with segment propagator $\cala[S]$ defined in \q{defwilsonline}.  Let us introduce an index $\nu(g)$, which equals $-1$ (resp.\ $+1$) if $S_i(g)$ has the same  (resp. reversed) orientation as $\frako$. Depending on $\nu$, $\cala$ is composed of a concatenation of $\{\cala_i\}$ as
\bal 
\cala[\frako]= \big(\;\cala_L\ldots \cala_2\cala_1\;\big)^{-\nu}.\la{decomposea}
														\end{align}
Note that we have arbitrarily chosen the base point of $\frako$ as $\bk_1$, but this choice does not affect the eigenvalues of $\cala[\frako]$, which enter the quantization conditions [cf.\ \s{sec:simple}]. \\

A particularization of \q{wilsonlineconstraint} implies that
\e{ &e^{i\bdelta\cdot(g\circ \bk_{j+1}-g\circ \bk_{j} )}\breve{g}_{j+1}K^{s}\cala_jK^s\breve{g}^{-1}_j = \cala_{j+1} \la{constraintaj} \\
 \iff& e^{i\bdelta\cdot(g\circ \bk_{j}-g\circ \bk_{j+1} )}\breve{g}_{j}K^{s}\cala^{-1}_jK^s\breve{g}^{-1}_{j+1} = \cala^{-1}_{j+1} \la{constraintaj2} \\
 \iff &e^{i\bdelta\cdot(g\circ \bk_{j} )}\breve{g}_{j}K^{s}\cala^{-1}_j= e^{i\bdelta\cdot(g\circ \bk_{j+1} )}\cala^{-1}_{j+1}\breve{g}_{j+1}K^s.\la{constraintaj3}
   }
Inserting \q{constraintaj} into \q{decomposea} for $\nu=-1$,
\e{ \nu(g)=-1,\as \cala[\frako] \eq \breve{g}_0K^s\cala_{L-1}\ldots \cala_1 \cala_L K^s \breve{g}^{-1}_0 \lin
\eq \breve{g}_0K^s\cala_0^{-1}\cala_L\cala_{L-1}\ldots \cala_1 \cala_0 K^s \breve{g}^{-1}_0\lin
\eq \tilde{g}K^s \cala[\frako] K^s\tilde{g}^{-1},}
where $\tilde{g}$ in the last line is defined as
\e{ [\tilde{g}K^s]=  [e^{i\bk_1\cdot \bdelta}\breve{g}_0K^s\cala_0^{-1}], \as \tilde{g}K^s \sim \cala[\frako]\,\tilde{g}K^s. \la{definecanonicalgtilde}}
Inserting \q{constraintaj2} into \q{decomposea} for $\nu=+1$,
\e{ \nu(g)=1,\as  &\cala[\frako] = \breve{g}_0K^s\cala_{0}^{-1}\cala_{1}^{-1}\ldots \cala_{L-1}^{-1} K^s \breve{g}^{-1}_0 \lin
\eq \big(\breve{g}_0K^s\cala_0^{-1}\big)\cala_{1}^{-1}\ldots \cala_{L-1}^{-1}\cala_{L}^{-1} \big(\cala_0 K^s \breve{g}^{-1}_0\big)\lin
\eq \tilde{g}K^s \cala[\frako] K^s\tilde{g}^{-1},}
utilizing the same definition of $\tilde{g}$ in \q{definecanonicalgtilde}.  In either case for $\nu$, $\tilde{g}$ satisfies
\bal  
&\big(\tilde{g}K^s\big)^N=  \big(\;e^{i(g\circ\bk_0)\cdot \bdelta}\breve{g}_0K^s\cala_0^{-1}\;\big)^N \lin
\eq \cala^{-1}_{1}\big(e^{i(g\circ\bk_1)\cdot \bdelta}\breve{g}_{1}K^s\cala^{-1}_{1}\big)^{N-1}e^{i(g\circ\bk_1)\cdot \bdelta}\breve{g}_{1}K^s \lin
\eq  \cala^{-1}_{1}\cala^{-1}_{2}\big(e^{i(g\circ\bk_2)\cdot \bdelta}\breve{g}_{2}K^s\cala^{-1}_{2}\big)^{N-2} \lin
&\times e^{i(g\circ\bk_2)\cdot \bdelta}\breve{g}_{2}K^s  e^{i(g\circ\bk_1)\cdot \bdelta} \breve{g}_{1}K^s \lin 
\eq \big(\cala^{-1}_1\cala^{-1}_{2} \ldots \cala^{-1}_N \big) \; \big(e^{i(g\circ\bk_N)\cdot \bdelta}\breve{g}_{N}K^s   \ldots \lin
&\times e^{i(g\circ\bk_2)\cdot \bdelta}\breve{g}_{2}K^s  e^{i(g\circ\bk_1)\cdot \bdelta} \breve{g}_{1}K^s \big) \lin
\eq \cala[\frako]^{\nu m}\;(-1)^{Fa(g)}.
\end{align}
The second to fifth equalities are derived by $N$ number of iterative applications of \q{constraintaj3}; in the last line, we have employed \q{decomposea}, and the fact that the g-orbit $\{\bk_i\}_{i=1}^N$ contains $m$   cycle(s) of length $L=N/m$, with $\bk_{L}=\bk_0$.


\section{Intraband breakdown} \la{sec:intraband}

\begin{figure}[ht]
\centering
\includegraphics[width=4 cm]{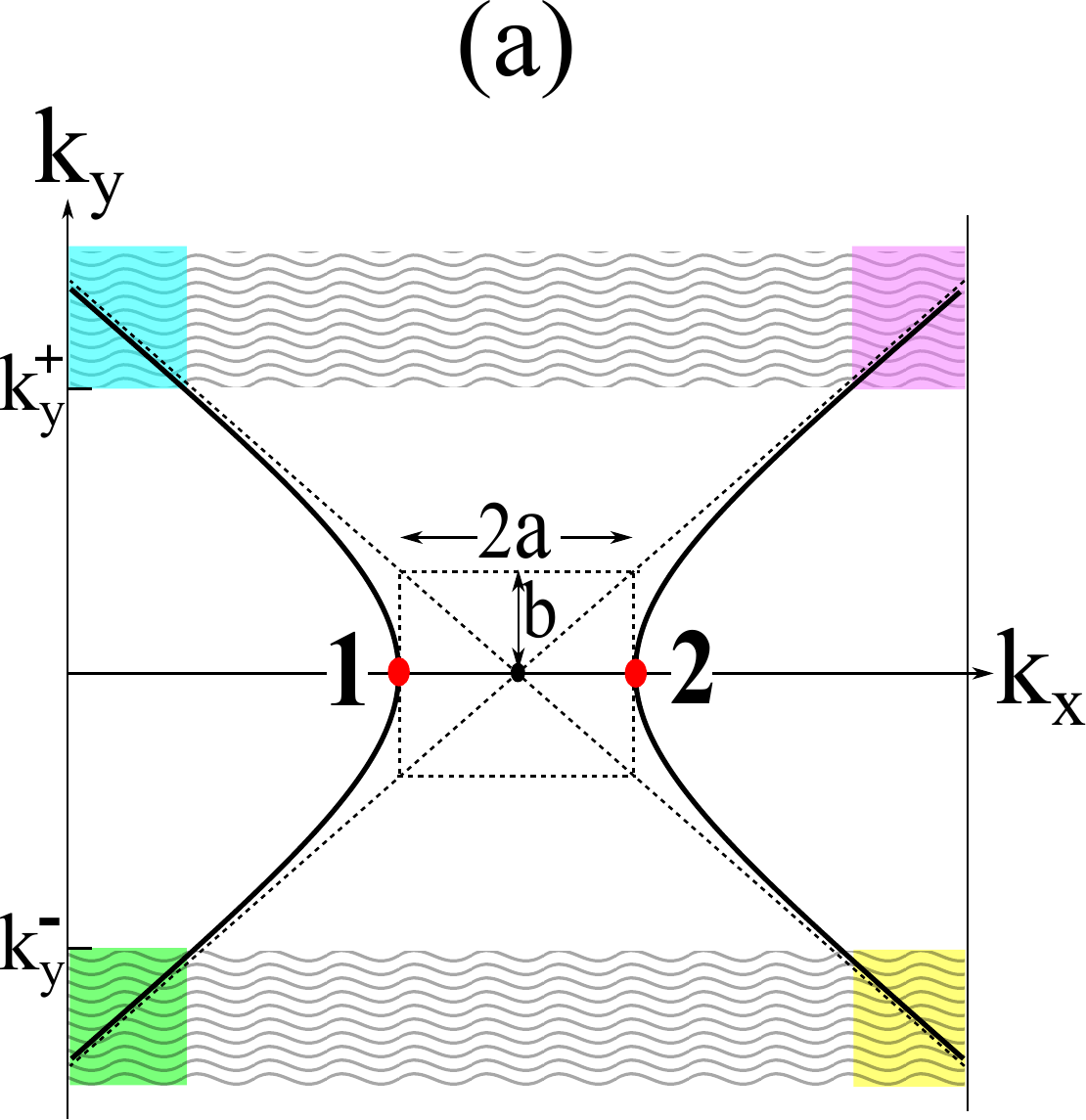}
\includegraphics[width=8 cm]{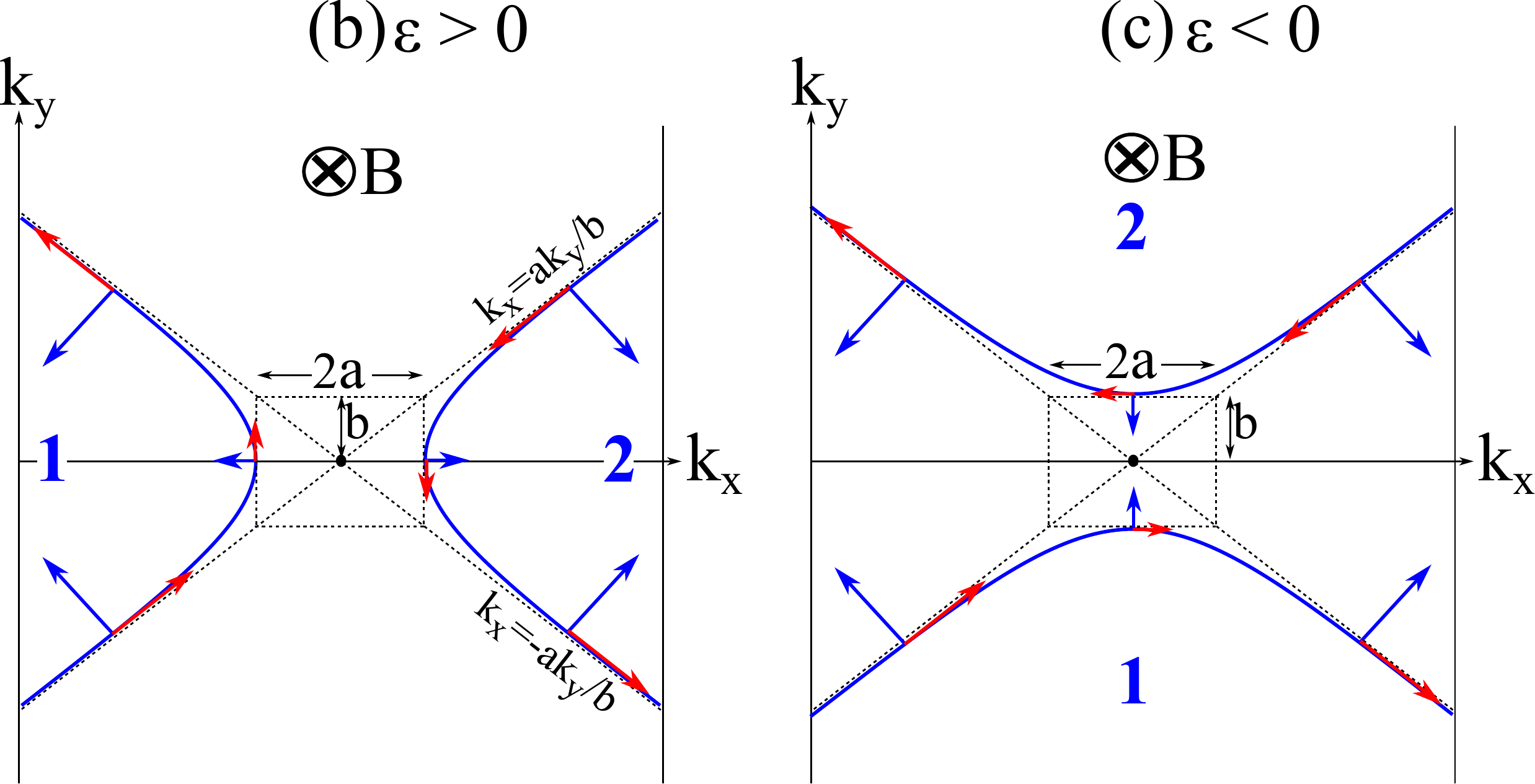}
\caption{(a) illustrates a region in $\bk$-space where quantum fluctuations are strong -- it shall be referred to as the breakdown region. The solid black lines are constant-energy band contours near a saddlepoint. The breakdown region overlaps with the semiclassical region (indicated by grey wavy lines). In (b-c), we representatively indicate the band and orbit velocities: respectively, $\nabla_{\bk}\var$ is indicated by blue arrows, and $\dot{\bk}$ (for a semiclassical wavepacket with negative charge in a magnetic field $\bB=-B\vec{z}$, $B>0$) is indicated by red arrows. (b) shows the velocities at positive $\var$, and (c) for negative $\var$.}\label{fig:saddlepoint}
\end{figure}

Intraband breakdown occurs in the vicinity of saddlepoints, which are the nuclei of Lifshitz transitions, i.e., changes in the topology of constant-energy band contours as a function of energy. In the neighborhood of a saddlepoint, the band contours approach each other as two arms of a hyperbola illustrated in \fig{fig:saddlepoint}. It is convenient to orient ourselves by parametrizing the zero-field, band dispersion as
\e{ \var_{\bk} \eq \f{k_x^2}{2m_1} -\f{k_y^2}{2m_2}, \la{saddlepointdispersion}}
 with $\bk:=(k_x,k_y)$ chosen so that both $m_j>0$. In 3D solids, $\var$ additionally depends on $k_z$ as 
\e{ \var_{k_x,k_y,k_z} \eq \f{k_x^2}{2m_1} -\f{k_y^2}{2m_2}+f(k_z). \la{saddlepointdispersion3D}}
Since $k_z$ remains a conserved quantity in the presence of a magnetic field along $\vec{z}$, we may as well define $\var_{k_x,k_y}:=\var_{k_x,k_y,k_z}-f(k_z)$ and work directly with \q{saddlepointdispersion}.\\

For a fixed energy $\var_{\bk}=E$, it is convenient to introduce the hyperbolic parameters
\e{ & \f{k_x^2}{a^2}-\f{k_y^2}{b^2} =\sgn[E], \lin
  &a(E) =  \sqrt{2m_1|E|}  ,\as b(E) =  \sqrt{2m_2|E|}, \la{defineab}}
such that the hyperbolic asymptotes are diagonal lines parametrized by $k_y =\pm (b/a)k_x$.  \fig{fig:saddlepoint}(b-c) illustrates a discontinuous change in the band contours at $E=0$.\\

A quantity of geometric significance is the area ($4ab$) of the rectangle inscribed between the two hyperbolic arms [see \fig{fig:saddlepoint}]; it is natural that the dimensionless parameter
\e{ \mu= \sgn[E]\f1{2}abl^2 = \sqrt{m_1m_2}El^2, \la{definemuintra}}
 determines the probability of tunneling between orbits; the exact form of $\mu$ will be motivated by the connection formula in \q{zerothscatteringmatrixsaddle}. 
	When $|\mu|=O(1)$, the minimal separation between two contours becomes of order $\lmo$, and tunneling between orbits -- intraband breakdown -- must be accounted for. One indication that the semiclassical equations of motion might fail is that the cyclotron mass $(\hbar^2/2\pi)\partial S/\partial E$ of the orbit diverges logarithmically as $E \rightarrow 0$;\cite{kosevich_topologyreview} a related symptom is that  both components of the band velocity $\nabk \var$ vanish at the saddlepoint, as illustrated in \fig{fig:saddlepoint}(b-c). Both symptoms suggest that a hypothetical, Hamilton-obeying wavepacket never reaches the saddlepoint in finite time.

\subsection{Connection formula with intraband breakdown}\la{sec:connectionintraband}

The method to determine energy levels is similar in spirit to the divide-and-conquer approach of \s{sec:turningpoint}. The vicinity of the saddlepoint is a region of quantum fluctuations where the Zilberman-Fischbeck (ZF) function [\q{zilbermanfischbeck}] loses its validity -- as may be inferred from the diverging prefactor of $1/|v^x|^{\sma{1/2}}$. What is needed is an approximate solution of the wavefunction in this breakdown region, with which to connect the two incoming ZF functions approaching along the $k_x=ak_y/b$ diagonal [see \fig{fig:saddlepoint}], with two outgoing ZF functions along the $k_x=-ak_y/b$ diagonal.  \\

The main goal of this section is to derive this connection formula. The first step is to derive an effective Hamiltonian that is valid in the breakdown region; we must then derive the eigenfunctions (of this effective Hamiltonian) to the same order of accuracy (in inverse powers of $l$) as the \zf function. For this purpose,  one must go beyond the Peierls substitution of \q{saddlepointdispersion}, which produces only the lowest-order, Peierls-Onsager Hamiltonian in the asymptotic expansion of \q{effhamasymptoticexpansion}. The Peierls-Onsager Hamiltonian  forms the basis of previous treatments\cite{azbel_quasiclassical,Wilkinson_criticalproperties,Davis_landauspectrum} of this problem, as briefly reviewed in \app{app:zerothorderconnectionintraband}. \\

Let us elaborate on how this connection is done. As illustrated in \fig{fig:saddlepoint}(a),  there exists an interval in $k_y \in [k_y^+,k_y^-]$, centered at the saddlepoint, where a semiclassical description  breaks down; we shall refer to $[k_y^+,k_y^-]$ as the breakdown interval. It is convenient to define four directed edges which meet in the breakdown interval, which we label by the directions of their semiclassical motion along the hyperbolic asymptotes: $\nu \in \{\nwarrow,\swarrow\}$ above the breakdown interval, and $\nu \in \{\nearrow,\searrow\}$ below. By `directed edge', we are utilizing graph-theoretic language that is reviewed in \s{sec:graph}.
Each edge is parametrized by two single-valued functions $k_x^{\nu}(k_y,E)$ and $k_y^{\nu}(k_x,E)$; it is convenient to define for each edge the coordinate of closest approach [$\bk_0^{\nu}(E):=(k_{x0}^{\nu}(E),k_{y0}^{\nu}(E))$] to the saddlepoint, as indicated by red dots in \fig{fig:saddlepoint}(a).\\

Above the breakdown interval, the general analysis of \s{sec:singlebandwkbwf} informs us that the wavefunction in the $(K_x,k_y)$ representation [\q{defineKyrepresentation}] is a linear combination of at least two ZF functions (corresponding to the two edges above a saddlepoint):
\e{ f^+_{\bk E} =c_{\sma{\nwarrow} E}\tilde{g}^{\sma{\nwarrow}}_{\bk E}+c_{\sma{\swarrow} E}\tilde{g}^{\sma{\swarrow}}_{\bk E}+\ldots \la{abovequantum}}
 $c_{\nu E}$ are edge-dependent constants which are to be determined. As denoted vaguely by $\ldots$, there might in general be more edges in the above sum which correspond to constant-energy band contours far away from the saddlepoint [and therefore not illustrated in \fig{fig:saddlepoint}], but they will not be important in the matching procedure. The ZF functions in \q{abovequantum} are defined (up to a $k_y$-independent phase) as eigenfunctions of the effective Hamiltonian in the semiclassical region (denoted $sm$):  
\e{ \ifor \bk \in sm,\as (H_0(\bK)+{H}_1(\bK) - {E})\tilde{g}^{\nu}_{\bk E}=O(\lmf).\la{zfsatisfies}}
Precisely, we define
\e{\tilde{g}^{\nu}_{\bk {E}}:\eq \f{e^{ik_xk_yl^2}}{\sqrt{|{v}^{\nu}_x|}}e^{-il^2\int^{k_y}_{k_{y0}^{\nu}(E)}\left({k}_x^{\nu}-{\tilde{H}^{\nu}_1}/{v^x_{\nu}}  \right)dz }\bigg|_{E \rightarrow \tilde{E}}; \la{lastlineiszf} \\
\as \tilde{E}:\eq E-H_1(\bze), \as \tilde{H}_1(\bk):=H_1(\bk)-H_1(\bze); \la{definetildeEH1}}
one may verify that $\tilde{g}$ indeed satisfies the eigenvalue equation \q{zfsatisfies}. Indeed, beginning from \q{zfsatisfies}, one may redefine the origin of the energy as in \q{definetildeEH1}, and utilize the known WKB solution from   \qq{eigenproblemsingleband}{zilbermanfischbeck}. The reader may wonder what is the point of the redefinition of energetic origin, i.e., why not directly use the simpler expression
\e{{g}^{\nu}_{\bk {E}}:\eq \f{e^{ik_xk_yl^2}}{\sqrt{|{v}^{\nu}_x|}}\exp\left\{-il^2\int^{k_y}_{k_{y0}^{\nu}(E)}\left({k}_x^{\nu}-\f{{H}^{\nu}_1}{v^x_{\nu}}  \right)dz \right\}\bigg|_{E}; \la{simplerwkb}}
which is also a solution of \q{zfsatisfies} in the semiclassical region. Indeed, $g^{\nu}_{\bk E}-\tilde{g}^{\nu}_{\bk E}=O(\lmt)$  in the semiclassical region, as proven in \app{app:gtildeeqg}. However, at the coordinate of the saddlepoint, the phase of $g$ (which includes a term proportional to $H_1(\bze)$log$|E|$) diverges logarithmically as $|E|\rightarrow 0$, while the phase of $\tilde{g}$ is continuous across $E=0$. For this reason, we will find that $\tilde{g}$ is a better WKB function to formulate quantization conditions that are valid in the vicinity of a saddlepoint.  \\

Below the breakdown interval, we analogously have
\e{ f^-_{\bk E} =c_{\sma{\nearrow} E}\tilde{g}^{\sma{\nearrow}}_{\bk E}+c_{\sma{\searrow}E}\tilde{g}^{\sma{\searrow}}_{\bk E}+\ldots \la{belowquantum}}
Assuming the non-WKB wavefunction in the breakdown region has been solved for, we may utilize this wavefunction as a bridge to coherently relate $\{c_{\sma{\nwarrow}E},c_{\sma{\swarrow}E}\}$ (defined above the breakdown region) to $\{ c_{\sma{\nearrow}E},c_{\sma{\searrow}E}\}$ (defined below). For the purpose of deriving quantization conditions in \s{sec:quantizationintraband}, we find it intuitive  to express this relation as a scattering-matrix equation connecting incoming to outgoing sections:
\e{ \vectwo{c_{\sma{\nwarrow}E}}{c_{\sma{\searrow}E}}= \bbs(E,k_z)\vectwo{c_{\sma{\nearrow}E}}{c_{\sma{\swarrow}E}}. \la{definescatteringmatrixintra}}
The scattering matrix in the Peierls-Onsager approximation is known to be:\cite{berry_mount_review,Wilkinson_criticalproperties,Davis_landauspectrum}  
\e{\bbs^{\sma{(0)}}(E,l^2) \eq \matrixtwo{\calt}{\calr}{\calr}{\calt}\bigg|_{El^2},\lin
   \calt(\mu) \eq e^{i\phi(\mu)} \,\f{e^{ \pi \mu/2}}{\sqrt{2 \,\text{cosh}(\pi \mu)}}, \lin
	\calr(\mu) \eq -i  \,  e^{i\phi(\mu)}\,\f{e^{ -\pi \mu/2}}{\sqrt{2 \,\text{cosh}(\pi \mu)}}, \lin
\phi(\mu) \eq  \arg[\Gamma(1/2-i\mu)]+\mu \log |\mu|-\mu,
	\la{zerothscatteringmatrixsaddle} }
with $\mu$ defined in \q{definemuintra} and $\Gamma$ the Gamma function. Alternatively stated, $\bbs^{\sma{(0)}}$ is the connection formula for Zilberman functions without higher-order corrections [i.e., \q{lastlineiszf} with $H_1(\bk)=0$]. \\

\begin{figure}[ht]
\centering
\includegraphics[width=8 cm]{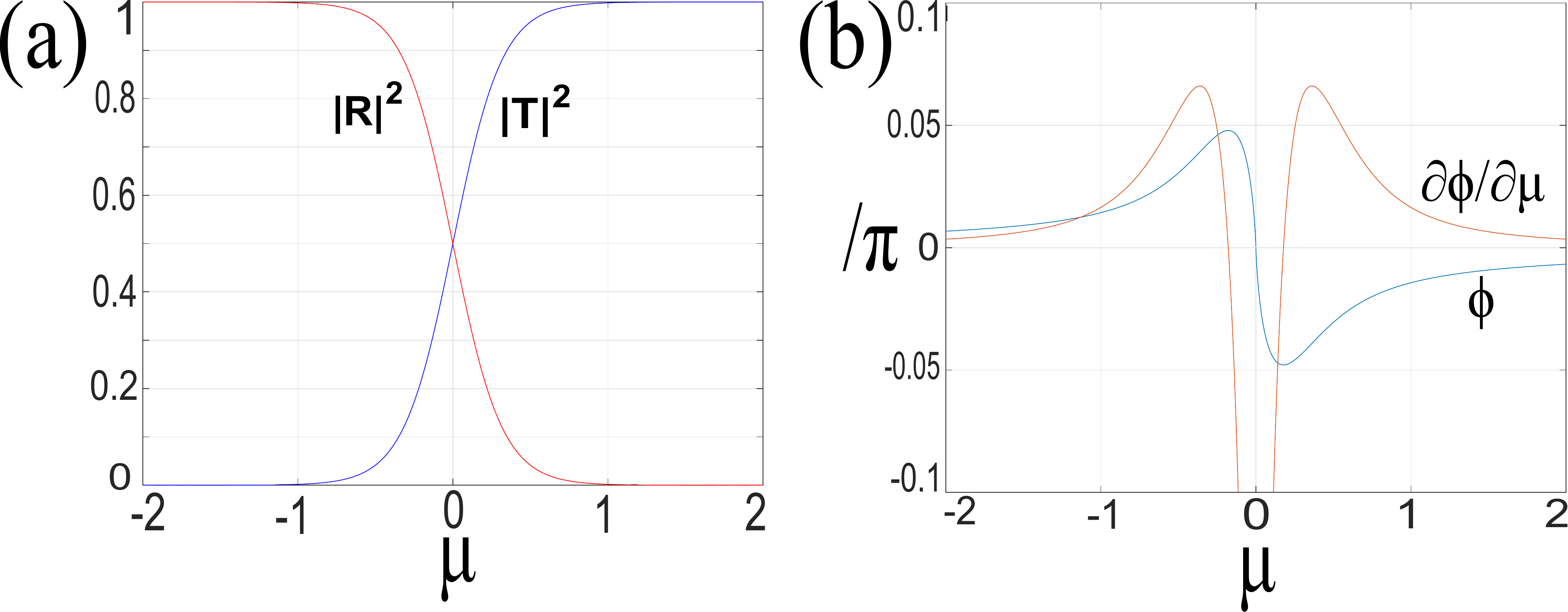}
\caption{Plots vs $\mu$ of (a) $|\calr|^2$ (red) and $|\calt|^2$ (blue), (b) the scattering phase $\phi$ (blue) and its derivative with respect to $\mu$ (red).}\label{fig:saddlescatteringmatrix}
\end{figure}

The derivation of \q{zerothscatteringmatrixsaddle} is reviewed in \app{app:zerothorderconnectionintraband}, where we elaborate on a useful analogy: magnetic tunneling of a Bloch electron near a saddlepoint is mathematically equivalent to a Schrodinger particle tunneling across an inverted parabolic barrier -- a problem first studied by Kemble.\cite{kemble} In particular, it is well-known\cite{kemble} that the tunneling probability at the barrier maximum is half of unity, which is reflected in \q{zerothscatteringmatrixsaddle} by $|\calt|^2=|\calr|^2=1/2$ for $\mu=0$ [see \fig{fig:saddlescatteringmatrix}(a)]; we shall refer to this as the Kemble limit. We refer to $\phi$ as the intraband scattering phase and plot it in \fig{fig:saddlescatteringmatrix}(b); $\phi$ has the following properties: (a) it is an odd function of $\mu$ that vanishes at zero and the limits $\pm \infty$, and (b) its first-order derivative diverges logarithmically as $\mu \rightarrow 0$. In all cases we have studied, property (b) does not lead to any irregularity in the Landau levels, owing to the cancellations of logarithmic divergences in $(\partial \phi/ \partial E)$ and the cyclotron mass $(\propto \partial S/\partial E)$; we will exemplify this cancellation in \s{sec:doublewell}. For quick reference in the future,
\e{  \mu \rightarrow +\infty, \as& \calt \rightarrow 1, \;\calr \rightarrow 0, \; \phi \rightarrow 0;  \la{muplus}\\
\mu \rightarrow 0, \as& \calt \rightarrow 1/{\sqrt{2}}, \;\calr \rightarrow -i/{\sqrt{2}}, \; \phi \rightarrow 0; \la{muzero}\\
\mu \rightarrow -\infty, \as& \calt \rightarrow 0, \;\calr \rightarrow -i, \; \phi \rightarrow 0.\la{muminus}} 
The transition from $\{|\calt|=1,|\calr|=0\}$ to $\{|\calt|=0,|\calr|=1\}$ reflects a Lifshitz transition of the band contours. While the change in the band contour is discontinuous across $E=0$, the scattering parameters are continuous in energy; there is, as noted, a first-order non-analyticity in $\phi$. The $\mu \rightarrow -\infty$ limit corresponds to the absence of tunneling (in the $\vec{y}$ direction) between the two semiclassical orbits drawn in \fig{fig:saddlepoint}(c). In this limit, $\calr=-i$ is the phase  acquired by a wavepacket as it approaches the saddlepoint and is reflected with unit probability -- the point of closest approach to the saddlepoint may therefore be identified as a turning point, just as discussed in \s{sec:turningpoint}. Indeed, we have demonstrated in \s{sec:turningpoint} that a wavepacket rounding a turning point with a clockwise orientation picks up a phase of $-i$, which we consistently identify with $\calr=-i$ here.\\

Let us argue generally that the scattering matrix, for any form of breakdown, should transform covariantly under gauge transformations within $P$. Viewed broadly, the scattering matrix describes the phase-coherent amplitudes for Feynman trajectories through the breakdown region. There is in general a phase ambiguity in how we define the incoming and outgoing scattering states, whose wavefunctions have the Zilberman-Fischbeck form in \q{lastlineiszf}; in particular, the phase difference between any two states connected by a tunneling trajectory has no gauge-invariant meaning. 
To appreciate this point, consider a phase redefinition of the cell-periodic function projected by $P$: $\ket{u_{\bk}}\rightarrow \ket{u_{\bk}}e^{i\phi(\bk)}$. The resultant non-covariant transformation of the Berry connection [$\bmx\rightarrow \bmx -\nabk \phi$] occurring in $\tilde{H}_1$ [cf.\ \q{lastlineiszf}] results in the scattering-state wavefunction transforming as 
\e{ g^{\nu \pm }_{\bk E} \rightarrow g^{\nu \pm }_{\bk E}e^{-i\phi^{\nu}(k_y)+i\phi(\bk^{\nu}_{0})},}
where in the last expression $\phi^{\nu}(k_y)$ equals $\phi(\bk)$ evaluated on the section $\nu$, and at the coordinate $k_y$. If hypothetically the scattering matrix were insensitive to phase redefinitions of the scattering states, as is $\bbs^{\sma{(0)}}$ [cf.\ \q{zerothscatteringmatrixsaddle}], one would conclude that the quantization condition depended on the phase difference $\phi(\bk^{\nu}_{0})-\phi(\bk^{\mu}_{0})$, which is generally nonzero for a tunneling trajectory connecting the edges $\nu$ and $\mu$ [note $\bk^{\nu}_{0} \neq \bk^{\mu}_{0}$]. The coeffients $c_{\nu E}$, defined in \q{abovequantum}-(\ref{belowquantum}), should transform with a \emph{cancelling} phase  factor
\e{ &c_{\nu E} \rightarrow c_{\nu E}e^{-i\phi(\bk_{0}^{\nu})} \imp \lin
&\bbs \rightarrow \diagmatrix{e^{-i\phi(\bk_{0}^{\nwarrow})}}{e^{-i\phi(\bk_{0}^{\searrow})}} \bbs \diagmatrix{e^{-i\phi(\bk_{0}^{\nearrow})}}{e^{-i\phi(\bk_{0}^{\swarrow})}}.\notag} 
Equivalently stated, the scattering matrix must transform gauge-covariantly. We see from this argument that the necessity of gauge covariance follows from the existence of tunneling trajectories, which is a characteristic feature of both intraband and interband breakdown -- but not of turning points. We believe that our argument should broadly apply to any quantum tunneling phenomenon within a subspace of states (bands, in our context) that is nontrivially embedded in a larger space of states; this point has been overlooked in conventional treatments\cite{berry_mount_review} of tunneling with scattering matrices. \\

Let us show that the next-order corrections to $\bbs^{\sma{(0)}}$ restores the essential gauge covariance. For this purpose, it is sufficient to consider the correction by the Berry term $H_1^B$ alone: 
\e{&\bbs(E,l^2) \condeq{H_1=H_1^B}  \lin
&\matrixtwo{\calt(\mu)\;e^{i \int^b_{-b}\mxy(0,k_y) dk_y}\as}{\calr(\mu)\;e^{-i \int^a_{-a}\mxx(k_x,0)dk_x}}{\calr(\mu)\;e^{i \int^a_{-a}\mxx(k_x,0)dk_x}\as}{\calt(\mu)\;e^{-i \int^b_{-b}\mxy(0,k_y) dk_y}}, \la{scatteringintraberry}}
neglecting terms of order $O(\lmt,(b/G)^2,(a/G)^2)$. $a(E)$ and $b(E)$ are the hyperbolic parameters defined in \q{defineab}, and $G$ is a typical reciprocal period. For positive $E$, $\int^{b(E)}_{-b(E)}\mxy(0,k_y) dk_y$ is the integral of the Berry connection along the shortest-length tunneling trajectory that connects $\bk_0^{\sma{\nearrow}}(E)$ to $\bk_0^{\sma{\nwarrow}}(E)$ through the classically forbidden region [e.g., the vertical dashed line in \fig{fig:necking}(b)]. That this tunneling trajectory is of the shortest length should  not be taken too seriously; a slightly deformed trajectory within the breakdown region gives a correction  [of $O(\lmt)$] that is beyond the accuracy of our calculation [detailed in \app{app:deriveintrascattering}]. Under a phase redefinition $\ket{u_{\bk}}\rightarrow \ket{u_{\bk}}e^{i\phi(\bk)}$,  the open-line Berry integral transforms as
\e{\int^b_{-b}\mxy dk_y \rightarrow \int^b_{-b}\mxy dk_y -\phi(\bk_0^{\sma{\nwarrow}})+\phi(\bk_0^{\sma{\nearrow}}), \la{transformopenline}}
which implies that the scattering matrix transforms covariantly as
\e{ \bbs \rightarrow \diagmatrix{e^{-i\phi(\bk_{0}^{\nwarrow})}}{e^{-i\phi(\bk_{0}^{\searrow})}} \bbs \diagmatrix{e^{-i\phi(\bk_{0}^{\nearrow})}}{e^{-i\phi(\bk_{0}^{\swarrow})}}. \la{definegaugecovari}}
Indeed, \q{scatteringintraberry} is minimally corrected from $\bbs^{\sma{(0)}}$ to ensure gauge covariance; our calculation shows that the minimally-corrected matrix completely accounts for corrections by $H_1^B$. In solids where the Roth ($H_1^R$) and Zeeman ($H_1^Z$) terms vanish by symmetry [e.g., $\rot_{2z}T$ symmetry; cf. \s{sec:symmetrysinglebandom}], there are no further leading-order corrections to the scattering matrix. \\

Unlike the Berry correction to the scattering matrix, the Roth and Zeeman corrections cannot be argued for from gauge covariance -- a calculation is necessary, which we detail in \app{app:deriveintrascattering}. When all three corrections are accounted for, we find that the scattering matrix takes the form
\e{ \bbs(E,l^2) \eq \matrixtwo{\calt(\tilde{\mu})e^{i \delta_y(\tilde{E})}\as}{\calr(\tilde{\mu})e^{-i \delta_x(\tilde{E})}}{\calr(\tilde{\mu})e^{i \delta_x(\tilde{E})}\as}{\calt(\tilde{\mu})e^{-i \delta_y(\tilde{E})}}\lin
& +O(\lmt,(b/G)^2,(a/G)^2),  \la{fullscattering} \\
\tilde{E} :\eq E-H_1(\bze), \as \tilde{\mu}:= \sqrt{m_1m_2}\tilde{E}l^2,   \la{definetildeEmu} \\
\delta_y(E):= 2 m_1&\,H_{1x}\,b({E})\,l^2, \;\delta_x(E):= 2 m_2\,H_{1y}\,a({E})\,l^2, \la{tunnelingphase}} 
where $H_1(\bze)$, $H_{1x}$ and $H_{1y}$ are defined as coefficients in the low-momentum expansion of $H_1=H_1^B+H_1^R+H_1^Z$ about the saddlepoint:
\e{H_1(\bk):\eq  H_{1}(\bze)+H_{1x}{k}_x+H_{1y}k_y+\ldots,  \lin
 H_1^B(\bk) \eq \lmt (\mx^y v^x -\mx^x v^y) \lin
\eq \lmt \bigg(\mx^y \f{k_x}{m_1} + \mx^x \f{k_y}{m_2}\bigg):= H_{1x}^Bk_x+H_{1y}^Bk_y, \la{H1Bexplicit}\\ 
H_1(\bze)\eq H^R_1(\bze)+H_1^Z(\bze). \la{approximateH1}}
Note that $H^B_1(\bze)=0$ because the saddlepoint is an extremum in the band dispersion, and the second line follows from particularizing the definition of $H_B^1$ [cf.\ \q{H1berry}] to the saddlepoint. \\

In \q{tunnelingphase}, we have defined $\delta_x$ and $\delta_y$ as phase corrections to the scattering matrix.  Their respective proportionality to $a(E)$ and $b(E)$ identifies them as phases acquired in the tunneling trajectories parallel to $\vec{x}$ and $\vec{y}$.
This tunneling phase includes the open-line Berry phase from our previous result in \q{scatteringintraberry}, e.g., we may identify
\e{ \delta_y= \int^{b}_{-b}\mxy(0,k_y) dk_y + 2m_1\,(H_{1x}^R+H_{1x}^Z)\,b\,l^2 +O(\tf{b^2}{G^2}),\la{tunnelingphasey}}
with aid from \q{H1Bexplicit}. Under a phase redefinition $\ket{u_{\bk}}\rightarrow \ket{u_{\bk}}e^{i\phi(\bk)}$, $\delta_y$ transforms just like \q{transformopenline} owing to the gauge invariance of  $H_1^R$ and $H_1^Z$ , and therefore $\bbs(E,l^2)$ in \q{fullscattering} transforms covariantly, just as in \q{definegaugecovari}. We might further motivate the form of \q{tunnelingphasey} by rewriting it completely in terms of $H_1$, $v^x$, and $b$: 
\e{ \delta_y \appr \int^{b}_{-b}\bigg\{ \overline{\f{H_1(\bk)-H_1(\bze)}{v^x(\bk)}} \bigg\}_{k_y} dk_y, \la{creativelic}} 
where $\{ \bar{\cdot} \}_{k_y}$ denotes the $k_x$-average of the quantity $\cdot$ over a fixed-$k_y$ cross-section of the classically-forbidden region. With some creative license, one might interpret \q{creativelic} as the Roth-Berry-Zeeman phase averaged over all possible tunneling trajectories parallel to $\vec{y}$. \\

As expressed in \ref{tunnelingphasey}, $\delta_y$ may be separated into gauge-dependent ($\delta_y^B$ with B for Berry) and gauge-invariant ($\delta_y^{RZ}$ for Roth and Zeeman) terms.  $\delta_y^{RZ}$ may be dropped if we are willing to accept an $O(1)$ accuracy for the scattering phase. Indeed, we make the following estimate for the size of $\delta_y^{RZ}$: since the tunneling trajectory has length $2b$ with $b$ the hyperbolic parameter [cf.\ \q{defineab}], $\delta^{RZ}_y$ is of order $O(b/G)$ with $G$ the reciprocal period. We might further bound $b \leq O(1/l)$, which is the width of the breakdown region. An analogous argument allows us to approximate 
\e{ &\delta_x = \delta_x^B+O(1/l), \as \delta^B_x:=\int^{a}_{-a}\mxx(k_x,0) dk_x, \lin
 &\delta_y = \delta_y^B+O(1/l), \as \delta^B_y:=\int^{b}_{-b}\mxy(0,k_y) dk_y.\la{definedeltaB}} 
Since $\delta_i^B$ is  gauge-dependent, there is no sense in which we might similarly conclude it is small.

\subsection{Quantization condition for closed orbits with intraband breakdown} \la{sec:quantizationintraband}


We summarize a few salient points from the previous subsection [\s{sec:connectionintraband}]: in the presence of intraband breakdown, we divide the Brillouin torus into overlapping subregions. A breakdown region is a strip centered at a saddlepoint in the energy-momentum dispersion, as illustrated in \fig{fig:saddlepoint}(a); wavefunctions  therein  are eigenfunctions of an approximate effective Hamiltonian in the $(K_x,k_y)$-representation. In the semiclassical subregions, the \zf wavefunctions [$\tilde{g}_{\bk E}$ in \q{lastlineiszf}] are asymptotically valid in the limit of weak fields. Both types of wavefunctions are matched where the breakdown and semiclassical subregions overlap; matching conditions are known as connection formulae, and may be expressed with the scattering matrix in \q{fullscattering}.  \\

\begin{figure}[ht]
\centering
\includegraphics[width=6 cm]{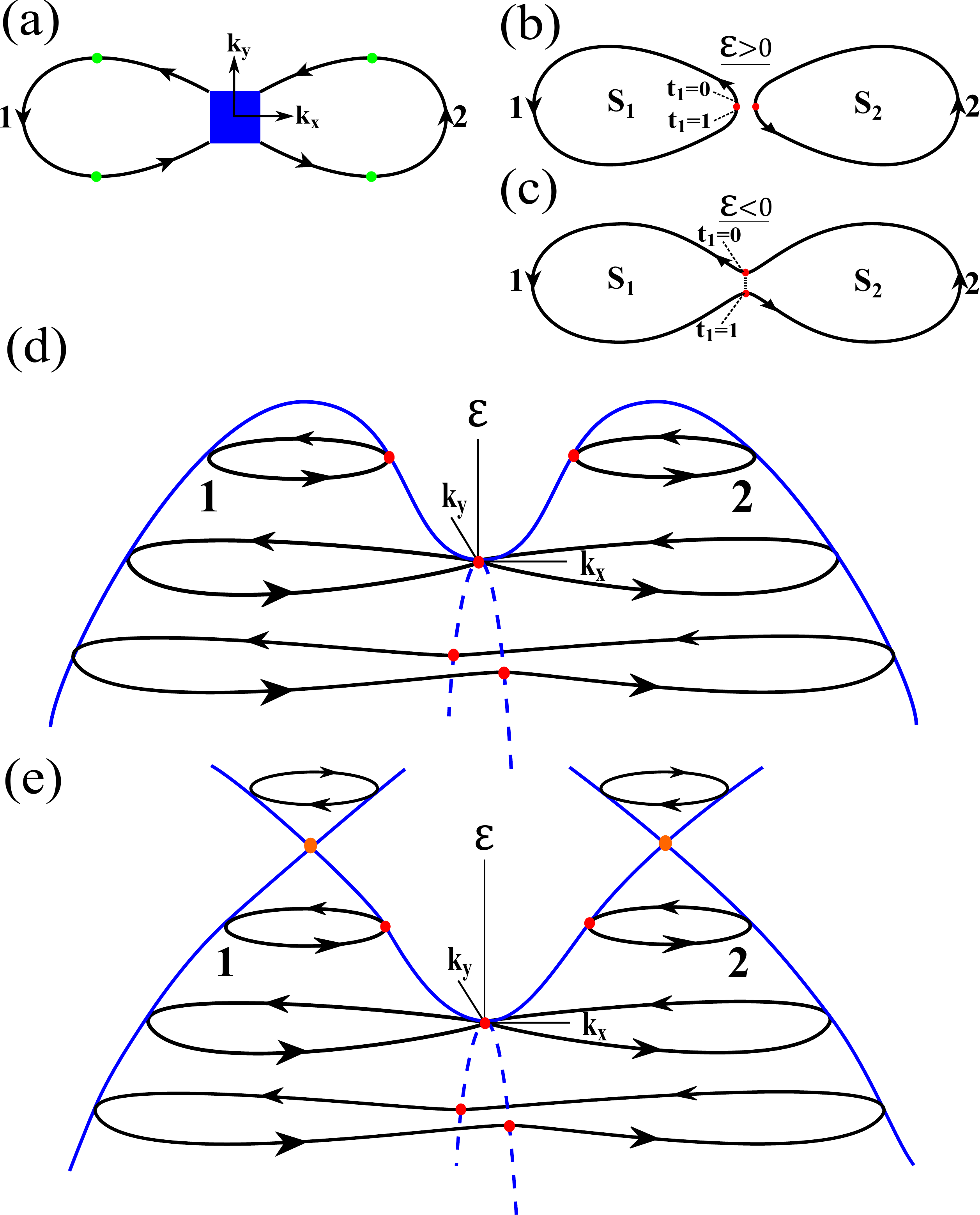}
\caption{(a) The double-well graph consists of two broken orbits (labelled $i=1,2$) linked by a single breakdown vertex.  Each broken orbit comprises three sections and two turning points. (b-c) illustrate the band contours at positive and negative energies respectively. (d-e) are possible realizations of the double-well graph: (d) depicts a band dispersion with two nearby peaks, and (e) illustrates two Dirac points (indicated by orange dots) in close proximity.}\label{fig:necking}
\end{figure}

The condition for an energy eigenstate at energy $E$ and wavevector $k_x$ is the continuity (with respect to $k_y$) of the wavefunction in the $(K_x,k_y)$-representation. This continuity condition has a simple graphical interpretation, which we will now develop. We view a closed-orbit configuration (which is presumably close to at least one saddlepoint) as a graph, which is composed of breakdown vertices and broken orbits. A breakdown vertex is region of dimension $1/l$ and centered at the coordinate of a saddlepoint, as illustrated in a blue patch in \fig{fig:necking}. A broken orbit is an orbit over a smooth trajectory that begins at a breakdown vertex and ends at a (possibly distinct) breakdown vertex [a precise definition is provided in \s{sec:graph}]. The continuity condition is conveniently expressed as a system of linear equations whose variables are scalar amplitudes ($\{A_{i E}\}$, defined in the next paragraph) which are associated to broken orbits (denoted $\{\frako_i\}$).  We will find it useful to parametrize each broken orbit ($\frako_i$) by a time-like variable $t_i \in [0,1]$, which increases along the orbit in a direction consistent with Hamilton's equation. $t_i=0$ corresponds to the point of closest approach to the saddlepoint of origin, and $t_i=1$  to the point of closest approach to the destined saddlepoint, as illustrated for the graph in \fig{fig:necking}(b) and (d). We caution the reader that: (i) these points of closest approach are zero-field band characterizations of each breakdown vertex, which is equipped with more internal structure than a point, and (ii) $t_i$ should be distinguished from $t_{\nu}$, which we introduced in \q{definescalarampsection2} to parametrize an edge $\nu$.  \\

To each point on the broken orbit $\frako_i$ we assign a scalar amplitude $A_{i,E}(t_i)$; while in principle we may specify its full functional dependence on $t_i \in [0,1]$, it is simplest in practice to just specify the ratio of the amplitudes at the end points:
\e{  e^{i\Theta_i(E,l^2)}:=\f{A_{i,E}(1)}{A_{i,E}(0)} := \prod_{p \in \frako_i}e^{i\phi_r^p} \prod_{\nu \in \frako_i}e^{i\tilde{\theta}_{\nu}}\bigg|_{E,l^2}.  \la{defineTheta}}
In the moving-wavepacket description [introduced in \s{sec:singlebandquant}], $A_{i E}(t_i)$ may be interpreted as the time-evolving amplitude for a wavepacket moving within $\frako_i$. $\Theta_i$ is then the net phase acquired by a wavepacket in traversing the full length of $\frako_i$. $\Theta_i$ includes the sum of semiclassical phases $\tilde{\theta}_{\nu}$ acquired along each edge $\nu \in \frako_i$:
\e{ e^{i\tilde{\theta}_{\nu}(E,l^2)} :=e^{-il^2\int_{k^{\nu}_{yi}}^{k^{\nu}_{yf}} \big(k_x^{\nu}-\tilde{H}^{\nu}_1(v^x_{\nu})^{\mo}\big)dk_y}\bigg|_{\tilde{E},l^2}, \la{ratioofamplitude2333} }
with $\tilde{E}$ and $\tilde{H}$ defined in \q{definetildeEH1}, and $k_{yi}^{\nu}$ [resp.\ $k_{yf}^{\nu}$] defined as the $k_y$-coordinate of the wavepacket as it enters [resp.\ leaves] the oriented edge $\nu$; precisely, if the edge $\nu$ is bounded by two turning vertices, $k_{yi}^{\nu}$ and $k_{yf}^{\nu}$ are coordinates of these two vertices; if the edge $\nu$ enters a breakdown vertex, $k_{yf}^{\nu}$ is the coordinate of closest approach to the saddlepoint. We further add to $\Theta_i$ a reflection phase $\phi_r^p$  for each turning vertex $p \in \frako_i$.  As discussed in \s{sec:turningpoint}, $\phi_r^p=\pm \pi/2$, with the sign depending on the sense of circulation of each turning vertex.  \\

The connection formula at each saddlepoint (labelled $s$) may be expressed as a scattering matrix (which in general depends on $s$) that  maps two incoming amplitudes  to two outgoing amplitudes: 
\e{ \vectwo{A_{i_{\sma{\nwarrow}},E}(0)}{A_{i_{\sma{\searrow}},E}(0)}= \bbs_s(E,l^2)\vectwo{A_{i_{\sma{\nearrow}},E}(1)}{A_{i_{\sma{\swarrow}},E}(1)}.  \la{definescatteringgeneral}}
The expression for $\bbs$ may be found in \q{scatteringintraberry} and (\ref{fullscattering}). $i_{\sma{\nearrow}}$ labels the broken orbit that is approaching the saddlepoint from the $\sma{\nearrow}$ direction, i.e., in the direction of increasing $k_x$ and $k_y$; take care that $\{i_{\sma{\nwarrow}},i_{\sma{\searrow}},i_{\sma{\nearrow}},i_{\sma{\swarrow}}\}$ do not necessarily correspond to four distinct broken orbits. \\

Combining \q{defineTheta} and \q{definescatteringgeneral} for all broken orbits in the graph, we obtain a system of linear equations with the variables $\{A_{i,E}(0)\}$, which is then solved by standard algebraic methods. A solution exists upon satisfaction of a determinantal equation that is parametrized by energy $E$ (and wavevector $k_z$ in 3D solids) -- this is the generalized Bohr-Sommerfeld quantization condition. For comparison, \q{multibandquantizationcondition} shows an analogous determinantal equation for a simple, closed orbit without breakdown.  Let us follow this algorithm to determine the quantization conditions for two case studies.


\subsection{Case study: the double-well graph, applied to conventional and topological metals}\la{sec:doublewell}

The simplest graph with a single breakdown vertex describes a Lifshitz transition where two orbits merge into one, as illustrated in  \fig{fig:necking}(a-c). Scattering from a saddlepoint is analogous to a Schrodinger particle scattering from an inverted parabolic potential. The semiclassical motion of wavepackets on either side of the saddlepoint is reminiscent of a Schrodinger particle in a double well, hence we shall refer to \fig{fig:necking}(a) as the double-well graph.\\ 

We offer two topologically distinct realizations of the double-well graph illustrated in \fig{fig:necking}(a): \fig{fig:necking}(d) illustrates a conventional metal whose band dispersion has two nearby maxima -- this has also been referred to as `necking' in \ocite{azbel_quasiclassical}; \fig{fig:necking}(e) illustrates two Dirac/Weyl points in close proximity, which materializes in topological metals near a metal-insulator transition. The double-well graph is a good description of both conventional and topological metals for an interval of energy centered at their respective saddlepoints; however, their difference in Berry phase leaves a signature in the Landau levels which we will investigate. The quantization condition for the double-well graph was first derived by Azbel in the Peierls-Osager approximation;\cite{azbel_quasiclassical} here, we derive also the subleading corrections to the quantization condition that encode the Berry phase, the orbital moment and the Zeeman effect. A particular expression of this corrected condition was presented previously in \ocite{AALG_breakdown} assuming certain crystalline point-group symmetries; here, we shall assume no such symmetries and derive the most general form of the quantization condition.


\subsubsection{Quantization condition for the asymmetric double well}\la{quantcondgenericwell}

The two broken orbits in the double-well graph are denoted by $\frako_i$, with $i=1,2$ indicated in \fig{fig:necking}(a). Corresponding to these orbits are two scalar amplitudes $(A_{1E},A_{2E})$, which are related by the scattering matrix as
\e{ &\vectwo{A_{1E}(0)}{A_{2E}(0)}= \bbs(E,l^2)\vectwo{A_{1E}(1)}{A_{2E}(1)} \lin
&\imp \det\left[\bbs \diagmatrix{e^{i\Theta_1}}{e^{i\Theta_2}}  -I  \right]\bigg|_{E,l^2}=0.
\la{necking}}
with $\Theta_j(E)$ defined in \q{defineTheta}. The above equation may be interpreted thus: a wavepacket that traverses the full length of $\frako_i$ accumulates a phase $\Theta_i$; as it passes through the breakdown region, the incoming wavepacket splits into two outgoing wavepackets with amplitudes determined by the scattering matrix. The determinantal equation in \q{necking} expresses the condition that these amplitudes are everywhere single-valued.\\   

Employing the expression for the scattering matrix [\q{fullscattering}] and the identity $\calt^2-\calr^2 = e^{i2\phi}$, the determinantal equation may be expressed trigonometrically as
\e{\cos\left[\f{\Omega_1+\Omega_2}{2}\bigg|_{E,l^2}+\phi(\tilde{\mu}) \right] = |\calt(\tilde{\mu})|\cos\left[\f{\Omega_1-\Omega_2}{2}\bigg|_{E,l^2} \right]. \la{quantizationnecking}}
$\tilde{\mu}$ has been defined in \q{definetildeEmu}, and  $\Omega_j \in \R$ may be expressed, modulo $2\pi$, as
\e{&\Omega_j(E,l^2) := \Theta_j(E,l^2)+ (-1)^{j+1}\delta_y(\tilde{E},l^2)\lin
\eq \pi+ \bigg\{l^2S_j +l^2\int_0^1 \f{\tilde{H}^{\nu(t_j)}_1}{v^x_{\nu(t_j)}} \f{dk_y}{dt_j} dt_j  +(-1)^{j+1}\delta^B_y\bigg\}_{\tilde{E},l^2}, \la{definenuj}}
with  $\bk(t_j,E)$ defined as the point on $\frako_j$ at time-like $t_j$ and energy $E$, and $\nu(t_j,E)$ labels uniquely the edge that contains $\bk(t_j,E)$. For $E>0$, $\Omega_j$ is simply the phase ($\Theta_j$)  acquired by a wavepacket as it traverses the full length of $\frako_j$ [cf.\ \q{defineTheta}]; $E<0$, $\frako_j$ is not closed [see \fig{fig:necking}(c)], and $\Omega_j$ includes an additional Berry phase ($\delta^B_y$) acquired in a tunneling trajectory that connects the two endpoints of $\frako_j$. In more detail, let us describe the four terms in \q{definenuj} in their order of appearance:\\


\noi{i} The $\pi$ term originates from the two turning vertices on each broken orbit, as indicated by green dots in \fig{fig:necking}(a). Each turning vertex has an anticlockwise circulation and contributes a $+\pi/2$ reflection phase, as discussed in \s{sec:turningpoint}.\\

\noi{ii}  We have previously employed $S(E)$ to denote the oriented area of a simple, closed orbit in \s{sec:simple}; $S$ is positive for clockwise-oriented orbits and vice versa. In the presence of intraband breakdown,  $S_j(E)$ denotes analogously the oriented area of a closed Feynman trajectory (denoted $\bar{\frako}_{j,E}$), which is a `minimally-modified closure' of the broken orbit $\frako_j$ at energy $E$. That is, we extend the broken orbit by the shortest possible path to form a closed loop. For $E>0$, $\frako_j$ is already closed [see \fig{fig:necking}(b)]; for $E<0$, we add an oriented vertical line [dashed line in \fig{fig:necking}(c)] of length $2b$ across the classically-forbidden region. We shall refer to this added line as a tunneling trajectory.  \\

\noi{iii} The two additional terms that contribute to $\Omega_j$ [in \q{definenuj}]  represent the leading-order corrections to the Peierls-Onsager approximation. The first corrective term is a phase acquired over $\frako_j$, and is generated by the Roth-Berry-Zeeman correction to the Peierls-Onsager Hamiltonian [$H_1$ in \q{H1}]. \\

\noi{iv} The second corrective term [$\pm \delta^B_y$ in \q{definenuj}] is defined to vanish for $E>0$, but for $E<0$ it is the Berry phase acquired over the tunneling trajectory that connects the boundary points of $\frako_j$ [cf.\ \q{definedeltaB}].  We may combine $\delta^{B}_y$ with the Berry contribution to $\int \tilde{H}_1dt$ to obtain an integral of the Berry connection over the closed loop $\bar{\frako}_{j,E}$. Thus, a gauge transformation of the type \q{gaugetransform} (with $D=1$) may modify $\Omega_j$ by any integer multiple of $2\pi$, but does not affect the quantization condition in \q{quantizationnecking}. This concretely exemplifies how the gauge-covariance of the scattering matrix (originating from $\delta^B_y$) results in the gauge-invariance of the quantization condition.\\

\qq{quantizationnecking}{definenuj} is the main result of this section. This quantization condition provides an algebraic approach to determine the Landau levels for any tunneling strength, and without recourse\cite{Serbyn_LandauofTCI,obrien_breakdown,koshino_figureofeight} to large-scale numerical diagonalization.   There are two limits ${\mu} \rightarrow \pm \infty$ where the Landau levels determined by \q{quantizationnecking} are  locally periodic in the sense of \q{orderedspectrum}. For $E>0$ and in the limit of weak field, we combine \q{muplus} and \q{quantizationnecking} to obtain
$\sin(\Omega_j/2)=0$, which are independent quantization conditions for two orbits with negligible tunneling, as illustrated in \fig{fig:necking}(b). Each condition may be cast more familiarly as
\e{ 2\pi(n+1/2)\appr  l^2S_j(E_n) +\oint_{\bar{\frako}_j}  (\orb+\bmx)\cdot d\bk \lin
&+ Z(\sigma^z/{v^{\sma{\perp}}}) |d\bk|\bigg|_{E_n},\la{anticlockwisequantcond}}
with $v^{\sma{\perp}}:=(v^x+v^y)^{\sma{1/2}}$; this expression is the anticlockwise-oriented analog of the single-band quantization condition in \q{rule3a} for simple closed orbits. The last three terms on the right-hand-side are the Roth, Berry and Zeeman contributions, as we have defined below \q{rule3a}. In deriving \q{anticlockwisequantcond}, we have employed a well-known expression for the cyclotron mass\cite{ashcroft_mermin} [$\partial S/\partial E=-\oint |d\bk|/v^{\sma{\perp}}$]
and the identity:  
\e{ &\text{for}\; |E|>0, \as \bigg\{S_j +H_1(\bze)\int_0^1 \f{dt_j}{v^x_{\nu(t_j)}} \f{dk_y}{dt_j} \bigg\}_{E-H_1(\bze)}\lin
\eq \bigg\{S_j -H_1(\bze) \int_{\bar{\frako}_j}  \f{|d\bk|}{v^{\sma{\perp}}} \bigg\}_{E-H_1(\bze)} =S_j(E)+O(\lmf). \notag}
A different, locally-periodic spectrum emerges in the weak-field limit for $E<0$: combining  \q{muminus} and \q{quantizationnecking}, we obtain a single quantization condition for the combined orbit illustrated in \fig{fig:necking}(c): $\cos(\Omega_1/2+\Omega_2/2)=0$. This condition is equivalent to \q{anticlockwisequantcond} with the replacements $S_j \rightarrow S_1+S_2$, and $\oint_{\bar{\frako}_j} \rightarrow \oint_{\bar{\frako}_1+\bar{\frako}_2}$.\\

For general $\mu$ and not assuming any symmetry, the spectrum of \q{quantizationnecking} is neither  locally periodic, nor completely random. Corresponding to the two distinct arguments of the cosine functions in \q{quantizationnecking}, there are generally two, distinct harmonics that competitively produce a quasirandom\cite{kaganov_coherentmagneticbreakdown} spectrum, i.e., a spectrum that is intermediate between that of an ordered and disordered system. Consequently, magnetic oscillatory patterns (e.g., of the de Haas-van Alphen type) are not completely smeared out, but retain a regularity that reflects the long-range correlations in a quasirandom spectrum.\cite{kaganov_coherentmagneticbreakdown} We will refer to linearly-independent arguments of trigonometric functions in the quantization condition as `trigonometric harmonics', to distinguish them from the related concept of dHvA harmonics in the magnetization. \\

While our quantization condition is valid for any tunneling strength, we may anyway gain some intuition about quasirandomness in a weak-tunneling parameter regime where one trigonometric harmonic is dominant over the other. The dominant harmonic determines a semiclassical Landau fan  in the absence of tunneling; to clarify, a Landau fan describes discrete energy levels $\{E_j^0(B)\}_{j\in \Z}$ whose separation $(E_{j+1}-E_j)$ increases with the magnetic field, i.e., the levels fan out. To leading order in a tunneling parameter (specified below), the tunneling correction to the fan $\delta E_j(B)$ oscillates with the frequency corresponding to the weaker harmonic. Such a perturbative treatment of quasirandom spectra is developed generally in \s{sec:perturbation}. As an example, let us perturbatively treat the regime $\mu \ll 0$, where $(\Omega_1+\Omega_2)/2$ dominates over $(\Omega_1-\Omega_2)/2$. The dominant harmonic determines the semiclassical Landau fan through $\cos(\Omega_1/2+\Omega_2/2)=0$; Landau levels  are indexed by $j \in \Z$ as
\e{ \f{\Omega_1+\Omega_2}{2}\bigg|_{E_j^0}= \f{\pi}{2}+j\pi. \la{fanintra}}
To leading order in $|\calt|$ and $\phi$, the correction to the Landau fan is  
\e{ \delta E_j(B)= \f{\phi +(-1)^j|\calt|\cos\big[\,(\Omega_1-\Omega_2)/2\,\big]}{ (-1/2)\big[\,\partial (\Omega_1+\Omega_2)/\partial E\,\big]}\bigg|_{E_j^0},\la{fanintra2}}
where the factor $(-1)^j$ originated from our evaluation of sin$[(\Omega_1+\Omega_2)/2]$ at $E_j^0$. The above equation is valid assuming $|\calt|$ and $\phi$ are small and slowly varying on the scale of $\delta E$. Indeed, the typical scale of variation for $|\calt(\mu)|$ and $\phi(\mu)$ is $\Delta \mu \sim 1$ [see \fig{fig:saddlescatteringmatrix}(a-b)], which implies an energy scale $\Delta E \sim  1/\sqrt{m_1m_2}l^2$ from the defining relation $\mu=\sqrt{m_1m_2}El^2$. It follows that 
\e{\f{\delta E_j}{\Delta E} \sim  \f{\sqrt{m_1m_2}}{\partial (S_1+S_2)/\partial E}\bigg(\phi +(-1)^j|\calt|\cos\f{\Omega_1-\Omega_2}{2}\bigg)\bigg|_{E_j^0},\notag}
which vanishes for small enough field or large enough $|E_j^0|$.


\subsubsection{Quantization condition for the symmetric double well}\la{sec:symmetricdoublewell}

Next, we discuss how certain (magnetic) point-group symmetries may simplify the quantization condition, and make contact with the simpler expressions found in \ocite{AALG_breakdown}. \\

\noi{i} Consider a time-reversal-symmetric ($T$), spin-orbit-coupled solid with a two-fold rotational axis ($\rot_{2z}$) parallel to the field, but lacking spatial inversion symmetry. The latter implies bands are nondegenerate at generic wavevectors. We shall assume the Weyl points and saddlepoints [\fig{fig:necking}(c-d)] lie on generic wavevectors in a plane (e.g., $k_z=0$) that is invariant under both rotation and time reversal. Weyl points in a rotationally-invariant plane are not uncommon, as exemplified by TaAs.\cite{TaAs,ChinadiscoversTaAs,Princeton_discovers_TaAs} The combined symmetry $T\rot_{2z}$ ensures that $H_1^R=H_1^B=0$ at any $\bk$ in this plane [cf.\ \s{sec:symmetrysinglebandom}], hence \q{definenuj} simplifies to 
\e{\Omega_j(E,l^2)=\pi+l^2S_j(E)+ \oint_{\bar{\frako}_j(E)}\bmx \cdot d\bk,}
with the right-hand-side evaluated at $E=\tilde{E}$ (recall that $\tilde{E}$ differs from $E$ by $H_1^R(\bze)+H_1^Z(\bze)$).\\

\noi{ii} Suppose a mirror symmetry ($x{\rightarrow}{-}x$) relates the two maxima in \fig{fig:necking}(c) and the two Weyl points in \fig{fig:necking}(d); the saddlepoint lies on the mirror line where $H_1^R=H_1^Z=0$ [cf.\ \s{sec:symmetrysinglebandom}], hence $E=\tilde{E}$ also. Note however that the Roth and Zeeman terms are not constrained to vanish at generic wavevectors away from the mirror line, thus 
\e{\Omega_j(l^2,E)\eq \pi+l^2S_j(E) +\oint_{\bar{\frako}_j(E)}\bmx\cdot d\bk \lin
&+\oint_{{\frako}_j(E)} \orb\cdot d\bk + Z(\sigma^z/{v^{\sma{\perp}}}) |d\bk|.}

\noindent The Landau levels and dHvA oscillations for both cases (i-ii) have been studied in \ocite{AALG_breakdown}.\\

In our next case study, we will apply the algorithm developed in \s{sec:quantizationintraband} to derive the quantization condition for a relatively more complicated graph.

\subsection{Case study of topological crystalline insulators: the butterfly graph}\la{sec:tci}

\begin{figure}[ht]
\centering
\includegraphics[width=8 cm]{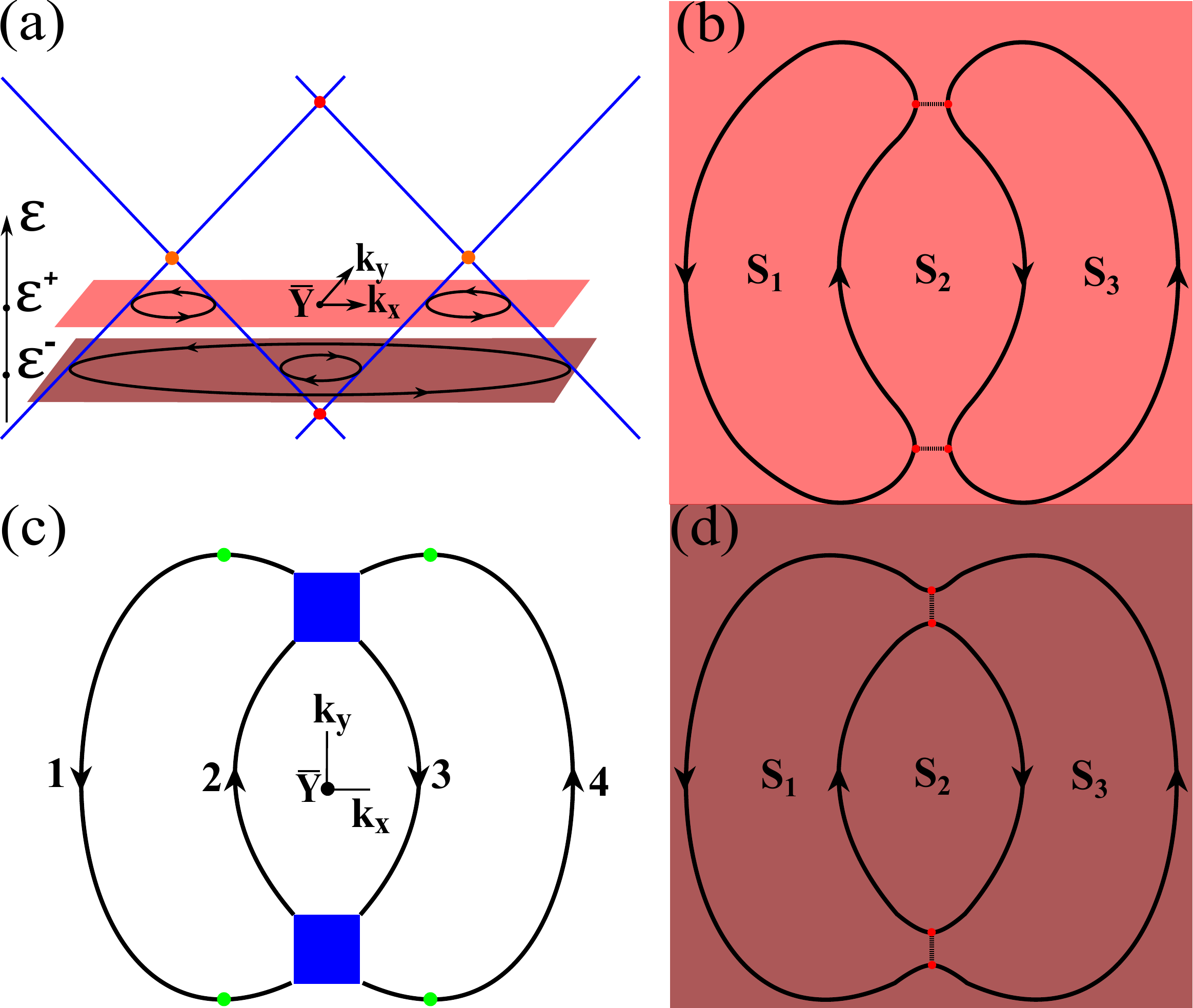}
\caption{Surface-state band contours in the SnTe-class of topological crystalline insulators. The corresponding graph consists of four broken orbits (labelled $i=1,2,3,4$) linked by two breakdown vertices. Orbits $1$ and $4$ each comprises three edges and two turning vertices; orbits $2$ and $3$ each comprise a single edge.}\label{fig:tci}
\end{figure}

The butterfly graph illustrated in \fig{fig:tci}(c) is materialized on the 001 surface of the SnTe-class of topological crystalline insulators,\cite{Hsieh_SnTe,Xu_observeSnTe,Tanaka_observeSnTe} which has the same symmetry as rocksalt. The 001 surface is symmetric under the point group $C_{4v}$,  which is generated by the four-fold rotation $\rot_{4z}$ and the reflection $\mir_x$. We focus on the vicinity of the $\rot_{2z}$-invariant wavevector $\bar{Y}$, which is an intersection of two orthogonal reflection-invariant ($\mir_x$ and $\mir_y$) lines. Along the $\mir_x$-invariant line, the dispersion of the surface states is plotted with blue lines in \fig{fig:tci}(a). The four surface bands intersect at four Dirac points, two of which (indicated by red dots) are robust due to Kramers degeneracy, and the other two (brown dots) are robust due to $\mir_x$ symmetry.  We shall distinguish them by calling the former $T$-Dirac points, and the latter $\mir_x$-Dirac points. At energy $\var^+$ just below the $\mir_x$-Dirac points, the band contours form two non-concentric circles (within the red plane); at energy $\var^-$ just above the lower $T$-Dirac point, the band contours form two concentric circles (within the brown plane). At an intermediate, critical energy, there is necessarily a  Lifshitz transition\cite{Hsieh_SnTe} facilitated by two saddlepoints, as illustrated in \fig{fig:tci}(b-d). \\

Following our algorithm to determine the quantization condition, we first identify four broken orbits and label them as $1,2,3,4$ in \fig{fig:tci}(c). Corresponding to these orbits are four scalar amplitudes, which are related by the scattering matrices as 
\e{ &\vectwo{A_{1E}(0)}{A_{3E}(0)}= \bbs(E)\vectwo{A_{2E}(1)}{A_{4E}(1)} , \lin
 &\vectwo{A_{2E}(0)}{A_{4E}(0)}= \bbs(E)\vectwo{A_{1E}(1)}{A_{3E}(1)} \lin
 \imp& \det\left[\bbs \diagmatrix{e^{i\Theta_2}}{e^{i\Theta_4}} \bbs \diagmatrix{e^{i\Theta_1}}{e^{i\Theta_3}}  -I  \right]\bigg|_{E}=0. \la{tci}}
Here, the scattering matrices corresponding to the two saddlepoints are identical owing to $\mir_y$ symmetry; we remind the reader that $\Theta_i$ is the semiclassical phase acquired by a wavepacket in traversing the full length of $\frako_i$, as defined in \q{defineTheta}. In spin-orbit-coupled systems with $\rot_{2z}T$ symmetry, both \emph{single}-band Roth and Zeeman terms vanish (i.e., $H_1^Z=H_1^R=0$); this follows from particularizing the general symmetry constraints in \q{gactsonMmultibandBz}-(\ref{transformszmatrix}). The Berry term is, however, non-negligible due to the Dirac cones present in this bandstructure. \\

Let us then insert the Berry-corrected scattering matrix [\q{scatteringintraberry}] into \q{tci} and perform the necessarily algebraic manipulations, with aid from the identity $\calt^2-\calr^2=e^{i2\phi}$. The result may be stated intuitively in this manner: let us define for each of the three delineated regions  in \fig{fig:tci}(c) a closed Feynman trajectory [a concept described below \q{definenuj}], which we denote respectively as $\bar{\frako}_{1,E},\bar{\frako}_{2,E}$ and $\bar{\frako}_{3,E}$. The semiclassical phase acquired from traversing each Feynman trajectory in a direction determined by Hamilton's equation is, respectively, 
\e{&\Omega_1(E,l^2)=l^2S_1(E),\as \Omega_2(E,l^2)=l^2S_2(E)+\pi,\lin
    &\Omega_3(E,l^2)=l^2S_3(E),}
with negative $S_1$ and $S_3$ (due to the anticlockwise orientations of $\bar{\frako}_{1,E}$ and $\bar{\frako}_{3,E}$) and positive $S_2$.  Each of $\{\bar{\frako}_{j,E}\}_{j=1}^3$ encircles a Dirac point (as illustrated in \fig{fig:tci}), and is therefore characterized by a Berry phase of $\pi$; once again, the robustness of $\pi$ is due to $\rot_{2z}T$ symmetry. There are two turning points on each of $\bar{\frako}_{1,E}$ and $\bar{\frako}_{3,E}$, as indicated by green dots in \fig{fig:tci}(c)  -- the resultant Maslov correction cancels the Berry-phase correction in $\Omega_1$ and $\Omega_3$. Finally, we should exploit that the area of left and right boundaries ($\bar{\frako}_{1,E}$ and $\bar{\frako}_{3,E}$) are identical due to $\mir_x$ symmetry, hence $\Omega_1=\Omega_3$. Putting all this together, the quantization condition may be expressed as a competition of two trigonometric harmonics:
\e{ 0 \eq e^{-i2\phi}+e^{i(\Omega_1+\Omega_3+2\phi)}+|\calr|^2\big[e^{i(\Omega_1-\Omega_2+\Omega_3)}+e^{i\Omega_2}\big]\lin
&-|\calt|^2\big[e^{i\Omega_1}+e^{i\Omega_3}\big] \lin
\imp 0\eq \cos\big(\Omega_1+2\phi\big)+|\calr|^2\cos\big(\Omega_1-\Omega_2\big)-|\calt|^2.\notag}
In the three limits of $\mu$ described in \q{muplus}-(\ref{muminus}),
\e{ \mu \rightarrow +\infty, \as& l^2S_1=2n\pi;\lin
\mu = 0,\as& 1=2\cos[l^2S_1]-\cos\left[l^2(S_1+S_2)\right]; \lin
\mu \rightarrow -\infty, \as& l^2(2S_1+S_2)=2m\pi, \as l^2S_2=2n\pi,}
with $m,n \in \Z$. There are two semiclassical limits of the quantization condition: for $\mu \rightarrow +\infty$ (resp. $\mu\rightarrow -\infty$), we obtain independent quantization conditions for two non-concentric (resp. concentric) simple orbits; in these cases, the Maslov and Berry corrections sum to zero modulo $2\pi$. Except in these two semiclassical limits, the spectrum is quasirandom, and may be analyzed with the perturbative techniques developed in \app{app:perturbinter}.




\section{Effective Hamiltonian for general band touchings}\la{sec:effhamgen}

Band touchings have long provided endless entertainment in condensed-matter physics.\cite{Herring_accidentaldeg,Herring_effectofTRS} There are two senses in which bands may robustly touch at a point  in $\bk$-space.  In one sense, the touching is movable, but alone it is unremovable. A 3D Weyl point exemplifies a linearly-dispersing touching between two bands which is free to move in the Brillouin torus,\cite{Nielsen_ABJanomaly_Weyl,Murakami2007B,wan2010,burkov2011} but can never be removed unless it meets a Weyl point with an opposite chirality.\cite{barrysimon_holonomy,Horava_stabilityofFSandKtheory} The freedom of one Weyl point to move but not to gap out may be understood from the following argument: in the absence of symmetry, a touching between two bands is described locally (in $\bk$ space) by a two-dimensional Hamiltonian having no constraints. For a generic two-by-two Hermitian matrix, three real parameters must be tuned to impose a degeneracy. In general, we refer to the number of real Hamiltonian parameters needed to tune a degeneracy as the {co-dimension} ($p$) of the Hamiltonian;\footnote{There exists a related notion of co-dimension that is defined differently in the literature.\cite{Horava_stabilityofFSandKtheory,burkov2011}}  the co-dimension depends on the symmetry class of the Hamiltonian, which in the present discussion is trivial. In 3D solids, the Brillouin torus affords us three parameters, hence perturbations of the $\bk$-dependent Hamiltonian of a Weyl fermion merely moves the Weyl point but cannot gap it out.\\

 Imposing a point-group symmetry (of both symmorphic and nonsymmorphic kinds), often in combination with time-reversal symmetry, may reduce the co-dimension. If  such symmetry exists in the groups of all wavevectors in the 3D Brillouin torus, then line nodes  are stable. More generally, the stable nodes form a $(d-p)$-dimensional submanifold of a $d$-dimensional manifold in $\bk$-space;  $p$ is the symmetry-dependent co-dimension of the Hamiltonian,  $d$ is the dimension of manifold where this symmetry acts locally (i.e., maps $\bk \rightarrow \bk$). $d$ may be less than the spatial dimension of the solid. For example, 3D Weyl points are stable in 2D submanifold ($d=2$) that is invariant under the composition of two-fold rotation and time reversal (which enforces $p=2$);\cite{TaAs} other examples where $d=p=1$ may be found in the literature.\cite{chen_multiweyl,zhijun_3DDirac,LMAA} \\


In the other sense of robustness, a band touching may be both immovable and unremovable. It occurs at high-symmetry points or lines, and is attributed to a high-dimensional irreducible representation of the little group at such a point. In time-reversal-invariant, spin-orbit-coupled systems, the possible dimensions of these irreducible representations are: $3,4,6,8$.\cite{Young_diracsemimetal2D,Bradlyn_newfermions} \\

The physical phenomena that are attributed to all these band touchings form an immense literature; much of this literature focuses on their unusual magnetic response.\cite{Nielsen_ABJanomaly_Weyl,son_chiralanomaly,xiong_discoverchiralanomaly,obrien_breakdown,AALG_breakdown} Any theoretical understanding of these magnetic phenomenon begins in the formulation of an effective Hamiltonian that is applicable to band degeneracies; however, this formulation is complicated by the discontinuity\cite{Zak_diracpoint} of the band eigenfunction at a touching point. The standard lore is to operationally implement the Peierls substitution in a $\bk \cdot \bp$ Hamiltonian. To our knowledge, such a lowest-order effective Hamiltonian has only been justified for a two-band touching with a linear dispersion,\cite{slutskin,chambers_breakdown} i.e., no justification exists for: (a) two-fold degeneracies with nonlinear dispersions (e.g., the multi-Weyl points in \ocite{chen_multiweyl}), and (b) higher-fold band degeneracies.\cite{Bradlyn_newfermions} \\

Moreover, there has been no attempt to derive higher-order (in $\lmt$) corrections to the Peierls-substituted Hamiltonian. A Peierls-subtituted Hamiltonian for a low-energy band subspace (that touch) accurately determines Landau levels if this subspace is from all other bands by an energy gap that is large compared to the cyclotron energy. However, in naturally-occurring solids, the band-touching subspace is typically embedded in a larger space of bands which disperse like spaghetti, and energy gaps between bands are typically small.\cite{Bradlyn_newfermions} In some cases,\cite{tiantianAA_doubleweylphonons,adrienbouhon_globalbandtopology} the band-touching subspace is connected (in the sense of a graph\cite{connectivityMichelZak,elementaryenergybands}) to a larger-rank elementary band representation.\cite{Zak_bandrepresentations,Evarestov_bandrepresentations,Bacry_bandrepresentations,TQC} Simply stated, symmetry enforces that there are other bands close by.   \\
 
This chapter addresses the above issues by presenting an effective Hamiltonian that is applicable to \emph{any} type of band touching, including all cases mentioned above. The lowest-order effective Hamiltonian confirms the standard lore that the Peierls substitution works, if correctly done. Motivated by applications to spaghetti bands, we also derive the subleading corrections to the Peierls-type Hamiltonian, which encode the band-degenerate generalization of the orbital moment and the geometric phase.


\subsection{Basis functions in the vicinity of a band degeneracy}\la{sec:interbandbasis}

In the rest of this section, we use $\bk:=(k_x,k_y)$ to denote a two-component wavevector, with the understanding that $k_z$ (for 3D solids) is a conserved quantity for a field aligned in $\vec{z}$.
The majority of band touchings occur at isolated wavevectors in the constant-$k_z$ plane -- these point degeneracies are 2D Dirac points. Even generically-dispersing line nodes may be viewed as a point degeneracy, when we restrict the line node to a constant-$k_z$ plane.\footnote{If the line node is protected by a mirror symmetry, we have assumed here that the field is not aligned orthogonal to the mirror plane.}\\


The effective Hamiltonian [cf.\ \q{effhamasymptoticexpansion}] in a basis of field-modified Bloch functions [cf.\ \q{zerothrothfunction}] is not applicable near a point degeneneracy $\bar{\bk}$. Indeed, \q{effhamasymptoticexpansion} is an asymptotic expansion in the parameter $\lmt$, and each power of $\lmo$ is accompanied with a derivative (with respect to $\bk$) of either the Bloch Hamiltonian, the band dispersion, or the cell-periodic energy eigenfunction ($u_{n\bk}$).\cite{rothI} The validity of this expansion thus relies on  $\lmo \nabk u_{n\bk}$ being of order $a/l$ (with $a$ a lattice period), lest there is no sense in which $H_{j+1}$ is smaller than $H_j$. However, this would not be true in the vicinity of the Dirac point, where the Berry connection (for a $\bk$-space derivative in the azimuthal direction) diverges.\cite{Blount,Zak_diracpoint} This directly invalidates the first-order Berry term [$H_1^B$ in \q{H1berry}] in the expansion. \\




The appropriate basis functions  near a band-touching point are  either field-modified Wannier functions (in a generalized sense\cite{chambers_breakdown,nenciu_review}) or field-modified Luttinger-Kohn functions.\cite{slutskin} Luttinger-Kohn functions  are well-known from the effective-mass theory,\cite{Luttinger_Kohn_function,foreman_effectiveHamfordegbands} and have been reviewed in \s{sec:reviewluttkohn}.  We will adopt the latter approach by Slutskin, which produces an effective Hamiltonian that acts on wavefunctions over quasimomentum space. \\

Previous derivations\cite{chambers_breakdown,slutskin} of the effective Hamiltonian have only been carried out to lowest order in the field, and only for a conventional Dirac point with a conical dispersion. Here, no assumptions will be made about the degeneracy or the band dispersion.  We will employ an ansatz for the wavefunction that is inspired by Slutskin:\cite{slutskin}
\e{ \Psi(\br) = \tf1{\sqrt{N}}\sum_{n\bk} e^{i\bk \cdot \br}{u}_{n,K_x,0}(\br)f_{n\bk}, \la{ansatzexpansion}}
with $K_x$ [the kinetic quasimomentum operator defined in \q{defineKyrepresentation}] acting on $f_{n\bk}$, which we refer to as the wavefunction in the $(K_x,0)$-representation; shortly we will derive an effective Hamiltonian for $f_{n\bk}$. In \q{ansatzexpansion} and henceforth, we suppress the spin index and assume $\bar{\bk}=\bze$ for notational simplicity. $u_{nK_x0}$ is defined by replacing $k_x$ in $u_{nk_x0}$ by the kinetic quasimomentum $K_x$. Explicitly, employing the Wannier-function expansion of $u_{nk_x0}$ in \q{expandcellperiodicfunc0},\footnote{If the band is not representable by Wannier functions, one may employ a Fourier-integral representation of $u_{nk_x0}$ instead [cf.\ \q{fourierinversion}-(\ref{replace})].} we replace  $k_x$ in  the exponent of \q{expandcellperiodicfunc0} by $K_x$:
\e{ u_{nK_x0}(\br) := \tf1{\sqrt{N}}\sum_{\bR}e^{-iK_x(x-R_x)}W_n(\br-\bR).\la{expandcellperiodicfunc}}
  The boundary conditions on $f_{n\bk}$ are determined [as detailed in \app{app:slutskinbasis}] by the condition that the expansion in \q{ansatzexpansion} is independent of the unit cell in $\bk$ space, i.e.,
\e{  \alpha(\bk,\br):=\sum_n  e^{i\bk \cdot \br}{u}_{nK_x0}(\br)f_{n\bk} =  \alpha(\bk+\bG,\br). \la{bcalpha}}
for any reciprocal vector $\bG$. For definiteness, we will choose $\sum_{\bk}$ to be an integral over the first Brillouin zone.\\

The main motivation for our ansatz is that $\{u_{nk_x0}\}$ can be chosen to be smooth with respect to $k_x$ (even at the band touching point), and we might therefore anticipate that the resultant effective Hamiltonian is well-behaved analytically. An example of a smooth basis would be the energy eigenfunctions of $\hH_0(k_x,0)$, with corresponding energy functions $\{\var_{nk_x0}\}$ that are smooth across $k_x=0$; we refer to this as the `energy basis'.\\

  Under certain formal assumptions, our ansatz in \q{ansatzexpansion} is equivalent to an expansion in Slutskin's basis functions\cite{slutskin} [cf.\ \app{app:slutskinbasis}]. An analogy can also be made with Roth's basis\cite{rothI} of field-modified Bloch functions [cf.\ \q{modifiedblochfunctions}-(\ref{zerothrothfunction})]. Indeed, Roth's ansatz is equivalent to \q{ansatzexpansion} with $u_{nK_x0}(\br)$ replaced by  $u_{n\bK}(\br)$, as we demonstrate in App.\ \ref{app:fieldmodBlochfunc}.\\





We will demonstrate that our basis functions are complete and orthonormal with respect to functions in $\mathbb{R}^d$; neither of these properties were proven in the previous works,\cite{chambers_breakdown,slutskin} and instead a variational argument was used. The question of completeness: can any function over $\R^d$ be written in the form of \q{ansatzexpansion}? We may make the following argument for the positive claim: if $\lmt$ is set to zero in \q{ansatzexpansion}, it reduces to an expansion over Luttinger-Kohn functions: $u_{nk_x0}\eikr$, which are known to form a complete and orthonormal set of basis functions.\cite{Luttinger_Kohn_function} For sufficiently small fields, it is plausible that the completeness and orthonormality relations are preserved; the latter property should presently be understood as an operator relation
\e{ \int d\br \; \dg{u}_{mK_x0}(\br) \emikr e^{i\bk'\cdot\br} {u}_{nK'_x0}(\br) = \delta(\bk-\bk')\delta_{mn}, \la{generalizeorthonormality}}
with $\bk$ and $\bk'$ restricted to the first Brillouin zone. Let us prove our claim.\\


\noindent Some well-known properties of Luttinger-Kohn functions will be useful, including the completeness and orthonormality of $\{u_{nk_x0}\}$ with respect to cell-periodic functions [reviewed in \q{cellperiodicbasis}]. These properties are simply generalized to the operator relations
\e{ \sum_n{u}_{n,K_x0}(\btau) \dg{{u}}_{n,K_x0}(\btau') \eq \delta(\btau-\btau'); \la{slutcomplete}\\
 \int d\btau \dg{{u}}_{m,K_x0}(\btau) {u}_{l,K_x0}(\btau) \eq \delta_{ml}. \la{slutortho}}
We remind the reader that $\btau$ is the cell-periodic position coordinate, and $\int d\btau$ is the integral over a unit cell; we will often decompose $\br=\btau+\bR$, with $\bR$ labelling a Bravais-lattice cell. The adjoint operation in \q{slutortho} is defined as $\dg{u}_{nK_x0}:=[u_{nk_x0}^*]$, i.e., we first complex-conjugate the symbol and then symmetrize it. We will employ that the Bloch functions (denoted $\{v_{n\bk}(\br)\eikr\}_{n\in \Z}$ with $v$ cell-periodic) are complete with respect to functions of $\br \in \R^d$, i.e., any $\Psi(\br)$ may be expressed as
\e{ \Psi(\br)=\tf1{\sqrt{N}}\sum_{n\bk}e^{i\bk \cdot \br} v_{n\bk}(\btau)g_{n\bk} \la{proofcomplete1}}
for some function $g_{n\bk}$. By expressing $v_{n\bk}(\btau)=\int d\btau'\delta(\btau-\btau')v_{n\bk}(\btau')$ and inserting \q{slutcomplete}, we arrive at
\e{ \Psi(\br)=\tf1{\sqrt{N}}\sum_{m\bk}e^{i\bk \cdot \br} u_{mK_x0}(\btau)\sum_n  \braket{u_{mK_x0}}{v_{n\bk}}g_{n\bk}, \la{proofcomplete2}}
from which we identify the wavefunction in the $(K_x,0)$-representation as $f_{m\bk}=\sum_n\braket{u_{mK_x0}}{v_{n\bk}}g_{n\bk}$. This proves completeness. To prove the orthonormality, we exploit the translational symmetry of $u_{nK_x0}(\br)=u_{nK_x0}(\br+\bR)$ to express the left-hand-side of \q{generalizeorthonormality} as
\e{\int d\btau\; \dg{u}_{mK_x0}(\btau) \bigg\{\sum_{\bR}e^{i(\bk'-\bk)\cdot \bR}\bigg\} e^{i(\bk'-\bk)\cdot \btau}   {u}_{nK'_x0}(\btau).}
The sum over $\bR$ produces $\delta(\bk-\bk')$; from \q{slutortho}, we derive that the integral over $\btau$ produces $\delta_{mn}$. The orthonormality condition implies that, given any $\Psi(\br)$, we may extract its wavefunction in the $(K_x,0)$-representation by
\e{f_{n\bk}= \tf1{\sqrt{N}}\int d\br \emikr \dg{u}_{nK_x0}(\br)\Psi(\br);\la{extractslutskinwf}}
here, we have assumed $\bk$ lies in the first Brillouin zone.



\subsection{Effective Hamiltonian in the vicinity of a band degeneracy}\la{sec:interbandhamiltonian}

Our goal is to derive an effective Hamiltonian in the $(K_x,0)$-representation, i.e., acting on the wavefunction $f_{n\bk}$ which we introduced in \q{ansatzexpansion}. Due to the periodicity of $\alpha(\bk,\br)$ [cf.\ \q{bcalpha}], the position operator acts in a simple manner:
\e{ \hat{\br} \Psi(\br) = \tf1{\sqrt{N}}\sum_{n\bk} e^{i\bk \cdot \br} (i\nabk){u}_{nK_x0}f_{n\bk}, \la{actionpositionopptbasis}}
and therefore the mechanical momentum acts as
\e{
\{\hbp+\ba(\hbr)\}\Psi(\br) = \tf1{\sqrt{N}}\sum_{n\bk}\eikr \{\hbp + \bK\} u_{nK_x0}f_{n\bk}}
with $\bK=\bk+\ba(i\nabk)$ the kinetic quasimomentum operator. It follows that the field-on Hamiltonian [\q{definefieldonhamiltonian}-(\ref{definefieldonpaulihamiltonian})] acts as
\e{\hH \Psi(\br) = \tf1{\sqrt{N}}\sum_{n\bk}\eikr \hH_0(\bK) u_{nK_x0}f_{n\bk}. \la{hha2}}
Applying the operation $(1/N)\int d\br \emikr \dg{u}_{\sma{mK_x0}}$ to the time-independent Hamiltonian equation: $(\hH-E)\Psi=0$, we obtain an effective Hamiltonian equation
\e{ \sum_{n}\big\{\tilde{\calh}_{mn}-E\delta_{mn}\big\}f_{n\bk}=0.\la{effhameqslutskin}}
The $E\delta_{mn}$ term in \q{effhameqslutskin} is simply obtained from the wavefunction extraction of \q{extractslutskinwf}; determining $\tilde{\calh}$ requires a calculation that we detail in \app{app:unsymmetrized} -- its complete form is
\e{ \tilde{\calh}\eq \bigg[\tilde{H}_0+ k_y\tilde{\Pi}^y+\f{k_y^2}{2m} {-}\f1{l^2}\bigg(\tilde{\mx}^x \tilde{\Pi}^y +\f{k_y}{m}\tilde{\mx}^x\bigg) \lin
&+ \f1{2ml^4}\bigg(\tilde{\mx}^x\tilde{\mx}^x- i\partial_{k_x}\tilde{\mx}^x\bigg)\bigg]_{K_x,0}.\la{effhamslutfirst} }
Here, $\tilde{H}_0$, $\tilde{\bPi}$ and $\tilde{\bmx}$ are matrices defined respectively in \q{definehamiltonianmatrix}, (\ref{definevelocitymatrix}) and (\ref{defineberryconnection}); the notation $[\tilde{H}_0]_{\sma{K_x,0}}$ is shorthand for the operator $[\tilde{H}_0(k_x,0)]$. To simplify the presentation, we assume that $\hH$ corresponds to the Schrodinger Hamiltonian minimally coupled to the electromagnetic field [cf.\ \q{definefieldonhamiltonian}]; the Pauli case [cf.\ \q{definefieldonpaulihamiltonian}] is a simple generalization of the present equations.  \\

While \q{effhamslutfirst} is formally an infinite-dimensional matrix equation that is valid over the entire Brillouin torus, we are pragmatically interested in a few-band, effective Hamiltonian that corresponds to a low-energy subspace projected by $P$; in the $\bk$-region of interest, it is assumed there are no band touchings between $P$ and its orthogonal complement. 
To achieve an effective few-band Hamiltonian, we need to transform $\calh$ as 
\e{ \dg{S}\tilde{\calh} S=\tilde{\calh}', \la{blockgoal}}
such that $\tilde{\calh}'$ is block-diagonal with respect to the decomposition $P\oplus Q$.  From the wavefunction perspective, we are modifying our ansatz in \q{ansatzexpansion} as
\e{\Psi'(\br) = \tf1{\sqrt{N}}\sum_{n\bk} e^{i\bk \cdot \br}{u}_{n,K_x,0}(\br)S(\bK)f_{n\bk}. \la{modifiedansatz}}
One aspect of the block-diagonalization is well-known: for any Luttinger-Kohn-type basis functions which are evaluated at $k_y=0$, we expect that any few-band, effective Hamiltonian should be valid only for small $k_y$. Consequently, we would treat $k_y/G_y$ (with $G_y$ a reciprocal period) as a small parameter. Using standard \low partitioning techniques which are well known in $\bk\cdot \bp$ theory,\cite{lowdin_partitioning,Luttinger_Kohn_function,winklerbook} the block-diagonalization may then be carried out perturbatively in $k_y$.\\


However, a nontrivial generalization of \low partitioning techniques is required, since every term in \q{blockgoal} is a function of noncommuting variables ($\bK$). The major difficulty lies in evaluating a product of matrix functions of $\bK$. To overcome this, we borrow an  insight from past constructions of effective Hamiltonians\cite{kohn_effham,blount_effham,rothI} -- namely, we will organize $\tcalh$ and $S$ in an expansion in powers of $\lmt$, such that each term in the series is a symmetrized function of $\bK$.  Once this organization is performed, we may then exploit well-known rules for the calculus of symmetrized operators. Of particular utility is the following product rule:\cite{rothI}
\e{ A(\bK)B(\bK) =\left[ e^{(i/2)\lmt \epsilon_{\ab}\nabk^{\alpha}\nabla_{\bk'}^{\beta}} A(\bk)B(\bk')\bigg|_{\bk=\bk'}\right], \la{multiplicationrule}} 
which we derive in \q{proofmultiplicationrule}. \q{multiplicationrule} is a particularization of a Moyal expansion, which is familiar from the correspondence between quantum and classical physics.\cite{moyal_QMasstatisticaltheory} As it stands, our expression for $\calh$ in \q{effhamslutfirst} is not organized in the above sense, but this will be rectified in \s{sec:symmetrized}. \\

The upshot of the last two paragraphs is that both $k_y$ and $\lmt$ should be taken as independent, small parameters. To our knowledge, partitioning the Hilbert space simultaneously with these two parameters has never been done. In \s{sec:blockdiag}, we formulate an algorithm for this partitioning, which may in principle be carried out to any order in $k_y$ and $\lmt$. When this algorithm is carried out to the lowest nontrivial order, we derive the following effective Hamiltonian:   
\e{ \calh \eq \calh_0 +\calh_1^{R}+\calh_1^{B} +O(k_y\lmt,k_y^2,\lmf) \la{effhambanddeg} \\
      \calh_0 \eq H_0(K_x,0)+ \f1{2}\big[\{k_y,{\Pi}^y(k_x,0)\}\big],\la{effhambanddegPeierls}\\
					\calh_1^{R} \eq  \f1{2l^2}\bigg[\{\tilde{\Upsilon}^y,\tilde{\Pi}^{x}\}-\{\mathring{\mx}^x,\tilde{\Pi}^y\}\bigg]_{K_x,0},\la{effhambanddegRoth}\\
					\calh_1^{B} \eq -\f1{2l^2}\{{\mx}^x,{\Pi}^y\}_{K_x,0}, \la{effhambanddegBerry}}
where $[\{A,B\}]:=[AB+BA]$. We have retained our convention that the infinite-dimensional matrices $\{\tilde{H}_0,\tilde{\mx}^x,\tilde{\bPi}\}$, when restricted to the $D$-dimensional vector space projected by $P$, are to be denoted by the same symbols without the tilde accent; cf.\ \q{restriction}. $\calh_1^{R}$ in \q{effhambanddegRoth} should be understood as the $D$-rank projection of infinite-dimensional matrices -- two of those matrices, which are both off-block-diagonal with respect to $P\oplus Q$, are defined for the first time here: (a) $\mathring{\mx}^x$ is the off-block-diagonal component of
\e{ &\tilde{\mx}^x = \mathring{\mx}^x+\dot{\mx}^x; \as \mathring{\mx}^x_{mn}=\mathring{\mx}^x_{\bar{m}\bar{n}}=0, \lin
& i\mathring{\mx}^x_{{m}\bar{n}}(k_x,0)= -\f{\tilde{\Pi}^x_{m\bar{n}}(k_x,0)}{\var_{mk_x0}-\var_{\bar{n}k_x0}},\la{splitbmx}}
while $\dot{\mx}^x$ is block-diagonal. (b) $\tilde{\Upsilon}^y$ is defined by its elements: for any $m,n$ (labelling bands projected by $P$) and $\bar{m},\bar{n}$ (labelling bands projected by $Q$),
\e{ &\tilde{\Upsilon}^y_{mn}=\tilde{\Upsilon}^y_{\bar{m}\bar{n}}=0, \as i\tilde{\Upsilon}^y_{\bar{m}n}(k_x)=-\f{\tilde{\Pi}^y_{\bar{m}n}(k_x,0)}{\var_{\bar{m}k_x0}-\var_{nk_x0}},\lin
& i\tilde{\Upsilon}^y_{{m}\bar{n}}(k_x)=-\f{\tilde{\Pi}^y_{m\bar{n}}(k_x,0)}{\var_{mk_x0}-\var_{\bar{n}k_x0}}.\la{defineUpsilon}}
These particular expressions for $\mathring{\mx}^x$ and $\tilde{\Upsilon}^y$ are valid in a certain basis for the cell-periodic functions -- namely, where  $\{u_{nk_x0}\}$ from our ansatz [cf. \q{modifiedansatz}] correspond to energy bands, and are also smooth in $k_x$; in this basis, $H_0(k_x,0)$ is a diagonal matrix, with diagonal elements equal to energy functions  $\{\var_{nk_x0}\}_{n=1}^D$ which are also smooth in $k_x$. For any line (at fixed $k_y=0$) that does not  form a loop (around the Brillouin torus), such an `energy basis' can always be found.\\


Let us discuss the possible bandstructures for which \q{effhambanddeg} may be applied. While we have motivated the choice of our basis functions [in \s{sec:interbandbasis}] by their utility in the vicinity of a point degeneracy, we should clarify that the derivation of the effective Hamiltonian [cf.\ \q{effhambanddeg}]  makes no assumptions about the presence of a point degeneracy, and is therefore also applicable to nondegenerate bands. If there exists multiple touchings between bands in the subspace of $P$, \q{effhambanddeg} is applicable if there exists an orthogonal coordinate system where all touchings occur on the straight line of fixed $k_y=0$.  The range of $k_x$ for which $\calh$ is valid is only restricted by the existence of a smooth (in $k_x$) `energy basis'; in some cases, this smooth basis may be found over the entire circle of fixed $k_y=0$.  For applications to a single point degeneracy, the essential physics is often captured by an effective Hamiltonian that is linearized in $k_x$ around said point, in which case \q{effhambanddegBerry} particularizes to
\e{ &\calh = H_0 +K_x\Pi^x + iK_x[\mxx,H_0]+ k_y\Pi^y +\f1{2l^2}\bigg(\{\tilde{\Upsilon}^y,\tilde{\Pi}^{x}\}\lin
&-\{\mathring{\mx}^x,\tilde{\Pi}^y\}-\{\mxx,\Pi^y\}\bigg) +O(k_i\lmt,k_i^2,\lmf), \la{effhambanddeglinearkxnondeg}}
where all matrices above are evaluated at $\bar{\bk}=\bze$. The above equation is derived by utilizing the identity in \q{idenderivativek}. In particular, if $\bar{\bk}=\bze$ is a point of degeneracy for all $D$ bands projected by $P(\bk)$, then $[\mxx,H_0]$ vanishes and \q{effhambanddeglinearkxnondeg} further simplifies to
\e{ &\calh =H_0 +K_x\Pi^x + k_y\Pi^y +\f1{2l^2}\bigg(\{\tilde{\Upsilon}^y,\tilde{\Pi}^{x}\}\lin
&-\{\mathring{\mx}^x,\tilde{\Pi}^y\}-\{\mxx,\Pi^y\}\bigg) +O(k_i\lmt,k_i^2,\lmf). \la{effhambanddeglinearkx}}

$\calh_0$ in \q{effhambanddegPeierls} [as well as the first three terms in \q{effhambanddeglinearkx}] shall be referred to as the Peierls-Onsager Hamiltonian in the $(K_x,0)$-representation; its form is closely analogous to the Peierls-Onsager Hamiltonian in the $(K_x,k_y)$-representation [cf.\ \q{Peierlsonsageroneband}]. Indeed, we may arrive at the first three terms in \q{effhambanddeglinearkx} by the Peierls substitution $\bk \rightarrow \bK$ of the Bloch Hamiltonian in the Luttinger-Kohn representation: $H_0(\bk)=H_0(\bze)+k_x\Pi^x(\bze)+k_y\Pi^y(\bze) +O(k_ik_j)$ [derived in \q{LKhamlinear}].  In the presence of a point degeneracy, this Peierls substitution is only valid for a Luttinger-Kohn basis that is smooth (in $k_x$) across the degeneracy. A case in point is the Peierls-Onsager Hamiltonian for the Dirac point in graphene: $\calh=  vK_x\tx+vK_y\ty$, where $\tau_j$ are Pauli matrices that span a vector space corresponding to the two sublattices. Here, we may identify $\tx=\pm 1$ as labelling the two Luttinger-Kohn functions ($u_{\pm,k_x,0}$), which depend smoothly on $k_x$ across the Dirac point. Going beyond two-band touchings with conical dispersions, we emphasize  that \q{effhambanddeglinearkx} proves the lowest-order validity of the Peierls-Onsager Hamiltonian for band touchings of any kind, including: (a) those with nonlinear dispersions, e.g., the double-Weyl point in \ocite{chen_multiweyl} disperses quadratically in two directions, as well as (b) higher-degeneracy touchings, e.g., the `spin-one Weyl' point described by $ K_xL_x+K_yL_y+k_zL_z$,\cite{berry1984,Bradlyn_newfermions} where $\bL$ are the generators of SO(3) in the spin-one representation.\\
 
Going beyond the leading-order Peierls substitution, we view $\calh_1^R$ [in \q{effhambanddegRoth}] as the direct generalization of the Roth orbital moment, and the $\calh_1^B$ [in \q{effhambanddegBerry}] as the direct generalization of the Berry term; their implications on the Landau levels will be investigated in a future work. In their original formulation,\cite{kohn_effham,blount_effham,rothI} the Roth and Berry terms describe the first-order corrections to the Peierls-Onsager effective Hamiltonian for either (i) a single, nondegenerate band [as reviewed in \s{sec:singlebandeffham}], or (ii) a subspace of degenerate bands [reviewed in \s{sec:multibandeffham}]. Here, we are claiming that $\calh_1^R$ and $\calh_1^B$ are applicable to multiple  bands, degenerate or nondegenerate, which disperse in any fashion -- possibly touching at isolated wavevectors. The broadness of our claim suggests that if we particularize \q{effhambanddeg} to cases (i) or (ii), we should be able to recover an analog of the previously-derived effective Hamiltonians -- we demonstrate this in \s{sec:recovery}.

\subsection{Derivation of symmetrized effective Hamiltonian in the $(K_x,0)$-representation}\la{sec:symmetrized}

As motivated in the paragraph containing \q{multiplicationrule}, the goal of this subsection is to derive \q{effhameqslutskin} with $\tcalh$ expressed in a power series in $\lmt$, such that each term is symmetrized with respect to $\bK$.  $\tcalh$ is defined implicitly through 
\e{\sum_n\tilde{\calh}(\bK)_{mn}f_{n\bk}= \f1{\sqrt{N}}\int d\br  \;\dg{u}_{mK_x0}(\br)\emikr \hH\Psi(\br),\la{extractslutskinham}}
with $\Psi$ having the ansatz form in \q{ansatzexpansion}, and $\bk$ assumed to lie in the integral domain of  $\sum_{\bk}$ in \q{ansatzexpansion}.  \\

In the first step, we would show that $\tcalh$ has the more explicit form: 
\e{\tcalh(\bK)_{mn}= \int d\btau \dg{u}_{mK_x0}(\btau)  \hH_0(\bK) u_{nK_x0}(\btau).\la{directlyfollows}}
Beginning from the right-hand-side of \q{extractslutskinham}, we employ \q{hha2} and the translational symmetry of the operator $u_{nK_x0}(\br)=u_{nK_x0}(\br+\bR)$ to derive
\e{ &\f1{\sqrt{N}}\int d\br\; \dg{u}_{mK_x0}(\br)e^{-i\bk \cdot \br}\hH \Psi(\br)\lin
\eq  \f1{N}\int d\br \sum_{n\bk'} \dg{u}_{mK_x0}(\br)e^{i(\bk'-\bk)\cdot \br}  \hH_0(\bK') u_{nK'_x0}(\br)f_{n\bk'} \lin
\eq  \f1{N}\int d\btau \sum_{n\bk'} \dg{u}_{mK_x0}(\btau)e^{i(\bk'-\bk)\cdot \btau} \left\{\sum_{\bR}e^{i(\bk'-\bk)\cdot \bR}\right\} \lin
&\times \hH_0(\bK') u_{nK'_x0}(\btau)f_{n\bk'} \lin
\eq  \int d\btau \sum_{n} \dg{u}_{mK_x0}(\btau) \hH_0(\bK) u_{nK_x0}(\btau)f_{n\bk},\la{fromtranssymm}}
from which \q{directlyfollows} directly follows. In the second equality, we have split the integral $\int d\br f(\br)$ as $\sum_{\bR}\int d\btau f(\btau+\bR)$, i.e., we integrate over the cell-periodic position coordinate $\btau$ and sum over all unit cells labelled by the Bravais lattice vectors $\bR$. \\

The right-hand-side of \q{directlyfollows} involves a triple product of symmetrized operators -- we evaluate it utilizing the product rule of \q{multiplicationrule}. We first consider the product of $\hH_0(\bK)$ with any other symmetrized operator -- owing to $\hH_0(\bk)$ being quadratic in $\bk$, the expansion of \q{multiplicationrule} continues at most to second order:
\e{ \hH_0(\bK)B(\bK) \eq [\hH(\bk)B(\bk)] +\f{i}{2l^2} \epsilon_{\ab}[\hPi^{\alpha}(\bk)\partial_{\beta}B(\bk)] -\f1{8ml^4}\epsilon_{\ab}\epsilon_{\alpha \nu}[\partial_{\beta}\partial_{\nu}B(\bk)] = \left[ \hH(k_{\alpha}+(i/2)\lmt\epsilon_{\ab}\nabk^{\beta}) B(\bk)\right].\la{quadraticink}}
To evaluate a triple product of the form $A(\bK)\hH_0(\bK)B(\bK)$, we may first evaluate $\hH_0B$ using \q{quadraticink}, then apply \q{proofmultiplicationrule} to $\{A\{\hH_0B\}\}$. In this manner, we derive the symbol of $\tcalh(\bK)$ as
\e{ \tcalh_{mn}(\bk)  \eq e^{(i/2)\lmt\epsilon_{\ab}\nabk^{\alpha}\nabla_{\bk'}^{\beta}} \int d\btau {u}_{mk_x0}^*(\btau)  \hH_0(k'_{\alpha} +(i/2)\lmt\epsilon_{\ab}\nabla_{\bk'}^{\beta}) u_{nk'_x0}(\btau) \bigg|_{\bk'=\bk} \lin
\eq \braopket{u_{mk_x0}}{\hH_0(\bk)}{u_{nk_x0}} +\f{i}{2l^2} \left\{\bra{\nabk^{x}u_{mk_x0}}\hPi^y(\bk)\ket{u_{nk_x0}} - \bra{u_{mk_x0}}\hPi^{y}(\bk)\ket{\nabk^{x}u_{nk_x0}} \right\} \lin
  & \as + \f1{2ml^4}\braket{\nabk^xu_{mk_x0}}{\nabk^xu_{nk_x0}},\lin
\eq \braopket{u_{mk_x0}}{\hH_0(\bk)}{u_{nk_x0}} -\f1{2l^2} \left\{\tilde{\mx}^{x},\bigg(\tilde{\Pi}^{y}+\f{k_y}{m}\bigg)\right\}_{mn;k_x0}+\f1{2ml^4}(\tilde{\mx}^x)^2_{mn;k_x0}. 	\la{abovequantity}
}
In the last equality, we employed $\hbPi(\bk)= \hbPi +{\bk}/{m}$, which follows from the definition of the velocity operator in \q{definevelocity0}-(\ref{definevelocity}). We might also express
\e{ \braopket{u_{mk_x0}}{\hH_0(\bk)}{u_{nk_x0}} = \big(\tilde{H}_0+ k_y\tilde{\Pi}^y\big)_{mn;k_x0}+ \f{k_y^2}{2m}\delta_{mn},\la{reorg2}}
using the identity in \q{expandBlochham}. The highest-order term in \q{abovequantity} is $O(\lmf)$ due to the following two reasons: (i) $e^{(i/2)\lmt\epsilon_{\ab}\nabk^{\alpha}\nabla_{\bk'}^{\beta}}$ acts on a function which depends quadratically on $k_y$ [through $\hH_0(\bk)$], and, (ii) the operator $\hH_0(k'_{\alpha} +(i/2)\lmt\epsilon_{\ab}\nabla_{\bk'}^{\beta})$ is at most of order $\lmf$.\\

As motivated towards the end of \s{sec:interbandhamiltonian}, we should consider $k_y/G_y$ as a small parameter, independent of and in addition to $\lmt$.  For any function of $\bk$ and $\lmt$, we may indicate its order in $k_y$ by a superscript: 
\e{ G_b^a(\bk,\lmt) = O(k_y^al^{-2b});\la{introducesuperscript}}
we retain our convention of indicating the order in $\lmt$ through the subscript; for matrices, we  would have additional subscripts to indicate the row and column indices: $\{G_b^a\}_{mn}$. For symmetrized operators $G(\bK)$, we may also label them as $G_b^a(\bK)$ if their corresponding symbols satisfy \q{introducesuperscript}. We are now ready to organize the effective Hamiltonian in \q{abovequantity} in a power series in the two small parameters, with the aid of \q{reorg2}:
\e{ \tcalh(\bK)\eq \tcalh_0^0+\tcalh_0^1+\tcalh_1^0+\tcalh_1^1+\tcalh_0^2+\tcalh_2^0, \la{reorgsum}\\
\tcalh_0^0(\bK)\eq  \tilde{H}_0(K_x,0), \la{reorg00}\\
\tcalh_0^1(\bK)\eq  \f1{2}[\{k_y,\tilde{\Pi}^y(k_x,0)\}],\la{reorg01}\\ 
\tcalh_1^0(\bK)\eq -\f1{2l^2}\{\tilde{\mx}^x(K_x,0),\tilde{\Pi}^y(K_x,0)\}, \la{reorg10}\\
\tcalh_1^1(\bK)\eq -\f1{2ml^2}[\{\tilde{\mx}^x(k_x,0),k_y\}], \la{reorg11}\\
\tcalh_0^2(\bK)\eq  \f{k_y^2}{2m},\la{reorg02}\\
\tcalh_2^0(\bK) \eq \f1{2ml^4}(\tilde{\mx}^x)^2\bigg|_{K_x,0}.\la{reorg20}}

\subsection{Block-diagonalization of effective Hamiltonian}\la{sec:blockdiag}

Our goal is to find a transformation ($S$) that block-diagonalizes the effective Hamiltonian [$\tilde{\calh}$ in \q{reorgsum}]
with respect to the decomposition $P\oplus Q$; recall \q{blockgoal}. We will carry out this transformation perturbatively in the two small parameters $k_y$ and $\lmt$; our approach thus marries the traditional \low partitioning in $\bk \cdot \bp$ Hamiltonians (which utilizes $\bk$ as a small parameter),\cite{lowdin_partitioning,Luttinger_Kohn_function,winklerbook} with the lesser-known block-diagonalization procedures of effective Hamiltonians (which utilize $\lmt$ as a small parameter).\cite{kohn_effham,blount_effham,rothI} \\

Let us expand $S$ in a series organized in powers of $k_y$ and $\lmt$, where each term in the series is a symmetrized function of $\bK$:
\e{&S(\bK) = I+{\sum_{i,j}}'S_j^i(\bK), \as S^i_j(\bK)=[S^i_j(\bk)],\lin
& \dg{{S^i_j}}(\bK) = [\dg{{S^i_j}}(\bk)], \as S^i_j(\bk)=O(k_y^il^{-2j}).}
By $\sum_{i,j}'$, we mean to sum over all nonnegative integers but exclude the single case of $i=j=0$. $S$ is formally an infinite-dimensional matrix operator, as are $\tcalh$ and $\tcalh'$.  We have chosen the lowest-order term in $S$ to be the identity operation, since $\calh_0^0$ is already block diagonal [cf.\ \q{reorg00}], and requires no further modification.\\

From a wavefunction perspective, we are modifying our ansatz $\Psi\rightarrow \Psi'$ as in \q{modifiedansatz}. 
Following essentially the same steps as outlined in \s{sec:interbandhamiltonian},  the modified, time-independent Hamiltonian equation [$(\hH-E)\Psi'=0$] is equivalent to
\e{  \{\dg{S}\tcalh S- E\dg{S}S\}*f_{\bk}=0,}
where, again, $A *B$ denotes a matrix multiplying a vector with implicit index summation. To maintain the structure of an eigenvalue equation, we insist on the unitary condition $\dg{S}S=I$. In practice, this unitarity condition will be imposed perturbatively. That is, from 
\e{ \dg{S}(\bK)S(\bK) \eq I+  {\sum_{i,j}}' \dg{{S^i_j}}(\bK)\lin
&+{\sum_{a,b}}'S^a_b(\bK)+{\sum_{i,j,a,b}}' \dg{{S^i_j}}(\bK)S^a_b(\bK), \notag}
we impose the conditions order by order, e.g.,
\e{ &S^0_1=-\dg{{S^0_1}},\as S_0^1=-\dg{{S_0^1}}, \lin
& 0=S_1^1+\dg{{S_1^1}}+\dg{{S^0_1}}{S^1_0}+\dg{{S^1_0}}{S^0_1},\lin
 & 0= \dg{{S^0_2}}+S^0_2+\dg{{S^0_1}}{S^0_1}, \as 0= \dg{{S^2_0}}+S^2_0+\dg{{S^1_0}}{S^1_0}, \; \ldots \la{unitarybyorder}}
We may also expand 
\e{ &\dg{S}\tilde{\calh} S =\tcalh_0^0 + \bigg(\tcalh_0^1 +[\tcalh_0^0,S_0^1]\bigg) + \bigg(\tcalh_1^0 +[\tcalh_0^0,S_1^0] \bigg)\lin
&+ \bigg(\tcalh_0^2 +\tcalh_0^0S^2_0+\dg{{S^2_0}}\tcalh_0^0+[\tcalh_0^1,S_0^1]+\dg{{S_0^1}}\tcalh_0^0S_0^1 \bigg)+\ldots \la{expandshs}}
The commutators in the above expansion were derived by utilizing the anti-Hermiticity of $S_0^1$ and $S_1^0$ [cf.\ \q{unitarybyorder}]. The commutator of two symmetrized operators is not just the symmetrized commutator of their corresponding symbols, e.g.,
\e{ [\tcalh_0^0(\bK),S_j^i(\bK)]\eq \big[[\tilde{H}_0(k_x,0),S_j^i(\bk)]\big]+{\sum_{n=1}^{i}}C^{i-n}_{j+n}. \la{commh00}}
The additional terms $\{C_{j+n}^{i+n}\}_{n=1}^i$ on the right-hand-side originate from a Moyal expansion, which we express in a more general form in \q{moyalcommutator}. The regularity of $C_{j+n}^{i+n}$ implies that for every unit increase in the power in $\lmt$, there is a corresponding unit decrease in the power of $k_y$. After all, the Moyal expansion is an expansion in $\lmt \epsilon_{\ab}\nabk^{\alpha}\nabk^{\beta}$. Each of $\{C^{i-n}_{j+n}\}$ renormalizes $\{{\tcalh^{i-n}_{j+n}}'\}$ in the block-diagonalization procedure.\\

In principle, we may block-diagonalize $\tcalh$ to any order in $k_y$ or $\lmt$; this will be demonstrated explicitly for $\tcalh^i_j$ with $(i,j)=(0,1)$ and $(1,0)$. Two identities will be useful for this purpose, which are particularizations of \q{commh00}:
\e{ [\tcalh_0^0(\bK),S_0^1(\bK)]\eq \big[[\tilde{H}_0(k_x,0),S_0^1(\bk)]\big]\lin
&+ \f{i}{2l^2}\big[\{ \tilde{\Pi}^{x}(k_x,0),\,\nabla_{\bk}^{y}S_0^1\}\big], \la{moyalcommutatorapplied1} \\
[\tcalh_0^0(\bK),S_1^0(\bK)]\eq \big[[\tilde{H}_0(k_x,0),S_1^0(\bk)]\big].\la{moyalcommutatorapplied2}}
\q{moyalcommutatorapplied1} has one more term than (\ref{moyalcommutatorapplied2}) because $S_0^1$ is linear in $k_y$ while $S_1^0$ is independent of $k_y$.  We will exemplify how the last term in \q{moyalcommutatorapplied1} renormalizes $\tcalh_1^0$. \\

Employing the identities in \q{moyalcommutatorapplied1}-(\ref{moyalcommutatorapplied2}), the first two brackets in \q{expandshs} may be expressed as
\e{  \tcalh_0^1 +[\tcalh_0^0,S_0^1]\bigg|_{\bK}\eq \bigg[\tcalh_0^1(\bk) +[\tilde{H}_0(k_x,0),S_0^1(\bk)]\lin
&+ \f{i}{2l^2}\{\tilde{\Pi}^{x}(k_x,0),\,\nabla_{\bk}^{y}S_0^1\}\bigg], \lin
     \tcalh_1^0 +[\tcalh_0^0,S_1^0]\bigg|_{\bK}\eq \bigg[\tcalh_1^0(\bk) +[\tilde{H}_0(k_x,0),S_1^0(\bk)]\bigg],}
		leading to
		\e{ {\tcalh'}_0^1\eq \bigg[\tcalh_0^1(\bk) +[\tilde{H}_0(k_x,0),S_0^1(\bk)]\bigg],  \la{tcalh01}\\
     {\tcalh'}_1^0\eq \bigg[\tcalh_1^0(\bk) +[\tilde{H}_0(k_x,0),S_1^0(\bk)]\lin
		&+ \f{i}{2l^2}\{\tilde{\Pi}^{x}(k_x,0),\,\nabla_{\bk}^{y}S_0^1\}\bigg]. \la{tcalh10}}
To block-diagonalize ${\tcalh'}_0^1$ with respect to $P\oplus Q$, we choose $S^1_0$ so that the off-block-diagonal elements of $[\tilde{H}_0,S_0^1]$ exactly cancel the off-block-diagonal elements of $\tcalh_0^1$. A simple expression for $S_0^1(\bk)$ exists if we employ a basis for cell-periodic functions $\{u_{nk_x0}\}$ that: (i) correspond to energy bands, and (ii) retains our initial assumption of smoothness in $k_x$.  The cancellation of off-block-diagonal elements in the energy basis leads to the following condition
\e{ &\{\tcalh_0^1(\bk)\}_{\bar{m}n} = -(\var_{\bar{m}k_x0}-\var_{nk_x0})\{S_0^1(\bk)\}_{\bar{m}n}  \lin
&\Rightarrow \{S_0^1(\bk)\}_{\bar{m}n}=-k_y\f{\tilde{\Pi}^y_{\bar{m}n}(k_x,0)}{\var_{\bar{m}k_x0}-\var_{nk_x0}}, }
with $\bar{m}$ labelling bands in $Q$ and $n$ the bands in $P$; clearly the above equality also holds with $\bar{m}$ and $n$ interchanged. In the last equality, we extracted $\tcalh_0^1$ from \q{reorg01}. The block-diagonal elements of $S_0^1$ may be any smooth function of $\bk$ that is linear in $k_y$; in practice we will set all of them to zero, such that 
\e{ S_0^1(\bk) = -ik_y\tilde{\Upsilon}^y(k_x),}
with $\tilde{\Upsilon}$ defined in \q{defineUpsilon}.\\
			
Inserting this equation, as well as \q{reorg10}, into \q{tcalh10} we obtain
\e{ {\tcalh'}_1^0\eq \bigg[ [\tilde{H}_0(k_x,0),S_1^0(\bk)]-\f1{2l^2}\{\tilde{\mx}^x,\tilde{\Pi}^y\}_{k_x,0} \lin
&+ \f{1}{2l^2}\{\tilde{\Pi}^{x},\tilde{\Upsilon}^y\}_{k_x,0}\bigg] \la{notmiddleterm}\\
\eq  \bigg[ [\tilde{H}_0(k_x,0),S_1^0(\bk)]+\f1{2l^2}\big(\{\tilde{\Upsilon}^y,\tilde{\Pi}^{x}\}\lin
&-\{\mathring{\mx}^x,\tilde{\Pi}^y\}\big)_{k_x,0} -\f1{2l^2}\{\dot{\mx}^x,\tilde{\Pi}^y\}_{k_x,0}\bigg]. \la{middleterm}}
In the last equality, we have separated $\mx^x$ into its  block-diagonal ($\dot{\mx}^x$) and off-block-diagonal ($\mathring{\mx}^x$) components [cf.\ \q{splitbmx}].
In similar fashion to the case of ${\calh_0^1}'$, we will cancel the off-block-diagonal elements of ${\calh_1^0}'$ by a judicious choice of $S_1^0$: 
\e{\{S_1^0(\bk)\}_{\bar{m}n} \eq \f{\big( \{\tilde{\Upsilon}^y,\tilde{\Pi}^{x}\}-\{\mathring{\mx}^x,\tilde{\Pi}^y\} -\{\dot{\mx}^x,\tilde{\Pi}^y\} \big)_{\bar{m}n;k_x0}}{-2l^2(\var_{\bar{m}k_x0}-\var_{nk_x0})}, \lin
\{S_1^0(\bk)\}_{mn}\eq\{S_1^0(\bk)\}_{\bar{m}\bar{n}}=0. }
This completes the block-diagonalization to order $k_y$ and $\lmt$; what remains is to define finite-dimensional matrices, having dimension $D$ equal to the rank of $P(\bk)$, to replace the infinite-dimensional matrices.  We then finally obtain the effective Hamiltonian of \q{effhambanddeg}.





\section{Interband breakdown}\la{sec:interband}

\begin{figure}[ht]
\centering
\includegraphics[width=8 cm]{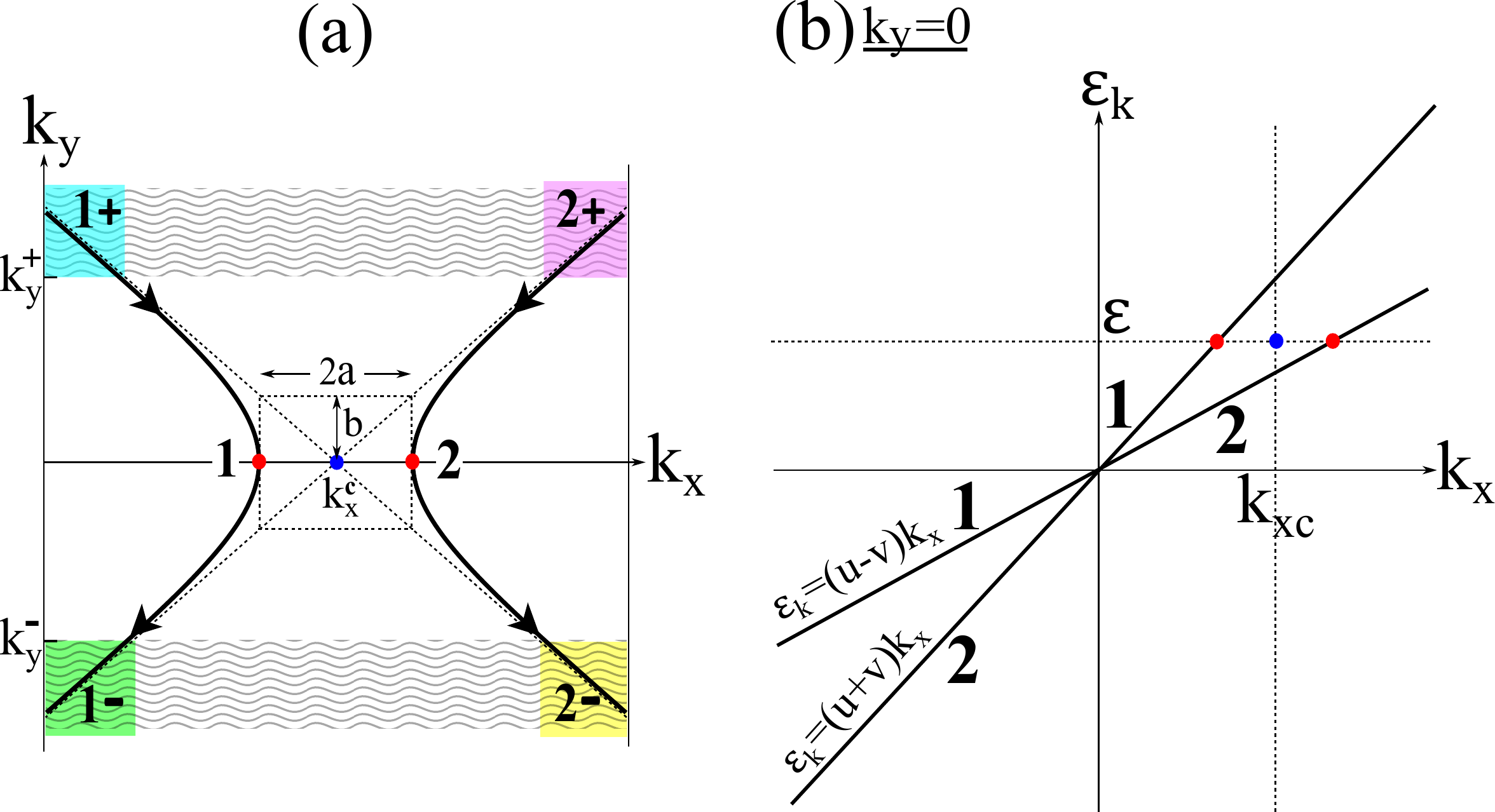}
\caption{(a) illustrates the interband breakdown region, which overlaps with the semiclassical region (indicated by grey wavy lines). Solid black lines illustrate the band contours at positive energy. The orbit velocity $\dot{\bk}$ is indicated by arrows, and determined by Hamilton's equation. (b) Band dispersion at fixed $k_y=0$ for the Hamiltonian in \q{minimalism}, with $u>v>0$ and $w>0$. The two bands are labelled by index $1$ and $2$.  }\label{fig:hyperbola_inter}
\end{figure}

Interband breakdown occurs where two constant-energy band contours -- belonging to distinct bands -- become anomalously close. As illustrated in \fig{fig:hyperbola_inter}(b), the two contours approach each other as two arms of a hyperbola, just as in the case for intraband breakdown. What distinguishes the two cases are the orientations\footnote{To define a local circulation for each hyperbolic arm, draw a line that closes the arm into a circle (in the sense of homotopy equivalence), e.g., $\curvearrowleft \rightarrow \circlearrowleft$, $\curvearrowright \rightarrow \circlearrowright$. We then assign the orientation by interpreting the circle as a clock face.} of travelling wavepackets (as determined by Hamilton's equation) on both arms: opposite for interband [\fig{fig:hyperbola_inter}(b)], and identical for intraband [\fig{fig:saddlepoint}(b-c)].    Another distinguishing feature is that for interband breakdown, only one of two in-plane components of the band velocity $(\nabk \var)$ becomes anomalously small, whereas this is true for both in-plane components in the intraband case.  \\ 


To systematize the derivation of quantization conditions, it will be useful to formalize the above discussion in the language of graph theory [cf.\ \s{sec:graph}]. Each region of strong tunnelling is identified with a degree-four, two-in-two-out vertex. For the intraband- (resp.\ interband-) breakdown vertex,  the two incoming edges are (resp.\ not) diametrically opposite to each other. To simplify notation, any `breakdown vertex' in the rest of this section should be understood as an interband-breakdown vertex.






\subsection{Symmetry analysis and Bloch Hamiltonian near a II-Dirac point}\la{sec:blochhamIIDirac}

Let us identify the symmetry classes which stabilize band touchings of the kind that leads to interband breakdown. To begin, how are band touchings (of any kind) stabilized in a Brillouin two-torus (\bt) parametrized by $\bk=(k_x,k_y)$; we shall again assume the field is aligned in $\vec{z}$. Our present discussion is restricted to \bt, but we will eventually comment on how \bt is embedded in a Brillouin three-torus. Applying the argument in the introduction of \s{sec:effhamgen}, robust band touchings occur on points or curves, if the co-dimension of the Hamiltonian is two or one respectively. We shall investigate the former case, and postpone the latter case to future studies. We shall also assume throughout this section that the point touching occurs between two nondegenerate bands; touchings between spin-degenerate bands are briefly discussed in \s{sec:outlook}. \\

We focus on point degeneracies which lie at the tip of a energy-momentum cone, i.e., the band degeneracy splits at linear order in $\bk$ (originating from the band touching), and the constant-energy contours intersect as an `X'. From a general classification of Fermi surfaces near conical band touchings,\cite{WTe2Weyl} this `X' must correspond to a type-II Dirac point.\cite{isobe_nagaosa_IIDirac,Bergholtz_typeII,LMAA}   A  II-Dirac point is minimally modelled by the following Hamiltonian in the Luttinger-Kohn representation [cf.\ \q{LKhamlinear}]:
\e{ &H_0(\bk) = (u+v\gamma_3)k_x+ wk_y \gamma_1, \lin
& \Pi^x(\bze)=u+v\gamma_3, \as \Pi^y(\bze) = w\gamma_1, \la{minimalism}}
with $\bk=(k_x,k_y)$  originating from the point of degeneracy; $\gamma_{j}$ are Pauli matrices spanning a (pseudo)spin-half basis.  The linearized band dispersion  is shaped as a `Dirac' cone which is rotationally invariant if $u=0$. If $u$ is continuously increased till $|u|>|v|$, the cone tilts over $E=0$ (the energy of the Dirac point), and the zero-energy band contour changes discontinuously from a point to an `X'. Precisely, for a finite energy window near zero, the corresponding contours form a family of  hyperbolic curves:
\e{  &\f{\left(k_x- k_{xc}\right)^2}{\bar{a}^2} - \f{k_y^2}{ \bar{b}^2} =1, \ins{with} k_{xc} := \f{uE}{u^2-v^2}, \lin
 &\bar{a}  := \f{vE}{u^2-v^2},\as \bar{b} :=  \f{vE}{w\sqrt{u^2-v^2}},\la{hyperbolainter}}
with $k_{xc}$ the center of the hyperbola; the `X' corresponds to the hyperbolic asymptotes: $k_y =\pm (\bar{b}/\bar{a})(k_x-k_{xc})$.  \\

 While \q{minimalism} is not the most general form of a II-Dirac Hamiltonian,\cite{WTe2Weyl} its simplicity manifests the physics we will describe. We may further motivate \q{minimalism}  as the most general Hamiltonian (up to unitary equivalence) satisfying the symmetry constraints:
\bal &[\hat{g}_1,H_0(\bk)]=0, \as \{\hat{g}_1,i\}=0, \as \hat{g}_1^2=+I, \la{symmg1}\\
 &\hat{g}_2H_0(\bk)\hat{g}_2^{\mo}=H_0(k_x,-k_y), \as  [\hat{g}_2,i]=0, \notag \\ 
& Tr[\hat{g}_2]=0,\as \hat{g}^2_2 =\begin{cases} +I, & [\hat{g}_1,\hat{g}_2]=0,  \\
                                    -I, & \{\hat{g}_1,\hat{g}_2\}=0.  
																			\end{cases}\la{symmg2}
\end{align}
${g}_1$ is a space-time transformation that maps $\bk \rightarrow \bk$ (within the plane), and has an antiunitary representation that squares to $+I$. For example,  $g_1$ could be $T\inv$ in an integer-spin representation; alternatively, $g_1=T\rot_{2z}$ in either half-integer or integer-spin representation, in which case \bt is identified with either of the high-symmetry planes: $k_z=0$ or $\pi$. In all these cases, $g_1$ lowers the co-dimension of the Hamiltonian to two, and hence stabilizes Dirac points in \bt. $g_2$ is an order-two, spatial symmetry which maps $\bk \rightarrow (k_x,-k_y)$. A touching between two orthogonal representations of $g_2$ occurs at the Dirac point; the commutation relations between $\hat{g}_1$ and $\hat{g}_2$ in \q{symmg2} imply that each representation of $g_2$ is invariant under $g_1$, so that the band touching may split away from $\bze$. Examples of $g_1$ include the reflection $\mir_y$, or the glide $\glide_{y,\vec{x}/2}$. Incidentially, $T\inv$  and $\glide_{y,\vec{x}/2}$  are the symmetries of the monolayer MTe$_2$ (M=W,Mo),\footnote{They are the symmetries of MTe$_2$ in the sense of a  group isomorphism.} which serves as a toy model for II-Dirac fermions.\cite{LMAA} For the pseudospin basis chosen in \q{minimalism}, $\hat{g}_1=K$ and $\hat{g}_2=\sz$. We clarify that  $\bk=\bze$ is \emph{not}  an inversion-invariant wavevector, but  a generic point on  either $g_2$-invariant line ($k_y=0$ or $\pi$). Lacking a symmetry (e.g., $\rot_{nz}$, $T$, $\mir_x$) that nontrivially transforms $k_x$ in the sense of \q{symmetryofH0}, the Hamiltonian term $uk_x$ is legal and tilts the Dirac cone in the direction parallel to the $g_2$-invariant lines.\footnote{In contrast, the Dirac cone of graphene is symmetric under $\rot_{3z}$ and cannot be tilted.} \\

There are two topologically distinct ways to embed \bt (containing a II-Dirac point) in a 3D Brillouin torus: (i) if the degeneracy splits away from \bt, it is a genuine 3D point degeneracy of the II-Weyl type; this may be modelled by adding $tk_z\gamma_2$ to the Hamiltonian in \q{minimalism}. (ii) If the degeneracy persists away from \bt, the II-Dirac point should be identified as a point on a line degeneracy; this may be modelled by adding $tk_z\gamma_3$ to \q{minimalism}, such that the line degeneracy lies on the intersection of two planes: $k_y=0$ and $tk_z+vk_x=0$. Interband breakdown in solids with line nodes was first studied by Slutskin,\cite{slutskin} but the conception of Weyl/Dirac points did not exist at his time (1967). \\


A quantity of geometric significance is the area ($4\bar{a}\bar{b}$) of the rectangle inscribed between the two hyperbolic arms [see \fig{fig:saddlepoint}]. It is natural that the dimensionless parameter
\e{ \bar{\mu}= \f1{2}\bar{a}\bar{b}l^2 = \f{v^2E^2l^2}{2w(u^2-v^2)^{3/2}} \geq 0, \la{definemuinter}}
 determines the probability of tunneling between orbits:  tunneling is negligible where $\bar{\mu}\gg 1$, and significant otherwise. The exact form of $\bar{\mu}$ will be motivated by the connection formula in \qq{abovequantumII}{definerhoandomega}, which is the key result of this section.\\


\subsection{Effective Hamiltonian for interband breakdown, and the Landau-Zener analogy}

 Following the divide-and-conquer strategy that we have employed for the turning point and the saddlepoint, we would likewise need to formulate an effective Hamiltonian that is valid at the interband-breakdown region, and solve for its wavefunction nonperturbatively. In the Landau electromagnetic gauge where $k_x$ is a good quantum number, the breakdown region is an interval $k_y\in [k_y^+,k_y^-]$  centered at the II-Dirac point where quantum tunnelling is significant, as illustrated in \fig{fig:hyperbola_inter}(a).\\

One complication for interband breakdown that did not occur in the previous two cases: a point degeneracy in the band dispersion invalidates the use of the field-modifed Bloch functions [cf.\ \q{zerothrothfunction}] as basis functions. As motivated in \s{sec:interbandbasis}, we will instead employ a set of field-modified\cite{slutskin} Luttinger-Kohn functions\cite{Luttinger_Kohn_function} which are analytic with respect to $\bk$ at the degeneracy -- these are basis functions in what we call the $(K_x,0)$ representation.  At energies where interband breakdown is relevant, our ansatz for the wavefunction is 
\e{ &\Psi_{k_x}(\br) = \f1{\sqrt{N}}\sum_{k_y \in [k_y^+,k_y^-]}\sum_{n=1}^2 e^{i\bk \cdot \br}\tilde{u}_{n,K_x,0}(\br)\tilde{f}_{n\bk}\lin
&+ \f1{\sqrt{N}}\sum_{k_y\notin [k_y^+,k_y^-]}\sum_{n=1}^2 e^{i\bk \cdot \br}{u}_{n,K_x,k_y}(\br)g_{n\bk} +O(\lmt), \la{ansatzexpansionrepeat}}
with $n=1,2$ labelling bands in the band-touching subspace, as illustrated in \fig{fig:hyperbola_inter}. $K_x$, the kinetic quasimomentum operator defined in \q{defineKyrepresentation}, acts on $\tilde{f}_{n\bk}$ and $g_{n\bk}$, which are wavefunctions in the $(K_x,0)$ and $(K_x,k_y)$ representations respectively. They are respectively valid in the breakdown and semiclassical intervals. An assumption (on the band parameters\cite{slutskin}) is made that the two domains of validity overlap. In this region of overap [indicated by wavy lines in \fig{fig:hyperbola_inter}(a)], the wavefunctions in the two representations may be matched as
\e{ \tilde{f}_{m\bk} = \sum_{n=1}^2\braket{\tilde{u}_{m,K_x,0}}{u_{n,K_x,{k}_y}}g_{n\bk} +O(\tf{k_y}{G_y},\lmt), \la{basischange}}
where $n=1,2$ are indices for the band-touching subspace. The proof of this relation is closely analogous to the proof of completeness in \q{proofcomplete1}-(\ref{proofcomplete2}); instead of employing Bloch functions which are complete with respect to functions of $r\in \R^d$, we would likewise use that field-modified Bloch functions are complete with respect to functions of $r\in \R^d$.\cite{nenciu_review,MISRA_completenessofroth} As denoted vaguely by $\ldots$ in \q{ansatzexpansionrepeat}, there  might generally be more contributions to $\Psi$ that are associated to edges far away from the II-Dirac point [and therefore not illustrated in \fig{fig:hyperbola_inter}]; these contributions  will not be important in the matching procedure described in \s{sec:connectioninter}.\\ 

As derived in \s{sec:interbandhamiltonian}, the effective Hamiltonian in the  $(K_x,0)$-representation is obtained from  \q{effhambanddeglinearkx} as
\e{ \calh_0(\bK) \eq H_0 +K_x\Pi^x +k_y\Pi^y, \la{effhambanddeglinearkx5666}}
where $H_0,\Pi_j$ are two-by-two matrices evaluated at the point of degeneracy $(\bk=\bze)$. When \q{effhambanddeglinearkx5666} is particularized to our minimal model in \q{minimalism}:
\e{ &\sum_{n=1}^2\big([\calh_0]_{mn}-E\delta_{mn}\big)\tilde{f}_{n\bk}, \iwith \lin
 &\calh_0(\bK) = K_x(u+v\gamma_3) +w k_y\gamma_1. \as  \la{effhambanddeglinearkx56}}
To simplify the notation, we would further assume $u,v,w$ are all positive. Our neglect of the first-order-in-$\lmt$ corrections in \q{effhambanddeglinearkx} is only justified if, near $\bk=\bze$, a large energetic gap separates the two-band subspace (involved in the degeneracy) from every other band.\footnote{The third $O(\lmt)$ term may be dropped explicitly if we work in a basis where $\bmx^x{=}0$, as further elaborated in \app{app:justifyWeber}.} \\

 Transforming the wavefunction as 
\e{ &\tilde{f}_{n\bk} = \alpha_{\bk E} \sum_{\bar{m}=\bar{1}}^{\bar{2}}\bar{T}_{n\bar{m}}\bar{f}_{\bar{m}\bk}, \iwith \lin
&\alpha_{\bk E} = e^{i(k_x-k_{xc}(E))k_yl^2}, \lin
& \bar{T} = \f1{\sqrt{2}}\matrixtwo{(u+v)^{\sma{-1/2}}}{(u+v)^{\sma{-1/2}}}{(u-v)^{\sma{-1/2}}}{-(u-v)^{\sma{-1/2}}}, \la{defTbar}} 
the effective eigenvalue equation describes the Landau-Zener dynamics of a two-level system: 
\e{ 0 = \left(\bar{a}\tx + \f{\bar{a}}{\bar{b}}k_y\tz + \f{i}{l^2}\p{}{k_y} \right)\bar{f}_{\bk}. \la{simplemodel}}
Here, $\tau_j$ are Pauli matrices, and \q{simplemodel} should be interpreted as a matrix differential equation acting on a two-component vector wavefunction $\bar{f}_{\bar{m}\bk}$. A more general transformation to a Landau-Zener dynamical equation is described in \app{app:justifyWeber}, which would apply to a larger class of matrix Hamiltonians than assumed for our minimal model. \\

\begin{figure}[ht]
\centering
\includegraphics[width=8.5 cm]{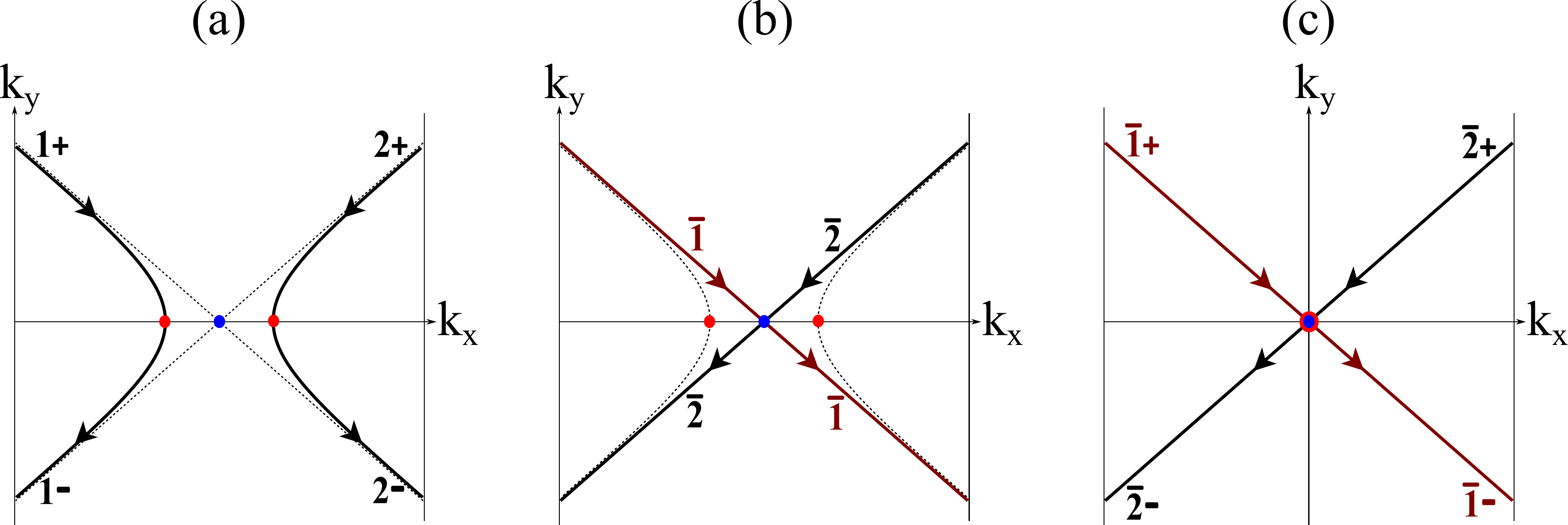}
\caption{ (a) Labels for the four edges that meet at an interband-breakdown vertex. Black solid lines correspond to constant-energy band contours at zero field and nonzero energy ($E=0$ being the energy of the II-Dirac point); we shall refer to energy bands for $E\neq 0$ as  the adiabatic basis. (b-c) illustrate the diabatic basis for $E\neq 0$ and $E=0$ respectively.  For $E\neq 0$, the diabatic basis coincides with energy bands only for $|k_y|\gg |\bar{b}|$.  For (and only for) $E=0$, the center of the hyperbola [indicated by blue dot in \fig{fig:didactic}(c)] coincides with the wavevector ($\bk=0$) of the II-Dirac point, and the diabatic basis coincides with the energy bands for all $k_y$.}\label{fig:didactic}
\end{figure}

In the Landau-Zener analogy, $k_y$ is interpreted as a time variable, and $\{k^n_x(k_y,E)\}_{n=1}^2$ as two `Landau-Zener energy' branches: 
\e{k_x^{n}(k_y,E)=k_{xc} +(-1)^n|\bar{a}| \sqrt{1+ \f{k_y^2}{\bar{b}^2}}, \as k_x^{n}(0,0)=0,\la{definekxn}}
with $k_{xc},\bar{a}$ and $\bar{b}$ being $E$-dependent  hyperbolic parameters defined in \q{hyperbolainter}. `Energy' in the Landau-Zener analogy  should not be confused with the actual energy ($E$) in the magnetic problem. We will refer to the zero-field energy bands labelled by $n=1,2$ as the $k_y$-dependent `adiabatic basis' in the Landau-Zener analogy -- as long as $E\neq 0$, there exists an adiabatic limit ($\lmt \rightarrow 0$) where the band is a conserved quantity. If $E=0$, such an adiabatic limit does not exist and the probability of tunneling is unity -- this has been described as a momentum-space analog of Klein tunneling.\cite{obrien_breakdown} At $E=0$, it is more convenient to employ a diabatic basis which  corresponds to the maximal-tunneling trajectories
\e{ k^{\bar{m}}_x(k_y) = (-1)^{\bar{m}}\big|\bar{a}/\bar{b}\big|k_y, \la{maximaltunneltraj}} 
as illustrated in \fig{fig:didactic}(c); the diabatic basis shall be labelled by $\bar{m}\in \{\bar{1},\bar{2}\}$, just as we have done for the two-component vector $\bar{f}_{\bar{m}\bk}$ in \q{simplemodel}. We see that the matrix $\bar{T}_{n\bar{n}}$ introduced in \q{defTbar} transforms between the adiabatic and diabatic bases.


\subsection{Connection formula and quantization condition for interband breakdown}\la{sec:connectioninter}


The Bohr-Sommerfeld quantization rule is the continuity condition on wavefunctions  in both representations [$\tilde{f}_{n\bk}$ and $g_{n\bk}$ in \q{ansatzexpansionrepeat}], with the understanding that $\tilde{f}$ and $g$ are related through \q{basischange} where the semiclassical and breakdown intervals overlap. In this section, we will derive a scattering-matrix formula that connects $g$ across the breakdown interval -- in effect, we may  forget about $\tilde{f}$ and impose continuity on $g$, which satisfies certain wavefunction-matching conditions.\\


Once $\tilde{f}$ is forgotten, we may proceed in close analogy to the graph-theoretic formulation of the quantization condition for intraband breakdown in \s{sec:graph}-\ref{sec:quantizationintraband}. We assume the reader has some familiarity with these sections -- we shall therefore avoid a lengthy exposition on similar-sounding generalities, in preference of a heuristic derivation of the quantization condition for a single II-Dirac point, as illustrated in \fig{fig:tilted_dirac} below. \\

A crucial ingredient to quantization conditions with interband breakdown is the connection formula for a single interband-breakdown vertex, which we will subsequently derive.  We will actually derive two connection formulae: (a)  the first formula, as summarized in \qq{abovequantumII}{definerhoandomega}, is applicable for $E\neq 0$ and   connects \zf (ZF) functions in the adiabatic basis, and (b) for $E=0$, the second formula [\qq{abovequantumIIII}{definescatteringmatrixinterEzero}] connects ZF functions in the diabatic basis.  These formulae extend a previous formula\cite{slutskin} to include the effect of the Berry phase.

\subsubsection{Connection formula for $E\neq 0$}\la{sec:connEneq0}

\begin{figure}[ht]
\centering
\includegraphics[width=7 cm]{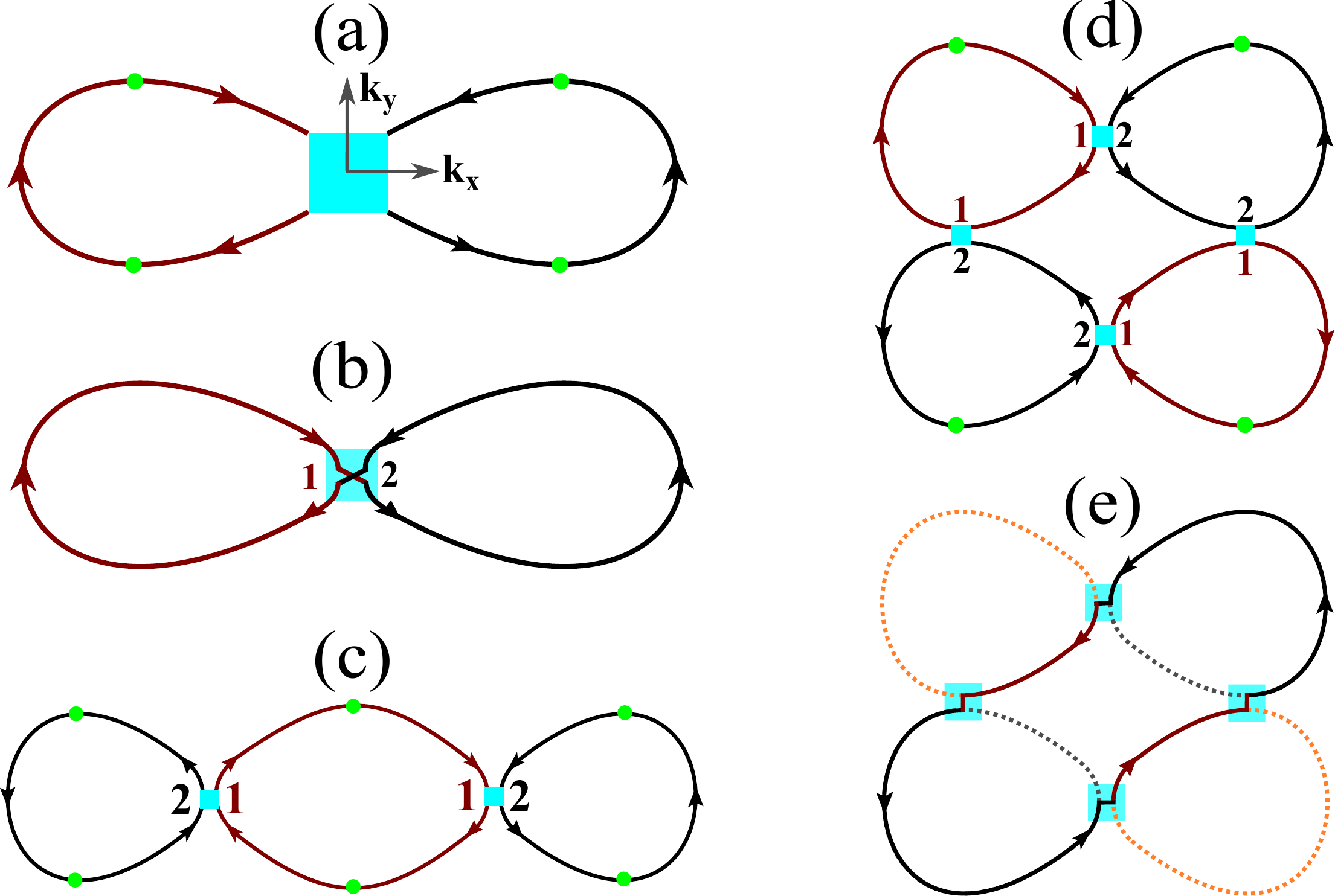}
\caption{(a) Graph for a single II-Dirac point. (b) Anti-crossing figure-of-eight trajectory.  (c) is a graph that typically occurs in band-inverted nonsymmorphic metals.\cite{LMAA} (b) and (e) illustrate closed Feynman trajectories. Brown and black pockets have opposite circulations.  }\label{fig:closed_feynman}
\end{figure}

As illustrated in \fig{fig:didactic}(a), the four edges which connect to the breakdown vertex are distinguished by the labels $(1+,1-,2+,2-)$. We will eventually formulate the connection formula  as a scattering matrix relating $(1+,2+)$ to $(1-,2-)$. Given an arbitrary graph with one or more breakdown vertices,   it is  important to correctly identify the four labels for each individual vertex. $1+$ and $2+$ label the two edges which are oriented toward the vertex, and $1-$ and $2-$ label the edges oriented away. We remind the reader that the orientation of each edge is the direction of a hypothetical wavepacket, which is determined by  Hamilton's equation with the convention $\bB=-|B|\vec{z}$. Let us set down local coordinates centered on each vertex, such that  $+$ lies to the north and $-$ to the south; we may then assign $1$ to the west, and $2$ to the east, as exemplified by the three graphs in \fig{fig:closed_feynman}(b-d). It is sometimes convenient to define (as we have already done) a right-handed coordinate system where $k_y$ increases in the direction from $-$ to $+$, and $k_x$ increases in the direction from $1$ to $2$, as exemplified in \fig{fig:closed_feynman}(a).   Since western and eastern edges also correspond to distinct bands, we might also view  $1,2$ as band indices which are locally-defined at each vertex.\\

For $E\neq 0$, the two edges belonging to band $n$ form a smooth curve given by $k_x^{n}(k_y,E)$ in \q{definekxn}. The corresponding \zf (ZF) wavefunction in the $(K_x,k_y)$ representation is 
\bal
w_{n\bk}= \f1{\sqrt{|v^x_{n}|}}e^{ik_xk_yl^2}e^{-il^2\int^{k_y}_{0} \big(k_x^{n}-H^{n}_1(v^x_{n})^{\mo}\big)dz}, \la{zilbermanfischbeckII}
\end{align}
following the general analysis of \s{sec:singlebandwkbwf}; $H_1^n$ (the Roth-Berry-Zeeman Hamiltonian) and $v^x_n$ (the band-diagonal velocity) are single-band quantities evaluated on the $n$'th band contour. The $H_1$-term in \q{zilbermanfischbeckII} is further simplified 
owing to the assumed space-time symmetry $g_1$ in \q{symmg1}: (i) the orbital moment vanishes, as may be deduced from \tab{tab:magnetization}. (ii) In spin-orbit-coupled systems, the Zeeman coupling also vanishes owing to $g_1$ [cf.\ \tab{tab:magnetization}]. (iii) In solids with negligible spin-orbit coupling, we work in the eigen-basis of the spin operator $\hat{\sigma}_z$, and the Zeeman splitting results in a constant term in $H_1$ which we will not write out explicitly. What remains of $\int H^n_1(v^x_n)^{\mo}$ is the integral of the single-band Berry connection $\bmx_n$ along the $n$'th band contour, i.e.,
\e{ w_{n\bk}\eq \f1{\sqrt{|v^x_{n}|}}e^{ik_xk_yl^2}e^{-il^2\int^{k_y}_{0} k_x^{n}dz}\;\W_{nk_y}, \la{zilbermanfischbeckIII}}
with $\W$ defined as 
\bal
\W_{nk_y}:\eq \exp\bigg[{i\int_{\sma{k_x^n(0),0}}^{\sma{k_x^n(k_y),k_y}} \bmx_n \cdot d\bk'}\bigg]\lin
\eq \begin{cases} \W_{n+}, &k_y \gg +|\bar{b}|, \\ \W_{n-}, & k_y \ll -|\bar{b}|.\end{cases} \la{definewilsonline}
\end{align}
$\bmx_n$ is only well-defined everywhere along $k_x^n(k_y,E)$ for $E\neq 0$. For $E=0$, the Berry connection (for a $\bk$-space derivative in the azimuthal direction) diverges at the II-Dirac point; it is appropriate here to employ different ZF functions in a diabatic basis, which will be described in \s{sec:connEeq0} below. We will henceforth refer to $w_{n\bk}$ defined in \q{zilbermanfischbeckII} as the adiabatic ZF functions, and restrict our attention to $E\neq 0$ in the remainder of this section.\\



\begin{figure}[ht]
\centering
\includegraphics[width=5 cm]{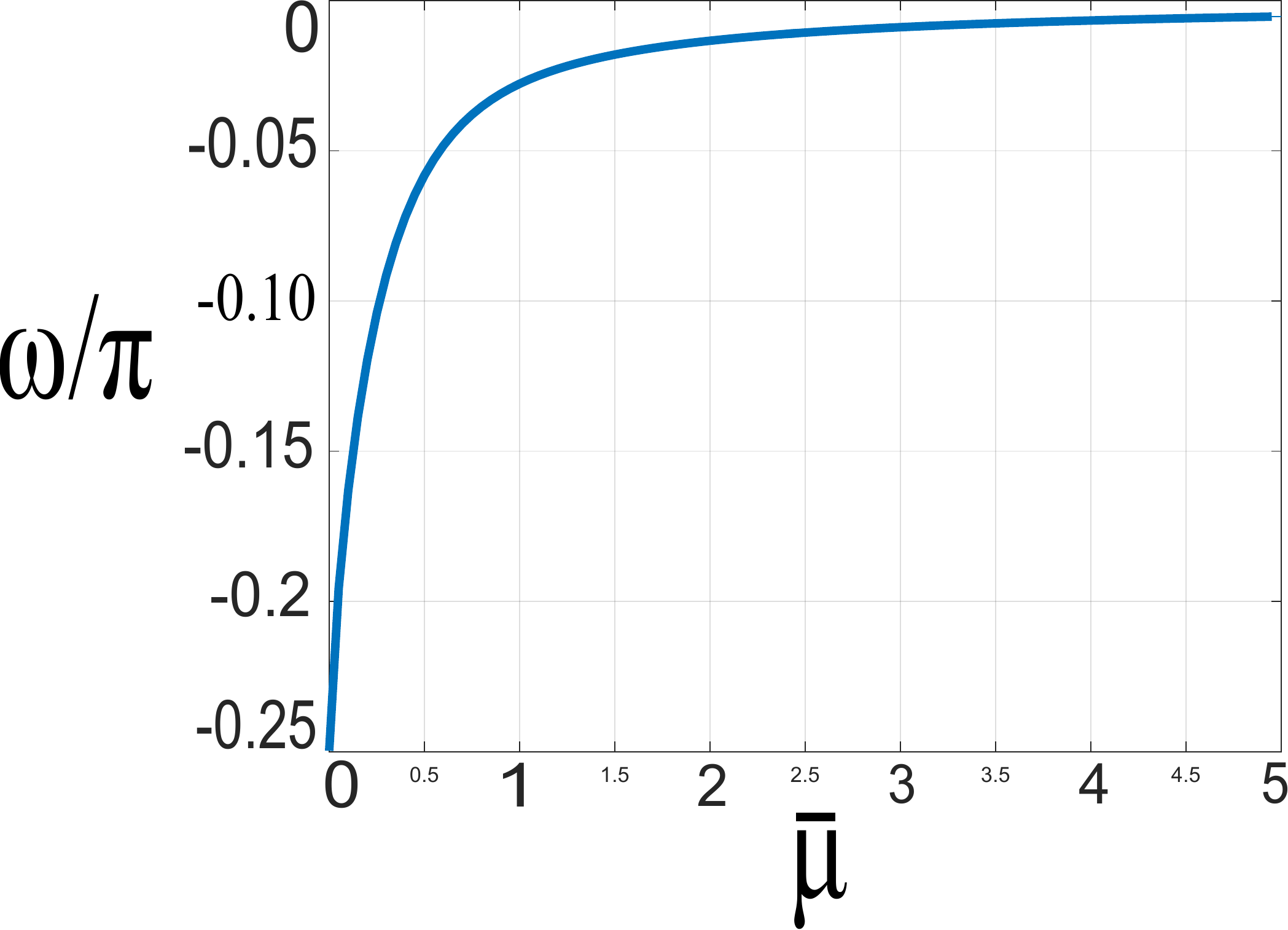}
\caption{Interband scattering phase $\omega$ vs $\bar{\mu}$.}\label{fig:scatteringphase_typeIIdirac}
\end{figure}

Since the $(K_x,k_y)$-representation is not valid for an interval of $k_y$ in the breakdown region [illustrated by the white region in \fig{fig:hyperbola_inter}(b)], we introduce a label to distinguish $(K_x,k_y)$-wavefunctions that are valid above ($+)$ and below ($-$) the II-Dirac point: 
\e{ \text{For}\;E \neq 0, \as g^{\pm}_{1\bk} =c^{\pm}_{1}w_{1\bk}+{\ldots},\as g^{\pm}_{2\bk}=c^{\pm}_{2}w_{2\bk}+{\ldots} \la{abovequantumII}} 
$\ldots$ indicates contributions by edges far away from the II-Dirac point; they will not play a role in deriving the connection formula. The derivation proceeds in three steps: (i) in the breakdown interval, we solve for the eigenfunction of the effective Hamiltonian  in the $(K_x,0)$-representation [cf.\ \q{effhambanddeglinearkx56}]. (ii) In the interval of overlap, we transform the eigenfunction of (i) to a wavefunction in the $(K_x,k_y$)-representation through \q{basischange}, and (iii)  match the resultant wavefunction to the WKB wavefunctions ($g^{\pm}_{n\bk}$) defined above.  (i) is elaborated in \app{app:justifyWeber}, and (ii-iii) in \app{sec:matching}.  In this manner, we obtain a scattering-matrix equation relating incoming (at positive $k_y$) to outgoing (negative $k_y$) amplitudes:
\e{ \ifor E\neq& 0,\as \vectwo{c^-_{1}}{c^-_{2}}=\mathds{S}\; \vectwo{c^+_{1}}{c^+_{2}}, \lin
   \mathds{S}(E,l^2) \eq  \matrixtwo{\sqrt{1-\rho^2}e^{i\omega} }{ -e^{i(\theta_1-\theta_2)}\rho }{e^{i(\theta_2-\theta_1)}\rho }{ \sqrt{1-\rho^2}e^{-i\omega}}, \lin
		\rho(\bar{\mu}) \eq  e^{-\pi \bar{\mu}}, \la{definescatteringmatrixinter}}
with $\bar{\mu}$ the dimensionless tunneling parameter defined in \q{definemuinter}.  $\omega$ is the interband scattering phase plotted in \fig{fig:scatteringphase_typeIIdirac}, and defined by
\bal
\omega(\bar{\mu}) \eq \bar{\mu} -\bar{\mu} \ln \bar{\mu} +\text{arg}\left[ \Gamma(i\bar{\mu})\right] +\pi/4 \lin
&\rightarrow \begin{cases} -\pi/4, & \bar{\mu} \rightarrow 0,\\ 0, & \bar{\mu} \rightarrow +\infty, \end{cases} \la{definerhoandomega}
\end{align}
with $\Gamma$ the Gamma function. In particular, $\omega= -\pi/4$ at the energy of the II-Dirac point, which may alternatively be derived by perturbation theory.\cite{wittig_landauzener} \\

The interband tunnelling amplitude ($\mathds{S}_{12}$) may be viewed\cite{kane_blount} as the exponentiated action of a tunneling trajectory that encircles a Kohn branch point\cite{kohn1959} in complex-$k_y$ space; $|\mathds{S}_{12}|^2=e^{-2\pi \bar{\mu}}$ is the famous  Landau-Zener tunneling probability. The unspecified phase $(\theta_1-\theta_2)$ in $\mathds{S}_{12}$ reflects an intrinsic phase ambiguity between two nondegenerate bands. This ambiguity was implicit in the expansion of \q{abovequantumII}, where we might have arbitrarily redefined $w_{n\bk}\rightarrow w_{n\bk}e^{i\theta_n}$ by a $\bk$-independent but band-dependent phase.  This arbitrary phase should not, however, affect the quantization condition for closed orbits, owing to the following argument: the quantization condition is a function of  phases acquired by wavepackets along closed Feynman trajectories. Let us consider those closed trajectories that involve interband tunneling. The two-in-two-out rule at an interband vertex guarantees that a wavepacket must traverse an even number  ($p\in 2\Z$) of pockets before forming a loop. For illustration, $p=2$ for the figure-of-eight trajectory in \fig{fig:closed_feynman}(b), and $p=4$ for the trajectory (indicated by solid lines) in \fig{fig:closed_feynman}(e). Each pocket corresponds to a single band, and a wavepacket that tunnels between two pockets (labelled $i_1$ and $i_2$) picks up  the inter-pocket phase difference $(\theta_{i_2}-\theta_{i_1})$. Since the wavepacket must eventually return to the pocket it originated, the sum of all inter-pocket phase differences acquired in a closed trajectory vanishes. This shall be made more explicit in our subsequent case study of \fig{fig:closed_feynman}(a-b) [\s{sec:casestudysingleIIDirac} below].



\subsubsection{Connection formula for $E= 0$}\la{sec:connEeq0}

As motivated in the discussion below \q{definewilsonline}, we would like to define a different set of ZF functions (henceforth referred to as diabatic ZF functions) which are applicable at the energy of the II-Dirac point. By requiring that these functions are continuous along the the maximal-tunneling trajectories [$k_x^{\bar{n}}(k_y)$, as defined in \q{maximaltunneltraj}], they assume the form 
\e{ w_{\bar{n}\bk}\eq \f1{\sqrt{|v^x_{\bar{n}}|}}e^{ik_xk_yl^2}e^{-il^2\int^{k_y}_{0} k_x^{\bar{n}}dz}\;\W_{\bar{n}k_y}, \la{zilbermanfischbeckIIIdiabatic}}
with  $\W_{\bar{n}}$ defined just as in \q{definewilsonline}, except the Berry connection is integrated over $k_x^{\bar{n}}(k_y)$. Following essentially the same argument as in \q{abovequantumII}, we define the coefficients $c^{\pm}_{\bar{n}}$ through 
\e{ g^{\pm}_{\bk, E=0} =c^{\pm}_{\bar{1}}w_{\bar{1}\bk}+c^{\pm}_{\bar{2}}w_{\bar{2}\bk}+\ldots \la{abovequantumIIII}}
These coefficients correspond to the relabelled edges illustrated in \fig{fig:didactic}(c), and are related simply as
\e{ \ifor E= 0,\as \vectwo{c^-_{\bar{1}}}{c^-_{\bar{2}}}=\bar{\mathds{S}}\; \vectwo{c^+_{\bar{1}}}{c^+_{\bar{2}}}, \as   \bar{\mathds{S}}(l^2) \eq  \matrixtwo{1}{0 }{0 }{1}, \la{definescatteringmatrixinterEzero}}
which states that Landau-Zener tunneling occurs with unit probability, independent of the strength of the field.
This connection formula may be derived from solving \q{simplemodel}, which  decouples (for $E=0$) to two scalar, first-order  differential equations.  \\

Let us compare this zero-energy connection formula [$\bar{\mathds{S}}(l^2)$] to the finite-energy formula [$\mathds{S}(E,l^2)$ in \q{definescatteringmatrixinter}] in the limit $E \rightarrow 0^{\pm}$, with $0^{\pm}$ a vanishingly small positive/negative quantity. Ignoring the $(\theta_2-\theta_1)$ phase (whose irrelevance was argued for in \s{sec:connEneq0}), $\mathds{S} \rightarrow -i\tau_2$, with $\tau_2$ a Pauli matrix -- this implies $c^{+}_{\bar{1}}=c^{-}_{\bar{1}}$ and $c^{+}_{\bar{2}}=-c^{-}_{\bar{2}}$, which differs from \q{definescatteringmatrixinterEzero} by a minus sign. This apparent discontinuity in the connection formula  does not  imply that the quantization condition is also discontinuous at $E=0$; we will see how this tension is resolved -- by the Berry phase -- in the following case study.


\subsection{Case study: single II-Dirac point} \la{sec:casestudysingleIIDirac}

\begin{figure}[ht]
\centering
\includegraphics[width=6 cm]{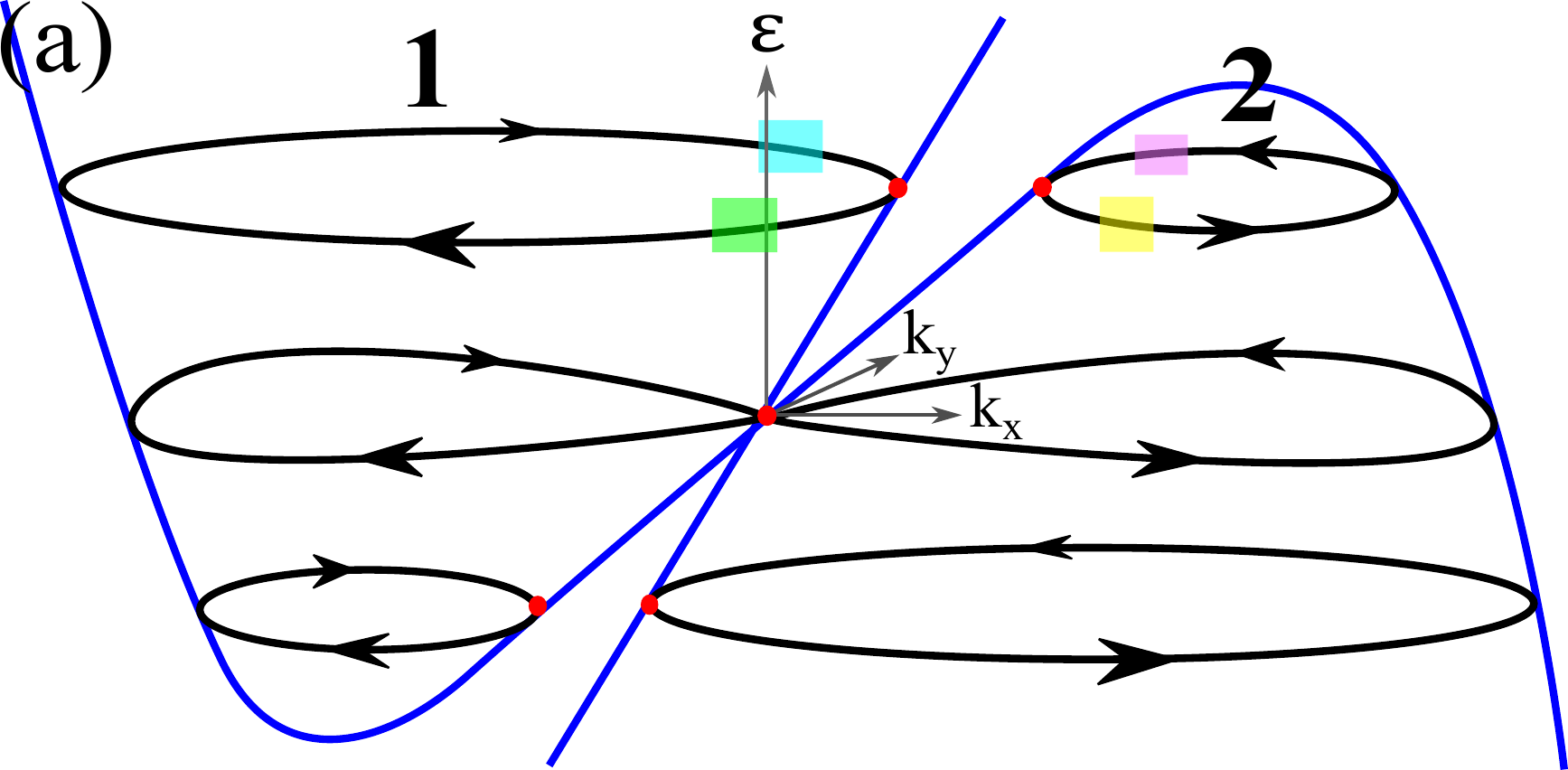}
\includegraphics[width=4 cm]{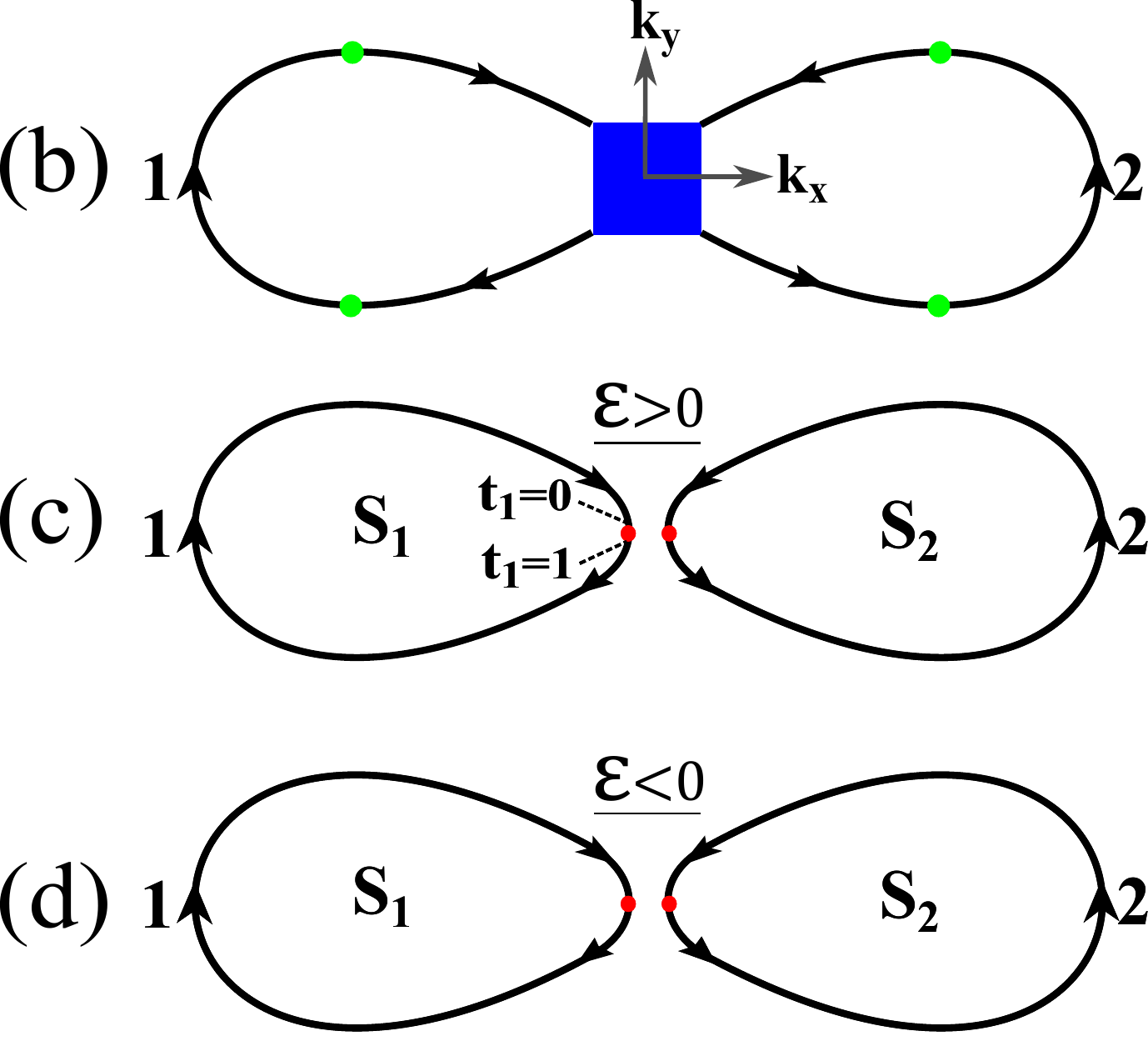}
\caption{(a) Energy-momentum dispersion of a single II-Dirac fermion. (b) shall be referred as the II-Dirac graph. }\label{fig:tilted_dirac}
\end{figure}

We study the simplest example of a single II-Dirac point, where the Fermi surface closes off as in \fig{fig:tilted_dirac}(a); this may be modelled by adding a cubic term to the II-Dirac Hamiltonian [cf.\ \q{minimalism}]
\e{ &H_0(\bk)=(u+v\gamma_3)k_x +wk_y\gamma_1-t(1-\gamma_3)k_x^3, \lin
 & u,v,t,w>0,\as u>v. \la{modelsingleII}}
This model has various realizations in the literature,\cite{YongXu_structuredWeylpoints,obrien_breakdown,koshino_figureofeight} 
 The corresponding graph in \fig{fig:tilted_dirac}(b) comprises four edges, four turning vertices and a single interband-breakdown vertex. The Landau levels of this model  were first studied in \ocite{obrien_breakdown} and \ocite{koshino_figureofeight} using a combination of  semiclassical analysis (for single-band transport) and large-scale numerical diagonalization; a quantization condition that determines this Landau levels for any tunneling strength was first formulated by us in \ocite{AALG_breakdown}. In this section, we derive the same quantization condition in greater detail -- we hope to equip the interested reader with the technical know-how to construct quantization conditions for other, possibly more complicated, graphs. It should be clarified at the onset that we are constructing  quantization conditions  that apply to homotopy classes of graphs [a definition of homotopy equivalence is provided in \s{sec:graph}], of which  \q{modelsingleII} merely describes one representative.

\subsubsection{Quantization condition for the II-Dirac graph}\la{sec:quantcondIIdirac}

The II-Dirac graph is similar to the double-well graph of \s{quantcondgenericwell} in having two broken orbits linked by a single degree-four vertex [these graph-theoretic terms have been defined \s{sec:graph}]. The major differences between inter- and intraband breakdown lie in (i) the scattering amplitudes [contrast \q{definescatteringmatrixinter} with \q{fullscattering}] , and in (ii)  the orientations of the four edges adjacent to the breakdown vertex [see \fig{fig:graphs}]. This orientation demonstrably affects the signs of the semiclassical phases acquired along the broken orbits.\\


To each of our broken orbits (labelled $\{\frako_i\}_{i=1}^2$) we assign a scalar amplitude $A_{i,E}(t_i)$, with $t_i\in [0,1]$ a time-like variable that increases along the $\frako_i$ in a direction consistent with Hamilton's equation. $t_i=0$ corresponds to the point of closest approach to the initial hyperbolic center $[\bk=(k_{xc}(E),0)]$, and  $t_i=1$ to the point of closest approach to the final hyperbolic center, as illustrated in \fig{fig:tilted_dirac}(c). In general, the initial and final hyperbolas may correspond to distinct II-Dirac points; in the present case study they are identical. \\


For $E\neq 0$, the Bloch functions can be made first-order differentiable with respect to $\bk \in {\frako}_j$, and consequently the Berry connection is well-defined; this is not true at $E=0$, where  Bloch functions are discontinuous at the cusp of $\frako_j$. Let us then define $\Omega_j$ for $E\neq 0$ as the net phase acquired by a wavepacket in traversing $\bar{\frako}_j$:  this has the form 
\e{\Omega_j(E,l^2)=l^2S_j(E)+\oint_{\bar{\frako}_j}\bmx \cdot d\bk+ \pi,}
where $S_j$ is the oriented area of $\frako_j$; note $S_1$ and $S_2$  have opposite signs. $\pi$ in the above equation corresponds to the Maslov correction for simple closed orbits, and the Berry phase contribution is fixed to $\pi$ or $0$, corresponding respectively to whether $\bar{\frako}_j$  encircles the II-Dirac point or not [\s{sec:symmetrycovariancepropag}]. The robust quantization of the Berry phase is a result of  the symmetry $g_1$ [cf.\ \q{symmg1}], which additionally ensures that the Roth and Zeeman contributions to $\Omega_j$ vanish for spin-orbit-coupled solids [cf.\ \s{sec:symmetrysinglebandom}]. Since the orbit ${\frako}_j$ that encloses the II-Dirac point changes discontinuous across $E=0$, $\Omega_j$ is also necessarily discontinuous:
\bal
\Omega_1(E,l^2) \eq \begin{cases} l^2S_1(E), & E>0, \\ 
                                l^2S_1(E)+\pi, & E<0,\end{cases} 
\lin
\Omega_2(E,l^2) \eq \begin{cases} l^2S_2(E)+\pi, & E>0, \\ 
                                l^2S_2(E), & E<0.\end{cases} \la{defineOmegaj}
\end{align}
Since loop integrals of the Berry connection are only uniquely defined modulo $2\pi$, some phase convention has been chosen in the above expressions; such a choice will not matter to the quantization condition, which is only a function of exp$[i\Omega_j]$ [as justified in \qq{IIdiracsinglevalued}{functionofthree} below].  \\

The following determinantal equation expresses the condition that the amplitudes $\{A_i\}$ are everywhere single-valued:
\e{ \vectwo{A_{1E}(0)}{A_{2E}(0)}\eq  \mathds{S}(E,l^2)\vectwo{A_{1E}(1)}{A_{2E}(1)}\lin
 \eq \mathds{S}(E,l^2) \diagmatrix{e^{i\Omega_1}}{e^{i\Omega_2}}\vectwo{A_{1E}(0)}{A_{2E}(0)}\lin
 \imp& \det\left[\mathds{S} \diagmatrix{e^{i\Omega_1}}{e^{i\Omega_2}}  -I  \right]\bigg|_{E,l^2}=0.
\la{IIdiracsinglevalued}}
Employing the expression for the scattering matrix [\q{definescatteringmatrixinter}] at nonzero $E$, the determinantal equation may be expressed as:  
\e{ 0=1+e^{i(\Omega_1+\Omega_2)}- \sqrt{1-\rho^2}\big(e^{i(\Omega_1+\omega)}+e^{i(\Omega_2-\omega)}\big).\la{functionofthree}}
The three phases occurring above may be identified with the phases acquired by a wavepacket in traversing three closed Feynman trajectories, e.g., $\Omega_1+\Omega_2$ corresponds to the figure-of-eight trajectory illustrated in \fig{fig:closed_feynman}(b). As we have argued generally in \s{sec:connEneq0}, the phase $(\theta_1-\theta_2)$ in the tunneling matrix element should not affect the quantization condition since it expresses an arbitrary phase difference between the electron and hole pocket. We should see this directly from our case study: the figure-of-eight trajectory includes an electron-to-hole tunneling trajectory [occurring with amplitude $\mathds{S}_{21}=\rho e^{i(\theta_2-\theta_1)}$], and also the reverse hole-to-electron tunneling trajectory [with amplitude $\mathds{S}_{12}=-\rho e^{i(\theta_1-\theta_2)}$]. \q{functionofthree} is equivalent to:
\e{\cos\f{\Omega_1+\Omega_2}{2}\bigg|_{E,l^2} = \sqrt{1-\rho^2}\cos\left[\f{\Omega_1-\Omega_2}{2}\bigg|_{E,l^2} +\omega(\bar{\mu})\right], \la{quantizationIIdirac}}
which we have previously analyzed in \ocite{AALG_breakdown}. \\



\begin{figure}[ht]
\centering
\includegraphics[width=5 cm]{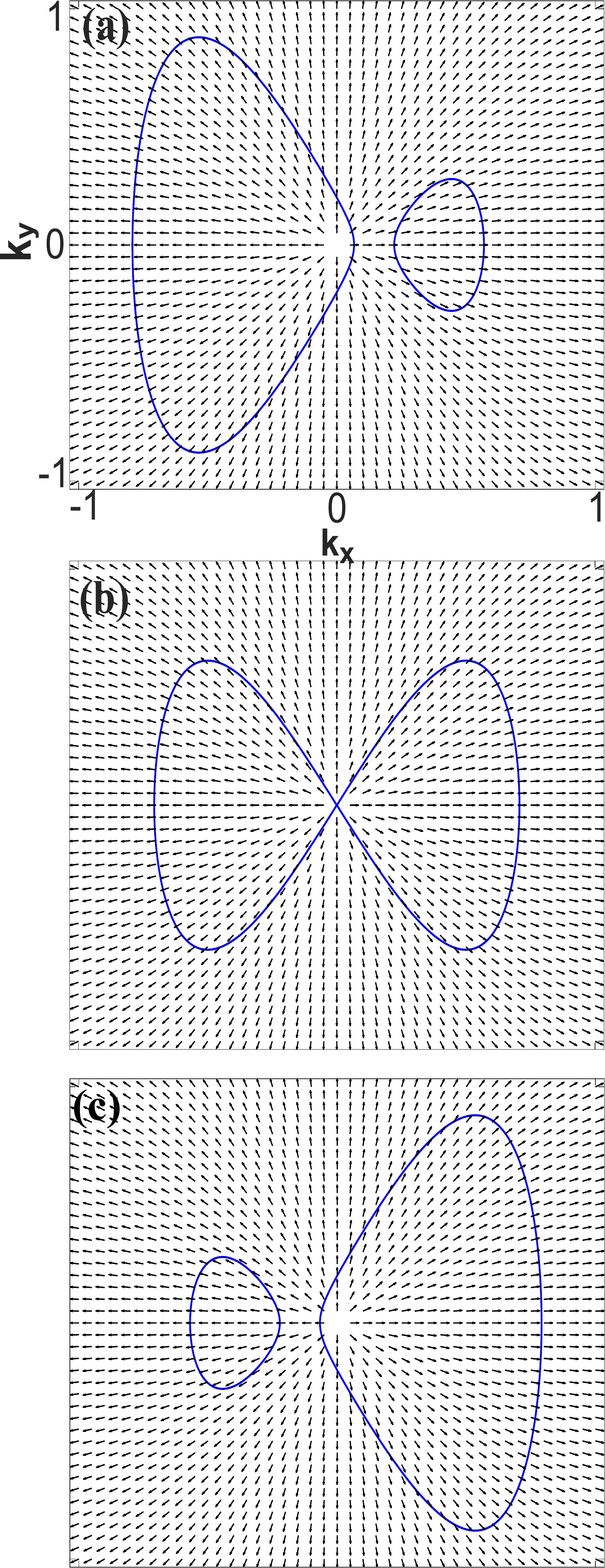}
\caption{ Expressing $H_0(\bk)=d_0(\bk)+d_1(\bk)\gamma_1 +d_3(\bk)\gamma_3$ from \q{modelsingleII}, we plot the two-vector [$\bd=(d_1,d_3)$]  as black arrows over $\bk$-space, for $E>0$, $E=0$ and $E<0$ respectively. The horizontal (resp.\ vertical) component of each arrow is proportional to $d_3$ (resp. $d_1$). The eigenfunctions of $H_0(\bk)$ are two pseudospinors which are parallel and anti-parallel to $\bd$. The parallel transport condition for a wavefunction is that its pseudospin remains parallel/anti-parallel to $\bd$ at all times.  The Berry phase ($\phi_B$) of an orbit may be deduced by evaluating the winding of the pseudospin over said orbit. (a)   The pseudospin winds by $2\pi$ for the left orbit (hence $\phi_B=\pi$ owing to Berry's argument\cite{berry1984}), but does not wind for the right orbit ($\phi_B=0$).  (b) The pseudospin does not wind for the crossing figure-of-eight orbit, hence $\phi_B=0$. Take care that as the wavefunction is parallel-transported across the II-Dirac point, $\bd$ flips sign but the pseudospin does not.   }\label{fig:figof8spin}
\end{figure}

Here, we focus on resolving the tension originating from a discontinuity of our connection formula at $E=0$ [see \s{sec:connEeq0}] -- the upshot is that a simultaneous discontinuity in the Berry phase ensures that the quantization condition remains continuous at $E=0$. We remind the reader that \q{quantizationIIdirac} has been derived utilizing the connection formula for nonzero $E$. In the limit $\bar{\mu} \rightarrow 0^{\pm}$ (equivalently, $E \rightarrow 0^{\pm}$ at finite field),  $\rho \rightarrow 1$, and \q{quantizationIIdirac}   simplifies to 
\e{ \f{l^2(S_1+S_2)}{2}\bigg|_{E_n^0}= n\pi. \la{faninter}}
Despite our proximity to a band degeneracy, the form of \q{faninter} is reminiscent of an Onsager-Lifshitz-Roth quantization condition for \emph{single-band} magnetotransport; the resultant Landau levels are also locally periodic. We may abscribe this emergent periodicity (in the Landau spectrum)  to the periodic motion of a wavepacket over the figure-of-eight illustrated in \fig{fig:closed_feynman}(b). Over one cyclotron period, the wavepacket accumulates: (a) a trivial Maslov phase from four turning points with vanishing net circulation [cf.\ \fig{fig:turning_point}(h)], (b) a net $\pi$-Berry phase of the two pockets [owing to a pseudospin argument in \fig{fig:figof8spin}], and (c) a net $\pi$ phase from two Landau-Zener tunnelings. The last phase is obtained by multiplying the two off-diagonal elements of the scattering matrix: $(\mathds{S}(0^{\pm})=-i\tau_2)$. \\

At strictly zero energy, exactly the same quantization condition [\q{faninter}] may be derived with the connection formula of \q{definescatteringmatrixinterEzero}. 
 Here,  the emergent periodicity (in the Landau specrum) is ascribed to periodic motion over a topologically-distinct figure-of-eight [illustrated in \fig{fig:figof8spin}]. The two  figures-of-eight differ in the vicinity of the II-Dirac point: bands cross at $E=0$, but anti-cross at $E=0^{\pm}$. For the crossing figure-of-eight, (a) the scattering matrix is trivially identity, (b) the Maslov phase vanishes [for the same reason described in \fig{fig:turning_point}(h)], and (c) the Berry phase is also trivial, owing to a pseudospin argument given in \fig{fig:figof8spin}. One practical implication of this discussion is that the two limiting values of \q{quantizationIIdirac}  as $E\rightarrow 0^{\pm}$ are equal, so one may as well extend the domain of \q{quantizationIIdirac} to include $E=0$; this extended quantization condition is then continuous in $E$.\\

Another aspect of the $|E|\rightarrow 0$ limit is worth discussing:  the second-order derivatives (with respect to $E$) of $l^2(S_1-S_2)/2$ [occurring in the right argument of \q{quantizationIIdirac}] diverges logarithmically. This divergence is a generic property of hyperbolic curves at the point of intersection;\cite{kosevich_topologyreview} physically stated, it originates from the transition from crossing to anti-crossing orbits at the II-Dirac point. This divergence does not lead to any irregularity in the quantization condition, due to a cancelling logarithmic divergence of the scattering phase $\omega$ [which also occurs in the right argument of \q{quantizationIIdirac}]. It is remarkable that an analogous cancellation of  divergences occurs for intraband breakdown. This is exemplified by the quantization condition for the double-well graph [cf.\ \q{quantizationnecking}], where the \emph{first}-order derivative of  $l^2(S_1+S_2)/2$ diverges logarithmically, but is also cancelled by the diverging scattering phase ($\phi$). These two case studies suggest that quantum tunneling, of both interband and intraband types, tends to smoothen out non-analyticities in the classical action function.



\subsection{Perturbative treatment of quasirandom spectrum}\la{sec:perturbation}

The typical spectrum of quantization conditions with tunneling is quasirandom, unless symmetry imposes commensuration of phases in the quantization condition.\cite{AALG_breakdown} The goal of this section is formulate a general perturbation theory to treat quasirandom spectra, and then apply it to our II-Dirac case study.\\

The general structure of the perturbation theory may be formulated in this manner. Let the quantization condition be expressed as
\e{ f(E,B;\tau(E,B))=0 \rightarrow E_n(B),}
which is an implicit equation for the discrete solutions $E_n(B)$; $\tau$ is a tunneling parameter whose functional form depends on the type of breakdown (whether inter- or intraband) and  the type of graph.  We consider a semiclassical limit of the quantization condition where $\tau(E,B)\rightarrow \tau_0$ (a constant), such that
\e{ f(E,B;\tau_0)=0 \rightarrow E^0_n(B);\la{generalfan}}
determines a locally-periodic spectrum that forms a Landau fan. Let $\tau=\tau_0+\delta \tau$ and consider a perturbative expansion in $\delta \tau$. To linear order, 
\e{0 = f(E,B;\tau_0) +\delta \tau(E,B)\, f_1(E,B) +O(\delta \tau^2);} 
the first-order-corrected energy levels are defined by $E_n^1 =E_n^0+\delta E_n^1+O(\delta \tau^2)$ with the assumption $\delta E^1=O(\delta \tau)$. Subtituting $E_n^1$ into the above equation, 
\e{ O(\delta \tau^2) \eq f(E_n^1,B;\tau_0) +\delta \tau f_1\bigg|_{E_n^0,B}\lin
\eq \bigg\{\delta E_n^1 \p{f}{E}\bigg|_{\tau_0}  +\delta \tau f_1\bigg\}_{E_n^0,B},\lin
\imp \delta E_n^1 \eq -\delta \tau \f{f_1}{(\partial f/\partial E)|_{\tau_0}}\bigg|_{E_n^0,B}. \la{generalfirstorder}}
This equation is valid assuming that $\delta \tau$ is small and slowly-varying on the scale of $\delta E_n^1$ -- this should be checked for self-consistency. \qq{generalfan}{generalfirstorder} have been exemplified for intraband breakdown in \qq{fanintra}{fanintra2}, and we shall now apply it to our case study of interband breakdown. One key equation [\q{faninter2}] in the subsequent section has been presented in \ocite{AALG_breakdown}, but the reader may benefit from a more detailed discussion. 






\subsubsection{Case study: quasirandom spectrum of the II-Dirac graph}\la{sec:perturbation_IIDirac}

Since no (magnetic) space-group symmetry relates an electron to a hole pocket, the two distinct arguments in the cosine functions of \q{quantizationIIdirac} competitively produce a quasirandom Landau spectrum. In the regime $\bar{\mu} \approx 0$, the dominant trigonometric harmonic $(\Omega_1+\Omega_2)/2$ determines a semiclassical Landau fan  indexed by $n \in \Z$ [cf.\ \q{faninter}]. The largeness of $|S_1|$ and $|S_2|$ relative to $\lmt$ justifies the semiclassical approximation; there is, however, no need for $|S_1+S_2|$ to be large. In particular, the zeroth ($n=0$) Landau level is non-dispersive and occurs at an energy ($E_0^0$) where electron and hole pockets are perfectly compensated: $(S_1+S_2)|_{E_0^0}=0$; this energy does not necessarily lie at the II-Dirac point.\cite{obrien_breakdown}\\

To leading order in $\sqrt{1-\rho^2}$, the correction to the Landau fan is derived in \app{app:casestudyperturbIID} to be  
\e{\delta E_n^1 \eq  2(-1)^{n+1}\text{sign}[E]\f{\sqrt{1-\rho^2}}{l^2(S_1+S_2)'}\lin
&\times \sin\left[ \omega+\f{l^2(S_1-S_2)}{2}\right]\bigg|_{E_n^0}, \la{faninter2}}
with  the shorthand $O'=\partial O/\partial E$. In particular, the correction to the zeroth Landau level is a sinuisoid enveloped by a function  $\propto B^{1/2}$:
\e{ \delta E_0^1\eq -2\sqrt{\pi}\f{v}{\sqrt{w}(u^2-v^2)^{3/4}}  \f{E_0^0}{l(S_1+S_2)'}\lin
&\times \sin\left[ \omega+\f{l^2(S_1-S_2)}{2}\right]\bigg|_{E_0^0} +O(\bar{\mu}^{3/2}\lmt).\la{firstorderzerothll}} 
For the non-zeroth Landau levels, we show in \app{app:casestudyperturbIID} that the envelop function grows as $B^{1/2}$ at weak field, but eventually crosses  over to a $B^{3/2}$ dependence at a scale that depends on the band parameters.\\

For \qq{faninter}{faninter2} to be consistent, $\sqrt{1-\rho^2}$ should be small and slowly varying on the scale of $\delta E_n^1$. Indeed, the typical scale of variation for $\sqrt{1-\rho^2}$ is $\Delta \bar{\mu} \sim 1$, which implies an energy scale 
\e{\Delta E \sim  \f{\sqrt{{w}}(u^2-v^2)^{3/4}}{v}\f1{l},} 
from the definition of  $\bar{\mu}$ in \q{definemuinter}. It follows that
\e{ \f{\delta E_n^1}{\Delta E} \sim \f{v^2}{w(u^2-v^2)^{3/2}}\f{E_n^0}{(S_1+S_2)'}}
vanishes for small enough $E_n^0$.  One additional remark is that the typical spacing of the Landau fan is small  compared to the energy interval where breakdown is significant:
\e{ \f{E_{n+1}^0-E_n^0}{\Delta E} \sim \f{2\pi}{l(S_1+S_2)'|_{E_n^0}}\f{v}{\sqrt{{w}}(u^2-v^2)^{3/4}} =O(1/l).}


\section{Discussion and outlook}\la{sec:outlook}

We have provided the recipe to cook up quantization rules for a large class of closed orbits: (i) in the absence of breakdown, our rules apply to band subspaces of arbitrary energy degeneracy. (ii) For band subspaces which are nondegenerate at generic wavevectors, we have accounted for intraband breakdown associated to saddlepoints, and interband breakdown associated to conical touching points between two bands (II-Dirac points).\\

 This certainly does not exhaust all types of band touchings: not all point touchings are conical, e.g., the band dispersion around a multi-Weyl point\cite{chen_multiweyl} is quadratic in $\bk$. Not all band touchings occur between two bands, e.g., the spin-one Weyl point\cite{Bradlyn_newfermions} is a touching of three bands. If bands are spin-degenerate at generic wavevectors, a touching point is minimally four-fold degenerate, e.g., an over-tilted 3D Dirac point.\cite{huaqing_IIDirac3D,tayrong_IIDirac3D} Spin-degenerate orbits may also intersect at four-fold-degenerate saddlepoints. Moreover, band touchings occur not just at isolated points, but also along lines. The connection formulae in all the above cases are unknown, but we hope that this work lays the groundwork for their future derivation. One necessary ingredient  would be an effective Hamiltonian that is valid at any type of band-touching point, as we have derived in \s{sec:effhamgen}. The connection formula should be derivable by matching the eigenfunctions of this effective Hamiltonian  to semiclassical WKB wavefunctions. For spin-degenerate bands, the matching should be performed for the multi-component WKB wavefunction derived in \s{sec:simple}. \\

The quantization rules in this work apply only to closed orbits, and include the complete subleading-in-$B$ correction. For an energy-nondegenerate band in the absence of breakdown, higher-order corrections to the quantization rule have been derived with various methods: beginning from the effective-Hamiltonian formalism, higher-order corrections may be obtained from an equation-of-motion method\cite{rothII} as well as with WKB methods;\cite{fischbeck_review} alternatively, these corrections may be derived from the zero-field, zero-temperature magnetic response functions.\cite{Gao_Niu_zerofieldmagneticresponse} However, a higher-order theory for energy-degenerate bands has not been developed. Finally, it would be interesting to generalize this work to open orbits, i.e., noncontractible orbits that extend across the Brillouin torus.


\begin{acknowledgments}
The authors are grateful to Nicholas Read, Meng Cheng, Titus Neupert, Lee Chinghua, Xi Dai, T. O'Brien, Lukas Muechler, Wang Zhijun, Kadigrobov, Qian Niu,  and Chen Fang for various discussions. We especially thank Judith H{\"o}ller for mathematical advice, and Wang Chong for a critical reading of the manuscript.  We acknowledge support by  the Yale Postdoctoral Prize Fellowship and NSF DMR Grant No.\ 1603243.
\end{acknowledgments}

\appendix

\section{Appendix to `Review of effective Hamiltonian in cases without interband breakdown'}

\subsection{Introduction to field-modified Bloch functions} \la{app:fieldmodBlochfunc}

We provide a pedagogical introduction to field-modified Bloch functions, and derive a few useful identities which will be used throughout the main text.\\

Let us first motivate the form of the field-modified Bloch functions in \q{zerothrothfunction} by an argument\cite{lifshitz_pitaevskii_statphys2} involving gauge invariance. Suppose at zero field the energy eigenfunctions are expressed in Bloch form: $\hat{H}_0\psi_{n\bk'}=\var_{n\bk'}\psi_{n\bk'}$. A zero field is expressible as the curl of a constant vector potential $\ba_0$,  hence by gauge invariance, 
\e{0=\left[\hat{H}[\ba_0]-\var_{n\bk'}\right]e^{i(\bk'-\ba_0)\cdot\br}u_{n\bk'}.}
We see that $\bk=\bk'-\ba_0$ is the quantity that determines the change in phase of the wavefunction under discrete translations, and $\eikr u_{n\bk+\ba_0}$ has energy eigenvalue $\var_{n\bk+\ba_0}$. In a weak field, i.e., for  $\ba(\br)$ that slowly varies in space, the appropriate basis functions to describe field-induced dynamics within the band $n$ is just $\eikr u_{n\bk+\ba(\br)}$ to leading order in the field [cf.\ \q{zerothrothfunction}]; to our knowledge, these types of basis functions were first proposed by Zilberman.\cite{zilberman_wkb} \\

Let us derive equivalent expression for the field-modified Bloch functions which is more amenable to algebraic manipulations:
\e{ &u_{n\bK^*}e^{i\bk \cdot \br}= \int d\br' \check{u}_{n\br'} e^{-i\bK^*\cdot \br'}\eikr 
\lin
\eq \int d\br' \check{u}_{n\br'} e^{-i[\bk + \ba(-i\nabk)]\cdot \br'}\eikr  \lin
\eq  \int d\br' \check{u}_{n\br'} e^{-i[\bk + \ba(\br)]\cdot \br'}\eikr = u_{n,\bk+\ba(\br)} \eikr.}
Here, $\check{u}_{n\br'}$ is the Fourier transform of $u_{n\bk}$, $\bK$ are the kinetic quasimomentum operators defined in \q{defineK}, 
the second-to-last equality is valid in the symmetric gauge, where 
$[\bk\cdot \br', \ba(-i\nabk)\cdot \br']=0.$ An arbitrary state may be expanded in field-modified Bloch functions as in \q{modifiedblochfunctions}, which is equivalently expressed as
\e{ \Psi(\br) =  \sum_{n\bk}(u_{n\bK^*}e^{i\bk \cdot \br})g_{n\bk} = \sum_{n\bk}e^{i\bk \cdot \br}(u_{n\bK}g_{n\bk}), \la{expandroth}}
where $\sum_{\bk}$ is really a continuous integral. After the above `integration-by-parts,' the basis functions effectively become operators acting on the wavefunction $g_{n\bk}$.\\

This `integration-by-parts' formula [\q{expandroth}] was proven in Ref.\ \onlinecite{rothI}. Here, we offer a more explicit proof for pedagogy.\\

\noindent \emph{Proof:} for a constant magnetic field, the vector potential can be written in the linear gauge as $\ba(\br) =  \bb^jr_j$, or equivalently $a_i(\br) = b_i^jr_j$. A useful identity in this context is then
\e{ e^{ (1/2)[\pm i\bv\cdot \bb^j\nabla_{k_j},\bv\cdot \bk ]} = e^{\pm (i/2) v_ib_i^jv_j}.}
By the Baker-Campbell-Hausdorff lemma,
\e{ e^{-i(\br-\bR)\cdot (\bk + \ba(\pm i \nabk))} \eq  e^{-i(\br-\bR)\cdot\bk}e^{-i(\br-\bR)\cdot\ba(\pm i \nabk)}\lin
&\times e^{\mp (i/2)(r_i-R_i)b_i^j(r_j-R_j)}. \la{bch2}  }
Sandwiching \q{bch2} in two different ways (also with opposite signs in the argument of $\ba$), we obtain an indentity
\e{  & e^{i\bk \cdot \br} e^{-i(\br-\bR)\cdot (\bk + \ba( i \nabk))} e^{-i\bk \cdot \bR} = e^{- (i/2)(r_i+R_i)b_i^j(r_j-R_j)} \lin
\eq e^{-i\bk \cdot \bR} e^{-i(\br-\bR)\cdot (\bk + \ba(- i \nabk))} e^{i\bk \cdot \br}, \la{sandt}}
which will be used in the following.
We apply the Fourier expansions
\e{   &g_{n\bk} = \sum_{\bR} \check{g}_{n\bR} e^{-i\bk \cdot \bR}, \ins{and} \lin
& u_{n\bK}= \sum_{\bR} W_n(\br-\bR) e^{-i\bK \cdot (\br-\bR)} , }
to express
\e{ &\Psi = \sum_{n\bk}g_{n\bk}(u_{n\bK^*}e^{i\bk \cdot \br}) \lin
\eq \sum_{n\bk,\bR,\bR'} \check{g}_{n\bR'} W_n(\br-\bR) e^{-i\bk \cdot \bR'} e^{-i\bK^* \cdot (\br-\bR)} \eikr \lin 
\eq \sum_{n\bk,\bR,\bR'} \check{g}_{n\bR'} W_n(\br{-}\bR) e^{{-}i\bk \cdot \bR'} e^{{-}i(\bk +\ba({-}i\nabk)) \cdot (\br{-}\bR)} \eikr \lin 
\eq \sum_{n\bk,\bR,\bR'} \check{g}_{n\bR'} W_n(\br-\bR) e^{-i\bk \cdot \bR'} e^{-i(\bk +\ba(\br)) \cdot (\br-\bR)} \eikr  \lin
&\propto \sum_{\bk} e^{i\bk \cdot (\bR-\bR')}.}
The delta function allows us to express the above equation as 
\e{ \Psi \eq N\sum_{n\bk,\bR} \check{g}_{n\bR'} W_n(\br-\bR) e^{-i\bk \cdot \bR} \lin
&\times e^{-i(\bk +\ba(-i\nabk)) \cdot (\br-\bR)} \eikr.}
We would like to show that this equals
\e{ \sum_{n\bk} \eikr u_{n\bK}g_{n\bk} \eq N\sum_{n\bk,\bR} \check{g}_{n\bR} W_n(\br-\bR) \eikr \lin
&\times e^{-i(\bk +\ba(i\nabk)) \cdot (\br-\bR)} e^{-i\bk \cdot \bR}.}
In deriving the above equality, we have reduced the double summation over $\bR$ to a single summation, by similar manipulations. Comparing the last two equations, and applying the identity \q{sandt}, we thus derive the desired relation.

\subsection{Equivalent expressions for the orbital magnetic moment} \la{sec:equivalentexpressionsorbitalmag}

\subsubsection{Single-band orbital moment} \la{sec:equivalentexpressionssinglebandorbitalmag}

The gauge-independent orbital moment, in the spatial direction $\vec{\alpha}$ ($\alpha=x,y,z$), for a band labelled $n$, is defined as [cf.\ \q{expressH1Osingleband}]
\e{ M(\bk)_n^{\alpha} = -\f{|e|}{2\hbar c}\epsilon^{\alpha \beta \gamma}\big[\mx^{\beta}(\Pi^{\gamma}-v^{\gamma})\big]_{nn}. \la{orbitalmagnetization}}
Applying the identity \q{idenvel},
\e{ \big(\mx^{\alpha}(\Pi^{\beta}-v^{\beta})\big)_{nn}=\sum_{l;\var_l\neq \var_n}\mx^{\alpha}_{nl}\Pi^{\beta}_{ln}=\sum_{l;\var_l\neq \var_n}\f{\Pi^{\alpha}_{nl}\Pi^{\beta}_{ln}}{i(\var_n-\var_l)}, \la{orbitalmagnetization2}}
we derive an equivalent expression 
\e{ M(\bk)_n^{\alpha} \eq i\f{|e|}{2\hbar c}\epsilon^{\alpha \beta \gamma}\sum_{l;\var_l\neq \var_n}\f{\Pi^{\beta}_{nl}\Pi^{\gamma}_{ln}}{\var_n-\var_l}\lin
\eq i\f{|e|}{2m^2\hbar c}\epsilon^{\alpha \beta \gamma}\sum_{l;\var_l\neq \var_n}\f{p^{\beta}_{nl}p^{\gamma}_{ln}}{\var_n-\var_l}; \la{orbitalmagnetization3}}
in the last equality, $\bp_{mn}(\bk)$ is the canonical momentum matrix: $\braopket{u_{m\bk}}{\hat{p}}{u_{n\bk}}$. \q{orbitalmagnetization3} coincides with the correction $(-\bM\cdot \bB)$ to the energy of a wavepacket in Ref.\ \onlinecite{chang_niu_hyperorbit}. We offer yet another equivalent expression which is identical in form (but carrying a different name) to that found in the WKB treatment of coupled-wave\cite{littlejohn_short,littlejohn_long} and coupled-channel equations:\cite{yabana_coupledchannel} 
\e{&M(\bk)_{n}^{\alpha} = -i\f{|e|}{2\hbar c}\epsilon^{\alpha \beta \gamma} \braopket{\partial_{\beta}u_n}{\hH_0(\bk)-\var_{n\bk}}{\partial_{\gamma}u_n} \lin
\eq -i\f{|e|}{2\hbar c}\epsilon^{\alpha \beta \gamma} \sum_m\bigg[ \braopket{\partial_{\beta}u_n}{\hH_0(\bk)}{u_m}\braket{u_m}{\partial_{\gamma}u_n}\lin
  &- \var_n\braket{\partial_{\beta}u_n}{u_m}\braket{u_m}{\partial_{\gamma}u_n} \bigg] \lin
\eq -i\f{|e|}{2\hbar c}\epsilon^{\alpha \beta \gamma} \sum_{m;m\neq n}(\var_m-\var_n)\braket{\partial_{\beta}u_n}{u_m}\braket{u_m}{\partial_{\gamma}u_n}\lin
\eq -i\f{|e|}{2\hbar c}\epsilon^{\alpha \beta \gamma} \sum_{m;m\neq n}(\var_m-\var_n)\mx^{\beta}_{nm}\mx^{\gamma}_{mn}\lin
\eq i\f{|e|}{2\hbar c}\epsilon^{\alpha \beta \gamma} \sum_{m;m\neq n}\f{\Pi^{\beta}_{nm}\Pi^{\gamma}_{mn}}{\var_n-\var_m}.
 \la{orbitalmagnetization4}}
Let us compare these expressions to the gauge-dependent moment corresponding to the Berry term in the effective Hamiltonian, i.e., we express $H_1^B=-\tilde{\bM}\cdot \bB$ with
\e{ \tilde{M}(\bk)_n^{\alpha} \eq -\f{|e|}{\hbar c}\epsilon^{\alpha \beta \gamma} \big[\mx^{\beta}\bv^{\gamma}\big]_{nn} = -\f{|e|}{\hbar c}\epsilon^{\alpha \beta \gamma} \mx_{nn}^{\beta}\partial_{\gamma}\var_n\lin
\eq -\f{|e|}{\hbar c}\epsilon^{\alpha \beta \gamma} \left[\partial_{\gamma}(\mx_{nn}^{\beta}\var_n) - i\braket{\partial_{\gamma}u_n}{\partial_{\beta}u_n}\var_n  \right]\lin
\eq -\f{|e|}{\hbar c}\epsilon^{\alpha \beta \gamma} \left[-\partial_{\beta}(\mx_{nn}^{\gamma}\var_n) + i\braket{\partial_{\beta}u_n}{\partial_{\gamma}u_n}\var_n  \right].\la{definemagnetizationberry}}
The total derivative (i.e., the first term in brackets in the above equation) cannot be ignored: it makes this quantity independent of the zero of energy. The sum of the two moments is then 
\e{ \big[{M}+\tilde{M}\big](\bk)_n^{\alpha}\eq-i\f{|e|}{2\hbar c}\epsilon^{\alpha \beta \gamma} \braopket{\partial_{\beta}u_n}{\hH_0(\bk)+\var_n}{\partial_{\gamma}u_n} \lin
&+\f{|e|}{\hbar c}\epsilon^{\alpha \beta \gamma} \partial_{\beta}(\mx_{nn}^{\gamma}\var_n). \la{totalmagnetizationateachk}}
For insulators with vanishing Chern number ($C_1$) in the Brillouin two-torus ($T_{\perp}$) perpendicular to the field, a first-order-differentiable basis for $u_{n\bk}$ may be found over $T_{\perp}$. This implies that $\mx_{nn}$ is continuous over $T_{\perp}$ ($\var_n$ clearly also satisfies this property), and therefore integrating the total moment over $T_{\perp}$:
\e{ &\int_{T_{\perp}} \f{d^2k}{(2\pi)^2} \big[{M}+\tilde{M}\big](\bk)_n^{\alpha} \as\substack{\sma{C_1=0}\\=}\lin
 &i\f{|e|}{2\hbar c}\epsilon^{\alpha \beta \gamma} \int_{T_{\perp}} \f{d^2k}{(2\pi)^2}\braopket{\partial_{\beta}u_n}{\hH_0(\bk)+\var_n}{\partial_{\gamma}u_n}. \la{chernless}}
The right hand side seems at first sight to depend on the zero of energy, but note that the effect of such a shift is proportional to $C_1$, which  vanishes by assumption. \q{chernless} is identical to the zero-temperature expression obtained for the orbital magnetization using various methods: (a) a Wannier representation for bands was used in  Ref.\ \onlinecite{thonhauser_orbmag} and \onlinecite{ceresoli_multibandorbmag}, (b) quantum-mechanical perturbation theory in Ref.\ \onlinecite{junren_orbmag},  and (c) a Green's function approach in Ref.\ \onlinecite{kuangting_orbmag}.




\subsubsection{Orbital magnetic moment for any number of bands}\la{app:equivalentmultibandmag}

Let us derive equivalent expressions for the single- and multi-band orbital magnetic moments, which manifest how they transform under basis changes of the form \q{basistransf}. The basis transformations we consider preserve both $P$ and $Q$ [recall \q{defineprojP} and \q{defineprojQ}], i.e., the unitary $V$ in \q{basistransf} is block-diagonal with respect to the decomposition into $P$ and $Q$. From the simple identities,
\e{ \partial_{\alpha} P \eq \sum_{n}\ketbra{u_n}{\partial_{\alpha} u_n}+\ketbra{\partial_{\alpha} u_n}{ u_n},\lin
(\partial_{\alpha} P) Q \eq \sum_{n,\bar{m}}\ket{u_n}\braket{\partial_{\alpha} u_n}{u_{\bar{m}}}\bra{u_{\bar{m}}},}
we derive, 
\e{[(\Pi^{\beta}-v^{\beta})\mx^{\alpha}]_{mn} \eq i\sum_{\bar{l}} \Pi^{\beta}_{m\bar{l}}\braket{u_{\bar{l}}}{\partial_{\alpha}u_n} \lin
\eq i[P\hPi^{\beta}Q\partial_{\alpha}P]_{mn}}
\e{[\mx^{\alpha}(\Pi^{\beta}-v^{\beta})]_{mn}\eq i\sum_{{\bar{l}}} \braket{u_m}{\partial_{\alpha}u_{\bar{l}}}\Pi^{\beta}_{{\bar{l}}n} \lin
\eq i[P(\partial_{\alpha}Q)\hPi^{\beta} P]_{mn}.}
For the single-band orbital moment for band $n$, the last equality reduces to $i\braopket{u_n}{(\partial_{\alpha}Q)\hPi^{\beta}}{u_n}$, which leads directly to \q{expressH1Osingleband}.

\section{Appendix to `Quantization conditions for orbits without breakdown'}\la{app:quantizationwobreakdown}

\subsection{Identities for Weyl-symmetrized operators}\la{app:moyalidentities}

The following identities may be generalized to nonperiodic functions of $\bk$ by replacing the Fourier sum with a Fourier integral.\\

Let $\check{A}_j(\bk)$ be the Fourier transform of  $A_j(\bk)=O(l^{-2j})$, and applying the definition of a Weyl-symmetrized operator  [cf.\ \q{replace}], 
\e{
&\sum_{\bR}\check{A}_{j}(\bR)e^{i\bK \cdot \bR}e^{-i\psi(k_y)}= \sum_{\bR}\check{A}_{j}(\bR)e^{ik_xR_x}e^{-(R_x/l^{2})\partial_y+ik_yR_y}e^{-i\psi(k_y)}.\la{H0sand}
}
Applying the Baker-Campbell-Hausdorff identity for a central commutator,
\e{ e^{A+B} \eq e^Ae^Be^{(1/2)[B,A]} \imp 
e^{-(R_x/l^{2})\partial_y+ik_yR_y}=  e^{ik_yR_y}e^{-(R_x/l^{2})\partial_y}e^{-iR_xR_y/2l^2}, \la{BCHapplied}}
and a identity valid for any function $f(k_y)$:
\e{e^{-(R_x/l^2)\partial_y}f(k_y)=f(k_y-R_x/l^2)e^{-(R_x/l^{2})\partial_y},\la{idengeneratortrans}}
we derive that \q{H0sand} equals
\e{ &\sum_{\bR}\check{A}_{j}(\bR)e^{i\bK \cdot \bR}e^{-i\psi(k_y)} \lin
\eq\sum_{\bR}\check{A}_{j}(\bR)e^{ik_xR_x+ik_yR_y-iR_xR_y/2l^2}e^{-i\psi(k_y-R_x/l^2)}e^{-(R_x/l^{2})\partial_y} \la{withhelpfromb}\\
  \eq \sum_{\bR}\check{A}_{j}(\bR)e^{ik_xR_x+ik_yR_y-iR_xR_y/2l^2}e^{-i\psi(k_y)+i\{\psi_{-1}'(k_y)+\psi_0'(k_y)\}R_x/l^2 - (i/2)\psi_{-1}''(k_y)R_x^2/l^4 +O(l^{-4})}e^{-(R_x/l^{2})\partial_y} \\
  \eq e^{-i\psi(k_y)}\sum_{\bR}\check{A}_{j}(\bR)e^{i(k_x+\psi_{-1}'/l^2)R_x+ik_yR_y-iR_xR_y/2l^2}e^{i\psi_0'R_x/l^2 - (i/2)\psi_{-1}''R_x^2/l^4 +O(l^{-4})}e^{-(R_x/l^{2})\partial_y} \\	
  \eq e^{-i\psi(k_y)}\sum_{\bR}\check{A}_{j}(\bR)e^{i(k_x+\psi_{-1}'/l^2)R_x+ik_yR_y-iR_xR_y/2l^2}\{1+i\psi_0'R_x/l^2 - (i/2)\psi_{-1}''R_x^2/l^4+O(l^{-4})\}e^{-(R_x/l^{2})\partial_y}. \la{finallycorrect}}
If $\psi=O(1)$, the above equation particularizes to
\e{ \sum_{\bR}\check{A}_{j}(\bR)e^{i\bK \cdot \bR}e^{-i\psi(k_y)} = e^{-i\psi(k_y)}\sum_{\bR}\check{A}_{j}(\bR)e^{i\bk\cdot \bR-iR_xR_y/2l^2}\{1+i\psi_0'R_x/l^2 +O(l^{-4})\}e^{-(R_x/l^{2})\partial_y}. \la{finallyorderone}}
Letting the operator \q{finallycorrect} act on the identity function:
\e{ &\sum_{\bR}\check{A}_{j}(\bR)e^{i\bK \cdot \bR}e^{-i\psi(k_y)}1\\
\eq e^{-i\psi(k_y)}\sum_{\bR}\check{A}_{j}(\bR)e^{i(k_x+\psi_{-1}'/l^2)R_x+ik_yR_y}\bigg\{1+i(iR_x)(iR_y)/2l^2+\psi_0'(iR_x)/l^2 +(i/2)\psi_{-1}''(iR_x)^2/l^4+O(l^{-4})\bigg\}\lin
\eq e^{-i\psi(k_y)}\bigg\{ A_j(\bk)+\f{i}{2l^2}\f{\partial^2A}{\partial k_x \partial k_y} + \f{\psi_0'}{l^2}\f{\partial A}{\partial k_x} +\f{i\psi_{-1}''}{2l^4}\f{\partial^2A}{\partial^2 k_x}+O(l^{-4-2j})\bigg\}_{\bk \rightarrow \bk+\vec{x}\psi_{-1}'/l^2}. \la{letitactonidentity}}
By similar manipulations, we may derive an identity that is closely analogous to \q{finallycorrect}:
\e{ f(k_y)=O(1),\as \sum_{\bR}\check{A}_{j}(\bR)e^{i\bK \cdot \bR}f(k_y)= \sum_{\bR}\check{A}_{j}(\bR)e^{i\bk\cdot \bR-iR_xR_y/2l^2}\big\{f(k_y)-\tf{R_x}{l^2}f'(k_y) +O(\lmf)\big\}e^{-(R_x/l^{2})\partial_y}. \la{Hactsonorderone}}
If we let the operator in \q{Hactsonorderone} act on $e^{-i\psi}$,
\e{&\sum_{\bR}\check{A}_{j}(\bR)e^{i\bK \cdot \bR}f(k_y)e^{-i\psi(k_y)}=\sum_{\bR}\check{A}_{j}(\bR)e^{i\bk\cdot \bR-iR_xR_y/2l^2}\big\{f(k_y)-\tf{R_x}{l^2}f'(k_y) +O(\lmf)\big\}e^{-i\psi(k_y-R_x/l^2)}\lin
\eq f(k_y)A_j(\bK)e^{-i\psi(k_y)}+ i\lmt f'(k_y) \sum_{\bR}\check{A}_{j}(\bR)\p{\eikR}{k_x} e^{-iR_xR_y/2l^2}e^{-i\psi(k_y-R_x/l^2)} +O(l^{-4-2j})\lin
\eq f(k_y)A_j(\bK)e^{-i\psi(k_y)}+ i\lmt f'(k_y)e^{-i\psi(k_y)} \sum_{\bR}\check{A}_{j}(\bR)\p{}{k_x}e^{i(k_x+\psi_{-1}'/l^2)R_x+ik_yR_y}  +O(l^{-4-2j})\lin
\eq f(k_y)A_j(\bK)e^{-i\psi(k_y)}+ i\lmt f'(k_y)e^{-i\psi(k_y)} \p{A}{k_x}\bigg|_{\bk \rightarrow \bk+\vec{x}\psi_{-1}'/l^2}  +O(l^{-4-2j}). \la{uluright}
}


\subsection{Appendix to subsection `Turning points'} \la{app:proofmaslov}

Here we derive the Maslov correction to the single-band quantization conditions from a WKB approach. After reviewing the solution of the Peierls-Onsager Hamiltonian at the turning point in \app{app:reviewturningpoint}, we derive the first-order-corrected effective Hamiltonian and its solution in \app{app:turningfirstorder}. By wavefunction matching with the Zilberman-Fischbeck functions, we may determine the `reflection phase' ($\phi_r$) at each turning point -- the sum of all reflection phases is the desired Maslov correction. We pay careful attention to assigning a sense of circulation to each turning point in \app{app:reviewturningpoint} -- this determines the sign of each $\phi_r$, which is important to keep track of when we perform the sum $\sum \phi_r$. Finally, in \app{app:estimatesize}, we estimate the size of the turning region where quantum fluctuations render the Zilberman-Fischbeck wavefunctions invalid.

\subsubsection{Review of solution to the Peierls-Onsager Hamiltonian at the turning point}\la{app:reviewturningpoint}

Let us review the Peierls-Onsager solution at the turning point, which was first derived by Zilberman.\cite{zilberman_wkb} We assume that the reader has some familiarity with the WKB theory of turning points, and shall keep the review brief. We will go one small step beyond \ocite{zilberman_wkb} by defining a sense of circulation for each turning point, which determines the sign of the relative phase between incoming and outgoing WKB solutions.\\ 

We assume that the field-free Hamiltonian may be approximated by 
\e{H_0(\bk)=E+u_y k_y+ \f{k_x^2}{2m_x}, \la{airyham}}
with momentum coordinates originating from the turning point at energy $E$. The constant-energy band contour in the vicinity of the turning point may be split into two sections that touch at the same point; we use $\nu =+$ ($-$) to denote the section to the right (left) of the point: 
\e{ {k}^{\pm}_x(k_y,E)= \pm \sqrt{-{2m_xu_yk_y}}. \la{signkxlr}}
The sign of $m_xu_y$ determines whether the classical region lies at positive or negative $k_y$, as illustrated in \fig{fig:turning_point}(a-d).\\

$H_0(\bk)$ is in Weyl correspondence with the Peierls-Onsager Hamiltonian $H_0(\bK):=[H_0(\bk)]$; we shall assume the Landau gauge $K_x{=}k_x+i\lmt (\partial/\partial k_y)$ and $K_y{=}k_y$. $H_0(\bK)$ becomes independent of $k_x$ after the the basis transformation $e^{ik_xk_yl^2}$:
\e{ e^{-ik_xk_yl^2}H_0(\bK)e^{ik_xk_yl^2}= E+u_y k_y- \f{1}{2m_xl^4}\p{^2}{k^2_y}.\la{Airygen}}
We shall separately tackle the two cases corresponding to different signs of $m_xu_y$. \\

\noi{i} $m_xu_y>0$\emph{; band contour is an inverted parabola $\frown$, i.e., }$k_y \sim -k_x^2$\\

\q{Airygen} is an Airy differential equation with the dimensionless variable $z=(2m_xu_yl^4)^{\sma{1/3}}k_y$. In the limit $z \ll 0$ (i.e., within the classical region, and sufficiently far from the turning point), and assuming a hard-wall boundary condition, the Airy function has the asymptotic form\cite{griffiths_introQM}
\e{\limit{z \ll 0} Ai(z) \eq \f1{|z|^{1/4}}\big( e^{i(2/3)|z|^{3/2}+i\pi/4}\lin
&-e^{-i(2/3)|z|^{3/2}-i\pi/4} \big)
 \la{airyclassical}}
which is then matched with the Zilberman functions [\q{zilbermanfischbeck} without the $H_1$ correction]; some assumption must be made on the band parameters and the field for this matching region to exist.\cite{zilberman_wkb}  The prefactor $|z|^{\minus 1/4}$ is proportional to $|v^x_{\nu}|^{\minus 1/2}$ for both $\nu=\pm$ . The phase factor in the Zilberman function is
\e{ e^{-il^2\int^{k_y}_0k_x^{\nu}(t,E)dt},}
with $k_y$ negative in the classical region; we remind the reader that this sign is determined by the sign of $m_xu_y$. From \q{signkxlr}, $k_x^-\leq 0$ and $k_x^+\geq 0$ , so we identify
\e{\limit{z \ll 0} Ai(z) \propto&\; \f1{|v^{x}(k_y)|^{1/2}}\big( c^{\frown}_+\; e^{-il^2\int^{k_y}_0k_x^{+}}\lin
 &+ c^{\frown}_-\; e^{-il^2\int^{k_y}_0k_x^{-}}\big), \lin
& c^{\frown}_+:=e^{i\pi/4},\as c^{\frown}_- :=e^{i3\pi/4}.}
 
From Hamilton's equation [\q{hamilton2}], $\hbar\dot{k}_x=\lmt u^y$; ($u_y>0,m_x>0)$ thus corresponds to a wavepacket circulating in the clockwise sense: $\curvearrowright$ [illustrated in \fig{fig:turning_point}(a)], and $(u_y<0,m_x<0)$ to  $\curvearrowleft$ [\fig{fig:turning_point}(b)]. For the locally-clockwise [resp. locally-anticlockwise] trajectory, the relative phase factor between outgoing and incoming WKB wave is then ${c^{\frown}_+}/{c^{\frown}_-} =-i$ [resp. ${c^{\frown}_-}/{c^{\frown}_+} =+i$]; this may be interpreted as the phase acquired by a wavepacket as it is reflected (in $k_y$) from the turning point.\\


\noi{ii} $m_xu_y<0$\emph{; band contour is an upright parabola $\smile$, i.e., }$k_y \sim +k_x^2$\\

\q{Airygen} is an Airy differential equation with the dimensionless variable $z=-(2|m_xu_y|l^4)^{\sma{1/3}}k_y$, which differs from the previous case in the sign of $z/k_y$. The Airy solution in the classical region ($k_y \gg 0, z \ll 0$) has the same asymptotic form as in \q{airyclassical}. However, now  that $k_y$ is positive in the classical region (with $k_x^-$ and $k_x^+$ retaining their original signs), we switch the identification of $\nu=\pm$  Zilberman functions in the Airy function:
\e{&\limit{z \ll 0} Ai(z) \lin
\eq \f1{|z|^{1/4}}\left( e^{i(2/3)|z|^{3/2}+i\pi/4}-e^{-i(2/3)|z|^{3/2}-i\pi/4} \right) \lin
\prop \f1{|v^{x}(k_y)|^{1/4}}\left( c^{\smile}_- e^{-il^2\int^{k_y}_0k_x^{-}} + c^{\smile}_+ e^{-il^2\int^{k_y}_0k_x^{+}}\right), \lin
& c^{\smile}_-:=e^{i\pi/4},\as c^{\smile}_+ :=e^{i3\pi/4} \la{airyclassical2}}
A wavepacket that circulates the turning point in the locally-clockwise sense ($u_y<0,m_x>0$) thus picks up a phase factor $c_-^{\smile}/c_+^{\smile}=-i$ [illustrated in \fig{fig:turning_point}(d)]; the locally-anticlockwise wavepacket ($u_y>0,m_x<0$) picks up $+i$ [\fig{fig:turning_point}(c)].

\subsubsection{First-order-corrected wavefunction at the turning point}\la{app:turningfirstorder}

To account for $H_1$ in the above matching procedure, we first need to derive a first-order-corrected effective Hamiltonian ($\calh=H_0+H_1$) in the turning region. Let us expand $H_1$ around the turning point as
\e{H_1(\bk) \eq  H_{1}(\bze)+H_{1x}{k}_x+H_{1y}k_y+H_{1xx}{k}_x^2+ \ldots. \la{approximateH12}}
We argue that only the terms which are written explicitly above are relevant to $\calh$ in the limit of small field. Indeed, the neglected terms ($\delta H_1$) are bounded by their values at the boundary of the turning region:  $\delta H_1(\Delta \bk) =O(\lmf)$, with our estimates of  $\Delta \bk$ in the above paragraph. One may verify that the explicit terms in \q{approximateH12}, when evaluated on the boundary, are greater than $O(\lmf)$. When these explicit terms are added to $H_0$, the result is an effective Hamiltonian that is identical in form to \q{airyham}:  
\e{ [\calh(\bq)]=\calh(\bQ) \eq E + \tilde{u}_yQ_y  + \f{Q_x^2}{2\tilde{m}_x}+O(l^{-4}),}
but is shifted in velocity $\tilde{u}_y = u_y+H_{1y}$,  mass $\tilde{m}_x = m_x-2m_x^2H_{1xx},$ and the momentum variables
\e{ &q_x=k_x+mH_{1x},\as q_y=k_y+H_{1}(\bze)/{u}_y \as \longleftrightarrow  \lin
& Q_x=K_x+mH_{1x}, \as Q_y=Q_y+H_{1}(\bze)/{u}_y \la{relateqtok}.}
We assume $m_xu_y>0$ in this derivation, which is simply generalized for the other sign. $\calh$ may be solved with the same techniques; the Airy eigenfunction may be expressed as a sum of Zilberman functions:
\e{ f_{\bk E}=e^{iq_xk_yl^2}\sum_{\nu =\pm}c^{\frown}_{\nu}\f1{\sqrt{|v^x_{\nu}|}}e^{-il^2\int^{q_y}_0{q}_x^{\nu}(z,E)dz}, \la{Airysumzil}}
with $c^{\frown}_+/c^{\frown}_-=-i$, and $q_x^{\nu}$ describes a section of $\calh$ at energy $E$: 
\e{ 0=\calh({q}^{\nu}_x(q_y,E),q_y)-E \imp   {q}^{\pm}_x = \pm \sqrt{ -2\tilde{m}_x\tilde{u}_yq_y}.}
This function is related to the zero-field band contour $k_x^{\pm}=\pm (-2m_xu_yk_y)^{\sma{-1/2}}$ by
\e{ q_x^{\nu}(z,E)-k_x^{\nu}(z,E)=-\f{H_{1y}k_y+H_{1xx}({k}^{\nu}_x)^2}{v^x_{\nu}}+O(\lmf).}
Inserting this, as well as the left-hand-side of \q{relateqtok}, into \q{Airysumzil}, we express $f$ in terms of the original $\bk$ coordinates and the zero-field band contour:
\e{&f_{\bk E} = e^{ik_xk_yl^2}e^{im_xH_{1x}k_yl^2}\sum_{\nu=\pm}c^{\frown}_{\nu}\f1{\sqrt{|v^x_{\nu}|}}\lin
&\times \exp\bigg[{-}il^2\int^{k_y}_0  k_x^{\nu} dz +il^2\int_0^{k_y} (H_{1y}z+H_{1xx}({k}_x^{\nu})^2)\tf{dz}{{v}^x_{\nu}}\lin
& -il^2H_1(\bze){k}_x^{\nu}(k_y,E)/u_y+  O(l^{-2})\bigg].}
This complicated expression may be simplified with the identification 
\e{\int_0^{k_y} \f{{H}^{\nu}_1}{{v}^x_{\nu}} dz
\eq   m_xk_y H_{1x} +\int_0^{k_y} \f{H_{1y}z+H_{1xx}({k}^{\nu}_x)^2}{{v}^x_{\nu}}dz \lin
&- {k}^{\nu}_x(k_y,E)H_1(\bze)u_y^{-1} +O(l^{-4}); }
here, our estimation of $O(\lmf)$ was made by evaluating the neglected terms at the boundary of the turning region. Therefore, we arrive at
\e{f_{\bk E} = \sum_{\nu=\pm}c^{\frown}_{\nu}\f1{\sqrt{|{v}_{\nu}^x|}}e^{-il^2\int^{k_y}_0dz \big({k}_x^{\nu}-k_x-{H}^{\nu}_1/{v}_{\nu}^x\big) +  O(l^{-2})}, \la{leadingasymptoticairymod}}
which implies that the incoming and reflected Zilberman-Fischbeck functions are related by the reflection phase factor $e^{i\phi_r}=c^{\frown}_+/c^{\frown}_-=- i+O(\lmt)$. For an analogous result in the coupled-channel equations in nuclear physics, we refer the reader to \ocite{yabana_coupledchannel}.

\subsubsection{Estimation of size of the turning region} \la{app:estimatesize}

It is useful to estimate the size of the region in $\bk$-space ($\Delta k_x \Delta k_y$), in the vicinity of the turning point, where the \zf wavefunctions are invalid; equivalently, this is where the asympotic limits of the Airy functions would not apply  -- we have called this the turning region. From $z =O(1)$, we obtain $\Delta k_y =O(l^{\sma{-4/3}})$. The two sections $s_{\pm}$ of the band contour that meet at the turning point are described by $k_x^{\pm}=\pm (-2m_xu_yk_y)^{\sma{-1/2}}$. Combining this with our estimate of $\Delta k_y$, we obtain $\Delta k_x=O(l^{\sma{-2/3}})$; note that $\Delta k_x \Delta k_y =O(\lmt)$. We may further estimate the length of the semiclassical orbit that lies within the turning region as 
\e{&2\int^{\Delta k_x}_0\sqrt{1+(dk_y/dk_x)^2 }dk_x\lin
\eq 2\int^{\Delta k_x}_0\sqrt{1+\f{k_x^2}{(u_ym_x)^2}}dk_x =O(l^{-2/3}).}

\subsection{Quantization condition for the simplest closed orbit, from conventional means}\la{app:justifyrules}

We review the conventional determination\cite{zilberman_wkb} of the quantization conditions without breakdown, through the simplest case study of the closed orbit $\frako$ in \fig{fig:turning_point}(e); it is composed of two edges (labelled $\nu =\pm$) that touch at two turning points. Let us define the wavefunction in the $(K_x,k_y)$-representation as $f_{\bk E}$; 
the quantization condition is the condition of continuity of $f_{\bk}$ with respect to $k_y$.\\

For the interval of $k_y$ within the classical region and sufficient far from the two turning points, $f$ is the sum of two Zilberman-Fischbeck (ZF) functions which correspond to the two edges: $f_{\bk,E} =\sum_{\nu=\pm }c_{\nu}g^{\nu}_{\bk E}$, with $g$ defined in \q{zilbermanfischbeck}. To impose continuity, it is convenient to introduce the gauge-transformed wavefunction 
\e{\tilde{f}_{k_y,E}:=e^{-ik_xk_yl^2}f_{\bk E}=\sum_{\nu=\pm}c_{\nu}\,|v^x_{\nu}(k_y)|^{-1/2} a_{\nu}(k_y), \notag}
where $a_{\nu}$ are scalar amplitudes which we define for each edge $\nu$ as 
\e{ a_{\nu}(k_y):=e^{-il^2\int \big(k_x^{\nu}-H^{\nu}_1(v^x_{\nu})^{\mo}\big)d{k}_y}. \la{definescalarampsection}}
As mentioned in \q{sec:singlebandquant}, the phase $k_xk_yl^2$ is trivially continuous over a closed orbit.\\

$\tilde{f}_{k_yE}$ may be analytically continued into the turning region, such that its domain extends up to but excludes the turning point -- here, the velocity prefactor diverges. The function that facilitates this continuation is the leading asymptotic term of the modified Airy wavefunction at the turning point, which we derive in \q{leadingasymptoticairymod}.\\  

By analytic continuation to the top turning point (at wavevector $k_{y1}$), we arrive at the following expression for
\e{ \tilde{f}_{k_yE}\eq c^{\frown}_-\f1{|v^x_-(k_y)|^{1/2}}\f{a_-(k_y)}{a_-(k_{y1})}+c^{\frown}_+\f1{|v^x_+(k_y)|^{1/2}}\f{a_+(k_y)}{a_+(k_{y1})}, \lin
 \f{c^{\frown}_+}{c^{\frown}_-}\eq-i. \la{analytictop}}
By analytic continuation to the bottom turning point (at wavevector $k_{y2}$), we obtain a different expression
\e{ \tilde{f}_{k_yE}\eq c^{\smile}_-\f1{|v^x_-(k_y)|^{1/2}}\f{a_-(k_y)}{a_-(k_{y2})}+c^{\smile}_+\f1{|v^x_+(k_y)|^{1/2}}\f{a_+(k_y)}{a_+(k_{y2})}, \lin
 \f{c^{\smile}_+}{c^{\smile}_-}\eq+i. \la{analyticbot}}
The continuity condition is then equivalent to the identity of \q{analytictop} and (\ref{analyticbot}). Equating the right-hand-side of these two equations and eliminating $c_{\nu}$, we derive
\e{ -1 = \f{a_+(k_{y2})}{a_+(k_{y1})}\f{a_-(k_{y1})}{a_-(k_{y2})}. }
By reparametrizing $a$ by the time-like parameters $t_{\pm}$ [cf.\ \q{definescalarampsection2}-(\ref{ratioofamplitude2})], the above condition may be identified with \q{singleeq}.


\section{Appendix to `Symmetry in the first-order effective Hamiltonian theory'}

\subsection{Symmetry in Bloch Hamiltonians}\la{app:symmetryinbloch}

The aim of this section is to expand on the review of symmetries in \s{sec:symmetryinBloch} and further derive some identities which will be useful in deriving symmetry constraints on the effective Hamiltonian. These identities all involve cell-periodic functions and their symmetry constraints [cf.\ \q{gactsonu4}, \q{gactsonu3} and \q{ginverseactsonu2}].\\ 

To begin, let us recall some notation from \s{sec:preliminaries}. A cell-periodic function may be expanded as
\e{ \braket{\alpha}{u_{n\bk}} \eq u_{n\bk}(\alpha), \as \ket{u_{n\bk}} = \sum_{\alpha}u_{n\bk}(\alpha)\ket{\alpha},\lin
   \braket{u_{n\bk}}{\alpha} \eq u_{n\bk}(\alpha)^*, \as \bra{u_{n\bk}} = \sum_{\alpha}u_{n\bk}(\alpha)^*\bra{\alpha},\la{expandcellperiodic}}
where $\alpha$ is a shorthand for $(\btau,s)$, with $s$ a spin index, and $\btau$ the cell-periodic position coordinate that is defined with the equivalence $\btau \sim \btau +\bR$ ($\bR$ being a Bravais-lattice vector). $\sum_{\alpha}$ should be interpreted as an integration of $\btau$ over the unit cell, in addition to a sum over the spin index $\sigma$. The overlap of bra with ket is defined as
\e{ \braket{u}{v} = \sum_{\alpha}u^*(\alpha)v(\alpha), \as \braket{\alpha}{\beta} = \delta_{\ab},}
where $\delta_{\ab}$ is a shorthand for the product of  a Dirac delta function in real space and a Kronecker delta function in spin space. 
We remark that the final results of this section, and the way they are derived, are essentially unchanged if we interpret $\alpha$ as a discrete label for a basis of \low orbitals\cite{lowdin1950,slater1954} in tight-binding methods.\\

Let a symmetry operation $g$ act on the cell-periodic variable as
\e{ \hat{g}\ket{\alpha} =\ket{\beta}[U_g]_{\beta \alpha}K^{s(g)}, \as U_g^{-1}= \dg{U}_g, \la{ghatacts}}
with $s(g)$ defined in \q{definesg}, repeated indices are summed, $K$ is the complex-conjugation operation that leaves the basis vector invariant:
\e{ Kz K =z^*,\as K\ket{\alpha}K =\ket{\alpha}, \as  K^2 =I. }
To clarify, \q{ghatacts} is shorthand for
\bal
\hat{g}\ket{\alpha} =\begin{cases}  \ket{\beta}[U_g]_{\beta \alpha}, &g \ins{unitary,}\\
\ket{\beta}[U_g]_{\beta \alpha}K, &g \ins{antiunitary}.\end{cases} \la{unitarydefined}
\end{align}
For example, consider $g=M_x$ as a reflection that maps $x\rightarrow -x$, in which case
\e{ \hat{M}_x\ket{\tau_x,\tau_y,\tau_z,s} = -i \ket{-\tau_x,\tau_y,\tau_z,-{s}}.}
Here, $s$ labels the eigenvalue of spin component $S_z$, we have used that $M_x$ is a product of a spatial inversion with a two-fold rotation about $\vec{x}$: $\hat{M}_x = \inv C_{2x} = \inv e^{-iJ_x\pi} =\inv e^{-iL_x\pi}(-i\sigma_x)$. 
If $g$ is the spatial translation by $\bR$, then $U_g$ is the identity operation, due to the just-mentioned equivalence $\btau \sim \btau +\bR$. The triviality of spatial translations imply that $\{U_gK^{s(g)}|g\in G\}$ forms a representation of the point group of the crystal, i.e., the quotient of the full space group $G$ (or magnetic space group) over the subgroup of discrete real-space translations.\\

Bear in mind that $\hat{g}$ acts on complex numbers as
\e{ \hat{g}z = K^{s(g)}zK^{s(g)}\hat{g}.}
We further define $\hat{g}^*$ by
\e{ \hat{g}^*\ket{\alpha} = \ket{\beta}[U_g]^*_{\beta \alpha}K^{s(g)},}
such that
\e{ K\hat{g}K=\hat{g}^*.}
The inverse operation is
\e{\hat{g}^{-1}\ket{\alpha} =K^{s(g)}\ket{\beta}\dg{[U_g]}_{\beta \alpha}, \la{defineinverseg}}
from which one may verify $\hat{g}\hat{g}^{-1}=\hat{g}^{-1}\hat{g}=I$. From  \q{gactsonposition},
\e{ \hat{g}e^{i\bk\cdot \hbr}\hat{g}^{-1}=e^{[(-1)^{s(g)}i]\bk\cdot [\check{g}^{-1}(\hbr-\bdelta)]}=e^{i[g\circ\bk]\cdot( \hbr-\bdelta)}.\la{gactsoneikr2}}
Consequently, a Bloch function at wavevector $\bk$, when operated upon by $g$, transforms with a possibly distinct wavevector 
\e{ \bk':=g\circ\bk, \as \p{k'_{\alpha}}{k_{\beta}}=(-1)^{s(g)}\check{g}_{\ab}, \la{gactsonk2}}
as may be ascertained from
\e{\hat{g}e^{i\bk\cdot \hbr}\ket{u_{n\bk}} = e^{i\bk' \cdot \hbr} \hat{g}(\bk) \ket{u_{n\bk}}. \la{gactsonBlochwave}}
Here, we have combined $\hat{g}$ and the nonsymmorphic phase factor in \q{gactsoneikr2} as 
\e{\hat{g}(\bk):=  e^{-i(g\circ \bk)\cdot \bdelta}\hat{g}.}
Combining \q{expandcellperiodic} with \q{ghatacts},
\e{ \braopket{\alpha}{\hat{g}}{u} \eq \bra{\alpha}\sum_{\beta}K^{s(g)}u(\beta)K^{s(g)}\ket{\delta}[U_g]_{\delta \beta}K^{s(g)} \lin
\eq \sum_{\beta}K^{s(g)}u(\beta)K^{s(g)}[U_g]_{\alpha \beta}K^{s(g)} .}
If $g$ is a symmetry of the Hamiltonian, then, applying \q{gactsoneikr2}, 
\e{ &\hat{g}(\bk)\hH_0(\bk)\hat{g}(\bk)^{-1} =\hat{g}e^{-i\bk\cdot\hbr}\hat{g}^{-1}\hH_0 \hat{g}e^{i\bk\cdot\hbr}\hat{g}^{-1}\lin
\eq  e^{-i[g\circ\bk]\cdot(\hbr-\bdelta)}\hH_0e^{i[g\circ\bk]\cdot( \hbr-\bdelta)} =\hH_0\big( \;g\circ\bk\;\big).}
This implies that if $\ket{u_{m\bk}}$ is an eigenstate of $\hH_0(\bk)$ with eigenvalue $\var_{m\bk}$, then ${\hat{g}(\bk)}\ket{u_{m\bk}}K^{s(g)}$ belongs to the eigenspace of $\hH_0( \;g\circ\bk\;)$ with the same energy $\var_{m\bk}$; the ambiguity in how we pick basis vectors within each energy eigenspace is expressed as 
\e{ {\hat{g}(\bk)}\ket{u_{m\bk}}K^{s(g)} \eq \ket{u_{n,g\circ\bk}}\breve{g}(\bk)_{nm},\la{gactsonu2}}
where $\breve{g}$ is a `sewing matrix' that is block-diagonal with respect to the energy eigenspaces, such that each distinct block corresponds to a distinct energy. \q{gactsonu2} is a shorthand for
\e{   &e^{-i(g\circ \bk)\cdot \bdelta}\sum_{\beta}K^{s(g)}u_{m\bk}(\beta)K^{s(g)}[U_g]_{\alpha \beta} \lin
\eq u_{n,g\circ\bk}(\alpha)\breve{g}(\bk)_{nm}. \la{shorthandfor}}
\q{gactsonu2} implies
\e{  \hat{g}(\bk)\ket{u_{m\bk}}K^{s(g)}\breve{g}^{\mo}(\bk)_{mn} \eq  \ket{u_{n,g\circ\bk}}, \la{gactsonu4}}
from which one obtains,
\e{ &\ket{\nabk^{\alpha}u_{n,\bk}}\bigg|_{\bk \rightarrow g\circ\bk}  =  \p{k_{\beta}}{k'_{\alpha}}\nabk^{\beta}\left(\hat{g}(\bk)\ket{u_{m\bk}}K^{s}\breve{g}^{\mo}(\bk)_{mn}\right) \lin  
\eq  (-1)^{s}\check{g}_{\ab} \bigg(\ket{\nabk^{\beta}u_{m\bk}}K^{s}\breve{g}^{\mo}(\bk)_{mn}\lin
+&\;\ket{u_{m\bk}}K^{s}\nabk^{\beta}\breve{g}^{\mo}(\bk)_{mn}\bigg) -i\delta^{\alpha}\hat{g}(\bk)\ket{u_{m\bk}}K^{s}\breve{g}^{\mo}(\bk)_{mn} .
\la{gactsonu3}}
In the last equality we substituted $(\partial k_{\beta}/\partial k'_{\alpha})$ with \q{gactsonk2}. Taking the complex conjugate of \q{shorthandfor},
\e{  & e^{i(g\circ\bk)\cdot \bdelta}\sum_{\beta}K^{s(g)}u_{m\bk}(\beta)^*K^{s(g)}\dg{[U_g]}_{\beta \alpha} \lin
\eq \dg{\breve{g}(\bk)}_{mn} u_{n,g\circ\bk}(\alpha)^*.}
This may be shortened, with \q{defineinverseg}, as
\e{ K^{s(g)}\bra{u_{m\bk}}\hat{g}^{-1}(\bk) = \dg{\breve{g}(\bk)}_{mn} \bra{u_{n,g\circ\bk}}, \la{ginverseactsonu}}
which implies
\e{  \breve{g}(\bk)_{ml}K^{s(g)}\bra{u_{l\bk}}=\bra{u_{m,g\circ\bk}}\hat{g}(\bk). \la{ginverseactsonu2}}
This identity, with \q{gactsonu4}, will be used to derive how the current operator transforms under symmetry in the next subsection.

\subsection{Symmetry constraint on the \omm}\la{app:symmetryom}

We detail the derivation of the symmetry constraint of the multi-band \om in \q{gactsonMmultiband}; we assume the reader is familiar with the outline of the proof sketched in \s{sec:symmetrysinglebandom}. As an intermediate step, let us derive \q{gactsonvelocitymatrixfinal}, which describes the symmetry constraint on the current operator. \\

\noindent \emph{Proof of \q{gactsonvelocitymatrixfinal}:} The current operator transforms as
\e{& \hat{g}\hbPi\hat{g}^{-1} = \hat{g}(-i)[\hbr,\hH]\hat{g}^{-1} = (-1)^{s(g)}(-i)[\check{g}^{-1}(\hbr-\bdelta),\hH]=(-1)^{s(g)}\check{g}^{-1}\hbPi.}
Combining this with \q{gactsoneikr}, we see that the operator, defined by
\e{ \hbPi(\bk) = e^{-i\bk \cdot \hbr} \hbPi e^{i\bk \cdot \hbr},}
transforms as
\bal
\hat{g}(\bk)\hbPi(\bk)\hat{g}^{\mo}(\bk) =(-1)^{s(g)}\check{g}^{-1}\hbPi\bigg(g\circ\bk\bigg).
\end{align}
The matrix elements of the velocity operator thus satisfies the following symmetry constraint:
\e{ \bPi\bigg(g\circ\bk\bigg)_{mn} = (-1)^{s(g)}\check{g}\braopket{u_{m,g\circ\bk}}{  \hat{g}(\bk)\hbPi(\bk)\hat{g}^{\mo}(\bk) }{u_{n,g\circ\bk}}.}
Inserting \q{gactsonu4} and \q{ginverseactsonu2} into this expression,
\e{ \bPi\big(\;g\circ\bk\;\big)_{mn} \eq (-1)^{s(g)}\check{g}\breve{g}(\bk)_{ml}K^{s(g)}\bra{u_{l\bk}}\hat{g}^{\mo}(\bk)  \hat{g}(\bk)\hbPi(\bk)\hat{g}^{\mo}(\bk) {\hat{g}(\bk)}\ket{u_{a\bk}}K^{s(g)}\breve{g}^{-1}(\bk)_{an} \lin
\eq (-1)^{s(g)}\check{g}K^{s(g)}[\breve{g}^* \bPi \breve{g}^{t}]_{mn}K^{s(g)}\bigg|_{\bk}. \blacksquare \la{gactsonvelocitymatrix}}

In the degenerate subspace projected by $P$, let us define $\var=\var_m$ for $m\in \{1,\ldots, D\}$; $\bk$-dependence is implicit in this and the following notations. Combining \q{gactsonvelocitymatrix} with \q{definemultibandom},
 \e{  (-i)\epsilon_{abc}\sum_{\bar{n}}\f{\Pi_{m\bar{n}}^b\Pi_{\bar{n}l}^c }{\var-\var_{\bar{n}}}\bigg|_{g\circ\bk} \eq  (-i)\epsilon_{abc}\check{g}_{bd}\check{g}_{ce}\sum_{\bar{n}}\f{K^{s(g)}[\breve{g}^*\Pi^d\breve{g}^T]_{m\bar{n}}[\breve{g}^*\Pi^e\breve{g}^T]_{\bar{n}l}K^{s(g)}\bigg|_{\bk} }{(\var-\var_{\bar{n}})\bigg|_{g\circ\bk} }. \la{lefthandsideequals}}
Applying the levi-cevita identity (for an orthogonal matrix satisfying $R^T=R^{\mo}$)
\e{ & \text{det}[R^T]\,\epsilon_{lmn} = \epsilon_{abc} \,R^T_{la}\, R^T_{mb}\,R^T_{nc}  \imp R_{al}\, \text{det}[R]\,\epsilon_{lmn} = \epsilon_{abc} \, R_{bm}\,R_{cn},}
and the reality of $\check{g}$, we derive that the left-hand-side of \q{lefthandsideequals} equals
\e{ (-1)^{s(g)}\det[\check{g}] \check{g}_{ab} K^{s(g)} (-i)\epsilon_{bde}\sum_{\bar{n}}\f{[\breve{g}^*\Pi^d\breve{g}^T]_{m\bar{n}}[\breve{g}^*\Pi^e\breve{g}^T]_{\bar{n}l}\bigg|_{\bk}} {(\var-\var_{\bar{n}})\bigg|_{g\circ\bk} }K^{s(g)}.  \la{gactsonmagnetization}}
Let us introduce new labels $\bar{n}:= (a',a'')$, such that $a'$ labels the \emph{distinct} energy eigenvalues, and $a''$ labels an arbitrarily chosen basis in the finite-dimensional subspace corresponding to energy $\var_{a'}$. We see that in this labelling that $\var_{n\bk}$ does not depend on $a''$, so we may shorten $\var_{\bar{n}\bk} \rightarrow \var_{a'\bk}$. Moreover, since the symmetry commutes with the Hamiltonian,
$\var_{a'\bk} = \var_{a',g\circ\bk}.$ Therefore, \q{gactsonmagnetization} simplifies to
\e{ (-1)^{s(g)}\det[\check{g}] \check{g}_{ab} K^{s(g)} (-i)\epsilon_{bde}\sum_{a';\var_{a'}\neq \var}\f1{(\var-\var_{a'})}\sum_{a''}[\breve{g}^*\Pi^d\breve{g}^T]_{m,(a',a'')}[\breve{g}^*\Pi^e\breve{g}^T]_{(a',a''),l}\bigg|_{\bk}K^{s(g)}. }
Since $\breve{g}$ is block-diagonal in the index $a'$, the above equation may be expressed as
\e{&  (-1)^{s(g)}\det[\check{g}] \check{g}_{ab} K^{s(g)} (-i)\epsilon_{bde}\sum_{a';\var_{a'}\neq \var}\f1{(\var-\var_{a'})}\sum_{a'',b'',c''}[\breve{g}^*\Pi^d]_{m,(a',b'')}\breve{g}^T_{(a',b''),(a',a'')}\breve{g}^*_{(a',a''),(a',c'')}[\Pi^e\breve{g}^T]_{(a',c''),l}\bigg|_{\bk}K^{s(g)}.\notag}
Since each block diagonal of $\breve{g}$, corresponding to an energy subspace, is unitary, the above equation reduces to
\e{&  (-1)^{s(g)}\det[\check{g}] \check{g}_{ab} K^{s(g)} (-i)\epsilon_{bde}\sum_{a';\var_{a'}\neq \var}\f1{(\var-\var_{a'})}\sum_{b'',c''}[\breve{g}^*\Pi^d]_{m,(a',b'')}\delta_{b'',c''}[\Pi^e\breve{g}^T]_{(a',c''),l}\bigg|_{\bk} K^{s(g)}\lin
\eq  (-1)^{s(g)}\det[\check{g}] \check{g}_{ab} K^{s(g)} (-i)\epsilon_{bde}\sum_{a';\var_{a'}\neq \var}\f1{(\var-\var_{a'})}\sum_{a''}[\breve{g}^*\Pi^d]_{m,(a',a'')}[\Pi^e\breve{g}^T]_{(a',a''),l}\bigg|_{\bk} K^{s(g)}.}
Restoring the usual labelling, we conclude that the left-hand-side of \q{lefthandsideequals} equals
\e{ (-1)^{s(g)}\det[\check{g}]\check{g}_{ab}K^{s(g)}(-i)\epsilon_{bcd}\sum_{\bar{n}}\f{[\breve{g}^*\Pi]_{m\bar{n}}^c[\Pi\breve{g}^T]_{\bar{n}l}^d }{\var-\var_{\bar{n}}}\bigg|_{\bk}K^{s(g)},}
from which follows \q{gactsonMmultiband}.

\subsection{Appendix to symmetry of the first-order effective Hamiltonian} \la{app:symmetryH1}

Let us analyze the symmetry constraints on the (a) Roth, (b) Zeeman and (c) Berry terms in the first-order effective Hamiltonian [recall their definitions in \q{H1multiband2}], in that order. The final goal is to derive \q{H1transformssymmetry}.\\
 
\noi{a} For Bloch electrons immersed in a field parallel to $\vec{z}$, $H_1^R=-B^zM^z$. For symmetries of semiclassical orbits [defined precisely in \s{sec:symmetryoforbits}], \q{kactionblockdiag} and (\ref{defineug}) inform us that $[\check{g}\bM]^z = (-1)^{t(g)} M^z =(-1)^{u(g)}\text{det}[\check{g}]M^z$, and therefore \q{gactsonMmultiband} particularizes to 
\e{ M^z\bigg|_{g\circ\bk} \eq (-1)^{s(g)+u(g)}\,\breve{g}\, K^{s(g)} \, M^z\, K^{s(g)} \,\breve{g}^{-1}\bigg|_{\bk}, \la{gactsonMmultibandBz}}
with $u(g)\in \Z_2$ defined in \q{defineug}.\\

\noi{b} For symmetries in spin-orbit-coupled systems, we would like to demonstrate that $H_1^Z \propto B\sigma^z$ is constrained similarly to \q{gactsonMmultibandBz}:
\e{ \sigma^z\bigg|_{g\circ\bk} \eq (-1)^{s(g)+u(g)}\,\breve{g}\, K^{s(g)}\,{\sigma}^z\,K^{s(g)}\,\breve{g}^{-1}\bigg|_{\bk},\la{transformszmatrix}}
where $(\hbar/2)\sigma_{mn}^z(\bk) =(\hbar/2)\braopket{u_{m\bk}}{\hat{\sigma}_z}{u_{n\bk}}$ is the spin-half matrix defined in \q{definespinhalf}. We already know how the cell-periodic functions transform under symmetry [cf.\ \q{gactsonu}], so what remains is to determine how $\hat{\sigma}_z$ transforms under a symmetry of the orbit. For this purpose, the decomposition in \q{symmetrydecomposition} is useful in deriving
\e{\hat{g}^{-1}  \hat{\sigma}^z \hat{g} =(-1)^{s(g)+u(g)}\hat{\sigma}^z. \la{transformsz}}
Indeed, among the factors written on the right-hand-side of \q{symmetrydecomposition}, only time reversal (if present) and $\mir_x$ (if present) flips the $z$-component of spin. Combining \q{transformsz} with \q{gactsonu}, we then obtain \q{transformszmatrix}.\\

\noi{c} The Berry term $H_1^B=\lmt\epsilon_{\ab}\mx^{\beta}v^{\alpha}$. Combining \q{symmetryconstraintberryconn} with the constraint on the band velocity in \q{constraintbandvelocity}, 
\e{ &\epsilon_{\ab}\mx^{\beta}v^{\alpha}\bigg|_{g\circ\bk} -\epsilon_{\ab}\delta^{\beta}v^{\alpha}\bigg|_{g\circ\bk}\lin
\eq (-1)^s\epsilon_{\ab}\check{g}_{\alpha\mu}\check{g}_{\beta\nu} \bigg( \breve{g}K^{s}\mx^{\nu}K^{s}\breve{g}^{\mo}+  i(-1)^{s}\,\breve{g} \nabk^{\nu}\breve{g}^{\mo}\bigg)v^{\mu}\bigg|_{\bk}  \lin
\eq  (-1)^{s+u}\epsilon_{\ab} \bigg( \breve{g}^{\mo}K^{s}\mx^{\beta}K^{s}\breve{g}^{\mo}+  i(-1)^{s}\,\breve{g} \nabk^{\beta}\breve{g}^{\mo}\bigg)v^{\alpha}\bigg|_{\bk}.  \notag}
 
\noindent The net result of (a-c) is \q{H1transformssymmetry}.

\subsection{Topological obstruction to symmetry covariance of $H_1$}\la{app:no}

Supposing $H_0$ is $g$-symmetric, does a  basis (for the cell-periodic functions)  exist where $H_1$ transforms covariantly under $g$, for all $\bk$ in the Brillouin torus? This section is a self-constained exposition on the possible obstructions to symmetry covariance in topologically-nontrivial band subspaces.
 The existence of a topological obstruction is suggested by the observation in paragraph (ii) of in \s{sec:symmetryH1}: the source of non-covariance is the Berry term. The Berry curvature is a measure of the `twisting' of the filled-band wavefunctions in $\bk$-space; it is known that topologically-nontrivial band structures exist whose curvature cannot be made to vanish.\cite{schnyder_classify3DTIandTSC}\\ 

We support this claim with a few case studies in the following subsections; a recurrent theme in these case studies is that the effective Hamiltonian of a symmetry-protected topological phase transforms anomalously (i.e., non-covariantly) under the symmetry in question. Our last case study  has only the $U(1)$ symmetry of charge conservation, and we would show that the effective Hamiltonian for a nontrivial Chern band transforms anomalously under a gauge transformation. \\

Before beginning properly, let us introduce a terminology. Supposing the second and third terms in \q{H1transformssymmetry} were absent, we say that $H_1$ transforms covariantly under the symmetry (resp.\ antisymmetry)  $g$  if $(-)^{s(g)+u(g)}=+1$  (resp.\ $-1$). In simple words, $g$ is referred to as an antisymmetry of $H_1$ if it inverts the sign of $H_1$.


\subsubsection{Wigner-Dyson class AII}

Let us exemplify this claim with gapped band subspaces in the symmetry class AII in two\cite{kane2005A,kane2005B,HgTe_bernevig,QSHE_Rahul,fu2006} or three\cite{fukanemele_3DTI,Inversion_Fu,moore2007,Rahul_3DTI} spatial dimensions -- spin-orbit-coupled, with the time-reversal symmetry satisfying $\hat{T}^2=-I$; no assumption is made presently about the spatial symmetries, however we will elaborate on their roles in the next two sections [\s{sec:AIIwithinversion} and \s{sec:spatialonly}]. We have used the word `gapped' liberally to describe band subspaces which are energetically separated from all other bands at each wavevector in the Brillouin torus; this would include indirect-gap systems with nonvanishing Fermi lines or surfaces. It is well-known that gapped band subspaces, in either 2D or 3D, are classified by a strong $\Z_2$ invariant;\cite{kitaev_periodictable} we shall refer to the nontrivial phase (in both 2D and 3D) as the $\Z_2$-topological band.\\

In this context, we would like to define the effective Hamiltonian $H_1(\bk)$ over the entire torus -- it is minimally a two-band Hamiltonian due to Kramers degeneracy at $\bk^{(i)}$. Let us  ask if $H_1$ may transform covariantly under the antisymmetry $T$; the non-covariant term on the right-hand-side of \q{H1transformssymmetry} vanishes if either (i) $\bv(\bk)$ can be made to vanish everywhere, i.e., the band(s) in $P$ have a flat dispersion, or (ii) the sewing matrix corresponding to $T$, as defined by
\e{ \breve{T}(\bk)_{mn} = \braopket{u_{m,-\bk}}{\hat{T}}{u_{n\bk}}K, \la{sewingT}}
can be made to be independent of $\bk$. We disregard the implausibly fine-tuned scenario where the non-covariant term vanishes without satisfaction of (i) or (ii). For a trivial band subspace, we argue that an adiabatic continuation exists to a lattice of inert atoms, where both $\bv(\bk)=0$, and  $\breve{T}$ reduces to a $\bk$-independent matrix which represents time reversal in the basis of \low orbitals. Let us then consider (i) and (ii) in the context of a topological band subspace.\\

\noi{i} For an $\Z_2$-topological band, $\bv(\bk)$ cannot everywhere be zero if the associated tight-binding Hamiltonian has local (strictly short-ranged) hoppings.\cite{read_compactwannier} In fact, the impossibility of a strictly short-ranged, flat-band Hamiltonian is more generally true for all of the strong topological band subspaces in larger-than-one spatial dimensions; this was first proven for class A in 2D,\cite{lichen_impossibility} and then extended to the tenfold symmetry classes.\cite{read_compactwannier} This rigorous result suggests that if a strictly-flat-band Hamiltonian exists for an $\Z_2$-topological band, it is likely to be a highly-optimized scenario\cite{chinghua_bandflatness} which is challenging to realize in both theoretical and experimental laboratories. We know only of one model\cite{Kapit_flatbandchern} (of a strong 2D topological insulator)  with exactly flat bands;\footnote{The model described in this reference describes a Chern insulator, but one can combine Chern insulators with opposite chiralities to obtain a class-AII strong topological insulator.} the hopping elements here decay as a Gaussian. We henceforth assume that $\bv(\bk)$ is a non-constant function, which is the case of interest in almost all applications.\\

\noi{ii}  Given that the band is not flat, we are led to investigate the momentum-dependence of the sewing matrix for a topological band. One expression for the strong $\Z_2$ invariant, in both 2D and 3D, involves the even-dimensional sewing matrix $\breve{T}$ defined in \q{sewingT}. Since this matrix is skew-symmetric at any inversion-invariant wavevector $\bk^{(i)}$, we might evaluate the quantity
\e{\delta_i={\sqrt{\det[\breve{T}(\bk^{(i)})]}}/{\text{Pf}[\breve{T}(\bk^{(i)})]}=\pm 1,}
with Pf$[\cdot]$ denoting the Pfaffian of $[\cdot]$. The product of $\delta_i$ over all $\bk^{(i)}$ (numbering four in 2D, and eight in 3D) is the strong $\Z_2$ invariant, which equals $+1$ ($-1$) in correspondence with the trivial (topological) phase;\cite{fukanemele_3DTI} this definition implicitly assumes the continuity of the cell-periodic functions over the Brillouin torus. If the sewing matrix were constant over this torus, an immediate implication is $\prod_i\delta_i=+1$; alternatively stated, for the $\Z_2$-topological band, there is an obstruction to defining a constant sewing matrix; consequently, a non-flatband $H_1$ must transform non-covariantly under the antisymmetry $T$.

\subsubsection{Class AII with spatial inversion symmetry}\la{sec:AIIwithinversion}

Even for $\Z_2$-topological band subspaces, it is generically not true that a topological obstruction exists for all symmetries of the system. To exemplify this claim, let us consider a $\Z_2$-topological band (in 2D or 3D) with spatial inversion symmetry $\inv$. 
The space-time inversion $T\inv$ ($s=1,u=0$) acts as an antisymmetry on the first-order effective Hamiltonian. For simplicity, we assume that bands are  two-fold degenerate everywhere on the Brillouin torus, and hence $H_1$ is a two-band Hamiltonian. We then ask if $H_1$ transforms covariantly under the antisymmetry $T\inv$, or equivalently, if the corresponding sewing matrix (denoted as $\breve{T}_{\inv}$) can be made constant over the torus. An algorithm for this has been proposed in \ocite{Inversion_Fu}, which is plausibly valid in the $\Z_2$-nontrivial phase (i.e., with $\prod_i\delta_i=-1$). Assuming such a gauge is found, $T\inv$-symmetry then imposes the covariant antisymmetry condition:
\e{ H_1(\bk) = -\breve{T}_{\inv}H_1^*(\bk)\breve{T}_{\inv}^{\mo}; \as \nabk\breve{T}_{\inv}=0,   \la{H1transformsTisymmetry}}
which follows from \q{H1transformssymmetry} for a constant sewing matrix. Evaluating the trace on both sides of \q{H1transformsTisymmetry}, we further deduce that $H_1$ is traceless: 
\e{\text{Tr}[H_1(\bk)]=0 \imp H_1(\bk)=-\mu_B\tau_iB_j\xi_{ij}(\bk); \la{effectivespin}}
$\tau_i$ here are Pauli matrices describing an effective spin,  that is generally distinct from the free spin due to spin-orbit coupling. A heuristic argument for the tracelessless of $H_1$ in \q{effectivespin} already exists in the literature.\cite{rothII,Mikitik_quantizationrule,lifshitz_pitaevskii_statphys2}
Here, we have clarified that $H_1$ is traceless only in a special basis where the sewing matrix for $T\inv$ is constant -- consequently, the unitary generated by $H_1$ [cf.\ \q{definenonabelianunitary}] has unit determimant, and the eigen-phases of this unitary satisfy $\lambda_1=-\lambda_2$ mod $2\pi$. We remind the reader that $\lambda_a$ enter the multi-band quantization conditions in \q{rule3b}, and also the condition for dHvA oscillations in \q{laj}. \\

The reader may be unsatisfied that the above conclusions relied on the existence of a special gauge. A more general proof of $\lambda_1=-\lambda_2$ is provided in \s{sec:symmetryunitarygenbyH1} [see the paragraph surrounding \q{Tinvconstraint}].


\subsubsection{Topological band subspaces protected by crystalline symmetries}\la{sec:spatialonly}

One next case study demonstrates that a topological obstruction may exist for band subspaces, where the obstruction is  protected solely by crystalline symmetries. 3D insulators having an improper spatial symmetry (i.e., det$[\check{g}]=-1$) have a quantized magneto-electric response,\cite{Chen_bulktopologicalinvariants} i.e., the $\theta$ angle occurring in the axion Lagrangian\cite{Wilczek_axion} is  symmetry-fixed to $0$ or $\pi$. For inversion-symmetric ($\inv$) bands, $\theta/2\pi$ may be expressed as half the winding number of $\breve{\inv}$ (the sewing matrix for $\inv$):\cite{turner_inversionsymmetricTI,hughes_inversionsymmetricTI}
\e{\f{\theta}{2\pi} = -\f1{48\pi^2}\int d^3\bk \epsilon_{\ab \gamma}\text{Tr}\bigg[ (\breve{\inv}\nabla_{\alpha}\breve{\inv}^{\mo})(\breve{\inv}\nabla_{\beta}\breve{\inv}^{\mo})(\breve{\inv}\nabla_{\gamma}\breve{\inv}^{\mo}) \bigg].}
$\theta=\pi$ thus implies a topological obstruction against $H_1$ transforming covariantly under $\inv$.

\subsubsection{Wigner-Dyson Class A}

Having described the anomalous symmetry transformation of $H_1$ for symmetry-protected topological phases, we might ask if there is an analogous topological obstruction for  charge-conserving band subspaces having no other symmetries -- they fall into Wigner-Dyson class A. Even though the Bloch Hamiltonian is completely unconstrained, one always has, at the basic level, a `gauge symmetry', which reflects the ambiguity in how we label our bands in $P$; a gauge transformation such as in \q{basistransf} might be viewed as a `do-nothing' symmetry operation.\\    

Class-A band subspaces in 2D are classified by the TKNN invariant,\cite{TKNN} or, equivalently, the first Chern number ($C_1$). We would like to show that a nonzero $C_1$ neccesarily implies that $H_1$ is not gauge-covariant. Indeed, it was already noted in  \q{h1bnotunique} that the Berry term $\lmt \epsilon_{\ab}\mx^{\beta}v^{\alpha}$ generally results in a loss of covariance for any band subspace, trivial or nontrivial; the Roth and Zeeman terms are gauge-invariant (resp.\ -covariant) in the one-band (resp.\ multi-band) case. We are led to ask if $H_1^B$ can be made to vanish by basis transformations within $P$. For a band with a generic dispersion, this amounts to asking if there exists a gauge where $\bmx(\bk)= 0$ at each $\bk$; this gauge does not exist if  the Berry curvature $\calf^z(\bk):=\nabk \times$Tr$[\bmx] \neq 0$. Since the net Berry curvature for a Chern band is nonzero, we conclude that $H_1$ (for a generically-dispersing Chern band) must transform non-covariantly.\footnote{The problem actually goes deeper: $H_1$ cannot even be defined continuously over the torus. For a related discussion, see \ocite{freund_peierlsmagnetic}}

\section{Appendix to `Intraband breakdown'}





\subsection{Derivation of the intraband scattering matrix}\la{app:deriveintrascattering}

\subsubsection{Review of connection formula in the lowest order}\la{app:zerothorderconnectionintraband}

Let us pedagogically review the derivation of the connection formula in the lowest order in $\lmt$, with the eventual goal of generalizing the formula to the next order [to be carried out in the next subsection]. The lowest-order problem was first studied by Azbel\cite{azbel_quasiclassical,azbel_energyspectrum} and has reappeared in similar contexts,\cite{Wilkinson_criticalproperties,Davis_landauspectrum,robbins_uniformquantization}as more generally reviewed in \ocite{berry_mount_review}. \\

The Hamiltonian in the breakdown region is approximated by the Peierls-Onsager Hamiltonian, which is in Weyl correspondence with \q{saddlepointdispersion}:
\e{H_0(\bK) =  \f{K_x^2}{2m_1} -\f{K_y^2}{2m_2}, \as (H_0(\bK)-E)f_{\bk E}=0. \la{Peierlshamsaddle}}
We have further defined $f_{\bk E}$ as the eigenfunction corresponding to eigenvalue $E$. Working in the Landau electromagnetic gauge [recall \q{defineKyrepresentation}], we perform a gauge transformation
\e{ f=e^{iq_xk_yl^2}\bar{f};}
the resultant differential equation for $\bar{f}$ becomes independent of $k_x$, and is equivalent to the time-independent Schrodinger equation for a  particle in an inverted parabolic potential, with coordinate $k_y$, as was first studied by Kemble.\cite{kemble} Introducing the dimensionless variable
\e{ z = e^{-i\pi/4}\,(k_yl) \left(\f{4m_1}{m_2}\right)^{1/4}, \la{zdimensionless}}
we obtain a Weber differential equation\cite{NIST}
\e{\left(\partial_z^2-\tf1{4}z^2 +i\mu \right)\bar{f} = 0, \la{changeofvariablessaddle}} 
with $\mu$ defined in \q{definemuintra}. From section 12.2.2 in \ocite{NIST}, the solutions are linear combinations of two independent parabolic cylinder functions (PCF's):
\e{ \bar{f}(z) = \bar{c}_{\sma{\nearrow}E} U(-i\mu,z) + \bar{c}_{\sma{\swarrow}E} U(-i\mu,-z),}
with newly-introduced coefficients $\bar{c}_{\nu E}$ that are to be determined. In the limit $z \gg 1$, the PCF's may be matched with the Zilberman functions [\q{simplerwkb} without the $H_1$ correction]. We will assume some conditions on the zero-field band parameters, such that beyond-quadratic terms in $H_0(\bK)$ can still be neglected in this matching region.
 Employing the asymptotic expansion in 12.9.1 and 12.9.3 of \ocite{NIST},
\e{ (i) \;k_y \rightarrow +\infty, \as &\bar{f} \rightarrow   \left[c^{\sma{(0)}}_{\sma{\nearrow}E}  e^{ \pi \mu/2}  -i\,c^{\sma{(0)}}_{\sma{\swarrow}E} \, e^{- \pi \mu/2}\right]\lin
&\times e^{i\mu \log|\mu|-i\mu}\,\f{\Gamma(1/2-i\mu)}{\sqrt{2\pi}}\,\varpi   +c^{\sma{(0)}}_{\sma{\swarrow}E}\,\varpi^* \lin
		(ii) \;k_y \rightarrow -\infty, \as &\bar{f} \rightarrow  \left[-i\,c^{\sma{(0)}}_{\sma{\nearrow}E} e^{-\pi \mu/2} +c^{\sma{(0)}}_{\sma{\swarrow}E} \,e^{ \pi \mu/2} \right]\lin
		&\times e^{i\mu \log|\mu|-i\mu}\,\f{\Gamma(1/2-i\mu)}{\sqrt{2\pi}}\,\varpi      +c^{\sma{(0)}}_{\sma{\nearrow}E}\, \varpi^* \la{finalpcfasymptoticssaddle}
		}
where we have introduced the coefficients $\{c^{\sma{(0)}}_{\nu E}\}$ which differ from $\{\bar{c}_{\nu E}\}$ only by a $\nu$-independent proportionality constant; we have additionally defined
\e{ \varpi(k_y,E) \eq e^{-z^2/4+i\mu/2} \left[  \f{2|k_y|}{b}\right]^{i\mu-1/2}, \lin
 \varpi^*(k_y,E) \eq  e^{z^2/4-i\mu/2}\left[  \f{2|k_y|}{b}\right]^{-i\mu-1/2}, \la{defvarpi}
		}
which may be identified with the Zilberman functions in the limit $|k_y| \gg |b|$ [recall the definition of $b$ in \q{defineab}]:
\e{ &g^{\sma{\nwarrow} +}_{\bk E} \condprop{k_y \gg |b|} e^{ik_xk_yl^2}\varpi(k_y,E), \lin
& g^{\sma{\searrow} -}_{\bk E} \condprop{k_y \ll -|b|} e^{ik_xk_yl^2}\varpi(k_y,E),\lin
&g^{\sma{\swarrow} +}_{\bk E} \condprop{k_y \gg |b|} e^{ik_xk_yl^2}\varpi^*(k_y,E), \lin
& g^{\sma{\nearrow} -}_{\bk E} \condprop{k_y \ll -|b|} e^{ik_xk_yl^2}\varpi^*(k_y,E).}
We emphasize that these identifications are made for uniquely-defined Zilberman functions, for which the lower limits of the classical action integrals are specified as in \q{simplerwkb} (with $H_1=0$). Following the discussion surrounding \q{abovequantum}-(\ref{belowquantum}), we may then identify
\e{ c^{\sma{(0)}}_{\sma{\nwarrow} E} \eq \left[c^{\sma{(0)}}_{\sma{\nearrow}E} \,e^{ \pi \mu/2}  -i\,c^{\sma{(0)}}_{\sma{\swarrow}E} \, e^{- \pi \mu/2}\right]\lin
&\times e^{i\mu \log|\mu|-i\mu}\,\f{\Gamma(1/2-i\mu)}{\sqrt{2\pi}} \lin
c^{\sma{(0)}}_{\sma{\searrow}E} \eq \left[-i\,c^{\sma{(0)}}_{\sma{\nearrow}E}\,e^{-\pi \mu/2} +c^{\sma{(0)}}_{\sma{\swarrow}E} \,e^{ \pi \mu/2} \right] \lin
&\times e^{i\mu \log|\mu|-i\mu}\,\f{\Gamma(1/2-i\mu)}{\sqrt{2\pi}}, \la{relatecnu}}
which can be expressed as a matrix equation relating  incoming to outgoing states 
\e{ \vectwo{c^{\sma{(0)}}_{\sma{\nwarrow}E}}{c^{\sma{(0)}}_{\sma{\searrow}E}}= \bbs^{\sma{(0)}}(E,k_z)\vectwo{c^{\sma{(0)}}_{\sma{\nearrow}E}}{c^{\sma{(0)}}_{\sma{\swarrow}E}}, \la{definezerothscatteringmatrixintra}}
with the lowest-order scattering matrix defined in \q{zerothscatteringmatrixsaddle}. To summarize the results of this review, the eigenfunctions of \q{Peierlshamsaddle} in the limit $k_y \rightarrow \pm \infty$ are
\bal
f^{\pm}_{\bk E} =e^{ik_xk_yl^2}{\sum_{\nu}}^{\pm}c^{\sma{(0)}}_{\nu E} \f1{\sqrt{|v^x_{\nu}|}}e^{-il^2\int^{k_y}_{k^{\nu}_{y0}(E)} k_x^{\nu}(z,E)dz},\la{summarizereview}
\end{align}
where the superscript on $f^{\pm}$ corresponds to the sign in $k_y \rightarrow \pm \infty$; $k_{y0}^{\nu}$ is the coordinate of closest approach to the saddlepoint for the edge $\nu$;  $\sum_{\nu}^{\pm}$ runs over $\nwarrow$ and $\swarrow$ for $f^+$ (the two edges above the breakdown interval), and over $\nearrow$ and $\searrow$ for $f^-$; the various $c^{\sma{(0)}}_{\nu E}$ are related as in \q{relatecnu}-(\ref{definezerothscatteringmatrixintra}) and (\ref{zerothscatteringmatrixsaddle}).

\subsubsection{Derivation of first-order-corrected connection formula}

It is useful to estimate the size of the region in $\bk$-space ($\Delta k_x \Delta k_y$), in the vicinity of the saddlepoint, where the Zilberman-Fischbeck wavefunctions are invalid; equivalently, this is where the asympotic limits of the PCF's would not apply -- we have called this the breakdown region. This is the region where $z$, the dimensionless variable entering the Weber differential equation [cf.\ \q{zdimensionless}], is of order one. Further assuming $m_1/m_2=O(1)$, we obtain $\Delta k_y =O(\lmo)$. Utilizing the hyperbolic asymptotes $k_y =\pm (b/a)k_x$ and assuming $(b/a)=O(1)$, we estimate $\Delta k_x =O(\lmo)$; note that $\Delta k_x \Delta k_y =O(\lmt)$. \\

Let us derive a first-order-corrected effective Hamiltonian ($\calh=H_0+H_1$) in the breakdown region. We first expand the symbol $H_1$ around the saddlepoint as in \q{approximateH1}. The terms ($\delta H_1$) which we neglect to write explicitly are bounded by their values at the boundary of the breakdown region as  $\delta H_1(\Delta \bk) =O(\lmf)$, with our estimates of  $\Delta \bk$ in the above paragraph. In other words, the explicit terms in \q{approximateH1}, when evaluated on the boundary, are larger in magnitude than $O(\lmf)$ and therefore expected to be relevant in the limit of small field. When these explicit terms are added to $H_0$, the result is an effective Hamiltonian that is identical in form to \q{Peierlshamsaddle}:  
\e{\calh= H_0(\bK)+H_1(\bK)  = H_0(\bQ)+H_1(\bze)+O(l^{-4}),}
but shifted by an energy constant $H_1(\bze)$, and with shifted momentum variables
\e{ &q_x=k_x+m_1H_{1x}, \as q_y =k_y-m_2H_{1y}  \as\longleftrightarrow  \lin
 &Q_x=K_x+mH_{1x}, \as Q_y=K_y-m_2H_{1y} \la{qvsk}.}
It is useful to know which of the Roth, Berry or Zeeman terms contribute to the effective Hamiltonian; let us individually expand $H_1^R$, $H_1^B$ and $H_1^Z$ as in \q{approximateH1}, keeping only the linear terms, which we define by $H_{1j}^R k_j,$ etc. For example, the Berry term is expanded as
\e{l^2H^B_1(\bk)  \eq  \mxy(\bk) v^x(\bk)-\mxx(\bk) v^y(\bk) \lin
\eq \mxy(\bze) \f{k_x}{m_1}+\mxx(\bze) \f{k_y}{m_2}+\ldots \lin
:\eq l^2(H_{1x}^Bk_x+H_{1y}^Bk_y)+\ldots,\la{expandberry}}
and vanishes when evaluated at the saddlepoint, where the band velocity $\bv^{\sma{\perp}}$ vanishes. Therefore the shift in the energy constant is only contributed by the gauge-invariant Roth and Zeeman terms: 
\e{H_1(\bze) \eq H_1^R(\bze)+H_1^Z(\bze).}
We further deduce from \q{expandberry} that the shifts in the momentum variables $k_x$ and $k_y$ are, respectively,
\e{ m_1H_{1x} \eq \lmt\mxy(\bze)+m_1(H_{1x}^R+H_{1x}^Z), \lin
 m_2H_{1y} \eq \lmt\mxx(\bze)+m_2(H_{1y}^R+H_{1y}^Z). \la{defineH1jsaddle}}
The similarity of $\calh$ with the inverted-harmonic-oscillator Hamiltonian implies that it may be solved with the same techniques, with some small modifications. We assume here the reader has some familiarity with the `same techniques', which we have reviewed in the previous subsection [\app{app:zerothorderconnectionintraband}
] and will presently extend.\\

Let us then define the eigenfunction of $\calh$, with the $O(\lmf)$ correction henceforth truncated, as
\e{ 0\eq (\calh(\bK)-E)f_{\bk E}=(H_0(\bQ)+H_1(\bze)-E)f_{\bk E}\lin
\eq (H_0(\bQ)-\tilde{E})f_{\bk E}; \as \tilde{E}:=E-H_1(\bze).}
Performing a gauge transformation:
\e{ f=e^{iq_xk_yl^2}\bar{f},}
we see that $\bar{f}$ satisfies the Weber differential equation in the modified variable $q_y$ and with modified eigenvalue $\tilde{E}$.  Let us define $f^{\pm}_{\bk E}$ to be the asymptotic limits of $f$ in the limit $k_y \rightarrow \pm \infty$. Utilizing results from our review in \app{app:zerothorderconnectionintraband}, especially \q{summarizereview}, we obtain
\bal
&e^{-iq_xk_yl^2}f^{\pm}_{\bk E} \lin
\eq {\sum_{\nu}}^{\pm}\left\{c^{\sma{(0)}}_{\nu E} \f1{\sqrt{|v^x_{\nu}|}}e^{-il^2\int^{k_y}_{k^{\nu}_{y0}(E)} k_x^{\nu}(z,E)dz}\right\}_{(k_y,E)\rightarrow (q_y,\tilde{E})}, \la{summarizereview2}
\end{align}
with $\{c^{\sma{(0)}}_{\nu E}\}$ related as in \q{relatecnu}-(\ref{definezerothscatteringmatrixintra}) and (\ref{zerothscatteringmatrixsaddle}); $\bk_0^{\nu}$ is the wavevector of closest approach to the saddlepoint for the edge $\nu$; ${\sum}^{\pm}_{\nu}$ runs over $\nwarrow$ and $\swarrow$ for $f^+$ (the two edges above the breakdown interval), and over $\nearrow$ and $\searrow$ for $f^-$. Upon substituting $k_y \rightarrow q_y=k_y-m_2H_{1y}$ in the curly brackets of \q{summarizereview2}, \q{summarizereview2} is expressible as
\e{&f^{\pm}_{\bk,E}= {\sum_{\nu}}^{\pm}c^{\sma{(0)}}_{\nu \tilde{E}}\, e^{im_1l^2H_{1x}k_{y0}^{\nu}(\tilde{E})+im_2l^2H_{1y}{k_{x0}^{\nu}(\tilde{E})}}\f{e^{ik_xk_yl^2}}{\sqrt{|{v}^{\nu}_x|}} \lin
& \times e^{{-}il^2\int^{k_y}_{k_{y0}^{\nu}(E)}\left({k}_x^{\nu}-\f{H^{\nu}_1-H_1(\bze)}{v^x_{\nu}}  \right)dz }\bigg|_{E \rightarrow \tilde{E}}+O(l^{-2}),\la{elegant55}}
as we prove at the end of this subsection. We may identify the last line of \q{elegant55} as the \zf function defined in \q{lastlineiszf}. Further defining
\e{ c_{\nu E}:\eq c^{\sma{(0)}}_{\nu \tilde{E}}\;\upsilon_{\nu E} \lin 
    \upsilon_{\nu E}:\eq e^{im_1l^2H_{1x}k_{y0}^{\nu}(\tilde{E})+im_2l^2H_{1y}{k_{x0}^{\nu}(\tilde{E})}}, \la{relatecnus}}
we cast \q{elegant55} in the simple form
\e{f^{\pm}_{\bk,E}\eq {\sum_{\nu}}^{\pm}c_{\nu {E}}\tilde{g}^{\nu}_{\bk E},\la{simpleform55}}
which may be identified with \q{abovequantum}-(\ref{belowquantum}). 
We are finally ready to derive the scattering matrix defined in \q{definescatteringmatrixintra}  and expressed in \q{scatteringintraberry} and (\ref{fullscattering}). Combining \q{relatecnus} with (\ref{definezerothscatteringmatrixintra}) and (\ref{zerothscatteringmatrixsaddle}),
\e{\bbs(E,l^2) \eq  \diagmatrix{\upsilon_{\sma{\nwarrow  E}}}{\upsilon_{\sma{\searrow E}}}\bbs^{\sma{(0)}}(\tilde{E},l^2)\diagmatrix{\upsilon^*_{\sma{\nearrow} E}}{\upsilon^*_{\sma{\swarrow} E}} \lin
\eq \matrixtwo{\calt(\tilde{\mu})\upsilon_{\sma{\nwarrow  E}}\upsilon^*_{\sma{\nearrow} E}\as}{\calr(\tilde{\mu})\upsilon_{\sma{\nwarrow  E}}\upsilon^*_{\sma{\swarrow} E}}{\calr(\tilde{\mu})\upsilon_{\sma{\searrow E}}\upsilon^*_{\sma{\nearrow} E}\as}{\calt(\tilde{\mu})\upsilon_{\sma{\searrow E}}\upsilon^*_{\sma{\swarrow} E}}. }
Inserting the integral expression for $\upsilon$ [from \q{relatecnus}], and further applying the definition of $\bk_0^{\nu}$, we obtain \q{fullscattering}.
If we neglect the Roth and Zeeman corrections, we find that the Berry term is sufficient to restore gauge covariance: 
\e{ &\bbs(E,l^2) \condeq{H_1=H_1^B} \lin
& \matrixtwo{\calt(\mu)e^{i \mxy(\bze) 2 b(E)}\as}{\calr(\mu)e^{-i \mxx(\bze) 2a(E)}}{\calr(\mu)e^{i \mxx(\bze) 2a(E)}\as}{\calt(\mu)e^{-i \mxy(\bze)2b(E)}} +O(\lmt) \lin
 \eq  \matrixtwo{\calt(\mu)e^{i \int^b_{-b}\mxy(0,k_y) dk_y}\as}{\calr(\mu)e^{-i \int^a_{-a}\mxx(k_x,0)dk_x}}{\calr(\mu)e^{i \int^a_{-a}\mxx(k_x,0)dk_x}\as}{\calt(\mu)e^{-i \int^b_{-b}\mxy(0,k_y) dk_y}}\lin
&+O(\lmt,(\tfrac{b}{G})^2,(\tfrac{a}{G})^2)}
In the first equality, we have made use of the expansion of $H_1^B$ in \q{expandberry}; the second equality follows from 
\e{  \int^b_{-b}\mxy(0,k_y) dk_y=2\mxy(\bze)b + O((b/G)^2),}
where the correction is of order $(b/G)^2$, with $G$ a typical reciprocal period. \\

Up to $O(\lmt,(b/G)^2,(a/G)^2)$, the $O(1)$ phases in \q{fullscattering} may be creatively interpreted as the Roth-Berry-Zeeman phase averaged over all possible tunneling trajectories in the classically-forbidden region. For example, the phase acquired for the tunneling trajectory in the $\vec{y}$ direction may be expressed as
\e{ e^{i 2 m_1H_{1x}b(\tilde{E})l^2} \appr \exp\left[{i\int^{b}_{-b}\bigg\{ \overline{\f{\tilde{H}_1}{v^x}} \bigg\}_{k_y} dk_y}\right],} 
where $\{ \bar{\cdot} \}_{k_y}$ denotes the $k_x$-average of the quantity $\cdot$ over a fixed-$k_y$ cross-section of the forbidden region:   
\e{&\bigg\{ \f{\tilde{H}_1}{v^x} \bigg\}_{k_y} = \f{l}{2}\int_{-1/l}^{1/l} \f{{H}_1(\bk)-H_1(\bze)}{v^x(\bk)} dk_x \lin
&\approx \f{m_1l}{2}\int_{-1/l}^{1/l} \f{{H}_{1x}k_x+H_{1y}k_y}{k_x} dk_x = m_1H_{1x}.}
In the last equality, we have used the Cauchy principal value for the integral $\int dk_x/k_x$.\\

\noindent \emph{Proof of identification of \q{summarizereview2} with \q{elegant55}}\\
 
\noindent From the exponent in the second line of \q{elegant55},
\e{ &\int^{k_y}_{k_{y0}^{\nu}(\tilde{E})} \f{H^{\nu}_1-H_1(\bze)}{v^x_{\nu}}dt =  m_1H_{1x}\big( k_y-k_{y0}^{\nu}(\tilde{E})\big)\lin
& + m_2H_{1y}\big({k}_x^{\nu}(k_y,\tilde{E})-k_{x0}^{\nu}(\tilde{E})\big)+O(\lmf), \la{tobeelegant1}
}
where, as a reminder, $k_x^{\nu}(k_y,E)$ as the $k_x$-coordinate of the section $s_{\nu}$ at wavevector $k_y$ and energy $E$. In deriving \q{tobeelegant1}, we employed $H^{\nu}_1(\bk)-H_1(\bze)=k_xH_{1x}+k_yH_{1y}+\ldots$ from \q{H1Bexplicit}, and the identity $dk_y/v^x=-dk_x/v^y$ along a constant-energy contour. The uncertainty $O(\lmf)$ in \q{tobeelegant1} is estimated by evaluating the neglected terms at the boundary of the breakdown region, where $\bk=O(\lmo)$.  Substituting \q{tobeelegant1} into \q{elegant55}, we obtain
\e{&f^{\pm}_{\bk,E}= e^{ik_xk_yl^2}{\sum_{\nu}}^{\pm}c^{\sma{(0)}}_{\nu\tilde{E}} \,e^{il^2m_1H_{1x}k_y + il^2m_2H_{1y}{k}_x^{\nu}(k_y,\tilde{E})} \lin
&\times \; \f1{\sqrt{|{v}^{\nu}_x|}}\exp\left\{-il^2\int^{k_y}_{k_{y0}^{\nu}(E)}{k}_x^{\nu}dz \right\}\bigg|_{E \rightarrow \tilde{E}}+O(l^{-2}).\la{elegant556}}
To complete the identification of this expression with \q{summarizereview2}, we apply the following three observations: 
\e{ (i)\; e^{ik_xk_yl^2}e^{il^2m_1H_{1x}k_y}= e^{iq_xk_yl^2},}
from the fundamental theorem of calculus,
\e{ (ii)\;&  m_2H_{1y}{k}_x^{\nu}(k_y,\tilde{E})+\int^{k_y}_{k^{\nu}_{y0}(\tilde{E})} k_x^{\nu}(z,\tilde{E})dz \lin
\eq \int^{q_y}_{k^{\nu}_{y0}(\tilde{E})} k_x^{\nu}(z,E)dz+O(\lmf),}
and finally, (iii) bearing in mind that the expressions are to be identified with an uncertainty of $O(\lmt)$, we might directly replace $|v^x_{\nu}(q_y,\tilde{E})| \approx |v^x_{\nu}(k_y,\tilde{E})|$ in the square-root prefactor.

\subsubsection{Equivalence of two Zilberman-Fischbeck functions}\la{app:gtildeeqg}

We would like to prove in the semiclassical region ($sm$) that
\e{\ifor \bk \in sm, \as g^{\nu}_{\bk E} = \tilde{g}^{\nu}_{\bk \tilde{E}} +O(\lmt).\la{gequalgtilde}}
We need the following three identities: (i) in the semiclassical region  where the \zf functions are valid, we may assume $k_x^{\nu}=O(1)$ and therefore
\e{2m_1H_1(\bze)/k_x^{\nu}(k_y,E)^2=O(\lmt); \la{theassumption}} 
combining this assumption with \q{zerothorderkxnu}, we derive
\e{ k_x^{\nu}(k_y,\tilde{E})=k_x^{\nu}(k_y,E)- H_1(\bze)/v^x_{\nu}+O(\lmf).\la{definedeltakx}}
(ii) The same assumption in \q{theassumption} implies, with $v^x_{\nu}=k_x^{\nu}/m_1$, that
\e{ v^x_{\nu}(k_y,\tilde{E})= v^x_{\nu}(k_y,{E})- H_1(\bze)/m_1v^x_{\nu} +O(\lmf).}
(iii) Lastly, applying the fundamental theorem of calculus,
\e{&\int^{k_y}_{k_{y0}^{\nu}(\tilde{E})}\left({k}_x^{\nu}(z,E)-\f{H^{\nu}_1}{v^x_{\nu}}  \right) dz \lin
\eq \int^{k_y}_{k_{y0}^{\nu}(E)}\left({k}_x^{\nu}(z,E)-\f{H^{\nu}_1}{v^x_{\nu}}  \right) dz\lin
& + \big(\,k_{y0}^{\nu}(\tilde{E})-k_{y0}^{\nu}(E)\,\big)k_{x0}^{\nu}(E)+O(\lmf)\lin
\eq \int^{k_y}_{k_{y0}^{\nu}(E)}\left({k}_x^{\nu}(z,E)-\f{H^{\nu}_1}{v^x_{\nu}}  \right) dz +O((a/G)\lmt,\lmf). \notag}
In the last equality, we applied that the coordinate of closest approach $k_{x0}=\pm a(E)$ (the hyperbolic parameter) for $E>0$ and is otherwise zero. Substituting (i-iii) into \q{lastlineiszf}, we derive \q{gequalgtilde} as desired.

\section{Appendix to `Effective Hamiltonian for general band touchings'}\la{app:effhamgen}

\subsection{Calculus with Weyl-symmetrized operators}\la{app:weylcalculus}

Here we collect several identities which are useful in the calculus of Weyl-symmetrized operators.\\

We are interested in kinetic quasimomentum operators with the noncommutative relation:
\e{[K_x,K_y]=i\lmt.\la{noncommutivity}}
It immediately follows that
\e{ [f(K_x),k_y]=i\lmt f'(K_x),\la{rreallybasi}}
with $f'$ denoting a derivative with respect to $K_x$. \\

We are very often interested in symmetrized functions of $\bK$. 
Beside our definition of symmetrization with the Fourier formula in \q{replace}, a more elementary definition exists for polynomials:\cite{Tao} given a monomial $k_x^mk_y^n$, its symmetrized form is obtained from extracting all terms with $m$ powers of $K_x$ and $n$ powers of $K_y$ in the noncommutative binomial expansion
\e{ \f{m!n!}{(m+n)!}(K_x+K_y)^{m+n}. \la{binomial}}
One may verify that this symmetrization preserves the structure of products:
\e{ &\big[(sk_x+tk_y+u)^v\big]=(sK_x+tK_y+u)^v, \lin
& s,t,u \in \C,\; v\in \Z, \la{preservestructureofproducts}}
which implies that the exponential structure is also preserved:
\e{  e^{i\bK \cdot \bR}= \sum_{n=0}^{\infty} \f{(i\bK \cdot \bR)^n}{n!}= \sum_{n=0}^{\infty} \bigg[\f{(i\bk \cdot \bR)^n}{n!}\bigg] = [\eikr].}
This identify underlies the Fourier definition of symmetrization in \q{replace}.\\

For any function of $K_x$: 
\e{[k_y,f(K_x)]\eq (1/2)(k_yf(K_x)+f(K_x)k_y)\lin
:\eq (1/2)\{k_y,f(K_x)\}, \la{symmetrizeky}}
as may be proven by Taylor-expanding $f$ and symmetrizing individual terms (e.g., $[k_yK_x^n]$) with the rule in \q{binomial}.\\

Symmetrization of a symbol commutes with addition:
\e{ [f(\bk)]+[g(\bk)]=[f(\bk)+g(\bk)].\la{symmcommuteswadd}}
Like many basic identities, it may be proven by Fourier analysis:
\e{  &\int d\bR\check{f}(\bR)e^{i\bK\cdot \bR} +\int d\bR' \check{g}(\bR')e^{i\bK\cdot \bR'} \lin 
\eq \int d\bR\bigg(\check{f}(\bR)e^{i\bK\cdot \bR} + \check{g}(\bR)e^{i\bK\cdot \bR}\bigg) \lin
\eq \int d\bR\bigg(\check{f}(\bR) +\check{g}(\bR)\bigg)e^{i\bK\cdot \bR}.\notag
}
The product rule for two symmetrized operators is described in \q{multiplicationrule}; its nontriviality  originates from the noncommutivity of \q{noncommutivity}.  We review the proof of the multiplication rule by Roth,\cite{rothI} which combines Fourier analysis, and the Baker-Campbell-Hausdorff identity $e^Ae^B=e^{A+B}e^{[A,B]/2}$:
\e{ &\int d\br \int d\br' \check{A}(\br)\check{B}(\br') e^{-i\bK\cdot \br}e^{-i\bK\cdot \br'}\lin \eq \int d\br \int d\br' \check{A}(\br)\check{B}(\br') e^{-i\bK\cdot (\br+\br')}e^{-i\lmt \epsilon_{\ab}r_{\alpha}r'_{\beta}/2} \lin
\eq  \left[\int d\br \int d\br' \check{A}(\br)\check{B}(\br') e^{-i\bk\cdot (\br+\br')}e^{-i\lmt \epsilon_{\ab}r_{\alpha}r'_{\beta}/2} \right] \lin
\eq \bigg[e^{(i/2)\lmt \epsilon_{\ab}\nabk^{\alpha}\nabla_{\bk'}^{\beta}} \int d\br \int d\br' \check{A}(\br)\check{B}(\br')\lin
&\times e^{-i\bk\cdot\br}e^{-i\bk'\cdot \br'}\bigg|_{\bk=\bk'} \bigg]. \la{proofmultiplicationrule}}
An application of this product rule to a commutator of two symmetrized operators leads to
\e{ [A(\bK),&B(\bK)]= \big[[A(\bk),B(\bk)]\big]\lin
&+ \f{i}{2l^2}\epsilon_{\ab}\big[\{ \nabk^{\alpha}A,\,\nabla_{\bk}^{\beta}B\}\big]+O(\lmf),\la{moyalcommutator}}
where $[[a,b]]= [ab-ba]$ and $[\{a,b\}]= [ab+ba]$.

\subsubsection{Symmetrized operators which are independent of $k_y$}

A particularization of the Roth product rule [cf.\ \q{multiplicationrule}] for functions independent of $k_y$ is
\e{ A(K_x)B(K_x) = [A(k_x)B(k_x)]_{k_x\rightarrow K_x} \la{reallybasic}}
An operator acting in $\br$-space (or more generally, an operator acting in both $\br$- and $k_x$-space) commutes with the operation $[\cdot]_{k_x\rightarrow K_x}$, i.e.,
\e{ \hat{F}(\hbr,\nabr)A(K_x,\br)\eq \left[ \hat{F}(\hbr,\nabr)A(k_x,\br)\right]_{k_x\rightarrow K_x}, \la{commuterbr} \\
\hat{G}(K_x,\hbr,\nabr)A(K_x,\br)\eq \left[ \hat{G}(k_x,\hbr,\nabr)A(k_x,\br)\right]_{k_x\rightarrow K_x},\la{commuterbrKx} }
which may also be proven from Fourier analysis.

\subsection{Relating our ansatz to Slutskin's function} \la{app:slutskinbasis}

To lowest order in $\lmt$, our ansatz for the wavefunction takes the form 
\e{\Psi(\br) \eq \f1{\sqrt{N}}\sum_{\bk}\alpha(\bk,\br), \lin
  \alpha(\bk,\br):\eq \sum_{n}\eikr u_{nK_x0}(\br)f_{n\bk},\la{ansatz2}}
with $\sum_{\bk}$ shorthand for a continuous integral over the Brillouin torus. We would like our ansatz to be 
independent of the choice of unit cell in $\bk$-space, i.e.,
\e{ \alpha(\bk,\br)=\alpha(\bk+\bG,\br), \la{periodicityalpha}}
for any reciprocal vector $\bG$. This is ensured if we impose the following boundary conditions on the wavefunction in the $(K_x,0)$-representation:
\e{f_{n\bk}\eq f_{n\bk+\bG_x}, \la{firstbc} \\
f_{m\bk} \eq \sum_n \tilde{S}_{mn}(K_x,0;\bG_y)f_{n\bk+\bG_y},  \la{secondbc}\\
\tilde{S}_{mn}(K_x,0;\bG):\eq \int d\btau [u_{mk_x0}^*(\btau)]e^{i\bG\cdot \btau}u_{nK_x0}(\btau). \la{defineSmatrixslut}
}
Here, $\tilde{S}$ is formally an infinite-dimensional matrix, $\int d\btau$ denotes an integration over the real-space unit cell, and $\bG_x$ and $\bG_y$ are the primitive reciprocal vectors of a rectangular lattice:
\e{ &\bG_x:=2\pi \vec{x}/a_x, \as G_x:=2\pi/a_x, \lin
    &\bG_y:=2\pi\vec{y}/a_y, \as G_y:=2\pi/a_y;}
the choice of a rectangular lattice is merely for notational simplicity.  $\alpha(\bk,\br)=\alpha(\bk+\bG_x,\br)$ follows from the periodicity in $k_x$ of both (i) the wavefunction [\q{firstbc}], and (ii) the operator  that acts on the wavefunction: 
\e{ \eikr u_{nK_x0} \eq e^{i(\bk+\bG_x)\cdot \br} u_{n,K_x+G_x,0};}
[cf.\ \q{expandcellperiodicfunc}]. The same operator is, however, not periodic in $k_y$, and therefore the corresponding boundary condition on the wavefunction is more complicated. To verify that this boundary condition produces the desired periodicity: $\alpha(\bk,\br)=\alpha(\bk+\bG_y,\br)$, apply the operation $\sum_m \eikr u_{mK_x0}(\br)$ on both sides of \q{secondbc} and apply the completeness relation in \q{slutcomplete}.\\ 

Our discussion about boundary conditions may seem more formal than practical, since in many applications we would only be interested in $f_{n\bk}$ for $\bk$ in the vicinity of a point -- the area of interest is typically much smaller than the Brillouin torus. On the other hand, assuming such formalities, we would show that our ansatz is equivalent to an expansion in Slutskin's basis functions\cite{slutskin} [denoted $\chi_{n\bk}$]:
\e{ \Psi(\br) \eq \f1{\sqrt{N}}\sum_{n\bk} e^{i\bk \cdot \br} {u}_{nK_x0}f_{n\bk}= \f1{\sqrt{N}}\sum_{n\bk} f_{n\bk} \chi_{n\bk},\la{integrationbyparts}\\
\chi_{n\bk}(\br) :\eq  u_{n, k_x+y/l^2,0}(\br)\eikr.\la{defineslutskinfunc}
}
While $\eikr{u}_{nK_x0}$ is a differential operator acting on $f_{n\bk}$, $\chi_{n\bk}$ acts on $f_{n\bk}$ by multiplication, and therefore has a more intuitive interpretation as a wavefunction over real space. \\

The first step to proving \q{integrationbyparts} is to equivalently express Slutskin's function as
\e{&u_{n, k_x+y/l^2,0}(\br)\eikr \lin
\eq  \f1{\sqrt{N}}\sum_{\bR}e^{-i[k_x-(i/l^2)(\partial/\partial k_y)](r_x-R_x)}W_n(\br-\bR)\eikr \lin
\eq {u}_{n,K^*_x,0}e^{i\bk \cdot \br}; \as K_x^* := k_x-\f{i}{l^2}\p{}{k_y},}
with help from the identity  \q{expandcellperiodicfunc}. What remains is to prove
\e{ \Psi(\br)=\sum_{n\bk}f_{n\bk}u_{nK_x^*0}\eikr = \sum_{n\bk}\eikr u_{nK_x0}f_{n\bk},  \la{idenslut}}
which is analogous to an integration by parts; for notational simplicity, we shall no longer write out normalization factors. As an intermediate step, we would further identify the above quantity as equal to an expansion 
\e{ \Psi(\br)\eq\sum_{R_x,k_y} h(x-R_x,y,k_y);\lin
 h(x-R_x,y,k_y):\eq \sum_n h_n(x-R_x,y,k_y) \check{f}_{n R_xk_y},\la{RHSfinal}}
in the basis functions
\e{ h_{n}(x-R_x,y,k_y):=e^{i\{k_y-(x-R_x)/l^2\}y}\sum_{R_y}W_n(\br-\bR),\la{hybridlkw}}
with expansion coefficients
\e{ \check{f}_{n R_xk_y} :=  \sum_{k_x}e^{ik_xR_x}f_{n,\bk}. }
We may identify these expansion coefficients as the Fourier coefficients of the periodic function $f_{n\bk}$ [cf.\ \q{firstbc}]. $h_n$ may be viewed as the magnetic analog of a `hybrid' function, which is spatially extended in $\vec{y}$ (as a Luttinger-Kohn function) but exponentially localized in $\vec{x}$ (as a Wannier function). Indeed, setting $\lmt=0$ in \q{hybridlkw}, 
\e{ h_{n}(x-R_x,y,k_y) \;\condeq{\lmt=0}&\;  e^{ik_yy}\sum_{R_y}W_n(\br-\bR)\lin
\eq\sum_{k_x}e^{-ik_xR_x}\chi^{\sma{(0)}}_{n\bk},\la{hybridlkw0}}
with the Luttinger-Kohn function defined as
\e{ \chi^{\sma{(0)}}_{n\bk}(\br) = \eikr u_{nk_x0}(\br).}
It is known that the Luttinger-Kohn functions form a complete orthonormal basis in which any function can be expanded\cite{Luttinger_Kohn_function} --  we thus expect for small fields that $\{h_n\}$ forms a linearly-independent basis, though we avoid assuming orthogonality. Furthermore, we insist that the expansion \q{RHSfinal} is independent of the choice of unit cell in the Brillouin circle parametrized by $k_y$, i.e., for each $R_x$,
\e{h(x-R_x,y,k_y)=h(x-R_x,y,k_y+G_y);} 
this imposes a boundary condition on the wavefunction $\check{f}_{nR_xk_y}$, in close analogy with \q{periodicityalpha}-(\ref{secondbc}). We may exploit this periodicity to express \q{RHSfinal}
as
\e{& \Psi(\br) = \sum_{R_x}\sum_{k_y} h(x-R_x,y,k_y) \lin
\eq \sum_{R_x}\sum_{k_y} h(x-R_x,y,k_y+(x-R_x)/l^2) \lin 
\eq \sum_{R_xk_y}e^{ik_yy}\sum_{nR_y}W_n(\br-\bR) \sum_{k_x}e^{ik_xR_x}f_{nk_xk_y+(x-R_x)/l^2}.\notag}
This quantity is equal to the RHS of \q{idenslut}, as we now demonstrate: 
\e{&\sum_{n\bk}\eikr u_{nK_x0}f_{n\bk} \lin
\eq \sum_{n\bk}\eikr \sum_{\bR}W_n(\br-\bR)e^{-iK_x(x-R_x)} f_{n\bk} \lin
\eq \sum_{n\bk}\eikr \sum_{\bR}W_n(\br-\bR)e^{-i(k_x+il^{-2}\partial_y)(x-R_x)} f_{n\bk} \lin
\eq \sum_{nk_y}e^{ik_yy} \sum_{\bR}W_n(\br-\bR)\sum_{k_x} e^{ik_xR_x} f_{nk_xk_y+(x-R_x)/l^2}. \la{sumbeta}
}
The LHS of \q{idenslut} may be expressed as
\e{& \sum_{n\bk}f_{n\bk}u_{nK_x^*}\eikr \lin
\eq \sum_{n\bk}f_{n\bk}\sum_{\bR}W_n(\br-\bR)e^{-i(k_x-il^{-2}\partial_y)(x-R_x)} \eikr \lin
\eq \sum_{n\bk}f_{n\bk}\sum_{\bR}W_n(\br-\bR)e^{-ik_x(x-R_x)} e^{-iy(x-R_x)/l^2}\eikr \lin
\eq \sum_{k_yR_x}e^{i\{k_y-(x-R_x)/l^2\}y}\sum_{nR_y} W_n(\br-\bR)\sum_{k_x}e^{ik_xR_x}f_{n\bk},\notag
}
which may be identified with \q{RHSfinal}.

\subsection{Alternative derivation of the infinite-band effective Hamiltonian}\la{app:unsymmetrized}

We offer a derivation of  \q{effhamslutfirst}  and its equivalent, symmetrized form in \q{reorgsum}; these are effective-Hamiltonian equations which formally act on the wavefunctions over all bands. 
\q{reorgsum} was previously derived in \q{sec:symmetrized} utilizing the Roth product rule of two symmetrized operators [cf.\ \q{multiplicationrule}]; the following, alternative derivation does not rely on this rule.\\

 From \q{fromtranssymm},
\e{ &\sum_n\tcalh_{mn}(\bK)f_{n\bk}\lin
\eq  \int d\btau \sum_{n} \dg{u}_{mK_x0}(\btau) \hH_0(\bK) u_{nK_x0}(\btau)f_{n\bk} \lin
\eq  \int d\btau \sum_{n} \dg{u}_{mK_x0}(\btau)  \left\{ [\hH_0(k_x,0)]+ k_y\hPi_y + \f{k_y^2}{2m}\right\}\lin
&\times u_{nK_x0}(\btau)f_{n\bk}.\la{threeterms}
}
To derive the last equality in \q{threeterms}, we need the following idenity:
\e{ \hat{H}_0(\bK) \eq [\hat{H}_0(\bk)]=\left[\hat{H}_0(k_x,0) + \hPi^yk_y+ \f{k_y^2}{2m}\right]\lin
\eq \left[\hat{H}_0(k_x,0)\right] + \hPi^yk_y+ \f{k_y^2}{2m}.}
The second equality follows from \q{expandBlochham}, and the last equality assumed the Landau gauge for the kinetic quasimomentum operators: $K_x=k_x+i\lmt \partial_y, K_y=k_y$.\\

We separately consider each of the three terms in the last line of \q{threeterms}. The first term is simply evaluated as
\e{ 
&\int d\btau  \dg{u}_{mK_x0}(\btau)  \hH_0(K_x,0) u_{nK_x0}(\btau)\lin
\eq\left[ \int d\btau  {u}^*_{mk_x0}(\btau)  \hH_0(k_x,0) u_{nk_x0}(\btau) \right]=\tilde{H}_0(K_x,0)_{mn}.\la{interfirst}}
Here we have made use of the basic identities \q{reallybasic} and  
\e{\hH_0(K_x,0) u_{nK_x0}(\btau)=\left[ \hH_0(k_x,0) u_{nk_x0}(\btau)  \right];}
the latter follows from \q{commuterbrKx}. It should be emphasized that the right-hand-side of \q{interfirst} corresponds to the symbol $\braopket{u_{mk_x0}}{\hat{H}_0(k_x,0)}{u_{nk_x0}}$, with cell-periodic functions which are smooth with respect to $k_x$; this assumption of smoothness is justified in \s{sec:interbandbasis}. \\

For the second and third terms, a few basic identities for noncommuting operators [cf.\ \q{rreallybasi}] are helpful:
\e{ [\dg{u}_{mK_x0},k_y]\eq i\lmt\sq{\partial_{k_x}u_{mk_x0}^*}, \la{commutatorky}\\
    [\dg{u}_{mK_x0},k_y^2]\eq 2i\lmt k_y\sq{\partial_{k_x}u_{mk_x0}^*}-\lmf\sq{\partial^2_{k_x}u_{mk_x0}^*}.\la{commutatorky2}}
We remind the reader that $[\cdot,\cdot]$ is a commutator, while $[\cdot]$ is a Weyl symmetrization of $\cdot$. We would also need the identity
\e{ &i\int d\btau \sq{\partial_{k_x}u_{mk_x0}^*(\btau) } u_{nK_x0}(\btau) \lin
\eq  \sq{ i\int d\btau \partial_{k_x}u_{mk_x0}^*(\btau)u_{nk_x0}(\btau) } = -\tilde{\mathfrak{X}}^x_{mn}(K_x,0),\notag
}
with $\tilde{\bmx}$ defined in  \q{defineberryconnection}. Employing the above identity, \q{commuterbr} and (\ref{commutatorky}), the second term is evaluated as 
\e{ &\int d\btau \sum_{n} \dg{u}_{mK_x0}(\btau)   \big(k_y\hPi^y\big)u_{nK_x0}(\btau)f_{n\bk}\lin 
\eq \sum_{n,o}\left(k_y\delta_{m,o}-l^{-2}\tilde{\mathfrak{X}}^x_{mo}(K_x,0)\right)\tilde{\Pi}^y_{on}(K_x,0) \;f_{n\bk}.\la{intersecond}}
Here, it was also necessary to insert a complete set of cell-periodic operators [cf.\ \q{slutcomplete}].\\

The third term in \q{threeterms} is evaluated with aid from \q{commutatorky2} and the orthonormality condition in \q{slutortho}:
\e{ &\int d\btau \dg{u}_{mK_x0}\f{k_y^2}{2m}u_{nK_x0}(\btau) = \f{k_y^2}{2m}\delta_{mn}\lin
&-\f{k_y}{ml^2}\tilde{\mx}^x_{mn}(K_x,0)-\f1{2ml^4}\bigg[\braket{\partial^2_{k_x}u_{mk_x0}}{u_{nk_x0}}\bigg]. \la{intermediatethirdterm}}
We remind the reader of our Dirac notation:
\e{    \braket{\partial^2_{k_x}u_{mk_x0}}{u_{nk_x0}}:= \int d\btau \sq{\big\{\partial^2_{k_x}u_{mk_x0}^*(\btau)\big\}\,u_{nk_x0}(\btau)}.}
Applying the identity
\e{ &\braket{u_{m}}{u_{n}}=\delta_{mn} \imp \lin
&\braket{\partial^2u_m}{u_n}+2\braket{\partial u_m}{\partial u_n}+\braket{u_m}{\partial^2 u_n}=0,}
we may express the symbol of the last term in \q{intermediatethirdterm} as proportional to
\e{ \braket{\partial_{k_x}^2u_{mk_x0}}{u_{nk_x0}}= i\partial_{k_x}\tilde{\mx}^x_{mn}-{\sum}_o\tilde{\mx}^x_{mo}\tilde{\mx}^x_{on}\bigg|_{k_x,0}.\la{idenpartialsquare}} 
Inserting \q{interfirst},	(\ref{intersecond}), (\ref{intermediatethirdterm}) and (\ref{idenpartialsquare}) into \q{threeterms}, we finally obtain
\e{ &\f1{N}\int d\br \dg{u}_{mK_x0}(\br)e^{-i\bk \cdot \br}\hH \Psi(\br)
\lin
\eq \sum_n \bigg\{\{\tilde{H}_0 +k_y\tilde{\Pi}^y-\lmt \tilde{\mx}^x\tilde{\Pi}^y  +\f{k_y^2}{2m}-\f{k_y}{ml^2}\tilde{\mx}^x\}_{mn} \lin
&-\f1{2ml^4}\bigg( i\partial_{k_x}\tilde{\mx}^x_{mn}-{\sum}_o\tilde{\mx}^x_{mo}\tilde{\mx}^x_{on}\bigg) \bigg\}_{K_x,0}f_{n\bk},\la{effhammine} }
from which we may identify the effective Hamiltonian acting on $f_{n\bk}$ as that of \q{effhamslutfirst}.\\ 

We may symmetrize the above Hamiltonian with respect to $\bK$ to obtain \q{reorgsum}. The identity in \q{symmetrizeky} is useful for this purpose. Let us tackle \q{effhammine} term by term:
\e{ k_y\tilde{\Pi}^y\eq \f1{2}\{k_y,\tilde{\Pi}^y\}-\f{i}{2l^2}\partial_{K_x}\tilde{\Pi}^y \lin 
\eq \f1{2}\{k_y,\tilde{\Pi}^y\}+\f{1}{2l^2}[\tilde{\mx}^x,\tilde{\Pi}^y].}
Therefore, the sum of following two terms is symmetric: 
\e{ k_y\tilde{\Pi}^y-\lmt \tilde{\mx}^x\tilde{\Pi}^y = \f1{2}\{k_y,\tilde{\Pi}^y\}-\f{1}{2l^2}\{\tilde{\mx}^x,\tilde{\Pi}^y\}.}
Consider another term in \q{effhammine}:
\e{ -\f{k_y}{ml^2}\tilde{\mx}^x \eq -\f{1}{2ml^2}\bigg(\{k_y,\tilde{\mx}^x\} +[k_y,\tilde{\mx}^x] \bigg) \lin
\eq -\f{1}{2ml^2}\{k_y,\tilde{\mx}^x\} +\f{1}{2ml^4}i\partial_{K_x}\tilde{\mx}^x.  }
The last term here cancels a term in \q{effhammine}. Finally, note that the $(\tilde{\mx}^x)^2$ is already symmetric, trivially.

\subsection{Comparison with the effective Hamiltonian in the representation of field-modified Bloch functions}\la{sec:recovery}

We have claimed that the effective Hamiltonian of \q{effhambanddeg} validly describes any band dispersion; when particularized to the case of (i) a single nondegenerate band, or (ii) a subspace of degenerate bands, we may make an instructive comparison with the effective Hamiltonians derived by Roth\cite{rothI} [reviewed in \s{sec:singlebandeffham} and \ref{sec:multibandeffham}].\\ 

In both cases (i-ii), the full velocity matrix $\tilde{\bPi}$ and its diagonal component $\tilde{\bv}$ [recall their definitions in \q{definevelocitymatrix} and (\ref{definediagonalvelocitymatrix})] satisfy $(\tilde{\bPi}-\tilde{\bv})_{mn}=0$, or equivalently  $\bPi= \bv$; this follows from \q{idenvel} and \q{definediagonalvelocitymatrix}. This property and the diagonality of $\tilde{\bv}$  imply that
\e{ 
\big( \mathring{\mx}^{\beta}\tilde{\Pi}^{\alpha}\big)_{mn} \eq \sum_{\bar{l}}  \tilde{\mx}^{\beta}_{m\bar{l}}\tilde{\Pi}^{\alpha}_{\bar{l}n}=  \sum_{\bar{l}}  \tilde{\mx}^{\beta}_{m\bar{l}}\big(\tilde{\Pi}^{\alpha}-\tilde{v}^{\alpha}\big)_{\bar{l}n}\lin
+\sum_{{l}} & \tilde{\mx}^{\beta}_{m{l}}\big(\tilde{\Pi}^{\alpha}-\tilde{v}^{\alpha}\big)_{{l}n}= \big( \tilde{\mx}^{\beta}\big(\tilde{\Pi}^{\alpha}-\tilde{v}^{\alpha}\big)\, \big)_{mn},   \lin
\big(\tilde{\Upsilon}^y\tilde{\Pi}^x\big)_{mn} \eq \big(\tilde{\Upsilon}^y\big(\tilde{\Pi}^x-\tilde{v}^x\big)\,\big)_{mn}. \la{justplain}}
Furthermore, the assumption of nondegeneracy in the band energies (for at least a local region in $\bk$) imply the existence of energy functions ($\var_{n\bk}$) and cell-periodic functions ($u_{n\bk}$) which are both smooth with respect to $\bk$. In such a smooth energy basis, $\mx^y(\bk)$ is well-defined, and its off-block-diagonal component $\mathring{\mx}^y(\bk)$ satisfies
\e{ i\mathring{\mx}^y_{{m}\bar{n}}(\bk)= -\f{\tilde{\Pi}^y_{m\bar{n}}(\bk)}{\var_{m\bk}-\var_{\bar{n}\bk}},}
which is, for $\bk=(k_x,0)$, also the defining relation for $\tilde{\Upsilon}(k_x)$ [cf.\ \q{defineUpsilon}].
When this identification is made in \q{effhambanddegRoth}, as well as those in \q{justplain}, we obtain 
\e{ \calh_1^R \eq  \f1{2l^2}\bigg[\epsilon^{\ab}\{\tilde{\mx}^{\beta},\big(\tilde{\Pi}^{\alpha}-\tilde{v}^{\alpha}\big)\}\bigg]_{K_x,0}, \la{calh10degdeg}}
which is almost identical to the original Roth term [cf.\ \q{H1orbmag} and $H_1^R$ in \q{H1multiband2}]; the sole difference is that $\calh_1^R$ is independent of $k_y$.  This difference originates from the different representations for the wavefunctions: $\Psi=\sum_{n\bk} \eikr u_{nK_x0}f_{n\bk}$ in the basis of field-modified Luttinger-Kohn functions, and  $\Psi=\sum_{n\bk} \eikr u_{n\bK}f_{n\bk}$ for field-modified Bloch functions. Finally, $\bPi=\bv$ also implies that
\e{ \calh_1^B= -\f1{2l^2}\{{\mx}^x,v^y\}_{K_x,0},}
which may be compared to the original Berry term [cf.\ \q{H1berry} and $H_1^B$ in \q{H1multiband2}].  Since the cell-periodic function is independent [resp. dependent] of $k_y$ in the $(K_x,0)$-representation [resp. $(K_x,k_y)$-representation], the Berry term proportional to $\mxy$ is absent in \q{calh10degdeg}, but present in \q{H1berry} and (\ref{H1multiband2}).

\section{Appendix to `Interband Breakdown'}\la{app:interbandbreakdown}

\subsection{Connection to Weber's differential equation and Landau-Zener dynamics}\la{app:justifyWeber}


Our aim is to derive Weber's differential equation from the effective Hamiltonian equation [\q{effhambanddeglinearkx56}]. To begin, let us elaborate on the basis of field-modified Luttinger-Kohn functions in which  \q{effhambanddeglinearkx56} is represented. We have presupposed  a basis where $\tilde{u}_{nk_x0}$ are energy bands along $k_y=0$; this fixes the basis up to $U(1)\times U(1)$ gauge transformations, i.e., each energy band may be multiplied by a $k_x$-dependent phase. This arbitrariness is partially removed by insisting that the diagonal elements of $\mx^x(k_x,0)$ (a two-by-two matrix) vanish -- this is the  parallel-transport condition within each band. The off-diagonal elements of $\mx^x(k_x,0)$ vanish because they represent a coupling between distinct representations of  $g_2$ [cf.\ \q{symmg2}].\cite{JHAA} The vanishing of $\mx^x(\bze)$ (as a two-by-two matrix) justifies the neglect of the third $O(\lmt)$ term in \q{effhambanddeglinearkx}, from which we have derived \q{effhambanddeglinearkx56}.\\

We remove the $k_x$-dependence of \q{effhambanddeglinearkx56} by the transformation 
\e{\tilde{f}_{n\bk} \eq e^{ik_xk_yl^2}\phi_{n\bk},\la{newbasis} \\
 0 \eq e^{-ik_xk_yl^2}\left(\calh_0(\bK)-E\right)*\tilde{f}_{\bk}\lin 
\eq  \left( \;\left[\var_{\bze}-E\right]I +k_y \Pi^y \right)*\phi + \f{i}{l^2}\Pi^x \p{}{k_y}*\phi. \la{forgotten}} 
Here, we introduce $\var_{\bze}$ as the energy at the II-Dirac point [$\var_{\bze}=0$ in the main text],  $\Pi^x$ is a diagonal matrix with elements $\Pi^x_{11}:=u+v$ and $\Pi^x_{22}:=u-v$.  Assuming that $u^2>v^2$, one can find a non-unitary transformation to a two-component wavefunction $\bar{f}$ which satisfies a differential equation that has been well-studied in the Landau-Zener scattering problem. Each component of $\bar{f}$ satisfies Weber's differential equation, which is solved by parabolic cylinder functions (PCFs). The transformation has the form
\e{ \tilde{f}_{n\bk} \eq \alpha(\bk,E) \,\beta(k_y)\,\sum_{\bar{m}=1}^2\bar{T}_{n\bar{m}}\bar{f}_{\bar{m}}(k_y) \la{transformtoweber} \\
 \alpha(\bk,E) \eq \exp{\left[i\left(k_x - \f1{2}{\text{sgn}[\Pi^x_{11}]} \,\text{Tr}[(\tilde{\Pi}^x)^{-1}]\,E\right)k_yl^2\right]}\lin 
\eq \exp{\left[i\left(k_x - k_{xc}\right)k_yl^2\right]} \la{definealpha}\lin
\beta(k_y) \eq \exp\bigg[i \f1{2}\text{sgn}[\Pi^x_{11}] \,\text{Tr}[(\tilde{\Pi}^x)^{-1}]\,\var_{\bze}k_yl^2 \lin
&+ \f{1}{4}\text{sgn}[\Pi^x_{11}]\bigg(\f{\Pi^y_{11}}{|\Pi^x_{11}|}+\f{\Pi^y_{22}}{|\Pi^x_{22}|}\bigg)k_y^2 l^2\bigg] \lin
 \bar{T} \eq (\tilde{\Pi}^x)^{-1/2} V ,\;\text{with}\; \tilde{\Pi}^x:=\text{sgn}[\Pi^x_{11}]\Pi^x, }
and $V \in SU(2)$. Note that $k_{xc}$ in the second line is the coordinate of the hyperbolic center, $\tilde{\Pi}^x$ (defined above) is positive-definite, and $\bar{T}$ is independent of $\{k_x,k_y,E\}.$ 
This transformation was first derived in Ref.\ \onlinecite{slutskin}, with the  assumption that $\Pi^y$  is real owing to spacetime-inversion symmetry.\footnote{See App. 1 of Ref.\ \onlinecite{slutskin}} The more general proof that is presented here demonstrates that solubility by PCFs does not require this symmetry.\\

\noindent \emph{Proof of transformation to Weber's differential equation}

\noi{0} In the non-unitarily transformed basis
\e{ \bar{\phi} = \tilde{v}_{x}^{1/2}\phi,}
the eigenvalue equation [\q{forgotten}] takes the form
\e{ 0 \eq \left( \;\left[\var_{\bar{\bk}}-E\right](\tilde{\Pi}^x)^{-1} +k_y (\tilde{\Pi}^x)^{-1/2} \Pi_y (\tilde{\Pi}^x)^{-1/2}\right)*\bar{\phi}\lin
& + \text{sgn}[\Pi^x_{11}]\f{i}{l^2} \p{}{k_y}*\bar{\phi}.  }
\noi{i} We can remove the terms proportional to identity by
\e{ \tilde{\phi}\eq  \exp\bigg\{-\text{sgn}[\Pi^x_{11}]\bigg[\f{i}{2}l^2(\var_{\bar{\bk}}-E)\text{Tr}[(\tilde{\Pi}^x)^{-1}] k_y \lin
&+ \f{i}{4}l^2\bigg(\f{\Pi^y_{11}}{|\Pi^x_{11}|}+\f{\Pi^y_{22}}{|\Pi^x_{22}|}\bigg)k_y^2\bigg]\bigg\}(\tilde{\Pi}^x)^{1/2}\phi.   }
In our model,
\e{ \f{E}{2}\text{Tr}[(\tilde{\Pi}^x)^{-1}]= \f{E(\Pi^x_{11}+\Pi^x_{22})}{2\Pi^x_{11}{\Pi^x_{22}}}=\f{Eu}{u^2-v^2} = k_{xc}.}
The operator on $\tilde{\phi}$ has the generic form
\e{ m\sz + k_y\bv \cdot \bsigma +\text{sgn}[\Pi^x_{11}]\f{i}{l^2}\p{}{k_y}.}
In the next steps, we would find a basis where the coefficient of $\sz$ is linear in $k_y$, and that of $\sx$ is independent of $k_y$. Indeed, given any two three-vectors $\ba$ and $\bv$ [in our context $\ba=(0,0,m)$], we can always find a basis where
\e{ \ba \cdot \bsigma + k_y\bv \cdot \bsigma}
is transformed to 
\e{ a_1'\sx + a_3'\sz + k_y|v|\sz.}
This follows from the homomorphism between $SU(2)$ and $SO(3)$.\cite{tinkhambook} From a geometrical perspective, we are looking for a plane in $\R^3$ that is spanned by two vectors $\ba$ and $\bv$; we parametrize this plane by ($x,z)$, such that $\vec{z} = \bv/|v|$. Let us show this explicitly:\\

\noi{ii} We rotate to a basis where the matrix multiplying $k_y$ is diagonal:
\e{ \boldsymbol{m} \cdot \bsigma + k_y|v|\sz +\text{sgn}[\Pi^x_{11}]\f{i}{l^2}\p{}{k_y}.}
\noi{iii} Shifting the origin of $k_y$ to absorb the $m_3$ term, 
\e{ m_1\sx+m_2\sy + k_y|v|\sz +\text{sgn}[\Pi^x_{11}]\f{i}{l^2}\p{}{k_y}.}
\noi{iii} Performing a rotation with $\exp{i\sz \theta}$,
\e{ \sqrt{m_1^2+m_2^2}\sx + k_y|v|\sz +\text{sgn}[\Pi^x_{11}]\f{i}{l^2}\p{}{k_y}.}
Henceforth assuming $\Pi^x_{11}>0$, we obtain the first-order matrix differential equation in \q{simplemodel}, which is expressed with the hyperbolic parameters $\bar{a}$ and $\bar{b}$ defined in \q{hyperbolainter}. \\

The general procedure outlined above, when applied to our minimal model, leads to the particular forms: $V=e^{-i\sy \pi/4}\sz$ and $\bar{T}$ of \q{defTbar}. \\

The case of $\bar{a}=0$ was previously discussed in \s{sec:connEeq0}. Henceforth assuming $\bar{a}\neq 0$, and changing variables as
\e{z = 2\sqrt{-i} \f{\sqrt{\bar{\mu}}}{\bar{b}} k_y,} 
with $\bar{\mu}$ defined in \q{definemuinter},
each component of $\bar{f}$ now satisfies a second-order differential equation: 
\e{ &\left[\;\partial_z^2 - \f{z^2}{4} + \f1{2} + i\bar{\mu}\;\right]\bar{f}_{\bar{1}} = 0, \lin
&\left[\;\partial_z^2 - \f{z^2}{4} + \f1{2} + (i\bar{\mu}-1)\;\right]\bar{f}_{\bar{2}}=0. \la{weber2}}
The above equations may be identified  with Weber's differential equation:
\e{ \left[\;\partial_z^2 - \f{z^2}{4} + \f1{2} + \nu\;\right]\psi= 0,}
which is solved generally by parabolic cylinder functions (or Weber-Hermite functions)
\e{\psi = p_1 D_{\nu}(z) + p_2 \bar{D}_{\nu}(z), \ins{where} \bar{D}_{\nu}(z) = D_{\nu}(-z).}
$D_{\nu}(z)$ is an entire function of both $\nu$ and $z$,\cite{NIST} and satisfies the recurrence relation
\e{&\partial D_{\nu}\bigg|_{z} +\f{z}{2} D_{\nu}(z) = \nu D_{\nu-1}(z).}
Note that \ocite{NIST} employs a different notation for the PCF: $D_{\nu}(z) = U(-1/2-\nu,z)$. $\bar{f}_{\bar{1}}$ and $\bar{f}_{\bar{2}}$ are related as
\e{\bar{f}_{\bar{2}} = -\f{\sqrt{i}}{\sqrt{\bar{\mu}}} \left( \partial +\f{z}{2}\right)\bar{f}_{\bar{1}},} 
which implies, via the recurrence relation, that
\e{\bar{f}_{\bar{1}}(z) \eq p_1 D_{i\bar{\mu}}(z)+p_2 \bar{D}_{i\bar{\mu}}(z),\lin
\bar{f}_{\bar{2}}(z) \eq \f{\sqrt{\bar{\mu}}}{\sqrt{i}}\big[\;p_1 D_{i\bar{\mu}-1}(z)-p_2 \bar{D}_{i\bar{\mu}-1}(z)\;\big].}
Combining these equations with the asymptotic limits of the PCFs [Eq.\ (12.9.1) and (12.9.3) of \ocite{NIST}], we obtain the leading-order terms for $\bar{f}$ in the limits $k_y \rightarrow \pm \infty$ (denoted by $\pm$ in the argument): 
\e{\ins{for} \bar{b}>0,\as &\bar{f}_{\bar{1}}(+) \approx \varphi(|k_y/\bar{b}|)\left[ p_1 Z+p_2\,Z^{-3} \right], \lin
&\bar{f}_{\bar{1}}(-) \approx \varphi(|k_y/\bar{b}|)\left[ p_1 Z^{-3}+p_2\,Z \right], \lin
&\bar{f}_{\bar{2}}(+) \approx -p_2 \,\varphi^*(|k_y/\bar{b}|)\,\calg\,Z^{-1}, \lin
&\bar{f}_{\bar{2}}(-) \approx p_1\,\varphi^*(|k_y/\bar{b}|)\,\calg\,Z^{-1}, \la{asymptoticformofbarf}}
where we have introduced the variables
\e{ \varphi(|k_y/\bar{b}|) :\eq  \exp\,i\bar{\mu} \left[{\f{k_y^2}{\bar{b}^2}+ \ln \left|\f{2k_y}{\bar{b}} \right| + \f1{2} \ln \bar{\mu} }\right], \la{definevarphi}\\
 Z :\eq e^{\pi \bar{\mu}/4},\iand \lin
\calg :\eq \f{\sqrt{-i} \sqrt{2\pi \bar{\mu} }}{\Gamma(1-i\bar{\mu})} = \sqrt{Z^4-Z^{-4}}e^{i\arg \Gamma(1+i\bar{\mu})-i\pi/4}. \la{definezdefinecalg}}

\subsection{Derivation of connection formula for $|E|>0$} \la{sec:matching}

The goal of this section is to derive the connection formula [cf.\ \q{definescatteringmatrixinter}]  for interband breakdown at finite energy away from the II-Dirac point.  \\

Let us  assume that the semiclassical interval [where we apply the $(K_x,k_y)$-representation] overlaps with the breakdown interval [the $(K_x,0)$-representation]; this overlap region is an interval in $k_y$ satisfying
\e{  \big|k_y\big| \gg l^{-1},|\bar{b}|,{l}^{-1}\sqrt{\f{\bar{b}}{\bar{a}}} \ins{and} \big|k_y\big| \ll G_y, \la{conditionsmatching2}}
with ${k}_y$ originating from the II-Dirac point, and $G_y$ the reciprocal period. In this overlap region, will apply the general transformation [\q{basischange}] that relates wavefunctions in the two representations. Combining \q{basischange} with the WKB-form of the $(K_x,k_y)$-wavefunction in \q{abovequantumII}, we obtain that
\e{\ifor k_y>0,\as \tilde{f}_{l\bk} \eq \sum_{n=1}^2 c^+_{n} w_{n\bk} M_{ln}(k_y) +O(l^{-2},\tf{k_y}{G_y}), \lin
\ifor k_y<0,\as \tilde{f}_{l\bk} \eq \sum_{n=1}^2 c^-_{n} w_{n\bk} M_{ln}(k_y) +O(l^{-2},\tf{k_y}{G_y}), \la{basischange4} } 
with $w_{n\bk}$ a Zilberman-Fischbeck function defined in \qq{zilbermanfischbeckIII}{definewilsonline}, and the overlap matrix defined as
\e{M_{ln}(k_y)=\braket{\tilde{u}_{l,k_x^n(k_y),0}}{u_{n,k_x^n(k_y),k_y}}. \la{defineoverlapM}}
Here, $\tilde{u}_{l,k_x,0}$ and $u_{n,k_x,k_y}$ are classical symbols of operators occurring in the basis functions of the $(K_x,0)$ and $(K_x,k_y)$ representations [cf.\ \q{expandroth} and \q{ansatzexpansionrepeat}]. Take care in defining $M$ that the band index $n$ appears in both bra and ket. \\

To make progress, we would need the asymptotic forms of the quantities $w,M,\tilde{f}$ as $k_y \rightarrow \pm \infty$:
\e{ \{ w_{n\bk}^{\pm},\,M^{\pm}(E),\,\tilde{f}_{l\bk}^{\pm} \}  := \limit{k_y\rightarrow \pm \infty} \{w_{n\bk},M(k_y,E),\tilde{f}_{l\bk}\}.\la{asymptoticoverlapM}}
We consider them in turn:
\e{ w_{n\bk}^{\pm} \eq \f1{\sqrt{|\bar{v}_x|}}\alpha(\bk,E)\big( \varphi(|k_y/b|) \lambda^{-1/2} \big)^{\mp(-1)^n}\W_{n\pm}(k_y), \la{asymptoticwkb} \\
\lambda :\eq e^{i\bar{\mu} (\ln \bar{\mu}-1)},\as \lambda^{\pm 1/2}=e^{\pm i\bar{\mu}/2 (\ln \bar{\mu}-1)}, \la{definelambda} 
 }
with $\alpha$ defined in \q{definealpha} and $\varphi$ in \q{definevarphi}; $\W_{n\pm}$ is the single-band Wilson line defined in \q{definewilsonline}. For both $n\in \{1,2\}$, $|{v}^n_x|$ approaches the same value (denoted as $|\bar{v}_x|$)  as $k_y \rightarrow \pm \infty$. This asymptotic form of the overlap matrix $M$ is derived in \app{app:deriveoverlap} to be
\e{\ifor E&>0,\lin
 &M_{l1}^+ = e^{-i\theta_1}\W_{1+}^{-1} \bar{T}_{l\bar{1}}, \as M_{l2}^+ = -e^{-i\theta_2}\W_{2+}^{-1} \bar{T}_{l\bar{2}}, \lin
&M_{l1}^- = e^{-i\theta_1}\W_{1-}^{-1}\bar{T}_{l\bar{2}}, \as M_{l2}^- = e^{-i\theta_2}\W_{2-}^{-1}\bar{T}_{l\bar{1}}, \lin
\ifor E&<0,\lin
 &M_{l1}^+ = e^{-i\theta_1}\W_{1+}^{-1} \bar{T}_{l\bar{1}}, \as M_{l2}^+ = e^{-i\theta_2}\W_{2+}^{-1} \bar{T}_{l\bar{2}}, \lin
&M_{l1}^- = -e^{-i\theta_1}\W_{1-}^{-1}\bar{T}_{l\bar{2}}, \as M_{l2}^- = e^{-i\theta_2}\W_{2-}^{-1}\bar{T}_{l\bar{1}}, \la{relateMtoT}}
with $\bar{T}$ a $\bk$-independent, non-unitary transformation matrix defined in \q{defTbar}; $\theta_n$ are band-dependent phases that should be present on principle [see discussion in  \app{app:deriveoverlap}] but does not ultimately affect the quantization condition. 

Inserting \qq{asymptoticwkb}{relateMtoT}  to the right-hand-side of \q{basischange4}, the cancellation of the Wilson lines lead to 
\e{ \ifor E>0,\as 
\sum_{n} c^+_{n} w^+_{n\bk} M_{ln}^{+} 
 \eq \f{\alpha(\bk)}{\sqrt{|\bar{v}_x|}}\bigg\{ c_1^+ (e^{-i\theta_1}\bar{T}_{l\bar{1}}) (\lambda^{-1/2}\varphi)  + c_2^+(-e^{-i\theta_2}  \bar{T}_{l\bar{2}}) (\lambda^{1/2}\varphi^*) \bigg\} \lin 
\sum_{n} c^-_{n} w^-_{n\bk} M_{ln}^{-} 
\eq \f{\alpha(\bk)}{\sqrt{|\bar{v}_x|}}\bigg\{ c_1^-(e^{-i\theta_1}\bar{T}_{l\bar{2}}) (\lambda^{1/2}\varphi^*) +c_2^-(e^{-i\theta_2}\bar{T}_{l\bar{1}}) (\lambda^{-1/2}\varphi) \bigg\}.\lin
\ifor E<0,\as \sum_{n} c^+_{n} w^+_{n\bk} M_{ln}^{+}  \eq \f{\alpha(\bk)}{\sqrt{|\bar{v}_x|}}\bigg\{ c_1^+ (e^{-i\theta_1}\bar{T}_{l\bar{1}}) (\lambda^{-1/2}\varphi)  + c_2^+(e^{-i\theta_2}  \bar{T}_{l\bar{2}}) (\lambda^{1/2}\varphi^*) \bigg\} \lin 
\sum_{n} c^-_{n} w^-_{n\bk} M_{ln}^{-} \eq \f{\alpha(\bk)}{\sqrt{|\bar{v}_x|}}\bigg\{ c_1^-(-e^{-i\theta_1}\bar{T}_{l\bar{2}}) (\lambda^{1/2}\varphi^*) +c_2^-(e^{-i\theta_2}\bar{T}_{l\bar{1}}) (\lambda^{-1/2}\varphi) \bigg\}. \la{finalexpressionwbk}
}
 From \q{transformtoweber}, and applying that $\beta=0$ in our minimal model (since $\Pi^y$ is off-diagonal and $\var_{\bze}=0$), 
\e{ \tilde{f}_{l\bk}^{\pm}=  \alpha(\bk)\sum_{\bar{m}=\bar{1}}^{\bar{2}}\bar{T}_{l\bar{m}}\bar{f}_{\bar{m}}^{\pm}(k_y),}
where $\bar{m}$ labels the diadactic basis vectors, and $\bar{f}$ satisfies the matrix differential equation in \q{simplemodel}. Let us relate $\bar{f}$ above and below the Dirac point. Both $\bar{a}$ and $\bar{b}$ change sign across $E=0$, but $\bar{a}/\bar{b}$ does not. We will exploit  a symmetry of \q{simplemodel}: if $\bar{f}$ is a solution for $\bar{a}>0$, $\bar{r}(k_y) = \bar{f}(-k_y)$ is a solution for $\bar{a}<0$. Therefore, the asymptotic forms above and below the Dirac point are related by $\bar{r}_{\bar{i}}^{\pm}(k_y) =\bar{f}_{\bar{i}}^{\mp}(-k_y).$ Further employing that $\bar{f}$ only depends on $k_y$ through $\varphi(|k_y/\bar{b}|)$,
\e{ \bar{r}_{\bar{i}}^{\pm}(|k_y/\bar{b}|) =\bar{f}_{\bar{i}}^{\mp}(|k_y/\bar{b}|). \la{quickway}}
Utilizing the asymptotic forms of $\bar{f}$ in \qq{asymptoticformofbarf}{definezdefinecalg}  for $E>0$ (recall that the sign of $E$ and $\bar{b}$ are identical with our assumption that $u,v,w>0$), in combination with \q{quickway}, we derive
\e{ \ifor E>0, \as \tilde{f}_{l\bk}^{+}\eq \alpha(\bk)\left\{ \bar{T}_{l\bar{1}}\varphi \big( p_1Z+p_2Z^{-3} \big) + \bar{T}_{l\bar{2}}\varphi^* \big( -p_2\calg Z^{-1} \big)\right\} \lin
\tilde{f}_{l\bk}^{-}\eq \alpha(\bk)\left\{ \bar{T}_{l\bar{1}}\varphi \big( p_1Z^{-3}+p_2Z \big) + \bar{T}_{l\bar{2}}\varphi^* \big( p_1\calg Z^{-1} \big)\right\},\lin
\ifor E<0, \as \tilde{f}_{l\bk}^{-}\eq \alpha(\bk)\left\{ \bar{T}_{l\bar{1}}\varphi \big( p_1Z+p_2Z^{-3} \big) + \bar{T}_{l\bar{2}}\varphi^* \big( -p_2\calg Z^{-1} \big)\right\} \lin
\tilde{f}_{l\bk}^{+}\eq \alpha(\bk)\left\{ \bar{T}_{l\bar{1}}\varphi \big( p_1Z^{-3}+p_2Z \big) + \bar{T}_{l\bar{2}}\varphi^* \big( p_1\calg Z^{-1} \big)\right\}.
\la{airygives}}
Comparing this with \q{finalexpressionwbk}, we identify
\e{ \f1{\sqrt{|\bar{v}_x|}}\vectwo{c^-_1e^{-i\theta_1}}{c^-_2e^{-i\theta_2}}\eq \matrixtwo{ \lambda^{-1/2} \,\calg\,Z^{-1} }{0}{\lambda^{1/2} \,Z^{-3}}{\lambda^{1/2} \,Z}\sx^{(1-\text{sign}[E])/2}\vectwo{p_1}{p_2} \lin
\f1{\sqrt{|\bar{v}_x|}}\vectwo{c^+_1e^{-i\theta_1}}{c^+_2e^{-i\theta_2}}\eq  \matrixtwo{\lambda^{1/2} \,Z}{\lambda^{1/2} \,Z^{-3}}{0}{\lambda^{-1/2} \,\calg\,Z^{-1} }\sx^{(1-\text{sign}[E])/2} \vectwo{p_1}{p_2},
\la{mmm2}}
with $\sigma_1^1:=\sigma_1$ a Pauli matrix, and $\sigma_1^0:=I$  the two-by-two identity. Since the expressions for positive and negative energy differ only in the relabelling of dummy variables $p_1 \leftrightarrow p_2$, the  scattering matrix is independent of the sign of the energy.
Removing the $p_1$ and $p_2$ from our equations, we finally relate $c^-$ to $c^+$ by the scattering matrix  in \q{definescatteringmatrixinter}. 

\subsection{Asymptotic form of overlap matrix $M$}\la{app:deriveoverlap}

The overlap matrix  $M$ defined in \q{defineoverlapM} may be viewed as a basis transformation between the Luttinger-Kohn and crystal-momentum representations, as we have reviewed in \s{sec:reviewluttkohn}. $M$ is determined, with an accuracy of $O(k_y/G_y)$,  from the following eigenvalue equation:
\e{ \left[-E+{k}_x^n(k_y,E)\,\Pi^x(\bze)  +k_y{\Pi}^y(\bze)\right]_{ml} M_{ln}(k_y,E) =0. \la{fixedenergyeigenproblem2}}
 It is assumed that $u_{n k_x^n(k_y,E) k_y}$ [occurring in \q{defineoverlapM}] are energy eigenfunctions of $\hat{H}_0(\bk)$ with eigenvalue $\var_{nk_x^nk_y}=E$.  The goal of this section is to derive the asymptotic form of $M$ [denoted by $M^{\pm}$ in \q{relateMtoT}] as $k_y \rightarrow \pm \infty$.  \\

Before a detailed proof of \q{relateMtoT}, we would argue for the form of $M^{\pm}$:\\

\noi{i} In the limit $k_y \rightarrow \pm \infty$, the adiabatic basis of energy bands (labelled by $n$) coincides with the diabatic basis (labelled by $\bar{n}$) up to a phase, as we have argued in the caption of \fig{fig:didactic}. Each column of $M^{\pm}$ is therefore proportional to a column of the matrix $\bar{T}_{n\bar{n}}$ defined in \q{defTbar}, which transforms between adiabatic and diabatic bases.\\

\noi{ii} What remains is to argue for the proportionality phase factors. If we ignore $\W$ and $e^{i\theta_j}$, there remains a $-1$ phase factor which reflects the $\pi$ Berry phase acquired in the $2\pi$ rotation of pseudospin-half. Indeed, we might view $\bar{a}\tx + \f{\bar{a}}{\bar{b}}k_y\tz$ as the Hamiltonian of a pseudospin coupled to a pseudo-magnetic field, and label the diabatic basis $\ket{\pm 1}$ according to its eigenvalue under $\tz$. The diabatic basis coincides, modulo a phase factor, with the adiabatic basis in the two limits $k_y \gg |\bar{b}|$ and $k_y \ll -|\bar{b}|$. As $k_y$ is varied from $+\infty$ to $-\infty$, the adiabatic basis is parallel-transported along $k_x^n(k_y,E)$ as $\ket{+}\rightarrow e^{i\phi_{+-}}\ket{-}$ and $\ket{-}\rightarrow e^{i\phi_{-+}}\ket{+}$. The product of the two phases $e^{i(\phi_{+-}+\phi_{-+})}=-1$ independent of phase redefinitions of the diabatic basis: $\ket{\pm} \rightarrow \ket{\pm}e^{i\varphi_{\pm}}$. To explain this independence, we may view the combined parallel transport of $\ket{+}\rightarrow e^{i\phi_{+-}}\ket{-} \rightarrow e^{i(\phi_{+-}+\phi_{-+})}\ket{+}$ as the adiabatic rotation of a pseudospin-half by $2\pi$ within a plane. $e^{i(\phi_{+-}+\phi_{-+})}=-1$ may then be identified with the Berry phase, which is half the solid angle\cite{berry1984} subtended by the rotation. \\

\noi{iii} In the Bloch problem, there is an intrinsic ambiguity in the definition of nondegenerate energy bands  $\ket{u_{n\bk}}$, which  may arbitrarily be redefined by a $\bk$-dependent phase. This phase ambiguity may be separated into two contributions: (ii-a) the single-band Berry connection $\bmx_n(\bk)$ encodes the phase relationship between infinitesimally separated wavevectors $\bk$ and $\bk+\delta \bk$. (ii-b) In addition, there remains, for each band labelled by $n=1,2$, a global phase ambiguity encoded by $e^{i\theta_n}$, which explains their presence in \q{relateMtoT}. That is, given a fixed connection, there remains a gauge freedom in redefining each band by a $\bk$-independent phase. The ambiguity described in (ii-a) is expressed in  \q{relateMtoT} as the integral of the connection along the constant-energy band contour in \q{definewilsonline}.
We remark that there is no sense in which $\W_{n\pm}$ asympotically converges to a unique phase factor; reassuringly, the final expression for the quantization condition involves only closed-loop integrals of $\bmx_n$.  \\

\noindent \emph{Proof of \q{relateMtoT}}\\

\noindent Let us make the argument of (i) precise. One may verify that
\e{0 =\left[ -k_{xc}+\bar{a}\tx +\bar{k}^n_x+\f{\bar{a}}{\bar{b}}k_y\tz \right]_{ml}[\bar{T}^{-1}M]_{ln}.\la{determinationM}
}
For large $|k_y/\bar{b}|\gg 1$, we may neglect $\bar{a}\tx$ relative to ${\bar{a}}k_y\tz/{\bar{b}}$:
\e{0\approx \left[ -k_{xc} +\bar{k}^n_x(k_y,E)+\f{\bar{a}}{\bar{b}}k_y\tz \right]_{ml}[\bar{T}^{-1}M]_{ln} \la{neglectatx}}
which determines the columns of $M^{\pm}$ up to a phase (denoted as $\gamma$ below):
\e{ &{M}_{l1}^+ =  e^{i\gamma_{1+}}\bar{T}_{l\bar{1}}, \as {M}_{l2}^- =  e^{i\gamma_{2-}}\bar{T}_{l\bar{1}}, \lin
&{M}_{l2}^- =  e^{i\gamma_{2+}}\bar{T}_{l\bar{2}}, \as {M}_{l1}^- =  e^{i\gamma_{1-}}\bar{T}_{l\bar{2}}. \la{relateMTbar}}
The first line follows from setting $[\bar{T}^{-1}M]_{ln} \propto \delta_{l,1}$ in  \q{neglectatx}, which leads to  $0\approx -k_{xc} +\bar{k}^n_x+\bar{a}k_y/\bar{b};$ for $k_y \gg |\bar{b}|$, this corresponds to the left band contour $(n=1)$, and for $k_y \ll -|\bar{b}|$ to the right $(n=2)$.\\

Our next step is to derive the phases $\gamma_{n\pm}$. As an intermediate step, we would determine $\gamma_{n\pm}$ in a special basis for the energy bands which we denote as:
\e{ \{\breve{u}_{n\bar{k}_x(k_y,E),k_y}|n \in \{1,2\}, \as k_y \in \R\}.}
Up to a relabelling of band indices, $\breve{u}$ and $\tilde{u}$ are defined to be continuous where their domains  (in $\bk$-space) overlap -- at the two hyberbolic vertices. Explicitly, if we define  the two-by-two overlap matrix:
\e{ &\breve{M}_{ln}(k_y,E) = \braket{\tilde{u}_{l,\bar{k}_x^n(k_y,E),0}}{\breve{u}_{n,\bar{k}_x^n(k_y,E),k_y}}, \lin
&\ins{then} \breve{M}(0,E)=\sx^{\sma{(1-\text{sign}[E])/2}}. \la{definebreveoverlap}}
The reason for this dependence on the sign of $E$: we have defined $\tilde{u}_1$ to have a larger velocity $(\partial \var/\partial k_x)$ than $\tilde{u}_2$, so $\tilde{u}_1$ corresponds to the left hyperbolic vertex ($\breve{u}_1$) at positive energy, and to the right hyperbolic vertex ($\breve{u}_2$) at negative energy. We further insist that $\breve{u}$ satisfies the parallel-transport condition, i.e., for any  segment of a hyperbolic arm, $0=$exp$({-}\int \braket{\breve{u}_n}{\nabk \breve{u}_n}  \cdot d\bk)$. This condition, combined with the reality of the pseudospin Hamiltonian ($\bar{a}\tx + \f{\bar{a}}{\bar{b}}k_y\tz$), ensures the reality of $\breve{M}$ for all $k_y$. It is simple to find a real function that interpolates between the known values of $\breve{M}$ at $k_y=0$ [cf.\ \q{definebreveoverlap}] and at $\pm \infty$ [each column of $\breve{M}$ must be proportional to a column of the real matrix $\bar{T}$]. The result is:
\e{ \ifor E>0, \as &\breve{M}_{l1}(+) =  \bar{T}_{l\bar{1}}, \as \breve{M}_{l2}(+) =  -\bar{T}_{l\bar{2}}, \lin
&\breve{M}_{l1}(-) = \bar{T}_{l\bar{2}}, \as \breve{M}_{l2}(-) = \bar{T}_{l\bar{1}}, \lin
 \ifor E<0, \as &\breve{M}_{l1}(+) =  \bar{T}_{l\bar{1}}, \as \breve{M}_{l2}(+) =  \bar{T}_{l\bar{2}}, \lin
&\breve{M}_{l1}(-) = -\bar{T}_{l\bar{2}}, \as \breve{M}_{l2}(-) = \bar{T}_{l\bar{1}}.
 \la{relateMtoTptbasis}}
Given that \q{relateMtoTptbasis} holds for the special basis $\breve{u}$, it follows that   \q{relateMtoT} holds for any basis 
\e{\{u_{n\bar{k}_x^n(k_y),k_y}| k_y \in \R   \}}
that is differentiable with respect to $k_y$. Indeed, we may define the phase mismatch between $u_n$ and $\breve{u}_n$ at the hyperbolic vertex $(\bar{k}_x^n(0),0)$ as
\e{ u_{n\bar{k}_x^n(0),0}=e^{-i\theta_n}\breve{u}_{n\bar{k}_x^n(0),0}.}
Finally, the Wilson line [$\W_{n\pm}$, as defined in \q{definewilsonline}] accounts for the additional phase mismatch between $u_n$ and $\breve{u}_n$, which originates from $u_n$ not satisfying the parallel-transport condition. To recapitulate,
\e{ \breve{u}_{n,{\bk}_x^{n},k_y\rightarrow \pm \infty} = e^{i\theta_n}\,\W_{n\pm}\, {u}_{n,{\bk}_x^{n},k_y\rightarrow \pm \infty},  \la{uWtilu}} 
which may be substituted into \q{relateMtoTptbasis} to derive \q{relateMtoT}.

\subsection{Perturbative treatment of quasirandom spectrum}\la{app:perturbinter}

\subsubsection{Case study of interband breakdown: single II-Dirac graph with $\bar{\mu} \approx 0$} \la{app:casestudyperturbIID}

From \q{quantizationIIdirac}, we identify
\e{f(E,B;\tau) \eq f(E,B;0) +\delta \tau(E,B) f_1(E,B) \lin
f_0(E,B):\eq f(E,B;0) = \cos\left[\f{\Omega_1+\Omega_2}{2}\bigg|_{E,l^2}\right]\lin
 f_1(E,B) \eq -\cos\left[\f{\Omega_1-\Omega_2}{2}\bigg|_{E,l^2} +\omega(\bar{\mu})\right]\lin
\delta \tau(E,B) \eq \sqrt{1-\rho^2}, \as \rho(\bar{\mu})=e^{-\pi \bar{\mu}}, \as \tau_0=0, \la{definef0f1}}
with $\bar{\mu} \propto E^2/B$ and $\omega$ defined in \q{definemuinter} and (\ref{definerhoandomega}) respectively.\\

In the semiclassical limit $\bar{\mu}\rightarrow 0$,  the Landau fan is determined by \q{faninter}, and the first-order correction to the Landau fan is given in \q{faninter2}. Just as in \q{faninter2}, we will employ the shorthand $O'=\partial O/\partial E$ throughout this appendix. Further assuming that $(S_1+S_2)$ is slowly varying on the scale of $E_n^0-E_0^0$ [i.e., $(S_1+S_2)'=O(1)$], and restricting ourselves to $n=O(1)$, \q{faninter} and (\ref{faninter2}) particularize to
\e{  E_n^0(B) \eq E_0^0  + \f{2n\pi}{l^2(S_1+S_2)'|_{E_0^0}} +O(\lmf), \la{zerothorderfan}\\
\delta E_n^1(B) \eq \f{2\sqrt{\pi}(-1)^{n+1}}{l(S_1+S_2)'|_{E_n^0}}\f{v}{\sqrt{w}(u^2-v^2)^{3/4}}\lin
&\times \bigg\{ {E_0^0} +\f{2n\pi}{l^2(S_1+S_2)'|_{E_0^0}}\bigg\} \lin
&\times  \sin\left[ \omega+\f{l^2(S_1-S_2)}{2}\right]\bigg|_{E_n^0} +O(\bar{\mu}^{3/2}\lmt,l^{-5}).\notag} 
The validity of the last expression rests on a double constraint on the field: it cannot be too large, as reflected in the $O(l^{-5})$ uncertainty; on the other hand for any nonzero energy, the field also cannot be too small, since $\bar{\mu}^{3/2}\lmt \propto E^3l$. With these caveats in mind, we observe that the amplitude of $\delta E_n^1$ goes as $B^{1/2}$ at weak field; for $n \neq 0$, this crosses over to a $B^{3/2}$ dependence at intermediate field. \q{faninter2} and the second line of \q{zerothorderfan} are derived at the end of this section [\s{app:derivefirstll}].

\subsubsection{Derivation of first-order correction \q{faninter2}} \la{app:derivefirstll}

As defined in \q{defineOmegaj}, the domain of $\Omega_j$ does not include $E=0$.  Due to the continuity of the quantization condition \q{quantizationIIdirac} across $E=0$ [as we had argued in \s{sec:quantcondIIdirac}], we may as well extend the domain by
$\Omega_j(0):=\Omega_j(0^+),$
with $0^+$ an infinitesimally-small positive quantity; our results will be unchanged if we had instead chosen $\Omega_j(0):=\Omega_j(0^-)$. We may then  express the extended functions concisely as
\e{ \f{\Omega_1-\Omega_2}{2} \eq \f{l^2(S_1-S_2)}{2}- \f{\pi}{2}+ \pi \Theta^+(E),\lin
  \f{\Omega_1+\Omega_2}{2}\eq \f{l^2(S_1+S_2)}{2}+\f{\pi}{2},\la{gaugechoice}}
with the step function defined by
\bal
\Theta^+(x)= \begin{cases}0  &\ins{for} x\geq 0, \\
 1 &\ins{for} x<0 . \end{cases}
\end{align}
Inserting \q{gaugechoice} into \q{definef0f1},
\e{ f_0 \eq   -\sin\left[ \f{l^2(S_1+S_2)}{2}\right], \la{f0now}\\
\p{f_0}{E}\bigg|_{E_n^0} \eq  \f{(-1)^{n+1}}{2} l^2(S_1+S_2)'\bigg|_{E_n^0}, \la{df0dvar}\\
 f_1\eq -\cos\left[\omega+ \f{l^2 (S_1-S_2)}{2}-\f{\pi}{2} + \pi \Theta^+(E) \right] \lin
\eq (-1)^{\Theta^+(E)+1}\sin\left[ \omega+\f{l^2 (S_1-S_2)}{2}\right].  \la{f1now}}
In the second equality, we applied that sin$[\Omega_1/2+\Omega_2/2]=(-1)^n$ when evaluated at $E_n^0$, as deducible from \q{faninter}. Inserting \qq{df0dvar}{f1now} into \q{generalfirstorder}, 
the first-order correction in energy is then
\e{ \delta E_n^1 \eq \f{-\delta \tau f_1}{\p{f_0}{E} }\bigg|_{E_n^0(B)} = 2(-1)^{n+1+\Theta^+}\f{\delta \tau}{l^2(S_1+S_2)' }\lin
&\times \sin\left[ \omega+\f{l^2 (S_1-S_2)}{2}\right]\bigg|_{E_n^0}.\la{deltaEwiththeta}}
Since we are in a parameter regime where $\bar{\mu}$ is small, $\delta \tau$ [defined in \q{definef0f1}] is approximated by
\e{ \delta \tau  \eq \sqrt{2\pi \bar{\mu}}+O(\bar{\mu}^{3/2})  \lin
\eq  \sqrt{\pi  }\f{v}{\sqrt{w}(u^2-v^2)^{3/4}}\, l\,|E|+O(\bar{\mu}^{3/2}), \la{expanddeltatau}}
where we have utilized the definition of $\bar{\mu}$ in \q{definemuinter}. Since 
\e{(-1)^{\Theta^+(E)}|E|=E=\text{sign}[E]|E|,} 
we may just as well replace $(-1)^{\Theta^+}$ in \q{deltaEwiththeta} by sign$[E]$, and finally obtain \q{faninter2} as desired. For Landau levels indexed by $n =O(1)$, we may substitute \q{expanddeltatau} and (\ref{zerothorderfan}) into \q{faninter2}, and derive the second line of \q{zerothorderfan}.

\bibliography{bib_Apr2018}

\end{document}


\begin{table}[ht]
	
\centering
		
\begin{tabular} {|r c|c|c|l|l|l|l|} \cline{3-8}
			
\multicolumn{1}{r}{}&\multicolumn{1}{c}{} &\multicolumn{1}{|c}{$u(g)$}&  \multicolumn{1}{|c}{$s(g)$} &  \multicolumn{1}{|c}{Constraint on $\cala$} & \multicolumn{1}{|c}{Spectrum of $\cala$} & \multicolumn{1}{|c}{Representation of $g$} & \multicolumn{1}{|c|}{Ex. of $g$} \\  \hline \hline 
		  
\multicolumn{1}{|l}{(I)}&  \multicolumn{1}{c|}{$\forall \; \bkp,$} & $0$& $0$ &  $\cala=\breve{g}\cala\breve{g}^{\mo}$   & $\sigma(\cala){=}\sigma_{\sma{+}}{\sqcup}\,\sigma_{\sma{-}}$ &  $\breve{g}^{\sma{2}}{=}(\minus 1)^{\sma{Fa}}$ & $\mir_z,\glide_{z,\vec{x}/2}$     \\ \cline{3-8}			 

&$\bkp{=}g{\sma{\circ}}\bkp$ & $0$& $1$ &  $\cala=\breve{g}\cala^*\breve{g}^{\mo}$   & $\sigma(\cala)=\sigma(\cala)^*$ & $(\breve{g}K)^{\sma{2}}{=}(\minus 1)^{\sma{Fa}}$  & $T\inv,T\rot_{2z}$     \\ \hline

\multicolumn{1}{|r}{(II-A)}& & $0$& $0$ &  $\cala=\tilde{g}\cala\tilde{g}^{\mo}$    & \multicolumn{1}{c|}{$-$} &     $\tilde{g}^{\sma{N}}{=}\cala^{p}(\minus 1)^{\sma{Fa}}$    & $\inv,\rot_{nz}$     \\ \cline{3-8}			 
			 
&$\bkp \in \frako,$  & $0$& $1$ &  $\cala=\tilde{g}\cala^*\tilde{g}^{\mo}$   &  $\sigma(\cala)=\sigma(\cala)^*$ &    $(\tilde{g}K)^{\sma{N}}{=}\cala^{p}(\minus 1)^{\sma{Fa}}$  & $T,T\rot_{6z}$     \\ 			 \cline{3-8}

&$|\frako| = |g{\sma{\circ}}\frako|$ & $1$& $0$ &  $\cala =\breve{g}\cala^{\mo}\breve{g}^{\mo}$   & $\sigma(\cala)=\sigma(\cala)^*$ &  $\breve{g}^{\sma{N}}{=}(\minus 1)^{\sma{Fa}}$  & $\mir_x,\mir_y$     \\ \cline{3-8}			 

& & $1$& $1$ &  $\cala =\breve{g}\cala^{t}\breve{g}^{\mo}$   & \multicolumn{1}{c|}{$-$} &  $(\breve{g}K)^{\sma{N}}{=}(\minus 1)^{\sma{Fa}}$   &  $T\mir_x,T\mir_y$     \\ \hline

\multicolumn{1}{|r}{(II-B)}& & $0$& $0$ &  $\cala_2=\breve{g}_{\sma{1}}\cala_1\breve{g}_{\sma{1}}^{\mo}$   & $\sigma(\cala_2)=\sigma(\cala_1)$ &  $\breve{g}_{\sma{N}}\ldots \breve{g}_{\sma{1}}{=}(\minus 1)^{\sma{Fa}}$   & $\rot_{nz}$     \\ \cline{3-8}			 
			 
&$\bkp \in \frako,$ & $0$& $1$ &  $\cala_2=\breve{g}_{\sma{1}}\cala_1^*\breve{g}_{\sma{1}}^{\mo}$   & $\sigma(\cala_2)=\sigma(\cala_1)^*$ &  $\breve{g}_{\sma{N}}K\ldots \breve{g}_{\sma{1}}K{=}(\minus 1)^{\sma{Fa}}$  & $T$     \\ \cline{3-8}			 

&$|\frako| \neq  |g{\sma{\circ}}\frako|$ & $1$& $0$ & $\cala_2=\breve{g}_{\sma{1}}\cala^{\mo}_1\breve{g}_{\sma{1}}^{\mo}$    & $\sigma(\cala_2)=\sigma(\cala_1)^*$ &     $\breve{g}_{\sma{N}}\ldots \breve{g}_{\sma{1}}{=}(\minus 1)^{\sma{Fa}}$  &  $\mir_x,\mir_y$     \\ \cline{3-8}			 

		&								& $1$& $1$ &  $\cala_2=\breve{g}_{\sma{1}}\cala^t_1\breve{g}_{\sma{1}}^{\mo}$   & $\sigma(\cala_2)=\sigma(\cala_1)$ &   $\breve{g}_{\sma{N}}K\ldots \breve{g}_{\sma{1}}K{=}(\minus 1)^{\sma{Fa}}$  &    $T\mir_x,T\mir_y$     \\ \hline

\end{tabular}
		
\caption{The first column distinguishes distinguishes between three topologically distinct mappings of $g:\bk \rightarrow g\circ \bk$, as summarized in \qq{caseequal}{casenotequal}. The second and third columns subdivide the three mapping classes according to two $\Z_2$ indices defined in \q{definesg} and (\ref{defineug}); this gives ten classes in total. Fourth column describes the algebraic constraints on the propagator, and the fifth column describes the constraint on the spectrum of $\cala$, which we denote by $\sigma(\cala)$. $\sigma(\cala)=\sigma(\cala)^*$ means the spectrum is invariant under complex conjugation, which implies also that det$\,\cala=\pm 1$. The sixth column lists some representative examples of the ten symmetry classes.
	\label{tab:tenfold2}}
\end{table}